\documentclass[a4paper,11pt,twoside]{article}
\usepackage{latexsym}
\usepackage{array}
\usepackage{amsmath}
\usepackage{longtable}
\usepackage{epsfig}
\usepackage{graphics}
\usepackage{color}
\usepackage{booktabs}
\usepackage{feynmp}
\usepackage{a4wide}

\definecolor{dblue}{rgb}{0.05,0.05,0.35}
\definecolor{ddblue}{rgb}{0.07,0.07,0.4}
\definecolor{dred}{rgb}{0.6,0.05,0.05}
\definecolor{dgreen}{rgb}{0.05,0.6,0.05}
\definecolor{dblugr}{rgb}{0,0.7,0.4}
\definecolor{lgrey}{rgb}{0.9,0.9,0.9}
\renewcommand{\Re}{\mbox{Re}}

\newcommand{\Tr}[1]{\ensuremath{\mbox{Tr}\{ #1 \}}}
\newcommand{\difd}{\mbox{d}}

\newcommand{\twovec}[2]{\left(\!\begin{array}{c}#1\\ #2\end{array}\!\right)}
\newcommand{\tn}[1]{\ensuremath{10^{#1}}}
\newcommand{\ttn}[1]{\ensuremath{\times 10^{#1}}}

\newcommand{\tket}[1]{\ensuremath{|\ \! #1\ \!\rangle}}

\newcommand{\neut}{\ensuremath{\ti{\chi}^0}}
\newcommand{\charg}{\ensuremath{\ti{\chi}^+}}

\newcommand{\MeV}{\mbox{\!\ MeV}}
\newcommand{\GeV}{\mbox{\!\ GeV}}
\newcommand{\TeV}{\mbox{\!\ TeV}}
\newcommand{\fm}{\mbox{\!\ fm}}
\newcommand{\fb}{\mbox{\!\ fb}}
\newcommand{\mb}{\mbox{\!\ mb}}
\newcommand{\scm}{\mbox{\!\ cm$^{-2}$s$^{-1}$}}

\newcommand{\LV}{\mbox{$L$\hspace*{-0.5 em}/\hspace*{0.18 em}}}
\newcommand{\BV}{\mbox{$B$\hspace*{-0.65 em}/\hspace*{0.3 em}}}
\newcommand{\RV}{\mbox{$R$\hspace*{-0.6 em}/\hspace*{0.3 em}}}
\newcommand{\ET}{\ensuremath{E_T\hspace*{-1.15em}/\hspace*{0.65em}}}
\newcommand{\ETs}{\ensuremath{E_T\hspace*{-1em}/\hspace*{0.65em}}}

\newcommand{\slashp}{\not \! p}
\newlength{\bredde}
\def\slash#1{\settowidth{\bredde}{$#1$}\ifmmode\,\raisebox{.15ex}{/}
\hspace*{-\bredde} #1\else$\,\raisebox{.15ex}{/}\hspace*{-\bredde} #1$\fi}

\newcommand{\ti}[1]{\ensuremath{\tilde{#1}}}
\newcommand{\pythia}{\texttt{PYTHIA}}
\newcommand{\spythia}{\texttt{SPYTHIA}}
\newcommand{\atlfast}{\texttt{AtlFast}}
\newcommand{\herwig}{\texttt{HERWIG}}
\newcommand{\isajet}{\texttt{ISAJET}}
\newcommand{\isasusy}{\texttt{ISASUSY}}
\newcommand{\mathematica}{{\texttt{Mathematica}}}

\newcommand{\comma}{,}

\newlength{\tfcapsep}
\begin{document}
\pagestyle{empty}

\vspace*{-0.5cm} \begin{center}
{\sc P\!\ e\!\ t\!\ e\! r\!\ \!\ \!\ Z\!\ e\!\ i\!\ l\!\ e\!\ r\!\ \!\ \!\
S\!\ k\!\ a\!\ n\!\ d\!\ s}\\ \ \\
\textbf{\huge {\color{ddblue}{\boldmath $L$-Violating}}
{\color{ddblue}Supersymmetry\vspace*{5mm}}}\\ {\large 
\mbox{\sc implementation in pythia and study of lhc discovery 
potential}}\vspace*{5.mm}\\\normalsize \setlength{\unitlength}{0.1cm} 
\vspace{4.4cm} 
\includegraphics*[scale=2.7]{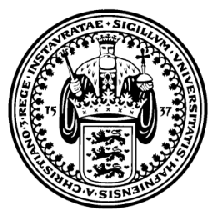}\vspace*{0.04cm}\\
{\normalsize\color{black}\sc thesis for the degree}
\vspace*{0.1cm}\\
{\normalsize\color{black}\sc candidatus scientiarum in physics}\vspace*{0.1cm}\\
\vfill 
{\color{black}\sc \normalsize{July 28, 2001}} \vspace*{0.1cm}\\ 
{\normalsize{\color{black}\sc NIELS BOHR INSTITUTE}}\vspace*{0.1cm}\\ 
{\normalsize{\color{black}\sc Department for Experimental High Energy Physics}}
\end{center}
\clearpage 

\noindent
This thesis is submitted for evaluation in accordance with the requirements
  for obtainment of the degree of Cand.\ Scient.\ 
in physics at the Niels Bohr Institute, University of Copenhagen. 
\vspace{0.3cm}

\noindent
I am very grateful to 
the L\o rup foundation, the Nordic Academy for Advanced
Study (NorFA), and the Niels Bohr Institute (the HEP group in particular) for
financial support. 
In addition, I would like to thank the 
University of Rostock and the Volkswagen Stiftung, the
University of Uppsala and the Nobel Comittee, the University of Oslo, 
the CTEQ/IPPP Summer School 2001, 
and the THEP group at Lund University. \vspace*{0.2cm}

\noindent \texttt{Mathematica} is a copyrighted program, trademark of Wolfram Research,
Inc. 

\noindent Version numbers for publically available programs used in this work
are -\pythia\ v.6.155, \isajet\ v.7.51, \herwig\ v.6.2, and \atlfast\ v.2.53.

\begin{center}
\vspace*{1.6cm}\setlength{\extrarowheight}{6pt}
\begin{tabular}{c} 
\textbf{Internal Supervisor (NBI):}\\ John Renner Hansen 
\\ Niels Bohr Institute (HEP), University of Copenhagen 
\\ Blegdamsvej 17 DK-2100 Copenhagen \O  
\\ renner@nbi.dk \end{tabular}\\
\vspace*{0.8cm}
\begin{tabular}{c} 
\textbf{External Supervisor (Lund):}\\
Torbj\"{o}rn Sj\"{o}strand\\
Dep.\ of Theoretical High Energy Physics, Lund University,  \\
P.O. Box 118 SE-221 00 Lund\\
Torbjorn.Sjostrand@thep.lu.se
\end{tabular}\vspace*{2.5cm}
\vfill
 \rule{2.in}{0.01in} \\ 
\mbox{Peter Z. Skands}\\
\end{center}

This document is set in \LaTeX{}.
\clearpage



\clearpage
\pagestyle{empty}
\vspace*{5cm}
\begin{abstract}
In the Minimal Supersymmetric Standard Model (MSSM), the simultaneous
appearance of lepton and baryon number
violation causes the proton to decay much faster
than the experimental bound allows. Customarily, a discrete symmetry known as
$R$-parity is imposed to forbid these dangerous interactions. This work
begins by arguing that there is no deep theoretical motivation for preferring
$R$-parity over other discrete symmetries and continues by 
adopting the MSSM with baryon number
conservation replacing $R$-parity conservation.
For the purpose of studying the influence of the 
consequent lepton number violating interactions, 1278 new decay channels of
supersymmetric particles into Standard Model particles 
have been included in the PYTHIA event generator.

The augmented event generator is then used in combination with the
\atlfast\ detector simulation to study the impact of
lepton number violation (\LV) on event topologies in the ATLAS detector, and
trigger menus designed for \LV-SUSY are proposed based on very general
conclusions. The subsequent analysis uses a combination of primitive cuts and
neural networks to optimize the 
discriminating power between signal and background events. In all scenarios
studied, it is found that a $5\sigma$ discovery is possible for cross sections
down to $10^{-10}\mb$ with an integrated luminosity of 30\fb$^{-1}$, 
corresponding to one year of data taking with the LHC running 
at ``mid-luminosity'', $L=3\ttn{33}\scm$.
\end{abstract}

\cleardoublepage
\setcounter{page}{1}
\tableofcontents
\listoffigures\listoftables
\clearpage
\vspace*{7cm}
\begin{center}
\parbox{0.8\textwidth}{\large \emph{
Once, this Earth was haunted 
by the flaming angels of elusive gods and by magic that seemed to move the
heavenly spheres. 
These mythic tales seem but superstitious fantasy,
despite their depth of colour, to the rational and sane of an enlightened
society. And yet this universe appears to the inquisitive mind still so
enigmatic, so full of the very mystery that breathed life and beauty into our
earliest imagination that man's greatest tragedy would be to dull his senses and
not still, despite the limits of his Earthly mind, seek to grasp the nature
of that which brought him forth.}}
\end{center}

\clearpage
\pagestyle{headings}
\pagestyle{headings}
\section{Introduction}
\pagestyle{headings}
The Herculean task of recapitulating the history of the science
of physics or even that small part of it which has direct connection to this
work  
is better left to other authors, yet for the inexperienced reader, 
I here will briefly set the stage on which the present work has been
performed. 

The essence of the 
school of thought initiated by Thales and his two contemporaries
in sixth century (BC) Ionian Miletus, Anaximander and Anaximenes, 
can be expressed as a belief that the world is not
governed by the wills and vanities of Gods but instead by immutable and
``natural'' laws arising from fundamental relationships between some basic
elements of Nature. 
This thought, formulated as the Democritian hypothesis that Nature, at the
most fundamental level, consists of a set of indivisible basic elements,
still dominates the science of physics. 
Today, the twilight of possible discoveries is found deep within the
atomic nucleus and even deeper, within the nucleons themselves, at the
smallest length scales yet probed by any experiment. In the continued quest
for Democritus' elements, we have reached objects which, if size they have,
are smaller than a billionth of a billionth of a meter across, around a 
billionth the
size of a normal atom. We do not yet know whether these objects have any
internal structure. They are the quarks and leptons (6 of each) which
constitute all matter known to us and the bosons which carry the three
fundamental forces of Nature known to us. The fourth force, gravity, is
still lacking a complete description. 

Combined, these particles and the interactions between them (excepting
gravity) are incorporated 
in a theoretical framework known as the Standard Model of particle physics
(the SM). We know that this model cannot be the whole story as it contains
some internal inconsistencies, and
therefore a number of speculations on the possible existence of more hitherto
undiscovered properties of Nature has arisen. That Nature could contain a
special symmetry known as supersymmetry is one of these speculations, and our
belief in it is spurred on by its providing an elegant solution to a
fundamental problem in the Standard Model connected with the nature of mass
and by its being a key ingredient in many of the attempts to provide a
quantum description of gravity. 
These theoretical speculations are not without
relevance to the world of experimental physics. There are many who believe
that the Large Hadron Collider (LHC) which is scheduled to commence
operations at CERN in 2006 will see the first experimental indications of
supersymmetry in Nature. This machine, the result of the combined efforts
of thousands  of engineers and physicists around the globe, will collide
protons against protons at 
hitherto unprecedented energies, allowing us to probe 
the extremely small length scales
at which supersymmetry, if it is indeed a property of Nature, should reveal
itself. The exact nature of supersymmetry and its
consequences are described in the main body of this text, yet first two very
important aspects of present day high energy physics deserve to be introduced.

\paragraph{1) The Experiment:}
The ATLAS detector is one of four detectors being constructed for the LHC. 
Its name is obtained by a rather contrived abbrevation of \emph{\textbf{A}
\textbf{T}oroidal \textbf{L}HC \textbf{A}pparatu\textbf{S}}. It will be
located in a newly excavated cavity in the tunnel that used to house the LEP
accelerator 140 metres below ground level at the CERN laboratory on the
French-Swiss border near Geneva. 
At the point where ATLAS will be located, the two beams of the
LHC, circulating in opposite directions at very close to a billion kilometers
per hour, will be brought to cross, and approximately 
a billion collisions of protons against protons will take place per
second. For someone interested in Supersymmetry or other rare physics, this
number is crucial. Even with this colossal event rate (and assuming of
course that Supersymmetry exists), less than \emph{once} per \emph{ten}
seconds will
the processes studied in this work occur. That means that we must somehow be
able to tell the difference between a ``supersymmetric event'' and a ``normal''
one to a very high precision. If we are wrong more than approximately 
1 time in 100 million, 
we will effectively be blind to supersymmetry. With
the aid of sophisticated computer simulations of both normal physics 
and supersymmetric physics, a
significant part of this thesis has been to study how well we can expect to
isolate the ``signal'' from the ``background'' processes.

From the technical point of view, the ATLAS detector itself follows the
by now classic ``cylindrical onion'' design for collider detectors,
beginning with a vertex detector closest to the beam pipe surrounded by
tracking detectors immersed in a strong magnetic field, these again being
surrounded by calorimeters and finally muon chambers. 
\begin{figure}[t]
\begin{center}\vspace*{5mm}
\textsf{\huge ATLAS}\\ \ \\
\includegraphics*[scale=3.0]{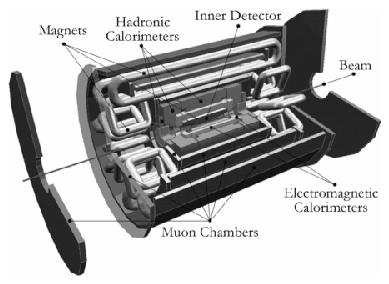}\\
\includegraphics*[scale=0.5]{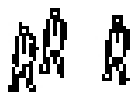}
\end{center}
\caption[\small Schematic of the ATLAS detector.]{Schematic of the ATLAS
detector. When completed, 
the contstruction will be about four stories in height and occupy
a ground area equivalent to half a football field.\label{fig:atlas1}}
\end{figure}

Briefly stated, the purpose of the vertex detector is to measure where the
produced particles are coming from. The tracking detectors then measure 
where the (charged) particles are \emph{going}. 
A strong magnetic field is applied to bend these tracks,
allowing the momenta of the particles making the tracks to be
measured. It is thereafter the
task of the calorimeters to completely stop the particles, neutral ones as
well as charged, and in the process give off a signal proportional to the
energy of each particle, completing the energy-momentum measurement for
the charged particles. The last particles to get caught are the muons. Their
high penetration ability makes special measures necessary, and so the
outermost part of the detector is the muon chambers. Another magnetic field
is applied here (see fig.~\ref{fig:atlas1}). The resulting bending of the
muon tracks yields a more accurate measurement of the muon momenta.

A last, quite remarkable thing about ATLAS is the extreme data rates which
will be output from the experiment with approximately 1 billion events per
second, each consisting of about 1MB of data \cite{atlastdr}. Using both
hardware close to the detector and software farther away from it (quite
possibly even on another continent), this tremendous 
data stream is reduced in real time
to a level where ``only'' 1 PetaByte,
or one million GigaBytes, are expected on disk per year. A very interesting
new type of network structure called ``GRID'' is being developed to handle
and distribute the data output from the detector over the internet to
local computing centers around the world.

\paragraph{2) Simulating the Experiment:} A multibillion dollar accelerator 
like the
LHC is not constructed without detailed prior studies of its capabilities.
In present-day high energy
experiments, an enormous amount of more or less well understood 
physical processes occur from the initial
state in time when two particles approach each other at ultra-relativistic
speeds inside the
beam pipe of some high-energy accelerator to the point, after the
interaction, when particle tracks, energy depositions, detector
status etc.\ are read our from the detector whose purpose it is to record
as detailed information as possible about the `event' that has taken place. 
Furthermore, the
probabilistic nature of quantum mechanics ensures that any given
initial state can result in a truly enormous number of possible
final states. It is therefore a central and highly non-trivial task for the
experimentalist to reconstruct what \emph{actually} happened in
each event recorded, and to design a detector that enables a
distinction between ``interesting'' and ``uninteresting'' events, i.e.\ events
that possibly contain traces of what is being searched for and events where
only ``trivial'' processes have occurred. Simply constructing a machine
capable of producing new particles is not enough - one must also
be sure that one can actually \emph{see} that the new particles
were there. 
The theorist faces similar problems. To enable a comparison with
experiment, the theory must deliver predictions of experimental
observables. It is therefore not enough simply to study and
calculate the elementary interactions contained in a given
theory; these are only very rarely directly visible to the
experiment. In most cases, a full, analytical calculation (using pen and
paper) of
everything that goes on between some initial state and some final
state is simply out of the question. The amount of physical processes that
would have to be calculated is simply too enormous, and for many of them, we
do not even have exact solutions of the underlying theory. 
It is here that Monte Carlo
generators come into the picture. The power of computers to
perform repetitive tasks according to specific rules combined
with the statistical Monte Carlo method has become an
indispensable tool in comparing theory with experiment.
It is the aim of Monte Carlo generators like \pythia\ and detector
simulations like \atlfast\ to provide
computer simulated `events' that mimic those which would be recorded in a
real detector, allowing us to bridge the gap between high energy theories and
high energy experiments.
 
In Monte Carlo generators, the elementary processes that
occur and about which one wishes to extract information are dressed
with all the `other things'\footnote{Of course, what is noise and background
to one physicist can be highly interesting phenomena to another.} 
that occur before, during, and after the
high-energy interaction, thus simulating a real interaction between physical
particles. Add to this a simulation of how the detector apparatus
works (\emph{no} detector catches everything), and the generated
event has all the likeliness of a real event, except that in
\emph{this} case we have full knowledge of what the fundamental
process was. It is then possible for the experimentalist to make his best
guess on what went on, based on the output from the detector simulation, and
compare that with what the generator says \emph{really} went on. This can be
done any number of times for any number of physics scenarios, and it is then
possible to test the efficiency of the detector and data analysis
before the experiment is constructed,
enabling an optimization of detector design, particle identification
algorithms etc. 
The limitation is, of course,
that it is impossible to include \emph{all} possible effects that
can occur in a physical interaction, and so one should never
blindly trust simulated results without a critical analysis of whether
all important effects have been taken into account. Exercising
caution on this point, Monte Carlo simulations can be extremely useful in
interpreting observed phenomena, in constructing a good detector and
in finetuning the analysis strategies used in the actual experiment. In this
work, the \pythia\ Monte Carlo generator is used to simulate a special class
of supersymmetric theories which were previously not available in the
generator and its output is combined with a crude simulation of the ATLAS
detector. The potential for ATLAS to find experimental evidence
for these theories is then analyzed for several specific
variants of the theories.

\subsection{Outline}
In section \ref{sec:susy}, a brief account of the theoretical
considerations fuelling 
the widespread belief in supersymmetry is given, together with an
introduction to that particular class of supersymmetric theories
called $R$-parity violating which comprises the subject of this
thesis. The details of the overall supersymmetric framework
adopted in this work and the $L$-violating decays of
supersymmetric particles together with their implementation in \pythia\ are
discussed in section \ref{sec:lspdecays}. In section 
\ref{sec:detector},
some technical aspects of the ATLAS detector are discussed, 
with some emphasis on the
detector and reconstruction parameters as they appear in \atlfast. 

Section \ref{sec:analysis} begins with a presentation of the mSUGRA points and
$L$-violating scenarios chosen for the analysis. The main part of the
analysis is then presented, beginning with proposals for trigger menus
optimized for \LV-SUSY at  ``mid luminosity'' running of the LHC 
(i.e.\ $L=3\times
10^{33}\mathrm{cm}^{-2}\mathrm{s}^{-1}$). Based on the events surviving these
triggers, an analysis of the
significance with which a discovery can be made at ATLAS for Lepton number
violating mSUGRA scenarios is then presented, assuming 30\fb$^{-1}$
integrated luminosity,
corresponding to one year of mid-luminosity running. 
This analysis 
begins with a
presentation of various kinematical and inclusive variables on which loose
cuts are made, designed to have a high acceptance for as many of the various
scenarios investigated as possible. Processing the remaining
events through neural networks trained to discriminate between SM and SUSY
events yields the final event numbers used to estimate the discovery
potential, i.e.\ the statistical
significance with which a discovery can be made, for each of the 50 separate
mSUGRA models investigated. These numbers, representing the main result
obtained in this work, can be found in table
\ref{tab:discoverypotential}. 

Finally, in section \ref{sec:conc}, we give an outlook on work remaining to
be done together with a summary of the main conclusions reached within the
body of this thesis. 


\clearpage
\section{Supersymmetry \label{sec:susy}}
For the past twenty years, high energy physics has experienced a unique state
of affairs. Following the revolutions of relativity and quantum mechanics in
the early parts of the 20'th century, Dirac's theory of the electron, 
the discoveries of antimatter and neutrinos, 
and the discovery that hadrons (e.g.\ protons
and neutrons) are composed of quarks, a unified framework describing the
electromagnetic, weak, and strong interactions emerged in the 70's, 
 the so-called Standard Model of particle physics. 

For the past twenty years, this
model has successfully explained and predicted an astounding amount of
experimental data. Only one piece is missing: the Higgs boson. In the
Standard Model, this boson must exist since the known particles would
otherwise be massless,  
and so the Standard Model would fall if the Higgs does not exist.
Due to the immense success of the SM, it is widely believed that
it is only a matter of time until this last piece of the puzzle is found, yet
such a discovery would as much mark the doom as it would mark the final
verification of the SM. If a fundamental Higgs boson exists, deeper problems
appear for which one must go beyond the SM to find a resolution. The most
severe of
these problems is known as the \emph{hierarchy problem}, and we shall discuss
it in detail in section \ref{sec:hierarchy}. For now, it suffices to note
that despite the great success of the SM, it contains many unexplained
parameters and has internal inconsistencies which can only be resolved by a
more general framework. The central question, 
if we are ever to identify the \emph{nature} of this ``New
Physics'', then becomes at what energy scale we will see the predictions of
the Standard Model begin to fail. The answer to this lies in the hierarchy
problem itself, and I shall summarize below why there is good reason to believe
that this energy scale cannot be much higher than 
the scales which will be probed by
high energy experiments over the next few decades. Several ideas have been
proposed for what the new physics could be. The possiblities range from
extending the Standard Model with new particles, new interactions, higher
symmetries of one form or another, or new spatial dimensions. 
Of these, Supersymmetry
has become by far the most extensively considered, and I shall present some
of the reasons for this in what follows.

\subsection{What \emph{is} Supersymmetry?}
Among the most
fundamental properties of any physical theory are the \emph{symmetries} which
the theory respects, since symmetries are so intimately connected with the
existence of conservation laws. One of the most fundamental theorems in both
classical as well as quantum mechanics is the famous 
\emph{Noether's Theorem} stating
that for each symmetry which nature respects there exists a corresponding charge
or current which must be conserved. Another reason that symmetries are of
prime interest is that they provide connections between one part of the
theory and another, namely the parts which the symmetry relates to each
other. For these reasons, a crucial task has been (and is) to discover
which symmetries are respected, or, if violated, to what extent, by
nature. In very general terms, there are two possible kinds of symmetries:
\emph{internal} symmetries and \emph{space-time} symmetries. The theory of
relativity, for example, tells us that Lorentz invariance is respected as a
symmetry in space and time. 
Rotational invariance, translational invariance, and temporal invariance are
other examples of space-time symmetries in that all of them imply a symmetry
between one spatio-temporal point and another. Dropping a ball today in
Copenhagen and
dropping it tomorrow in Tokyo, having first turned yourself by 90 degrees, 
should not affect the speed at which it hits the ground.
In contrast, the symmetries which we
believe to be responsible for the natural forces are examples
of internal symmetries. A familiar example is the invariance of the performance
of an electrical circuit to a global change of phase. 
As can be gleaned from this example, internal symmetries have to do
with degrees of freedom associated with the fields themselves and not
space-time, c.f.\ the phase of
the electric field. As is well known, 
the theoretical framework on which the Standard Model is
built is quantum field theory where particles are viewed as \emph{quanta} of a
\emph{field}. These fields possess internal degrees of freedom not unlike the
phases just mentioned, and the existence of internal symmetries require that
the physics be invariant under changes to these degrees of freedom. Such
internal degrees of freedom which have no effect on physical observables are
known as gauge degrees of freedom, and it was with the realization that gauge
symmetries can give rise to the fields associated with the natural forces
that quantum electrodynamics was first conceived. Turning now to
Supersymmetry, it will be useful to keep the distinction
between space-time and internal symmetries in mind.

The story of Supersymmetry began in 1967 when 
Coleman and Mandula \cite{coleman67} proved a very important theorem
which now bears their name. It deals with what \emph{kind} of space-time 
symmetries are at all possible in any quantum field theory, at least any
theory where the particles interact. They found that our theories already
contain their full share of bosonic space-time symmetries, 
the combination of which is described by the so-called Poincar\'{e} 
symmetry group. I shall come back to the significance of the word
\emph{bosonic}. Effectively, the existence of an additional symmetry
of this type would constrain the degrees of freedom in any interaction 
so much that the equivalent to no interaction at all takes place.
Such a non-interacting theory obviously has very little to do with the real,
very interacting universe, and thus it ceases to be
 physically interesting. The Poincar\'{e} group already contains its full
share of (bosonic) symmetries. But in the context of supersymmetry, the word
 \emph{bosonic} becomes all-important. That a symmetry is
bosonic is technical slang for a symmetry for
which \emph{the order} in which two transformations are performed cannot be
distinguished, i.e.\ a symmetry whose generators commute. It is, for
example, quite impossible to tell the difference between two sheets of paper,
one first moved 1 cm and then moved 2 cm, the other first moved 2 cm and then
moved 1 cm (in the same direction). They both end up in exactly the same
place. Thus, translational invariance is an example of a bosonic
symmetry. To get on to supersymmetry, note that the word bosonic is
customarily used
about one of the two fundamentally
different types of spin a particle can have. Bosons have integer spins
(0,1,2,...) while
fermions have half-integer spins (1/2,3/2,5/2,...). This difference causes
the two types of particles to behave in very different ways, and their
mathematical description is, likewise, very different. Bosons can be described
using ordinary, commuting complex numbers while fermions require a
description based on anti-commuting numbers, 
so-called Grassmann numbers for which $BA=-AB$. 

The case of fermionic symmetry
generators in quantum field theory (i.e.\ anticommuting operators carrying
half-integer spin)
was considered in 1975 by Haag, Lopuszanski, and Sohnius \cite{haag75}. 
Their discovery was that any number of \emph{fermionic}
symmetry generators could be introduced in the theory without
ruining it, provided they satisfied a rather constrained algebra;
an algebra they called the supersymmetry algebra\footnote{As usual, an
  ``algebra'' is simply a set of commutation (and anticommutation) rules
  between operators, in this case the generators of the usual space-time
  symmetries and the supersymmetry generators.}.

These considerations found their motivation simply in understanding
what symmetries were at all possible in a general, physical
quantum field theory, and nothing was said as to whether these
symmetries were actually realized in nature or not, yet since that
time, supersymmetry (SUSY) has risen to become the most
extensively considered extension of the current theory, the
so-called Standard Model (SM) of particle physics. The
motivations for believing that supersymmetry does indeed belong
among the basic properties of Nature are many, and there is some reason
to expect that we shall soon possess the ability to test this
belief in high energy accelerator experiments around the globe, most notably
at the future Large Hadron Collider (LHC) at CERN, scheduled to commence
operations around 2006. 

Supersymmetry has been extensively considered in
the literature, and a wealth of good textbooks and reviews on the
subject are readily available. Thus, for detailed discussions,
the reader is referred to e.g.\
\cite{kane98} or, for the more algebraically minded,
\cite{lykken96,cahill99}. It is understood that 
this by no means represents an exhaustive list. In the present
writing, I merely outline the most important motivations and
properties of supersymmetry. For completeness, the most important
alternatives are mentioned as well. 

\subsection{Properties of SUSY}
As already mentioned, the operators generating SUSY transformations
 carry spin $\frac12$. It should then not be
surprising that they transform a spin state $\tket{S}$ into
$\tket{S\pm\frac12}$, implying that a fermion is transformed into a
boson and vice versa. Supersymmetry is thus a symmetry between
fermions and bosons, and unbroken supersymmetry implies (among other things) that
these states have equal masses. 
The non-observation of
supersymmetric particles to date therefore implies that
supersymmetry must be a broken symmetry at our energy scales, if it
exists at all. Otherwise we would have seen these ``spin-partners'' of the
known particles long ago. 
How supersymmetry may be broken will be discussed in section
\ref{sec:susybreaking}, but first we must get slightly technical. 
Postulating for the moment that this symmetry really exists,
 which fields and interactions should the SUSY Lagrangian contain? 

The first
question to be addressed in answering this question is \emph{how
many} anticommuting SUSY generators one should introduce. To this
there is a natural limitation. In 4-dimensional space-time, only one
supersymmetry is allowed if the theory is to have parity
 violation and chiral fermions. Violation of parity symmetry, or
 space-inversion, 
 was first established by Wu et al.\ in 1957 with the famous measurement 
that the electrons emitted in $\beta$ decay of
 Cobalt-60 nuclei are correlated with the nuclear spin direction, a parity
 non-invariant situation \cite{wu57}. Chiral fermions means fermions that
 come with definite \emph{handednesses} as we know it from the SM. 
Theories with only one supersymmetry are known as 
$N=1$ SUSY, and in the rest of this text, I let $N=1$ be
implicit when talking about supersymmetry. We are now left with \emph{one}
supersymmetry whose spin $\frac12$ generators are customarily
 denoted $\mathcal{Q}$ and $\mathcal{Q}^\dagger$ \cite{lykken96}, i.e.\
\begin{equation}
\mathcal{Q}\tket{\mbox{boson}} = \tket{\mbox{fermion}}
\hspace*{1cm}\mbox{and}\hspace*{1cm}
\mathcal{Q}\tket{\mbox{fermion}} = \tket{\mbox{boson}}
\end{equation}
That a theory is supersymmetric simply means, in all generality, that these
operators leave its Lagrangian invariant -- that the physics is the same
before and after particles and superpartners are interchanged. 

The next question to consider is what the particle content of the theory 
looks like; what kind of multiplets should we include? A
``multiplet'' of a symmetry simply denotes a collection of 
any number of fields which are
transformed into each other by the symmetry such that one never goes
``outside'' the multiplet by acting with the symmetry operators. If there is
only one field (which is then, itself, invariant to the symmetry operation),
one speaks of a singlet. If there is a pair of fields which are transformed
into each other, one speaks of a doublet, etc. For brevity, and to
avoid a (technical) repetition of a well-known issue, the reader is
referred to \cite{martin98} for details regarding
supermultiplets. For our purposes, it suffices to note that ($N=1$)
supersymmetric theories can be built up from just two types of fundamental
multiplets: \emph{gauge} (or \emph{vector}) multiplets  and \emph{chiral}
 (or \emph{matter}) multiplets. 
\begin{figure}[t]
\center
\begin{fmffile}{supermultiplets}
\begin{tabular}{ccc}
\begin{fmfgraph*}(160,80)
\fmfforce{0w,0.2h}{fl}
\fmfforce{0.35w,0.2h}{fr}
\fmfforce{0.2w,0.4h}{fu}
\fmfforce{0.8w,0.4h}{su}
\fmfforce{0.65w,0.2h}{sl}
\fmfforce{w,0.2h}{sr}
\fmf{fermion,label=$f$}{fl,fr}
\fmf{scalar,label=$\ti{f}$}{sl,sr}
\fmf{plain,left=0.5,label=SUSY $Q$\comma $Q^\dagger$}{fu,su}
\fmfv{d.sh=triangle,d.siz=0.03w,d.ang=-28}{su}
\fmfv{d.sh=triangle,d.siz=0.03w,d.ang=28}{fu}
\end{fmfgraph*} & 
& 
\begin{fmfgraph*}(160,80)
\fmfforce{0w,0.2h}{fl}
\fmfforce{0.35w,0.2h}{fr}
\fmfforce{0.2w,0.4h}{fu}
\fmfforce{0.8w,0.4h}{su}
\fmfforce{0.65w,0.2h}{sl}
\fmfforce{w,0.2h}{sr}
\fmf{boson,label=$b$}{fl,fr}
\fmf{fermion,label=$\ti{b}$}{sl,sr}
\fmf{plain,left=0.5,label=SUSY $Q$\comma $Q^\dagger$}{fu,su}
\fmfv{d.sh=triangle,d.siz=0.03w,d.ang=-28}{su}
\fmfv{d.sh=triangle,d.siz=0.03w,d.ang=28}{fu}
\end{fmfgraph*}\vspace*{-4mm}\\
a) & \hspace*{15mm}& b) 
\end{tabular}
\end{fmffile}
\caption[\small A graphical illustration of supermultiplets]{A graphical
illustration of supermultiplets. a) The fermion and the scalar in a chiral supermultiplet. 
b) The vector boson and the fermion in a vector supermultiplet. Both are
related to each other by a supersymmetry
transformation. \label{fig:supermultiplets}} 
\end{figure}
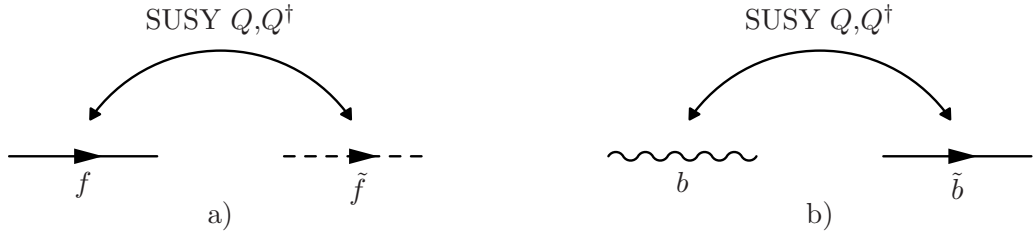
A chiral multiplet contains a chiral (i.e.\ left-handed or right-handed)
fermion together with a complex scalar. This is illustrated in
figure \ref{fig:supermultiplets}a. Thus, the fermions of the SM can be
accomodated in chiral supermultiplets, each receiving a scalar SUSY partner
called \emph{squarks}, \emph{sleptons}, and \emph{sneutrinos} 
where the `\emph{s}' is short for
scalar. A gauge multiplet contains a
 vector boson together with a chiral fermion (fig.~\ref{fig:supermultiplets}b). 
The gauge bosons of the SM can thus be
accomodated in gauge supermultiplets with the requirement that they each have
a supersymmetric fermion partner (the so-called \emph{gauginos}). So far,
each SM field (excepting the Higgs - see below) 
has been put inside a multiplet with one other field, and so we
have arrived at
 an exact doubling of the SM particle spectrum. More supermultiplets could of
course exist where \emph{both} fields have not been seen experimentally, yet
I here restrict the attention to
the minimal model where only those multiplets that are
strictly necessary are included.

The Higgs sector is slightly more complicated and deserves some
mention. Due to the Higgs being a scalar, 
it can only reside in a \emph{chiral}
supermultiplet together with a fermion which would have hypercharge
 $Y\!=\!\frac12$ or $Y\!=\!-\frac12$. 
This might sound very innocent, but in fact, it is a
disaster. Together with the third component of weak isospin, $T_3$, 
the hypercharge assignments of the SM particles uphold a delicate balance
where the traces $\Tr{Y^3}$ and $\Tr{T_3T_3Y}$ are both zero when taken over
all fermions. Introducing new, non-zero hypercharge 
spoils that relation, giving rise to an anomaly \cite{weinberg82} which 
would lead to unphysical divergences 
in the theory. If, however, \emph{two} Higgs supermultiplets are
introduced with different hypercharge signs, the cancellation is
reestablished, and so there must be at least two Higgs (super-) doublets for
SUSY to work. Indeed,
the requirement of two Higgs doublets finds its motivation from two other
sources as well; if the Lagrangian is to respect supersymmetry and gauge
invariance of hypercharge, the coupling of the ordinary Higgs doublet to
up-type quarks is forbidden. Thus, two Higgs multiplets are
needed to provide  
mass terms for both up- and down-type quarks (and leptons)
\cite{dimopoulos81}. Also, two
Higgs doublets are needed to provide all the SU(2) gauge bosons and their
superpartners with masses. 

This completes the specification of the minimal particle content
for a supersymmetric extension of the Standard Model. In this
work, we shall consider exclusively that version of SUSY known as the
Minimal Supersymmetric Standard Model (MSSM) (albeit with the modifications
that \RV\ introduces) which only contains the SM
particles (plus one additional Higgs doublet) together with their
superpartners. This choice is made due to the relative simplicity of the MSSM
and due to the lack of any real motivation for introducing more multiplets than
absolutely needed. The total particle content of this model (with the
superfields to which each particle belongs) is shown in table \ref{tab:mssm}.
\begin{table}
\setlength{\extrarowheight}{0.5pt}
\begin{center}
\begin{tabular}{ccccc}\toprule
\sc Name        & \sc Spin 0 & \sc Spin 1/2 & \sc Spin 1 & \sc\begin{tabular}{c}
Superfield\\
Notation
\end{tabular}\\\cmidrule{1-5}
squarks \& quarks & ($\ti{u}_L\hspace*{1.5mm}\ti{d}_L$) &
($u_L\hspace*{1.5mm}d_L$) & - & $Q$ \\
(3 families) & $\ti{u}_R^*$ & $u_R^\dagger$ & - & $\bar{u}$ \\
 & $\ti{d}_R^*$ & $d_R^\dagger$ & - & $\bar{d}$ 
\\ \cmidrule{1-5}
sleptons \& leptons & ($\ti{\nu}\hspace*{1.5mm}\ti{e}_L$) &
($\nu\hspace*{1.5mm}e_L$) & - & $L$ \\
(3 families) & $\ti{e}_R^*$ & $e_R^\dagger$ & - & $\bar{e}$ \\
\cmidrule{1-5}
Higgs \& higgsinos & ($H_u^+\hspace*{1.5mm}H^0_u$) &
($\ti{H}_u^+\hspace*{1.5mm}\ti{H}_u^0$) & - & $H_u$ \\
 & ($H_d^0\hspace*{1.5mm}H^-_d$) &
($\ti{H}_d^0\hspace*{1.5mm}\ti{H}_d^-$) & - & $H_d$
\\\cmidrule{1-5}
gluino \& gluon & - & $\ti{g}$ & $g$ & - \\\cmidrule{1-5}
winos \& $W$ bosons & - & $\ti{W}^\pm \hspace*{1.5mm} \ti{W}^0$ &
$W^\pm\hspace*{1.5mm} W^0$ & - \\\cmidrule{1-5}
bino \& $B$ boson & - & $\ti{B}^0$ & $B^0$ & - \\ 
\bottomrule
\end{tabular}
\caption[\small Supermultiplets and superfields in the MSSM]{Supermultiplets
 in the MSSM. The states appearing in each supermultiplet are collected 
in ``superfields'' in section \ref{sec:lspdecays} when Lagrangians are written down.\label{tab:mssm}}
\end{center}
\end{table}

\subsubsection{Sparticles and Mixing\label{sec:mixing}}
As is familiar from the SM, the spectrum of physical particles (mass
eigenstates) is not necessarily identical to the set of states
appearing in the interaction Lagrangian (current eigenstates). In general,
particles whose conserved quantum numbers are identical will mix with each
other. The well-known phenomenon
of quark mixing (see e.g.\ \cite{buras96}) such as causes $K -
\overline{K}$ and 
$B - \overline{B}$ mixing provides an important example of
this. When extending the SM with SUSY particles, more mixing appears, and it
is important that the conventions and nomenclature are firmly established
before calculations are carried out. The conventions employed in \spythia,
\herwig, and \isasusy\ can be found in Appendix \ref{app:A} for reference.

\paragraph{Scalars:}
Both the left-handed and right-handed versions of the SM 
quarks and leptons have scalar SUSY partners. Yet due to their scalar
nature, it is impossible for these particles to possess any intrinsic
`handedness' themselves. The only difference between e.g.\ the
$\tilde{e}_L$ and the $\tilde{e}_R$ is that they belong to different
supermultiplets. Apart from that, they have exactly the same quantum
numbers, and so they can mix with each other. In the language of ordinary
quantum mechanics, we are dealing with a 2-fold degenerate subspace of
states. Of course, any two orthonormal states will provide us a basis for
this space, yet two physically motivated choices suggest themselves:
the \emph{L-R} basis and the \emph{mass} basis, the former being a
description in terms of current eigenstates and the latter in terms of the
mass eigenstates, $\tilde{e}_1$ and $\tilde{e}_2$. 
Assuming for simplicity that there is no 
mixing between sfermions of different generations (see below), there appears
a $2\times 2$ mixing matrix between the mass and current eigenstates for each
sfermion pair, $\tilde{s}$ (in the convention of \cite{baer94,dreiner00}):
\setlength{\extrarowheight}{2pt} 
\begin{equation}
\left( \! \begin{array}{c}
\tilde{s}_L \\ \tilde{s}_R
\end{array}\! \right) = 
\left[ \begin{array}{cc}
\cos\theta_s & +\sin\theta_s\\ 
-\sin\theta_s & \cos\theta_s
\end{array} \right]
\left( \! \begin{array}{c} 
\tilde{s}_1 \\ \tilde{s}_2
\end{array}\! \right)
\label{eq:sfmix} 
\end{equation}  
 In a model without right-handed 
(SM) neutrinos, there is of course only one $\tilde{\nu}$ for each generation
(i.e.\ no sneutrino mixing).

As mentioned, it is the \emph{current} eigenstates (the chiral SM fields and
their superpartners, $\tilde{s}_{L,R}$) which appear in the interaction 
Lagrangian. Since it is not these states but rather the mass eigenstates
(being eigenstates of the Hamiltonian and thus of the time-development
operator) which propagate through spacetime, one is most often interested in
calculating production and decay 
properties of \emph{mass} eigenstates\footnote{An important exception is in
neutrino physics where one stays in the current state basis and mixing is
replaced by oscillation. See e.g.\ \cite{diracmajorana}.}, resulting in the appearance of mixing
factors in the matrix elements in section \ref{sec:lspdecays}. A more general
calculation 
taking inter-generational mixing into account can be found in \cite{baltz98}, 
yet the magnitude of such mixing is highly constrained by the non-observation
of large Flavour Changing Neutral Current rates
and so we shall discount this possibility here (Flavour Changing Neutral Currents are
quark transitions where the quark flavour changes, but the charge does not. The
lowest order SM contributions are loops containing virtual, heavy
particles, and so new physics effects (new loop particles) would enter at the
same order in the couplings and thus could be extremely visible). 
\paragraph{Neutralinos:}
There are four neutral fermions in the MSSM: the partners of the neutral U(1)
and SU(2) gauge bosons, $\tilde{B}^0$, $\tilde{W}^0$ (or, 
equivalently, $\tilde{\gamma}$, $\tilde{Z}$), and the partners of the
two neutral Higgs scalars,
$\tilde{H}_u^0$, and $\tilde{H}_d^0$. These states mix to form the four 
mass eigenstates
$\tilde{\chi}_1^0$, $\tilde{\chi}_2^0$, $\tilde{\chi}_3^0$,
$\tilde{\chi}_4^0$. 
In analogy with the scalar case, the mixing is parametrized by a $4\times 4$
mixing matrix, $N_{ij}$ \cite{haber85}:
\begin{equation}
\left( \!\begin{array}{c}
\tilde{\chi}^0_1\\ 
\tilde{\chi}^0_2\\ 
\tilde{\chi}^0_3\\ 
\tilde{\chi}^0_4\end{array}\! \right)
= N_{ij} \left(\!\begin{array}{c}
-i\tilde{B}^0\\ -i\tilde{W}^0\\
\tilde{H}_u^0\\ \tilde{H}_d^0\end{array}\!\right)
\end{equation}
Note that \cite{haber85} uses the notation $\lambda'=B^0$ and $\lambda^3=W^0$. 
The analogous mixing matrix, $N'$, for the
$(\tilde{\gamma}, \tilde{Z},\tilde{H}_u^0, \tilde{H}_d^0)$ basis can be
obtained simply by rotating the $\tilde{B}^0, \tilde{W}^0$ components by the
Weinberg angle (see e.g.~\cite{gunion86}) so that:  
\begin{eqnarray}
N'_{j1} = N_{j1}\cos\theta_W + N_{j2}\sin\theta_W \\
N'_{j2} = -N_{j2}\sin\theta_W + N_{j2}\cos\theta_W
\end{eqnarray}
Depending on what coupling is considered one may see either of these used in
the literature. This should not lead to confusion when keeping in mind
that $N$ and $N'$ express the same mixing, albeit on different bases.
\paragraph{Charginos:}
The charginos, $\tilde{\chi}_i^\pm$ ($i=1,2$), are defined as the mass
eigenstates arising from the mixing of the charged Winos and
Higgsinos. Naturally, states of different electric charge cannot mix, and so
two separate mixing sectors appear \cite{haber85}:
\begin{equation}
\twovec{\ti{\chi}_1^+}{\ti{\chi}_2^+} = V \twovec{-i\ti{W}^+}{\ti{H}_u^+}
\hspace*{1.5cm}
\twovec{\ti{\chi}_1^-}{\ti{\chi}_2^-} = U \twovec{-i\ti{W}^-}{\ti{H}_d^-} 
\end{equation}

\subsubsection{Supersymmetry Breaking}
\label{sec:susybreaking}
It was stated above that supersymmetry must be broken since unbroken
supersymmetry implies an unoberserved 
mass degeneracy between each SM particle and its
respective superpartner. An important part of constructing a viable 
supersymmetric model is therefore to specify 
some model for \emph{how} supersymmetry is broken. At
present, this issue is unresolved and there is no outlook to convergence on
any one model in the near future. 
At first glance, a spontaneous breaking mechanism would probably seem most
natural, yet it is 
generally very difficult to obtain an acceptable mass spectrum in
such models. The problem
is that while the superpartner of \emph{one} of the chiral states of the SM
fermions is heavier, the superpartner of the other chiral state tends to be 
\emph{lighter} than its SM counterpart, again 
in contradiction with the experimental
non-observation of SUSY particles. Several more or less
well-motivated alternatives exist: Explicit (soft) breaking, 
SuperGravity (SUGRA), Anomaly-mediated SUSY breaking (AMSSB), 
and Gauge-mediated SUSY breaking (GMSSB). We shall here
focus on the most popular model, the minimal Supergravity model (mSUGRA).

In Supergravity models,
supersymmetry is imposed \emph{locally} rather than globally. The resulting gauge
field is interpreted as the gravitational one - an economic solution since it
introduces only a field we already know might exist, the gravitational one, to
break supersymmetry. The model is an example of a
more general class of breaking mechanisms called ``hidden sector breaking
models'' where supersymmetry is broken by some unobservable (called ``hidden'')
physics which only interacts with ``the visible sector'' (i.e.\ you, me, and
all fields which we can interact with) through some ``messenger field'', 
in this case gravity. This keeps the breaking flavour-blind as required from
the absence of low energy Flavour Changing Neutral Currents and, since
gravity is weak, it keeps the degree of breaking small which is
generally required if Supersymmetry is to solve the hierarchy problem (see
below). Adopting this scenario in phenomenological applications, 
a number of further assumptions are usually made, as will also 
be the case in the present work. They find their inspiration  
in Grand Unified Theories (GUTs) by which one 
justifies the plausibility of setting classes of
masses and couplings equal at the GUT scale, and after the inclusion of which
one prepends a small ``m'' (for  \emph{minimal}) to the SUGRA name. Some
theorists argue against this practice due to generally declining interest in
GUTs, yet it is often the only practically feasible thing in view of the 
the several hundred dimensonal parameter space
one would otherwise face. 
What is needed in order to specify a complete mSUGRA model is 
just five parameters: 
\begin{enumerate}
\item $m_0$ : A common mass for all scalars in the theory, i.e.\ sfermions and
higgs scalars, at the GUT scale.
\item $m_\frac12$ : A common mass for all fermions in the theory, i.e.\ the
SM fermions, the gauginos, and the higgsinos, at the GUT scale.
\item $A_0$ : A common trilinear coupling (trilinear just meaning that three
fields are coupling together).
\item $\tan\beta$ : The ratio of vacuum expectation values for the two Higgs
doublets. Its value is constrained between approximately 5 and 40.
\item sign$(\mu)$ : $\mu$ is the mass parameter for the higgsinos. In
general, this parameter could have an arbitrary complex value
\cite{martin98}, but a non-vanishing imaginary part generally results in
in large, unobserved low energy $CP$ violating effects, the most severe being 
 an electric
dipole moment for the neutron. It is therefore customary to assume $\mu$
real, with only the sign remaining to be specified since the magnitude 
is given by the other parameters.
\end{enumerate}
One should note that the choice of basis is not unique, and so one often sees
slightly different quantities parametrizing the space. It remains, however,
that the degrees of freedom are inherently four real numbers and one
sign. Graphically speaking, these parameters describe the physics we would
see if we could probe the ultra-short GUT length scales. At lower energies,
the physics ``looks different'', in analogy with the
shielding of the electron charge by quantum fluctuations very near the
charge. The physics at different length/energy scales is related by the
procedure of renormalization, specifically the famous Renormalization Group
Equations (RGE's). In mSUGRA (and in any other high-energy field theory),
these equations are used to evolve the parameters from the input scale down to
the observable scale. In
physical language, the quantum fluctuations (loops) which are too highly
energetic to be resolved by the probing energy ($\mathcal{O}(1\TeV)$ for the
LHC) are ``put inside'' the masses and couplings, causing each parameter of
the theory to be a function of the energy scale. This is the reason one
speaks of ``running'' couplings and masses. In this work,
the minimal supergravity scenario will be assumed with the additional
condition that $A_0=0$, mainly to limit 
the dimensionality of the SUSY parameter space, yet a small study of the
consequences of this assumption is also performed (see section
\ref{sec:analysis}). For each of the scenarios
studied, the input parameters along with the mass spectrum at the
electroweak scale are given in table \ref{tab:sugrapoints} in section
\ref{sec:analysis}. 

\subsection{Motivations for SUSY}
\subsubsection{The Hierarchy Problem \label{sec:hierarchy}}
When attempting to lend credence to an extension of the Standard
Model, it is perhaps of importance to note that the SM by itself
cannot be quite right. Although the theory is in truly
astonishing agreement with experiments at the hitherto accessible
energy scales, the SM Higgs sector poses important problems when
extrapolating the theory to higher energy scales. Virtual
particles coupling to the Higgs field yield corrections to the
Higgs mass squared which depend quadratically on the momentum
cutoff used to regulate the theory. The only physical interpretation
that can give meaning to such a cutoff is that it represents the energy
scale at which some new physics steps in to halt the divergent behaviour.
At the Planck scale, \emph{some} new physics
  \emph{must} step in, as this scale by definition is the scale at
  which quantum gravity effects become sizeable (at least in 4 dimensions),
but if the ultimate 
validity scale of the Standard Model is interpreted as either the 
Planck scale
($10^{19} \GeV$) or some other very large energy scale, then,
naturally, the quantum corrections to the Higgs mass would make it
nowhere near the electroweak scale. However, the Higgs
mass \emph{must} be of order the electroweak scale: 
if the theory is to be perturbative for $W$ and $Z$ bosons, $m_H$ must
be less than about $1\TeV$ (see e.g.\ \cite{okun99}), and from below,
the theoretical limit from vacuum stability, $m_H>1\GeV$ \cite{okun99}, has
been superceded by the 
LEP limit, $m_H>113\GeV$ \cite{ellis00}. The discrepancy between these
two fundamental scales of Nature is known as the \emph{hierarchy
problem}. How can the electroweak mass scale be so low 
when the Planck scale is so high? 

In a supersymmetric scenario, the hierarchy problem would
exist no more as long as the mass differences between SUSY particles and SM
particles do not exceed $\mathcal{O}(1\TeV)$ \cite{witten81_2}. 
Since the supersymmetric particles by construction have the same couplings as
their SM counterparts,
every correction to the Higgs mass from an SM particle would receive an
identical contribution from a supersymmetric particle, but with
opposite relative sign (due to the sign difference between fermions and
bosons), and thus the quadratic divergence is cancelled. When SUSY is broken,
this cancellation is not exact (the masses are not the same), 
and so the cutoff mentioned above is effectively
replaced by the SUSY breaking mass scale. 
If this scale is close to the
electroweak scale ($M_{SUSY}\le\mathcal{O}(1\TeV)$), 
then so would the Higgs mass be, in accordance with
experiment. Note that it is only the technical part of the hierarchy problem
which is solved by this. The electroweak scale can be made stable, but there
is no explanation for its actual size. Supersymmetry only began to gain
widespread acceptance around 1981/82 
when a mechanism was discovered \cite{ibanez82}
which could generate the large gap between $M_{X}$ and $m_{EW}$ in a natural
way: radiative breaking of 
electroweak symmetry. The Higgs mass parameter squared which enters the Higgs
potential starts out positive at the GUT (or Planck) scale, but 
via radiative (= loop and therefore suppressed) 
corrections it is gradually driven smaller and
smaller until it finally drops below zero, triggering the symmetry
breaking. If this mechanism is the one used by Nature, then the
two fundamental scales are connected by an exponential relation
(see e.g.~\cite{drees95}), giving a natural explanation for
their wide separation.

Some recent alternative solutions to the hierarchy problem which
should be noted are the brane world scenarios in which the SM
fields, excepting gravity, are confined to a 4-dimensional
subspace (a ``3-brane'') in a higher-dimensional world. Two
dominant models exist. The first \cite{arkani-hamed98} forces the
Planck scale down by simply introducing large extra dimensions.
The high effective Planck scale $M_P=\sqrt{hc/G}$ is then
interpreted as being due to the spreading of gravity in more
dimensions, causing the $G$ we measure to be lower than the
fundamental gravitational coupling, and so the fundamental Planck
scale could be much smaller, possibly of the order of the
electroweak scale, nullifying the hierarchy problem. Another
important scenario has a single small extra dimension
\cite{randall99} where an exponentially decaying scale factor
(the \emph{warp factor}) in the extradimensional geometry is used
to explain $M_P/M_W$. With two branes (one of which contains the
SM fields) spaced some distance apart in this ``fifth
dimension'', the warp factor would cause an exponentially
different length scale between the two, generating the weak scale
and the Planck scale in a natural way. One could raise the objection
 that what is done in this model is simply to state that 
while $M_P/M_W$ is a very large number, the logarithm of that number is much
smaller, and so more ``natural''. 
This may not be a wrong approach, but in this
model, one still needs to justify why the slightly unnatural 
exponential appears in the extradimensional geometry, and so the hierarchy
problem seems to be only partly solved by this model, in close analogy with
what was the case for
SUSY before radiative symmetry breaking was invented. References to
other extradimensional models for explaining the hierarchy
problem can be found in \cite{kanti00}. It should be noted that
while these scenarios do not as they stand require SUSY,
there is good reason to believe that branes and SUSY do go together.
The brane scenarios do not themselves include a description of
quantum gravity, and thus they must be embedded in a larger, more
fundamental framework. The most plausible of these at present
being superstrings (see below), it does seem that branes are
likely to indirectly require SUSY. For recent (legible) discussions
on embedding branes in superstring scenarios, see
\cite{randall99_2,antoniadis98}. In this work, 
the effects of an eventual higher dimensionality
of the underlying theory are assumed negligible. 

Yet another alternative solution to the hierarchy problem is provided by
the so-called Technicolour models which nullify the problem by simply not
having a Higgs scalar at all (see e.g.~\cite{gunion90}). 
These models are rather ill-favoured at present, and so I omit a presentation
of them here.

\subsubsection{Other Motivations}
In GUT scenarios, one imagines that the gauge groups of the SM 
unify at some high energy scale. Extrapolating the SM itself to these scales
does not result in a unification of the gauge couplings whereas the
introduction of supersymmetry can produce exact coupling unification. 
On the experimental side, SUSY GUTs 
received a great amount of attention when the measurement of the
weak mixing angle, $\theta_W$, at LEP was found to be in
agreement with predictions based on supersymmetric SU(5) GUTs, but  not with
predictions from SM-based GUTs, as shown in table \ref{tab:electroweak}.
\begin{table}[t]
\begin{center}
\begin{tabular}{llrr}
World Average & 0.2312 & $\pm$ 0.0002 & \cite{europhys} \\
SM GUT & 0.206 & $\pm$ 0.01 & \cite{llewellyn81} \\
SUSY GUTs & \multicolumn{2}{c}{0.196 -- 0.262} & \cite{ibanez81} 
\end{tabular}
\caption{\small $\sin^2\theta_W$ -- measured value and predictions from
supersymmetric and SM GUTs. \label{tab:electroweak}}  
\end{center}
\end{table}
Thus, from
a GUT viewpoint, supersymmetry is well motivated. One must keep in mind,
though, that the simple MSSM (with $R$-parity conservation) 
gets in conflict with proton decay when embedded in the most general GUTs. 
This is not the case for SUSY with e.g.\ baryon number conservation, as 
will be discussed in depth in section \ref{sec:assumptions}. Going one
step higher in energy (and one step deeper into speculation), conventional
string theory was found to contain unphysical anomalies which are cancelled 
when the theory is made supersymmetric (hence the ``super'' in superstrings),
and thus also in string theory, supersymmetry finds a possible motivation.

From a purely theoretical viewpoint, SUSY finds one more motivation. The
essence of the Haag-Lopuszanski-Sohnius theorem \cite{haag75} is that the
largest symmetry possible in an interacting field theory is the combination
of some internal gauge symmetries and (possibly local) supersymmetry (where
supersymmetry here includes the usual Lorentz group). The gauge symmetries
can of course be arbitrarily complicated, but the Lorentz group can only be
extended with supersymmetry. It is then possible to make a (very) loose
argument that the theory, in the absence
of evidence to the contrary, should be of the most general type possible and
thus include supersymmetry.

The last indication I shall be concerned with is the light Higgs scenario
which has received an explosion of interest caused by a few events in the LEP
2 data sample. Assuming (a discussion of the justification of this
assumption is beyond the scope of this work) 
that the LEP results of $m_H = 115\GeV$ are confirmed or that some other
not much greater Higgs mass is found, the
effective SM Higgs potential becomes unstable at $\mathcal{O}(10^6)\GeV$,
resulting in an unobserved instability of the electroweak vacuum
\cite{ellis00_2}. When supersymmetry is introduced, a light Higgs boson does
not lead to this instability, and so a light Higgs would be suggestive of the
existence of supersymmetry. This scenario is presently awaiting confirmation
from the Tevatron and the LHC.

\subsection[$R$-breaking Supersymmetry]{\boldmath $R$-breaking Supersymmetry}
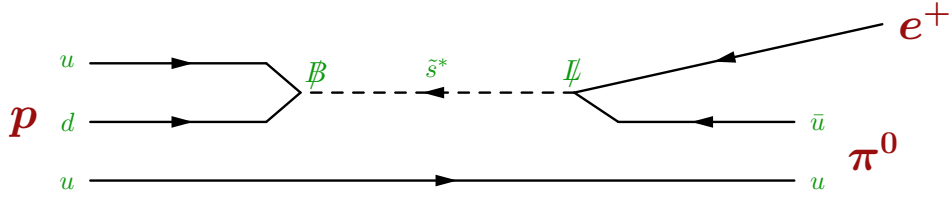
\begin{figure}[t]
\begin{center}
\color{dgreen}
\setlength{\unitlength}{0.65mm}
\begin{fmffile}{protondiags}
\begin{fmfgraph*}(180,40)\fmfset{arrow_len}{3mm}
\fmfforce{0.1w,0.8h}{ll1}
\fmfforce{0.1w,0.5h}{ll2}
\fmfforce{0.1w,0.2h}{ll3}
\fmfforce{0.3w,0.8h}{l1}
\fmfforce{0.3w,0.5h}{l2}
\fmfforce{0.3w,0.2h}{l3}
\fmfforce{0.65w,0.65h}{v2}
\fmfforce{0.7w,0.8h}{r1}
\fmfforce{0.7w,0.5h}{r2}
\fmfforce{0.7w,0.2h}{r3}
\fmfforce{1w,1h}{rr1}
\fmfforce{0.9w,0.5h}{rr2}
\fmfforce{0.9w,0.2h}{rr3}
\fmf{plain}{l1,v}
\fmf{fermion}{ll1,l1}
\fmf{plain}{l2,v}
\fmf{fermion}{ll2,l2}
\fmf{fermion}{ll3,rr3}
\fmf{scalar,tension=0.25,label=\small $\ti{s}^*$}{v2,v}
\fmf{plain,tension=0.55}{r2,v2}
\fmf{fermion}{rr2,r2}
\fmf{fermion,tension=0.2}{rr1,v2}
\fmfv{label=\LARGE\boldmath\textcolor{dred}{$e^+$}}{rr1}
\fmfv{label=\small$\bar{u}$}{rr2}
\fmfv{label=\small $u$}{ll1}
\fmfv{label=\small $d$}{ll2}
\fmfv{label=\small $u$}{ll3}
\fmfv{label=\small $u$}{rr3}
\fmfv{label=\normalsize$\BV$,label.angle=45,label.dist=2}{v}
\fmfv{label=\normalsize$\LV$,label.angle=90,label.dist=2}{v2}
\fmf{phantom,label=\LARGE\boldmath\textcolor{dred}{$p$},label.dist=20}{ll1,ll3}
\fmf{phantom,label=\LARGE\boldmath\textcolor{dred}{$\pi^0$},label.side=left,label.dist=20}{rr2,rr3}
\end{fmfgraph*}
\end{fmffile}
\color{black}
\caption[\small Proton decay via $L$ and $B$ violating SUSY vertices.]{Proton
decay, $p\to \pi^0 e^+$, via one baryon number violating vertex and one
lepton number violating vertex through the propagation of a (heavy)
sparticle. On a more general note, it is not difficult to realize that
something baryonic has to change
into something leptonic for proton decay to be kinematically allowed, and so
$L$ and $B$ \emph{must} both be broken for this to happen.
\label{fig:protondecay}}
\end{center}
\end{figure}
In the above, I have essayed to outline the most important
motivations for supersymmetry. There is, however, a snake in this
apparent paradise which immediately makes its presence known; the
most general SUSY Lagrangian is utterly incompatible with experiment. It
contains renormalizable Lepton and Baryon number violating interactions
\cite{weinberg82} which, being suppressed only
by {\tiny $\frac{1}{\mathrm{(sparticle\ mass\ scale)}^2}$} (from the
propagation of heavy SUSY particles, see fig.~\ref{fig:protondecay}), 
result in a proton lifetime much 
lower than the experimental limit of $\tau_{\mbox{\tiny proton}} \ge
10^{31}$yr \cite{europhys}. The bounds on the $B$ and $L$
violating couplings from this measurement are so strict that at
least one of them must be almost exactly zero 
\cite{dreiner98}. The only natural way
for this to come about is if there were a symmetry in the theory that would 
exactly forbid these reactions.
Thus, in a supersymmetric scenario, there must be some extra
symmetry to protect the proton, but \emph{which} symmetry? A very
sober discussion of this question can be found in
\cite{ibanez92}. Generally, there are three types of discrete
symmetries which draw special attention to themselves; $R$-parity, $B$-parity
(Baryon parity), and $L$ parity. They will be discussed in detail in 
 section \ref{sec:assumptions}. The latter two are slightly
favoured since $B$ and $L$ violating operators of dimension 5 from
supersymmetric GUT- and string-inspired models must be extremely 
suppressed relative to their ``natural'' values in the $R$-parity conserving
models \cite{hinchliffe93}, 
requiring the introduction of additional symmetry, whereas both
$B$ and $L$ parity automatically forbid all relevant baryon or lepton 
number violation, respectively.
However, $R$-parity is considered much more often in the
literature due to its providing a natural dark matter candidate
and due to its comparatively much simpler phenomenology.

In order to outline the difference, the dominant aspects of
$R$-conserving phenomenology are now briefly described. As the
name suggests, the Lagrangian is required to be invariant under a
discrete symmetry known as $R$-parity \cite{farrar78}:
\begin{equation}
\hat{R}\mathcal{L}=\mathcal{L}
\end{equation}
where $\hat{R}$ is an operator acting on products of fields by returning the
product of the $R$-parities of each field (i.e.\ $R$-parity is a
multiplicative quantum number) defined by:
\begin{equation}
R = (-1)^{3B+L+2S}
\end{equation}
Here, $B$ ($L$) is baryon (lepton) number, and $S$ is the spin.
Due to the $\frac12$ spin difference between SM and SUSY
fields, the
SM fields have $R=1$ whilst the SUSY particles have $R=-1$. With
the requirement that the Lagrangian be invariant under $R$, all
combinations of fields which change sign under this operation are
forbidden. This clearly implies that SUSY fields must always come
in pairs (since $-1\times -1=+1$). 
Furthermore, as the $B$ and $L$ violating terms
appearing in the general SUSY Lagrangian are without exception
\emph{odd} under $R$ (see e.g.~\cite{dreiner98}), they are
forbidden by this symmetry, and thus the proton is saved. In
addition to this, a dark matter candidate is gained: the decay
chain of any sparticle ends with an odd number of the so-called
Lightest Supersymmetric Particle (LSP) which is stable (and
neutral, for cosmological reasons). Being heavy, it is an ideal
candidate for the dark matter type known as Weakly
  Interacting Massive Particles (WIMPs). The experimental searches for $R$
conserving scenarios thus rely on pair production and missing energy (from
LSP's escaping the detector) as strong signals. 

The most important differences that should be noted between $R$ conserving
and $R$ breaking phenomenologies have to do with the LSP. In \RV\ the LSP can
decay, often with a lifetime short enough for the decay to proceed inside the
detector, and thus the missing energy signal can be 
greatly diminished. In addition, there is no longer any requirement from cosmology
that the LSP should be 
neutral, and so the list of possible LSP candidates becomes larger. It should
 be mentioned, though, that while $R$-breaking generally results in the loss
of the LSP as a dark matter candidate, viable $R$-breaking models 
exist in which the LSP is a gravitino having a
lifetime longer than the age of the Universe, thus restoring its
validity as a dark matter candidate \cite{takayama00}. This open question
must be resolved by experiment.

\subsubsection{Experimental Indications of $R$-Violation?}
\paragraph{Neutrino Masses:} 
Combining the measurements of the Super-Kamiokande detector in Japan
\cite{superK99} with the very recently published (June 2001) results of the
Sudbury Neutrino Observatory in Canada \cite{sno01}, one now has conclusive 
evidence for
the existence of neutrino oscillations, implying non-zero
neutrino masses. This would mean that lepton family number and/or total lepton
number is not conserved (Lepton family number is broken by Dirac mass
terms while total lepton number is broken by Majorana mass terms. For a
description of the difference, see \cite{diracmajorana}). 
The exact nature of the mass (Dirac or Majorana)
is difficult to determine. The most commonly used technique is to search
for neutrinoless double $\beta$ decay of nuclei, since this process is
forbidden unless there is a Majorana neutrino.
\begin{figure}[h]
\begin{center}
{\small\vspace*{3mm}
\begin{fmffile}{nucbdcy}
\begin{fmfgraph*}(114,76)
\fmfforce{0.78w,0.4h}{wen1}
\fmfleft{d1,l1,l2,d2}
\fmfright{u1,e1,e2,u2}
\fmf{fermion,tension=2,label=$d$,label.side=left}{d2,duw2}
\fmf{boson,label=$W^-$,label.side=right,label.dist=1.4}{duw2,wen2}
\fmf{fermion,label=$u$,label.side=left}{duw2,u2}
\fmf{fermion,label=$e^-$,label.side=left}{wen2,e2}
\fmf{fermion,label=$\bar{\nu}$,label.dist=3.,label.side=left}{wen1,wen2}
\end{fmfgraph*}\hspace*{5mm}\raisebox{1.4cm}{\huge\boldmath$+$}\hspace*{-15mm}
\begin{fmfgraph*}(114,76)
\fmfforce{0.78w,0.6h}{wen2}
\fmfleft{d1,l1,l2,d2}
\fmfright{u1,e1,e2,u2}
\fmf{fermion,tension=2,label=$d$,label.side=right}{d1,duw1}
\fmf{boson,label=$W^-$,label.side=left,label.dist=1.4}{duw1,wen1}
\fmf{fermion,label=$u$,label.side=right}{duw1,u1}
\fmf{fermion,label=$e^-$,label.dist=1.4,label.side=right}{wen1,e1}
\fmf{fermion,label=$\bar{\nu}$,label.dist=3.,label.side=right}{wen2,wen1}
\end{fmfgraph*}\hspace*{10mm}\raisebox{1.4cm}{\huge\boldmath$\stackrel{?}{=}$}\hspace*{10mm}
\begin{fmfgraph*}(114,76)
\fmfleft{d1,l1,l2,d2}
\fmfright{u1,e1,e2,u2}
\fmf{fermion,tension=2,label=$d$,label.side=right}{d1,duw1}
\fmf{fermion,tension=2,label=$d$,label.side=left}{d2,duw2}
\fmf{boson,label=$W^-$,label.side=left,label.dist=1.4}{duw1,wen1}
\fmf{boson,label=$W^-$,label.dist=1.4,label.side=right}{duw2,wen2}
\fmf{fermion,label=$u$,label.side=right}{duw1,u1}
\fmf{fermion,label=$u$,label.side=left}{duw2,u2}
\fmf{fermion,label=$e^-$,label.dist=1.4}{wen1,e1}
\fmf{fermion,label=$e^-$}{wen2,e2}
\fmf{plain,label=$\nu_M$,label.dist=2.}{wen1,wen2}
\end{fmfgraph*}
\end{fmffile}}
\caption[\small Neutrinoless double $\beta$ decay]{
Combining two diagrams for 
ordinary $\beta$ decay \emph{cannot} result in
neutrinoless double $\beta$ decay since the arrow on the neutrino propagator
cannot simultaneously point in both directions \emph{unless} the neutrino is
its own antiparticle (= a Majorana particle). 
Note: the two $d$ quarks in the last
diagram are imagined taken from two separate neutrons, and the ``+'' between
the two first diagrams should not be taken too literally.}
\end{center}
\end{figure}
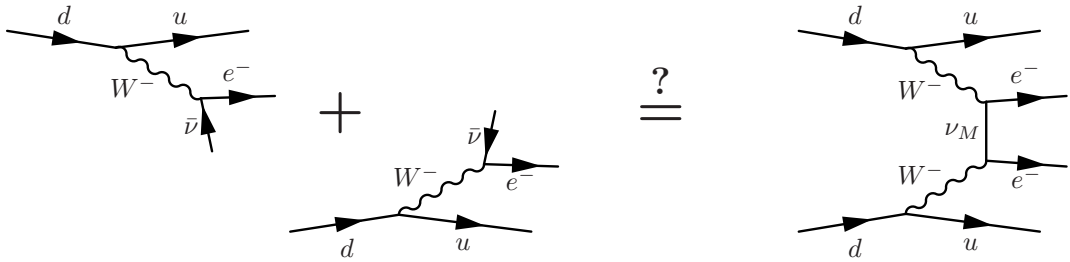
These experiments are not easy, however, and the determination of the nature
of the netrino mass will most likely have to wait for at least another
generation of neutrinoless double $\beta$ decay experiments. 
On the theoretical side,
neutrino mass models are enormously abundant at
present, and there is no outlook to consensus in the near future. In
connection with $R$-parity violation, I merely note that Majorana masses
can arise from one-loop self-energy diagrams containing SUSY particles with
\RV\ couplings. The resulting mass hierarchy is not well constrained, but at
least a neutrino mass pattern consistent with both atmospheric and solar
neutrinos can be generated (see e.g.\ \cite{drees98,abada00}).
There are, however, no strong arguments either for or against this
hypothesis. Naturally small neutrino masses can arise in GUT- and
string-inspired scenarios without breaking $R$-parity (see section
\ref{sec:assumptions}). However, \emph{if} there are Majorana neutrinos
present (i.e.\ total lepton number broken), then Baryon parity 
still seems a good candidate for the proton-protecting symmetry (proton decay
occurs only when $B$ and $L$ are violated \emph{simultaneously}). 

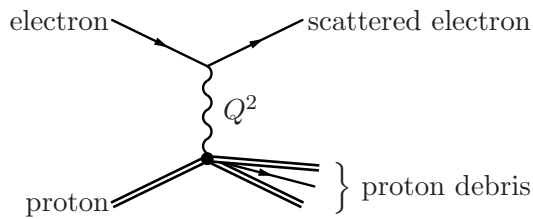
\begin{figure}[h]
\begin{center}\vspace*{2mm}
\begin{fmffile}{qsq}
\begin{fmfgraph*}(90,70)
\fmfset{arrow_len}{2mm}
\fmfleft{p,e}
\fmfright{r2,r1}
\fmfforce{.95w,.10h}{rm1}
\fmfforce{.97w,.2h}{rm2}
\fmfforce{.96w,.095h}{rml}
\fmf{fermion}{e,v1}
\fmf{double}{p,v2}
\fmf{boson,label=$Q^2$}{v2,v1}
\fmf{double}{v2,r2}
\fmf{fermion}{v1,r1}
\fmffreeze
\fmf{fermion}{v2,rm1}
\fmf{double}{v2,rm2}
\fmfdot{v2}
\fmfv{label=$\Big\}$ proton debris,label.ang=0}{rml}
\fmfv{label=proton,label.dist=1,label.ang=180}{p}
\fmfv{label=electron,label.dist=1,label.ang=180}{e}
\fmfv{label=scattered electron,label.dist=2,label.ang=0}{r1}
\end{fmfgraph*}\vspace*{2mm}
\end{fmffile}
\caption[\small Deep Inelastic Scattering at HERA]{Illustration of $Q^2$ at
HERA. As can be seen, this process is similar to ordinary scattering except that
the proton here breaks up, hence the name Deep Inelastic
Scattering. \label{fig:dis}}
\end{center}
\end{figure}

\paragraph{The HERA High $Q^2$ Anomaly:} Part of the experimental programme
for the H1 and ZEUS experiments at the HERA $ep$ collider (which is currently
under upgrading) is to search for so-called leptoquarks, meaning 
particles carrying quark as well as lepton quantum numbers. Although SUSY
does not imply the existence of such particles, \RV-SUSY can produce similar
signatures in the detector due to the violation of lepton or baryon
number. Before describing the anomaly, let us first explain what $Q^2$
means. This is the standard symbol for the momentum transfer squared in
so-called Deep Inelastic Scattering (DIS, see fig.~\ref{fig:dis}). 
At HERA, $Q$ denotes the size of the momentum exchange 
ocurring between the incoming
electron or positron and the proton constituent it interacts with.
The HERA High $Q^2$ Anomaly was an excess of
events with a positron in the final state at high (previously unexplored)
momentum transfers reported by the H1 and ZEUS collaborations at HERA
\cite{adloff97,breitweg97} based on $e^+p$ data taken during 1994-1997. An
unexpected bump ($\approx\!10$ events) 
was seen in the jet-positron invariant mass around 200\GeV, with a small
difference in the peak position between the two experiments.
One interpretation of this excess was to ascribe it to the resonant
production of a $\ti{u}$ squark in the process $e^+ d \to \tilde{u}$ (see e.g.\
\cite{altarelli97_1}) with the $\ti{u}$ decaying to a final state including a
positron, and thus the signal could be interpreted as an
indication of $R$-Violation. While this interpretation is still not completely
excluded, further data did not support the hypothesis, and
related processes which should occur in $e^- p$ collisions
(produced by HERA from 1997 till now) were not seen \cite{perez00}, 
and so the HERA High $Q^2$ Anomaly is at present relegated to one of 
history's many statistical fluctuations. 

\paragraph{The KARMEN Time Anomaly:} 
The KARMEN experiment, designed
to search for electron neutrinos appearing in 
$\nu_\mu\to\nu_e$ and $\bar{\nu}_\mu\to\bar{\nu}_e$ oscillations,
while having found no evidence for oscillations, has found something else --
evidence \cite{oehler00} for an unexpected anomaly in the time-of-flight
distributions of pion decay products.
The experiment works by shooting
protons on a stationary target, giving pions which are quickly stopped in
the target material. The decay $\pi^+\to\mu^+\nu_\mu$ then gives a source of
mono-energetic muon neutrinos, and the subsequent muon
decay $\mu^+\to e^+\nu_e\bar{\nu}_\mu$ provides the muon
antineutrinos, these neutrinos on average being produced a muon lifetime, 2.2
$\mu$s, later than the neutrinos from the pion decay. 

In the time distribution, one thus expects to see an
initial pulse (the $\nu_\mu$'s) at close to zero time after the proton pulse
hits target, due to the short lifetime of the pion,  
followed by a smooth exponential falloff (the $\bar{\nu}\mu$'s and $\nu_e$'s) 
with a time constant
charactaristic of muon decay, 2.2 $\mu$s. The suprise is that there seems to
be sitting a small gaussian on top of this exponential falloff at around 3.6
$\mu$s after the beam hits target. This effect was first observed in 1995 and
was since reconfirmed by a new run with improved shielding to
protect against cosmic rays. It is thus not the case for this anomaly that
its statistical significance has decreased. The first run resulted in $83\pm
28$ events in the gaussian fit, and the second in a total of $103\pm 34$
events \cite{zimmerman00}.

The interpretation of the KARMEN collaboration is that the bump is due to the
process $\pi^+ \to \mu^+ X^0$, where $X^0$ is some unknown particle which must
be neutral since it survives through 7 metres of steel. 
Based on the measured time of
flight, its mass must be 33.91 \MeV, just below the kinematical threshold to
produce it in pion decay at all. 
The likelihood analysis results in less than $1$ to
$10^4$ chance that this is a statistical
fluctuation. Several potential solutions can be excluded (see
\cite{choudhury00} for a brief review), yet two simple possibilities remain:
sterile neutrinos and three-body neutralino decay through
$R$-Violation, specifically the $\lambda'_{211}$ coupling. 
In models like SUGRA in which the neutralino and chargino
masses are related, the latter is already excluded 
from very stringent limits on chargino masses
\cite{choudhury00}, but it \emph{is} allowed in unconstrained supersymmetric
models and is consistent with all experimental data as well as cosmological
nucleosynthesis \cite{choudhury96}. The experimental 
bounds are nicely reviewed in \cite[chp.4]{richardson00} 
where both the limit from
the invisible $Z$ width and limits from precision electroweak measurements
are discussed at a pedestrian level.
Other collaborations have explored and excluded parts of the KARMEN 
signal region, yet ``No experiment currently approved is likely to have
sufficient sensitivity to confirm or rule out the $X^0$''\cite{zimmerman00}.
This is where the case rests.  
\vspace*{1.5ex}

To conlude in as fair a manner as possible, there are in my opinion 
at present no significant experimental indications of either a conserved or a
broken $R$-parity. \vspace*{1.5ex}

Having made these remarks, it is
deemed likely that $R$-breaking supersymmetry will be of
interest to experiments planning to explore SUSY at second
generation accelerators such as the LHC. With the \pythia\ Monte
Carlo generator being one of the most widely used, the motivation
for the present work also becomes clear. 


\clearpage
\section{L-violating Decays of SUSY Particles \label{sec:lspdecays}}
\subsection{The Model \label{sec:assumptions}}
A brief list of the assumptions that underlie the specific scenario of
\RV-SUSY 
adopted in this work is now given, together with attempts to explain and
justify each of them in turn. A considerable emphasis is given to the
theoretical motivations for and against the various proton-protecting
symmetries one can introduce, and several
important yet often understated arguments are presented.
\paragraph{The particle content is that of the MSSM.} In the absence of any
experimental indications to guide us, the choice of minimal particle content
is made strictly for reasons of simplicity.
\paragraph{Lepton number conservation is violated.} 
Since this assumption is
in some sense the crucial and somewhat controversial step taken in this 
thesis, I will present the arguments for it in some detail. Stated briefly, 
the advantage of
abandoning $R$-parity in favour of e.g.\ baryon parity 
is that the MSSM then no longer comes in direct experimental conflict with
proton decay when embedded in GUT scenarios.

As discussed above,
simultaneous violation of both $B$ and $L$ would lead to fast proton decay,
and so \emph{some} additional symmetry must be introduced. 
The choice of taking lepton number violated and baryon number conserved in
the supersymmetric interactions
has been made for the following reason: in theories where a discrete
symmetry (such as $B$, $L$, or $R$) arises from a broken, continuous gauge
symmetry, there have been put forth strong 
indications \cite{ibanez92} that only $B$ and $R$
do not to lead 
to so-called `discrete gauge anomalies' \cite{ibanez91}. 
Here, a brief digression to explain the
concept of an anomalous symmetry is probably in order.

 We know from classical field
theory that every symmetry of the Lagrangian implies the conservation of a
corresponding current, the Noether current.  When this conservation is not
respected by quantum corrections, i.e.\ when the divergence of the Noether
current is non-zero when calculating e.g.\ loop corrections, we speak of an
anomaly.  The fact that $L$-symmetry contains anomalies in some models
simply means that even if
we take the Lagrangian invariant under transformations generated by Lepton
number, quantum corrections will not respect the conservation of this
generator, and thus proton decay clearly \emph{will} take place (this is
still the case when $L$ is a discrete symmetry \cite{ibanez92}).
We can then
conclude that $B$ and $R$ are the strongest candidates \emph{if} the
symmetry that protects the proton is indeed the discrete remnant of some
broken gauge symmetry. That this is most likely the case can be seen when
taking quantum gravity effects into account. 

It has been pointed out \cite{gilbert88} that for any discrete symmetry
which is \emph{not} of gauge origin, the effect of taking wormholes into
account is to produce an
effective action at low energy which does not respect the selection rule
arising from the symmetry.
An intuitive example is that a
group of particles in an $R$-odd state may be engulfed by a baby universe,
causing a measurement in the parent universe to register a deficit of $R$. In
more precise language,
 quantum mechanical tunneling, represented by the formation of a wormhole,
can transport an $R$-odd state into a disconnected space (the baby universe) 
where it will not be measured. On the other hand, a very compact and elegant
argument (see e.g.\ \cite{skands01}) can be used to show
that discrete \emph{gauge} 
symmetries are absolutely stable under
quantum gravity effects \cite{wilczek89},
i.e.\ proton decay \emph{will not} be mediated even by quantum gravity
effects if and only if the symmetry has a gauge origin.
From these arguments, taking the uncertainties
involved in treating quantum gravity\footnote{In \cite{gilbert88}, the author
mentions that his method (summation over wormholes) may or may not be
correct.} and my own lack of superior judgement on
this point into account, I restrict myself to the conclusion that the
discrete symmetry protecting the proton is \emph{most likely} of gauge
origin, and thus $B$ and $R$ parity are the preferred candidates. 

That both $R$-conservation, $B$-conservation, and $L$-conservation
should be considered is indisputable, but I will make the point that
$B$- and $L$-parity seem to be more favoured from a GUT and string theoretical
viewpoint: In quantum field theory, only operators of
mass dimension $\le 4$ are renormalizable, and thus they are the only ones
considered in ordinary applications. 
In theories which go beyond the SM where superheavy particles 
appear at some high mass scale (such as the GUT or string scale), 
non-renormalizable $B$ and $L$ violating operators, i.e.\ operators 
of higher mass dimensions,
appear in the effective low-energy Lagrangian much like in the early
EW theory of four-fermion interactions (see
e.g.~\cite{weinberg82,dimopoulos81}). These terms have forms like 
e.g.~$\psi^2\varphi^2$, where the
fermionic operator, $\psi$, has dimension $(\mbox{mass})^\frac32$ and the
scalar operator $\varphi$ is of mass dimension 1. On dimensional grounds,
noting that the terms in the Lagrangian must be of dimension 4 to give a
dimensionless action, these
operators will be suppressed by $d-4$ powers of the superheavy mass scale
\cite{weinberg79}, and so they can  safely be neglected in 
ordinary applications, yet due to the very strict bounds from proton decay,
$B$ and $L$ violating operators of $d$=5 that appear in supersymmetric
GUTs \cite{murayama98} are not sufficiently suppressed by
even the Planck scale alone, and they will generate nucleon decay at an
`unacceptable rate' \cite{ibanez92,hinchliffe93} 
unless they are suppressed by several (7--8) orders of
magnitude from their ``natural'' values,
as defined by \cite[eq.~(17)]{hinchliffe93}.

Early arguments ran along the line of \cite{dimopoulos81} that
any supersymmetric GUT with $R$-parity will require a natural, but
very accurate adjustment of parameters. In later analyses
\cite{ibanez92,hinchliffe93} it is explicitly argued
that both $B$ and $L$ conservation, \emph{but not} $R$-parity conservation,
can save the proton from higher-dimension $B$- and
$L$-violating terms in Supersymmetric GUTs. It should be mentioned that in
the $B$-parity case (where Lepton number is violated), 
decays such as $\mu\to e \gamma$ would not be expected to
occur at rates above current experimental limits \cite{hinchliffe93}. 

The main conclusion is that 
in any supersymmetric extension of the SM where 
baryon and lepton number violation can occur
at \emph{some} scale below the Planck scale with $d=5$ operators in the
effective Lagrangian that are not additionally suppressed ($d=5$ operators
are very hard to avoid in a supersymmetric scenario due to the many scalars),
conservation of baryon number (and violation of lepton number)
or conservation of lepton number in the supersymmetric interactions 
are possible, 
and $R$-parity is excluded. This means that practically the only chance for
Supersymmetry to provide a natural Dark Matter candidate in such models is if
the gravitino has a lifetime longer than the age of the universe. 

From this discussion, it also becomes clear that $R$-conserving models can
generate naturally small neutrino masses through
higher-dimension operators suppressed by some large mass scale. 
\paragraph{Inter-generational mixing of squarks and sleptons is neglected.} 
The only
mixing matrices that appear for the SUSY particles are the
$\tilde{f}_L-\tilde{f}_R$ mixing matrices discussed above. This choice has
been made due to the observed absence of large Flavour Changing Neutral
Currents (see \ref{sec:mixing}) and 
due to the \spythia\ generator in its present form not being able to handle
this kind of mixing. 
\vfill
\paragraph{It is assumed that all SUSY decays happen on resonance.}
This is a more
technical assumption to do with the decay treatment in \pythia\ (see section
\ref{sec:resdec}). It means that all
decay widths and consequent branching fractions of sparticles are only
evaluated once (at the pole mass of the decaying particle)
during program execution, saving considerable CPU time. This has the drawback
that finer features of the decay distributions are integrated out and care
has to be taken to ensure energy conservation when thresholds are present
near the resonance.
\paragraph{
In the analysis performed in section \ref{sec:analysis}, minimal 
Supergravity
will be assumed} (see section
\ref{sec:susybreaking}), and the evolution equations of \isasusy\ are used to
evolve couplings and masses from the GUT scale to the electroweak. We make
one additional simplifying assumption. The common trilinear coupling, $A_0$,
which  controls the sfermion mixing sector is assumed vanishing. A study of the
effects of non-zero $A_0$ was performed at points 2, 9, and 12 (see table
\ref{tab:sugrapoints} for the definitions of these points). Varying $A_0$
between 0 and 500\GeV\ 
did not result in significant changes of the semi-inclusive
branchings (e.g.\ BR$(\neut\to qq\nu)$) 
for the \RV\ modes, and so the influence of
this parameter on the frequency with which sparticles decay in $L$-conserving
channels versus $\LV$-channels is concluded minimal. Since this is primarily
what determines whether we will be able to see $\LV$-SUSY in our detectors,
the impact of this assumption on the analysis presented in section
\ref{sec:analysis} is likewise concluded minimal. It cannot be guaranteed,
however, that this parameter may not shift the conventional SUSY discovery
signals around between clean and less clean modes. It is clear that such
effects cannot be taken into account by the present analysis.

\subsection{The \boldmath$L$-Violating Lagrangian}
There are three lepton-number
violating terms in the superpotential, one which is bilinear in the
superfields and two which are trilinear. The bilinear term may be
rotated away by a simple change of basis for the fields with no consequences
on the physics, the choice of basis being arbitrary. This is
advantageous for calculating lowest order decay rates such as we are
interested in here. In other applications, such as loop calculations and
renormalization group evolution, other bases are more appropriate (see e.g.\
\cite{carlos96}) to better keep track of the parameters or to obtain
expressions for which experimental bounds are in more direct correspondence. 
In our basis, the superpotential takes on the form
\cite{dreiner00}:  
\begin{equation} 
W_{\LV} = {\textstyle\frac12}\lambda_{ijk} \varepsilon^{ab}
L^i_aL^j_b\bar{e}^k + 
\lambda'_{ijk}\varepsilon^{ab}L^i_aQ^j_b\bar{d}^k \label{eq:LLV}
\end{equation}
where $L$ and $Q$ are the lepton and quark superfields respectively (see
table \ref{tab:mssm}), $i,j,k$
are  generation indices (e.g.\ $e$, $\mu$, or $\tau$ for leptons) 
and $a,b$ are $SU(2)_L$ indices (up-type or down-type, neutrino or charged
lepton)\footnote{Thus, e.g.\ $L^1_1=(\ti{\nu}_e,\nu_e)$, $L^1_2=(\ti{e}_L,e_L)$, 
$Q^2_2=(\ti{s}_L,s_L)$, and $Q^3_1=(\ti{t}_L,t_L)$.}.
From this potential, the following terms in the Lagrangian (corresponding to
the Lepton number violating vertices in the theory) are obtained (see
e.g.~\cite{lykken96,carlos96}):
\begin{eqnarray}
\mathcal{L}_{\LV} & = & {\textstyle\frac12}\lambda_{ijk}\left(
\bar{\nu}_{Li}^c e_{Lj}\tilde{e}_{Rk}^* -
e_{Li}\bar{\nu}_{Lj}^c\tilde{e}_{Rk}^* +\nu_{Li}\tilde{e}_{Lj}\bar{e}_{Rk} -
\tilde{e}_{Li}\nu_{Lj}\bar{e}_{Rk}
+ \tilde{\nu}_{Li}e_{Lj}\bar{e}_{Rk} - e_{Li}\tilde{\nu}_{Lj}\bar{e}_{Rk} 
\right)  \nonumber \\ & & +
\lambda'_{ijk}\left( \bar{\nu}_{Li}^c d_{Lj}\tilde{d}^*_{Rk}-\bar{e}_{Ri}^c
u_{Lj}\tilde{d}_{Rk}^* +\nu_{Li}\tilde{d}_{Lj}\bar{d}_{Rk} -
e_{Li}\tilde{u}_{Lj}\bar{d}_{Rk} + \tilde{\nu}_{Li}d_{Lj}\bar{d}_{Rk} -
\tilde{e}_{Li}u_{Lj}\bar{d}_{Rk} \right) \nonumber\\ & & + \mbox{ hermitian
conjugate (h.c.)}
\label{eq:LVLagrangian}
\end{eqnarray}
All of these terms contain two (SM) fermions and one (SUSY) scalar, and
thus the oddness under $R$ is here directly visible. Also,
it is clear that each scalar can decay in a number of ways to two SM fermions
via these couplings. The neutralino and chargino decays (and for that sake
gluino) must proceed via intermidate scalar particles, resulting in
three-particle decays and interference between several contributing diagrams.

We begin by considering diagrams corresponding to the terms multiplied
by $\lambda_{ijk}$ (the first line in the above equation).  These terms
involve only non-coloured fields, and so they have more 
clean signatures relative to the $\lambda'$ terms (the second line) where
the strong interaction is at play. 
These terms will be treated in section \ref{sec:lambdaprime}.

\subsection{Purely Leptonic Decays}
One property that should be noticed about the $\lambda$ coupling is that it
must be antisymmetric in its first two indices, $i$ and $j$. 
If $\lambda$ is decomposed
into a symmetric and an antisymmetric part, then it is easy to see that the
symmetric part multiplied by $L^i_a L^j_b$ will be symmetric in $a$ and $b$
and thus will give zero when multiplied by $\varepsilon^{ab}$. 
This 
antisymmetry puts some restrictions on the number of possible decay channels.
\subsubsection{Scalar Decays}
The antisymmetry of the $\lambda$ coupling can quickly
be used to rewrite the first line of eq.~(\ref{eq:LVLagrangian}) as:
\begin{eqnarray}
\mathcal{L}_{\LV}^\lambda & = & {\textstyle\frac12}\lambda_{ijk}\left(
\bar{\nu}_{Li}^c e_{Lj}\tilde{e}_{Rk}^* +
e_{Lj}\bar{\nu}_{Li}^c\tilde{e}_{Rk}^* +\nu_{Li}\tilde{e}_{Lj}\bar{e}_{Rk} +
\tilde{e}_{Lj}\nu_{Li}\bar{e}_{Rk} + 
e_{Lj}\tilde{\nu}_{Li}\bar{e}_{Rk} + \tilde{\nu}_{Li}e_{Lj}\bar{e}_{Rk} 
\right) \nonumber \\ 
& & + \mbox{h.c.} 
\end{eqnarray}
The purely leptonic decay modes allowed by the couplings in the above
Lagrangian can then be read off: 
\begin{enumerate}
\item $\ti{e}^-_{j\alpha} \to \bar{\nu}_i\ell_k^-$ (terms 3 and 4)
\item $\ti{e}^-_{k\alpha} \to \nu_i\ell_j^-$ (h.c.\ of terms 1 and 2)
\item $\ti{\nu}_j \to \ell^+_i\ell^-_k$ (terms 5 and 6)
\end{enumerate}
where roman indices are generation indices and $\alpha$ runs over the two
mass states of the sleptons.

\begin{figure}[ht]
\vspace*{4mm}
\input{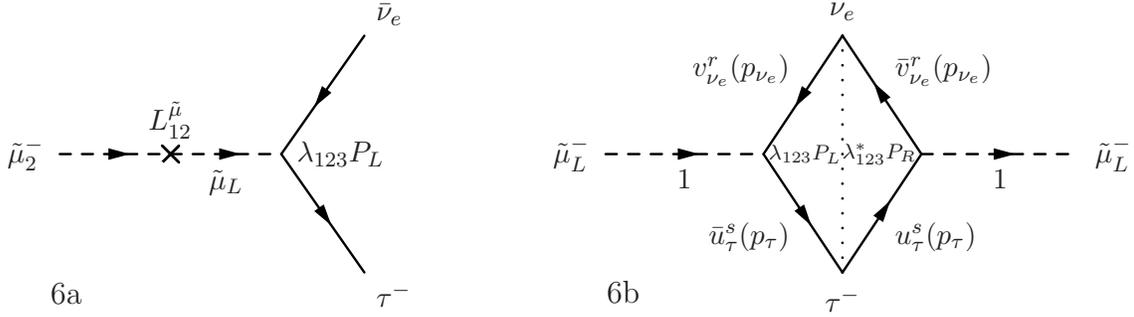}
\vspace*{5mm}
\caption[\small \LV\ decay of a slepton to two leptons]{\small An example: a)
the \LV\ decay of a slepton mass state to two leptons with the mixing shown symbolically as
a cross. b) the folding of the
diagram, for simplicity shown without the mixing
factor. \label{fig:scalardecay}} 
\end{center}
\end{figure}

It is well known from kinematics that the simple
shape of the $1\to 2$ phase space 
allows us to express the width of $\tilde{a}$ decaying to $b$ and $c$ (see
e.g.~\cite{europhys}) as: 
\begin{equation}
\Gamma (\tilde{a}\to bc) = \frac{|\overline{M}(\tilde{a}\to bc)|^2
                           p_{CM}}{8\pi M_{\tilde{a}}^2}
\end{equation}
with the final-state momentum in the decay rest frame
\begin{equation}
p_{CM}^2 =
\frac{1}{4M_{\tilde{a}}^2}\left(M_{\tilde{a}}^2-(m_b+m_c)^2\right)
                          \left(M_{\tilde{a}}^2-(m_b-m_c)^2\right) 
\end{equation}
The leading order matrix elements for these decays are excessively simple to
obtain. An example showing a decay proceeding via the third term in
eq.~(\ref{eq:LVLagrangian}) is depicted in fig.~\ref{fig:scalardecay}a. The
slepton mixing matrix giving the `left-handed' part of the mass state
$\tilde{e}_{j2}$ is denoted by $L^{\tilde{e}_j}_{12}$ (see section
\ref{sec:mixing}). A convenient and intuitive way of writing $|M|^2$
graphically which I would like to advocate is shown in 
fig.~\ref{fig:scalardecay}b with Feynman factors written explicitly for each
particle and vertex (I follow the conventions of \cite{peskin95}, using $u$
for spinors representing particles and $v$ for spinors representing
antiparticles).

\noindent Taking a look at figure \ref{fig:scalardecay}b, we can simply write
down $|M|^2$ for the process $\tilde{e}_{j\alpha}^-\to\bar{\nu}_ie^-_k$:
\begin{equation}
|M(\tilde{e}_{j\alpha}\to \bar{\nu}_ie_k)|^2 =
 |\lambda_{ijk}|^2|L^{\tilde{e}}_{1\alpha}|^2
 \left(\bar{u}_{e_k}^s(p_{e_k})(1-\gamma_5)v_{\nu_i}^r(p_{\nu_i})\right)
 \left(\bar{v}_{\nu_i}^r(p_{\nu_i})(1+\gamma_5)u_{e_k}^s(p_{e_k})\right) 
\end{equation}
where $i,j,k \in (e,\mu,\tau)$ and $\alpha \in (1,2)$.  Summing over the spin
states of the outgoing fermions (and keeping the neutrino mass for
generality) yields:
\begin{eqnarray}
|\overline{M}(\tilde{e}_{j\alpha}\to\bar{\nu}_ie_k)|^2 
& = & |\lambda_{ijk}|^2|L^{\tilde{e}}_{1\alpha}|^2
      \Tr{(\slashp_{\nu_i}-m_{\nu_i}){\textstyle\frac{1-\gamma_5}{2}}
(\slashp_{e_k}+m_{e_k}) 
      {\textstyle\frac{1+\gamma_5}{2}}} \nonumber \\    
& = & {\textstyle\frac12}|\lambda_{ijk}|^2|L^{\tilde{e}}_{1\alpha}|^2 \left(p_{\nu_i}^\mu
      p_{e_k}^\nu\Tr{\gamma_\mu\gamma_\nu}\right)\nonumber\\
& = & 2 |\lambda_{ijk}|^2|L^{\tilde{e}}_{1\alpha}|^2 \left(p_{\nu_i}\cdot
      p_{e_k}\right)
\end{eqnarray}
Noting that $p_e\cdot p_\nu =\frac12 (M_{\tilde{e}}^2-m_e^2-m_\nu^2)$ we get
the final form:
\begin{eqnarray}
|\overline{M}(\tilde{e}_{j\alpha}\to\bar{\nu}_ie_k)|^2
 & = & |\lambda_{ijk}|^2|L^{\tilde{e}}_{1\alpha}|^2 \left(
        M_{\tilde{e}_j}^2-m_{e_k}^2-m_{\nu_i}^2\right)
\end{eqnarray}
Estimating the other matrix elements of this kind proceeds along exactly the
same lines (also for \LV\ decays involving (s)quarks), resulting in the
general form: 
\begin{equation}
|\overline{M}(\tilde{a}\to bc)|^2 = C^a_{bc} \left(
 M_{\tilde{a}}^2-m_{b}^2-m_{c}^2\right)
\end{equation}
where $\tilde{a}$ can be any SUSY scalar, and the coefficients $C^a_{bc}$ are
given in \cite{dreiner00}. This form of the matrix agrees with
the result obtained in \cite{dreiner00}. 
 
\subsubsection{Kinematics and Phase Space Topologies}
\begin{figure}[ht]
\vspace*{4mm}
\begin{center}
\input{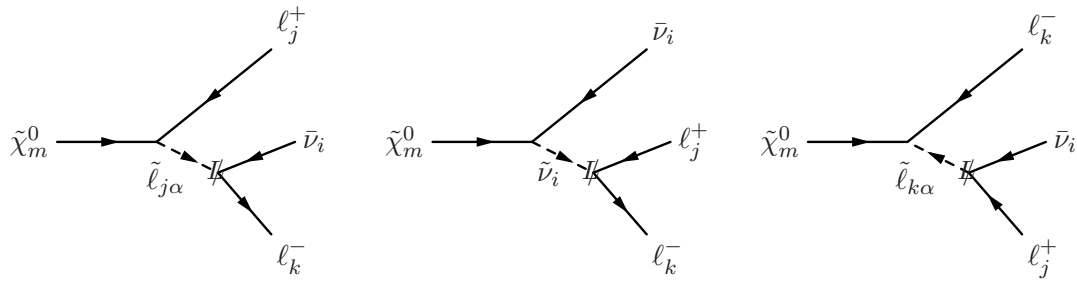}
\vspace*{4mm}
\caption[\small Contributions to $\tilde{\chi}^0_m\to
\bar{\nu}_i\ell^+_j\ell^-_k$]{\small Feynman diagrams contributing to
$\tilde{\chi}^0_m\to \bar{\nu}_i \ell^+_j\ell^-_k$. The index $\alpha$ runs
over the slepton mass eigenstates, $\tilde{\ell}_{1,2}$. An explicit \LV\
symbol adorns the vertices where the lepton number is
violated.\label{fig:chidecay}} 
\end{center}
\end{figure}

As none of the SUSY fermions have \LV\ couplings themselves, the leading
order contributions to the
\LV\ decays of neutralinos, charginos and gluinos proceed via intermediate scalar states 
with (for the purely leptonic processes) 
one $L$-conserving vertex where the SUSY fermion decays to a
slepton-lepton pair and one \LV\ vertex where the slepton decays to two
leptons. An example, showing all contributions to the neutralino decay
$\tilde{\chi}^0_m\to \bar{\nu}_i\ell^+_j\ell_k^-$ is given in figure
\ref{fig:chidecay}.
With $\alpha$ running over the two mass eigenstates of the sleptons, a total
of five interfering diagrams enter the calculation of the squared
amplitude. Combined with the more complicated phase space, the treatment
becomes much more cumbersome than for the scalar decays in the previous
section. The full calculations have been performed in \cite{dreiner00}, but one
must take care when performing numerical integrations over kinematical
variables as is typically the case in event generators.  Therefore,
a brief digression to define and comment kinematical variables is now 
appropriate.   

Any quantum mechanical transition probability depends sensitively on the
(energy-momentum) spectrum of the final states available for the system. In the
continuous limit of free particle final states, this dependence is cast in
the form of the Lorentz Invariant Phase Space, and we must now consider
how to integrate over this space, as we wish to obtain the total widths of
the decays we shall presently consider. In a general three-body
decay ocurring between unpolarized states, there are two kinematical degrees of
freedom. For example, one can specify the
sharing of (four-) momentum between two of the three final states, the third then
being given by momementum conservation. 
A commonly used parametrization is the Dalitz parametrization (see
e.g.~\cite{europhys}) where two parameters with dimensions of
$(\mbox{mass})^2$ are used to parametrize the degrees of freedom:
\begin{equation}
(0\to 1,2,3):\hspace*{2.5cm}m^2_{12} \equiv (p_1+p_2)^2 \hspace*{2cm}
m_{23}^2 \equiv (p_2+p_3)^2
\end{equation}
This is also the parametrization chosen by \cite{dreiner00} where the widths
of the decays we are presently interested in have been calculated. However,
$|\overline{M}|^2$ cannot in general 
be expected to be a smooth function of either of
these variables due to resonant peaks caused by the presence of the intermediate
scalar particles, and so we must understand the possible topologies we can
face in our way across phase space.
\begin{figure}[t]
\vspace*{4mm}
\input{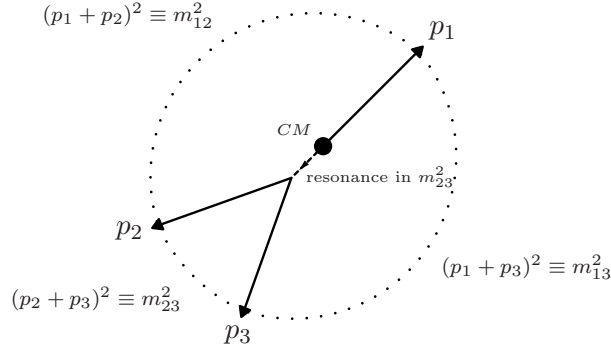}
\caption[\small Kinematics of a three-body decay containing a
resonance]{\small A
three-body decay proceeding through an intermediate state, as seen in the
plane spanned by the momentum vectors of the outgoing particles in the CM of
the decaying particle. \label{fig:threebodydecaykinematics}}
\end{center}
\end{figure}

In the CM of the decaying
particle, the momentum vectors of the decay products must lie in a plane, due
to momentum conservation. Figure
\ref{fig:threebodydecaykinematics} shows a decay proceeding through a
short-lived resonance in this plane, along with specifications of Dalitz
variables. When $m_{23}^2$ approaches the mass of the intermediate particle,
it is almost on its mass shell (i.e.\ it is almost a real particle and not a
quantum fluctuation). This fact results in a great
enhancement of the decay width in the region of phase space where $m_{23}^2
\approx m_{res}^2$, the shape of the enhancement being the familiar
Breit-Wigner profile\footnote{Also known as the Lorentz profile, the Cauchy
curve or even the Witch of Agnesi.}.

\begin{figure}[t!]
\begin{center}
\setlength{\extrarowheight}{7pt}
\begin{tabular}{ccc}
\multicolumn{3}{c}{\large \sc The orientation of resonances in Phase Space}
\\ & & \\ \input{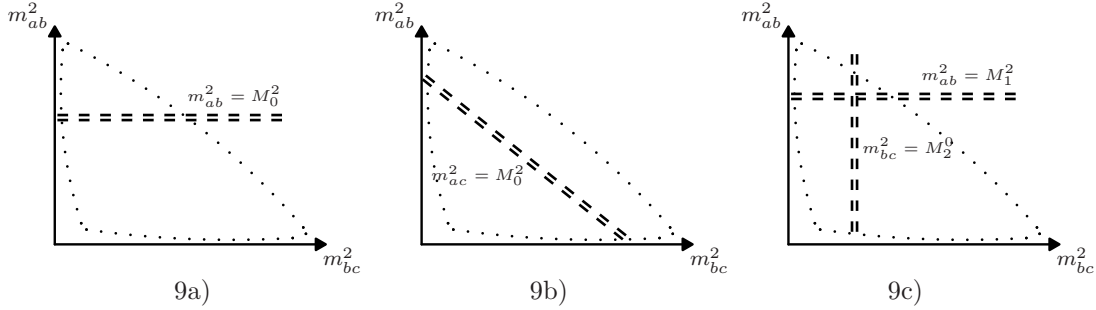}\\
\small \ref{fig:phasespacetopologies}a) 
& \small \ref{fig:phasespacetopologies}b) 
& \small \ref{fig:phasespacetopologies}c) 
\end{tabular}
\caption[\small Phase space topologies]{\small Various three-body phase space
topologies. 
The dotted lines denote the kinematical limits for the integration, and the
dashed lines show the placement of the resonances (it is here assumed that
the resonances lie inside the integration region). The
numerical integration over $m_{bc}^2$ can be performed trivially in the first
case only. In the second, one of the integration variables must be rotated to
$m_{ac}^2$.\label{fig:phasespacetopologies}} 
\end{center}
\end{figure}

In this case, we can use our knowledge of which variable contains the
resonance to deduce that $|\overline{M}|^2$ should be a smooth function of
$m_{12}^2$ whereas $m_{23}^2$ would require a more careful 
(and CPU-time consuming)
treatment. This procedure works perfectly well for decays where only one
diagram contributes, but in the cases considered here, multiple diagrams
contribute with different resonances in different variables and interference
between them, making the game more complicated. If, for example, one of the 
resonances is in $m_{13}^2$ and we are integrating over $m_{12}^2$ and
$m_{23}^2$, then due to the general relation for Dalitz variables
\begin{equation}
M_{0}^2 = m_{12}^2+m_{23}^2+m_{13}^2-m_1^2-m_2^2-m_3^2
\label{eq:dalitzrelation} 
\end{equation}
the resonance in $m_{13}^2$ will show up as a diagonal band in a plot of
$m_{12}^2$ versus $m_{23}^2$, possibly cutting across
the integration interval. Fortunately, by using
eq.~(\ref{eq:dalitzrelation}), 
we can simply change integration variables and then again have a resonance which
appears in only \emph{one} of the integration variables. This is important,
as the resonance can now be dealt with using a one-dimensional technique
rather than a two-dimensional one (see Appendix \ref{app:numint}). This
treats all but a few contributions. The remaining ones are terms where two
different resonances in two different variables interfere with each
other. Rewriting the integral so that the resonant variables become the
integration variables results in a ``resonance cross'', possibly inside
 the integration region. The various 
topologies are illustrated in figure \ref{fig:phasespacetopologies}. 

The standard expression for the width of an unpolarized three-body decay
expressed in Dalitz variables is \cite{europhys}:
\begin{equation}
\Gamma(0\to 1,2,3)
 = \frac{1}{32M_{0}^3}
\int_{m_{12,\mbox{\scriptsize min}}^2}^{m_{12,\mbox{\scriptsize max}}^2}
\hspace*{-2mm}\difd m_{12}^2\int_{m_{23,\mbox{\scriptsize
min}}^2}^{m_{23,\mbox{\scriptsize max}}^2} \difd m_{23}^2 |\overline{M}|^2
\end{equation}
with the kinematical limits\footnote{The corresponding expression in
\cite{dreiner00} for the $m_{23}^2$ limits suffers from a minor typographical error.}
\begin{equation}
\begin{array}{lcl}
m_{12,\mbox{\scriptsize min}}^2 = (m_1+m_2)^2 
& \hspace*{1cm} & 
m_{23,\mbox{\scriptsize min}} = (E_2^*+E_3^*)^2 -
  \left(\sqrt{E_2^{*2}-m_2^2}+\sqrt{E_3^{*2}-m_3^2}\right)^2 \\
m_{12,\mbox{\scriptsize max}}^2 = (M_0-m_3)^2 
& & 
m_{23,\mbox{\scriptsize min}} = (E_2^*+E_3^*)^2 -
  \left(\sqrt{E_2^{*2}-m_2^2}-\sqrt{E_3^{*2}-m_3^2}\right)^2
\end{array}
\end{equation}
This expression has the drawback that the variables to be integrated over and
their order is already specified. We must be more general if we are to
navigate through phase space safely and with a minimum of effort. Therefore,
a brief derivation of all expressions for phase space integrations in
any combination and order of Dalitz variables is given in Appendix
\ref{app:phasespace}, along with details on the numerical integrations
performed in \pythia. 

As we are now fully equipped to select the most advantageous path through
phase space, the discussion of the decay matrix elements can proceed.

\subsubsection{Neutralino Decays}
\label{sec:neutralino}
In models where supersymmetry is broken by supergravity, the Lightest
Supersymmetric Particle (LSP) is most often the lightest neutralino. It goes
without saying then that a major impact of $R$-violating SUSY is that this
particle can now decay into Standard Model particles. Neutralino decays are
highly relevant to accelerator searches for $R$-parity violation, and they
form the most important part of the analysis presented in section
\ref{sec:analysis}. 
 
Considering the couplings in the Lagrangian, there is only one type of purely
leptonic decay mode possible: 
$\ti{\chi}^0_m\to\bar{\nu}_i\ell^+_j\ell^-_k$. 
Due to the Majorana nature of the neutralinos (the neutralinos are their own
antiparticles), the charge
conjugate channels are equally possible, i.e.\
$\ti{\chi}^0_m\to\nu_i\ell^-_j\ell^+_k$. Since there are five
contributing diagrams (see 
fig.~\ref{fig:chidecay}), the amplitude can symbolically be written in the
form: 
\begin{equation}
A(\ti{\chi}^0_m\to \bar{\nu}_i\ell^+_j\ell^-_k) \propto \lambda_{ijk}\left(
\mathcal{A}(\ti{\nu}_i) + \sum_{\alpha=1}^2\mathcal{A}(\ti{\ell}_{j\alpha}) +
\sum_{\alpha=1}^2 \mathcal{A}(\ti{\ell}_{k\alpha}^*)
\right) 
\end{equation}
\begin{eqnarray}
\implies |A(\ti{\chi}^0_m\to \bar{\nu}_i\ell^+_j\ell^-_k)|^2 & \propto &
 |\lambda_{ijk}|^2 \left( |\mathcal{A}(\ti{\nu}_i)|^2 + \sum_{\alpha=1}^2|\mathcal{A}(\ti{\ell}_{j\alpha})|^2 +
\sum_{\alpha=1}^2 |\mathcal{A}(\ti{\ell}_{k\alpha}^*)|^2 +
 \right. \nonumber \\
 & & \left.\hspace*{-4.cm}+ 2\Re\left\{ \sum_{\alpha=1}^2
\mathcal{A}(\ti{\nu}_i)\mathcal{A}^*(\ti{\ell}_{j\alpha}) +
\sum_{\alpha=1}^2\mathcal{A}(\ti{\nu}_i)\mathcal{A}^*(\ti{\ell}_{k\alpha}^*) +
\sum_{\alpha=1}^2\sum_{\beta=1}^2\mathcal{A}^*x(\ti{\ell}_{j\alpha})
\mathcal{A}(\ti{\ell}_{k\beta}^*)    
\right\}\right)\label{eq:chiA2}
\end{eqnarray}
The first line of eq.~(\ref{eq:chiA2}) contains the pure resonance terms
whereas the second line contains interference terms which are expected to be
numerically small relative to the terms in the first line\footnote{This is
due to the important triangle inequalities for complex numbers (see
any standard text on complex analysis).}. 

The matrix elements for the decays considered in this work, including $L-R$
sfermion mixing, have been presented in \cite{dreiner00}, 
and there is little point in presenting
explicit formulae here, except to note that they all follow 
the general layout of eq.~(\ref{eq:chiA2}). 

\subsubsection{Chargino Decays}
The chargino decays are comparatively simpler, due to a generally smaller
amount of contributing diagrams, but on the other hand, there are more
possible channels. For the purely leptonic modes, there are three types of
processes, all of which have been implemented in \pythia:
\begin{enumerate}
\item $\ti{\chi}^+_{m}\to\bar{\nu}_i\ell^+_j\nu_k$
\item $\ti{\chi}^+_{m}\to\nu_i\nu_j\ell^+_k$ 
\item $\ti{\chi}^+_{m}\to\ell^+_i\ell^+_j\ell^-_k$
\end{enumerate}
See Appendix \cite{dreiner00} for detailed formulae. In all cases, the
type of decay is the same as for the neutralinos: an $L$-conserving vertex
where the chargino splits into a lepton-slepton pair (the slepton can be
off mass shell) and an $L$-violating vertex where the slepton decays to two
leptons. 
\subsection{\LV\ Decays Involving (s)quarks.}
\label{sec:lambdaprime}
Several additional channels arise from the terms involving
the $\lambda'$ coupling in eq.~(\ref{eq:LVLagrangian}), but besides involving
different (s)particles, the matrix elements are similar to the 
ones previously discussed. The relevant mixing factors, colour factors,
and complete formulae are given in \cite{dreiner00}.
These channels are:
\begin{enumerate}
\item $\ti{\nu}_i \to\bar{d}_jd_k$
\item $\ti{u}_{j\alpha}\to e^+_id_k$ 
\item $\ti{e}^-_{i\alpha}\to\bar{u}_jd_k$ 
\item $\ti{d}_{k\alpha}\to\nu_id_j$
\item $\ti{d}_{j\alpha}\to\bar{\nu}_id_k$ 
\item $\ti{d}_{k\alpha}\to e_i^-u_j$
\end{enumerate}
where the indices have been chosen so that the coupling in all cases is
$\lambda'_{ijk}$. Roman indices are generation indices and $\alpha$ runs over
mass states.

For the neutralinos and charginos, there are 108 decay modes involving the
$\lambda'$ coupling for each neutralino/chargino, including the conjugate
modes in the neutralino case. These decays are:
\begin{enumerate}
\item $\ti{\chi}^0_m \to \bar{\nu}_i\bar{d}_jd_k$
\item $\ti{\chi}^0_m \to \ell^+_i\bar{u}_jd_k$
\item $\ti{\chi}^+_m \to \bar{\nu}_i\bar{d}_ju_k$
\item $\ti{\chi}^+_m \to \ell^+_i\bar{u}_ju_k$
\item $\ti{\chi}^+_m \to \ell^+_i\bar{d}_jd_k$
\item $\ti{\chi}^+_m \to \nu_iu_j\bar{d}_k$
\end{enumerate}
where $m$ is the neutralino/chargino number and $i,j,k$ are generation
indices. Again, the contributing diagrams are of the same type as before, and
the matrix elements \cite{dreiner00} are almost identical. 

\subsubsection{Gluino Decays}
Since gluinos tend to be heavy in most supersymmetric scenarios, there is
almost no chance that the LSP is a gluino. \RV\ decays of gluinos are
therefore not at present implemented in \pythia. I estimate that this does
not severely 
affect the analysis of the ATLAS \LV-SUSY discovery potential presented in
section \ref{sec:analysis}. The gluino, being heavy, has a large number of
other, unsuppressed, decay channels. On this ground, the total branching into
\LV\ modes can be expected to be small. The impact on the present analysis is
therefore also likely to be small, at least in the part of SUSY parameter
space where the gluino is so heavy that it has a number of unsuppressed
$R$-conserving decay channels at its disposal.

\subsection{The Monte Carlo method and {\sc PYTHIA}}
\vspace*{-4mm}
\textsf{
\begin{quote}
The necessity to make compromises has one major implication: to
write a good event generator is an art, not an exact science.
\emph{\hspace{2.7cm}T.\ Sj\"{o}strand \cite{pythia5.7}}.
\end{quote}}
%
\noindent 
The original work performed in this thesis has noot been to present an
introduction to supersymmetry or to review the motiviations (or lack of same)
for the discrete symmetries thought to protect the proton. Nor has it
been to calculate Feynman diagrams to obtain the decay matrix elements. When
these calculations and arguments for and against have now been presented, it
has been with one specific goal in mind -- to argue for what 
I believe is the necessity of studying \LV-SUSY scenarios,
and to present enough phenomenological
as well as computational background to facilitate this study. 
The original work, then, that has
so far been contained within these pages is the 
F77 source code for the subroutines which
have been added to \pythia\ in the course of this work. 
Since this material is intended for general use, a few comments are necessary
regarding the parameters that an eventual user would have to be familiar
with. These comments can be found in Appendix \ref{sec:pythia}. Here, I
constrain my attention to a brief introduction to MC generators in general.

The name ``Monte Carlo'' is associated with a particular statistical
method which, due to its inherent win-or-loose technique,
 in turn received its name from the famous Monegasque 
gambling house on the French Riviera 
(for a review of the MC technique, see \cite{james80}). Turning immediately
to the practical implementation of the technique in particle physics
applications, it is worth to note that most generators on the market 
today include the following
(here listed in the order they occur in an experiment):
\begin{enumerate}
\item \textsf{Pre-Interaction Era}. 
The colliding particles (i.e.\ the incoming protons for the LHC) 
are described by parton distribution functions, $f_{i/h}(x,\mu_F^2)$, 
giving the probability of finding, inside hadron $h$, a parton of type $i$
carrying momentum fraction $x$ of the hadron momentum. The dependence on
$\mu_F$ is slightly trickier to explain. When thinking of parton
distributions, it is important to bear in mind that a hadron is not a static
thing. Rather, it is a continually
evolving dynamical system with interaction time scales between its
consituents of order the inverse of the hadronic size (1 fm). As seen by a
high-momentum probe, this time scale is very long, and so the hadron
will effectively \emph{seem} static during a hard interaction. This is what
allows us to describe it in terms of parton distributions, i.e.\ 
as an essentially frozen object consisting of free,
individual quarks and gluons. Recalling, however, that each of these is
subject to quantum fluctuations, and that as we increase the wavelength (or,
equivalently, the energy) with which we probe the hadron, we resolve more and
more of these fluctuations, we begin to run into trouble. What, exactly,
should we think of as being part of the hadronic structure, and what should
we think of as short-distance phenomena associated with the hard interaction?
This is exactly what the \emph{factorization scale}, $\mu_F$, defines. 
Partons which are softer (= less energetic) than $\mu_F$ are defined (by us)
to be part of the hadron structure and should be part of the parton
distributions while partons with higher energies belong in the hard
interaction cross section. The reason $\mu_F$ is called the
\emph{factorization} scale is now easy to see. We have effectively factorized
the total cross section for whatever process we are considering into a soft
part and a hard part:
\begin{equation}\hspace*{-8mm}
\begin{array}{lllllll}
\sigma & = &\sum_{i,j} &\int \difd x_1 \difd x_2 & f_{i/1}(x_1,\mu_F^2)
f_{j/2}(x_2,\mu_F^2) & \times & 
\sigma_{\mbox{\tiny hard}}^{ij}(x_1p_1,x_2p_2,Q^2/\mu_F^2) \\
& & \mbox{\tiny sum over} & \mbox{\tiny integration over}  & \mbox{\tiny
probability of finding partons $i$ and $j$}&& \mbox{\tiny cross section for hard}\vspace*{-3mm}\\
& & \mbox{\tiny possible} & \mbox{\tiny momentum fractions}&\mbox{\tiny
inside hadrons 1 and 2, respectively}&&\mbox{\tiny interaction, including}\vspace*{-3mm}\\
& & \mbox{\tiny partons} & \mbox{\tiny of the partons}     &\mbox{\tiny
with momentum fractions $x_1$ and $x_2$,}&&\mbox{\tiny anything going on}
\vspace*{-3mm}\\
& & & & \mbox{\tiny respectively, at scale $\mu_F$}& &\mbox{\tiny above $\mu_F$}
\end{array}
\end{equation}
Note that the factorization scale is not really a physical
parameter. In a perfect world, the cross section would be independent of it.
Usually, it is set equal to $Q^2$, and the same is true for the 
renormalization scale used for $\alpha_s$ in the calcuation. One
therefore most often sees parton distributions expressed as $f(x,Q^2)$.

Also note that parton distributions are inherently non-perturbative quantities
which cannot be calculated from first principles in QCD. 
Rather, some
plausible functional forms are assumed for them with the exact parameters
(coefficients and exponents) being fitted to data. 

What is left to describe is now the hard scattering cross section, $\sigma_{\mbox{\tiny hard}}^{ij}$. This can
easily involve multiple emissions of gluons and/or photons from the incoming
partons, collectively referred to as Initial State Radiation (ISR). Since
matrix elements are usually only available for small numbers of in- and outgoing
particles with complexity rapidly increasing as a function of the number of
particles involved, 
this radiation is described in MC generators by a
succession of $1\to 2$ splittings, thus not taking interference effects
between successive emissions into account (including higher order splittings 
is a tricky issue and still lacks a satisfying solution). An
extremely useful technical trick which is used in all MC generators today is
to evolve this shower backwards in time, starting from the hard interaction
and ending with the proton constituents. 

\item \textsf{Hard Interaction Era}. 
This is in some sense the simplest step in the MC generation, described by
(typically lowest order) matrix elements giving the 
differential cros sections for the interaction of two partons to produce a
given final state. As an example, \pythia\ sports all kinds of conceivable
processes with 1 or 2 in- and outgoing partons. 
At more than 2, only select processes
are implemented, and it is a rare occurence indeed 
to see a hard matrix element with 5 or
more in- or outgoing particles in an MC generator.
\item \textsf{Post-Interaction Perturbative Era / Parton Shower Era}. 
In analogy with Initial State Radiation, the partons coming out from the
hard interaction radiate off gluons and photons through bremsstrahlung. In
the case of gluons being radiated, they themselves will also radiate. This
process, in MC language, is known as the parton shower. These radiations 
decrease the average parton energy while
increasing multiplicity until the parton virtualities reach a cutoff value of
$Q_0 = 1-2\GeV$. Idealistically, perturbation theory applies until the
energies become comparable to the
hadronic scale given by $\Lambda_{QCD}\approx 200 \GeV$, equivalent to the
length scale being about the hadronic size (confinement scale) of
1\fm. However, it would clearly be nonsense to trust first order calculations
all the way to this scale. A reasonable choice, then, is to stop the
perturbative phase when the strong coupling becomes larger than approximately
0.5, exactly at around 1 \GeV.
\item \textsf{Post-Interaction Non-Perturbative Era / Hadronization
Era}. When the partons have become sufficietly soft, hadronization (or
\emph{fragmentation}) takes place. From the border of the perturbative region
and down, we enter the
never-land of not analytically solvable QCD. In one end, we have the perturbative
QCD of quark states that are treated as more or less free, and in the other
 confined quarks inside bound hadronic states. Several phenomenologically
inspired models have been proposed to bridge this gap, in turn giving us an 
understanding of the
physics that takes place, one of the more successful being the Lund ``string
fragmentation'' model used in \pythia. 
In the string picture, coloured objects (basically quarks
and gluons) are colour-connected by strings to each other, the strings being
inspired by narrow flux tubes behaving somewhat like a spring or a rubber
band. As such, a string carries a potential energy of its own, and this
energy can be used e.g.\ to create new quark pairs from the vacuum. In
graphical language, when the string is stretched too hard, it breaks. It is,
of course, the colour connection which breaks, since the newly created quarks
are also coloured. What ultimately happens is that
more or less collinear\footnote{Collinear: going in the same direction.} 
partons with similar momenta join to form
colour singlets, and the non-perturbative strong coupling at this soft scale 
ensures that we can now only talk about bound systems of quarks and gluons 
inside colourless hadrons. 
\item \textsf{Hadronic Era}. Having now fully crossed the quark-hadron
duality boundary, the description changes from the partonic one 
to one involving only hadronic states, in addition of course to any charged
leptons, neutrinos, and photons that may have been produced. 
Unstable particles (mostly hadrons and tauons) decay to lighter
particles, decreasing the average particle mass while increasing multiplicity.
\item \textsf{Visible Era}. Long-lived hadrons (pions and kaons), 
electrons, muons, 
neutrinos, and photons traverse the detector (possibly also the SUSY LSP) and
are recorded by the detector apparatus to the best of its ability, 
this ability depending on where the particles
hit, how energetic they are, and how strongly they interact. 
This step takes place outside the event generator itself, in the detector
simulation. 
\end{enumerate}
In this work, the five first points are handled by a slightly modified
version of \pythia6.1 combined with the more precise supergravity evolution
of \isasusy\ (in ISAJET-7.51). This includes full simulation of the standard
($R$-conserving SUSY + SM) production cross sections (\spythia) as well as shower
descriptions, fragmentation etc. 
The novelty is that all SUSY particles (excepting the
gluino) can now decay to SM particles via Lepton number violating
couplings. Without $R$-Violating production cross sections, the amount of
$R$-Violation going on is necessarily underestimated. Since the major
difference is that single sparticle production could now take place, it is
reasonable to suppose that this 
underestimation is significant only in the very heavy parts of parameter
space. 


\clearpage
\section{The ATLAS Detector and ATLFAST\label{sec:detector}}
\begin{figure}[h!]
\begin{center}
\includegraphics*[scale=3]{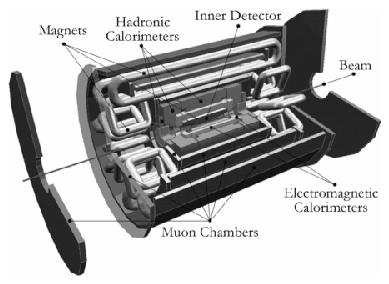}\\
\includegraphics*[scale=0.5]{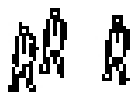}
\end{center}
\end{figure}
\noindent 
Among the most important factors shaping the general design of ATLAS
are the beam particles (protons), the energy (14\TeV), the luminosity
($10^{34}\scm$), and the cost (2.2 billion DKR). The composite nature of the
proton makes it impossible to predict what the CM energy of any given hard
interaction will be. Even though the proton-proton CM energy is known, the
parton-parton CM energy can be anything below that number, and so one 
must
reconstruct it on an event-by-event basis. If many particles escape
detection, 
the CM energy will be poorly determined, and 
so a significant effort goes into making ATLAS as
hermetically sealed as possible. 
The high energies of the collisions determine to some extent 
the physical dimensions of the detector, 
the other limiting factors being cost and
material resistance to the radiation damage caused by the high luminosity. 
Another technical factor is the astounding amount
of data which will be produced by the detector. With approximately 
100 million interactions per second, the amount of data generated 
corresponds to having to handle each and every person on earth talking in a
dozen telephones simultaneously. The so-called ``triggers'' play an essential
role in bringing this data stream down to a manageable level. The triggers
are cuts
which are applied (in several stages, depending on complexity) before events
are written to disk, i.e.\ non-triggered events are irrevocably lost. Thus,
the triggers must stay well below the physics cuts so that potentially
interesting events are kept. 

The main features of
the design and the physics goals of the detector are clearly established. 
ATLAS is first and foremost a Higgs-machine. Second,
it is a \TeV-scale explorer with the emphasis on Supersymmetry. 
For a general overwiev of 
the detector and its design, see the ATLAS Physics TDR\footnote{TDR:
Technical Design Report} \cite{atlastdr}.
I shall here be mainly concerned with the coarse, overall
features of the design such as play a role in the \atlfast\ detector 
simulation: angular coverage, resolution and electron/muon/jet identification.
Where comments are not explicitly made to the contrary, 
it is the \atlfast\ v.2.53 default parameters which have been used in this work.

\subsection{The Beam}
The LHC operates with two proton beams accelerated to 7 \TeV\ each. At these
energies, beam energy loss from synchrotron radiation is completely 
inhibitive for building circular $e^+e^-$ colliders, at least in as ``small'' 
a ring as the LEP/LHC tunnel, and so the choice of protons is unavoidable for
this machine. The design luminosity of the machine is
$10^{34}\scm$. With a total cross section of about
$100\mathrm{mb}$, this corresponds to an event rate of 1GHz, i.e.\ a billion
interactions per second. This ``design'' luminosity was initially foreseen to
be reached after three years of low-luminosity running, starting at 
$10^{33}\scm$. Due to postponements of the
scheduled startup, however, the low-luminosity period has now been replaced
by a shorter period of ``mid-luminosity'' running, starting at
$3\times 10^{33}\mathrm{cm}^{-2}\mathrm{s}^{-1}$, i.e.\ with approximately
300 million interactions per second. 
Presumably, ATLAS will still have collected a total of about 30$\fb^{-1}$ of
integrated luminosity after that period. Following this, the machine will
run at full luminosity, collecting a planned total of 300$\fb^{-1}$ by
2010/2011. These numbers, of course, play a
crucial role in what kind of physics (how low cross sections) the detector
will be sensitive to. Their implications are discussed more closely in
section \ref{sec:trigger} on cross sections and trigger selections.

\subsection{Inner Detector}
\begin{figure}[h!]
\begin{center}
\includegraphics*[scale=2]{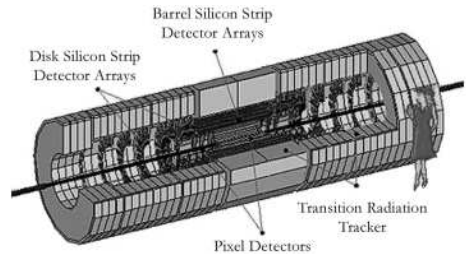}
\caption[\small The ATLAS Inner Detector system]{Schematic drawing showing
the placements of the Inner Detector Components. A female physicist is
shown for size comparison. \label{fig:innerlayout}}
\end{center}
\end{figure}
\noindent The primary functions of the 7.8 metres long and 1.1 metres radius
Inner Detector are track reconstruction for charged
particles and precise reconstructions of both the interaction vertex (where
the hard interaction ocurred) and
secondary vertices (where a particle has decayed) in the event.  
Three distinct subsystems make up the design,
all of them with a high tolerance against the tough radiation levels \cite{atlas_inner}:
Silicon pixel detectors, Silicon strip detectors (called the \emph{SCT:
Semi-Conductor Tracking}), and Transition Radiation Trackers (TRT). The total
cost for these components lies around 400 million DKR, to be compared with
the total cost for ATLAS of 2.2 billion DKR.

Particle tracks down 
to $\approx9.4^\circ$ from the beam pipe can be measured by the Inner
Detector system. Often, one sees \emph{pseudorapidity} used rather than the
angle itself:
\begin{equation}
 \eta\equiv -\ln\tan(\theta/2)
\end{equation}
where $\theta$ is the angle from the beam axis to the radius vector of the
point considered.
For massless particles, this quantity is equal to the rapidity and transforms
additively under Lorentz boosts along the beam direction (by convention, the
$z$ direction): 
\begin{equation}
\mathcal{L}_z(\Delta \eta)\eta = \eta + \Delta \eta
\end{equation}
where $\mathcal{L}_z$ symbolically expresses a Lorentz boost in the $z$
direction. For hadron colliders, where the CM of the interacting
partons is related to the CM of the incoming hadrons by exactly a $z$-boost
(neglecting the small transverse momenta of the partons), a quantity with
such simple transformation properties is
more convenient to work with. In this unit, the beam directions are given by
$\eta = \pm \infty$ and the direction transverse to the beam by $\eta = 0$.
The above statement that particle tracks down
to $9.4^\circ$ are measured translates to $|\eta| < 2.5$.
In reality, only particles
with transverse momenta, $p_T^2 = p_x^2+p_y^2$, above a given threshold
can be reconstructed (\atlfast\ default: $p_T>0.5\GeV$). 
Momenta are measured (SCT and TRT) with an accuracy of approximately
$\sigma(p)/p\approx6\times\tn{-4}p[\GeV]$, and vertex positions (pixels) with
radial resolution $\approx 14\mu\mbox{m}$ and longitudinal resolution $\approx
87\mu\mbox{m}$ \cite{atlas_inner}.   

\subsection{Calorimeters}
Weighing around 4000 tons, there is ample material in the ATLAS calorimeter
system to accomplish its purpose: to completely stop particles 
and measure the energy they give off as they are decelerated. 
This is done using arbsorbing materials in
which the particles loose energy through collisions interlaced
with scintillating media or other signal devices where the particles give off
a signal proportional to their energy, enabling energy loss (d$E$/d$x$) 
and shower profile measurements, highly discriminating variables in particle
identification. 

\subsubsection{EM Calorimeters} 
Electrons and photons are stopped rather more easily than hadrons, 
and so the innermost
calorimeter in any detector is the Electromagnetic Calorimer. In cylindrical
detectors like ATLAS, this is divided into a barrel part and an end-cap part.
In ATLAS, the barrel calorimeter, being 6.8\!\ m long and 4.5\!\ m 
in outer diameter, covers
$|\eta| < 1.475$ i.e.\ angles larger than $26^\circ$ from the beam-pipe while
the end-cap region covers $1.375<|\eta|<3.2$ ($\implies 4.5^\circ < \theta <
28^\circ$) \cite{atlas_calo}. 
Lead and liquid Argon are used as absorber and
scintillator, respectively. The individual cells (numbering in the hundred
thousands)  
are constructed in varying sizes (as measured in cartesian
coordinates) and are mounted facing the center of the detector 
such that their $\eta\times\phi$ areas remain constant
throughout the subdivisions of the detector with the best resolution
($\Delta\eta\times\Delta\phi = 0.025\times 0.025$ corresponding to about 4 by
4 centimetres at $\eta=0$) in the
$|\eta|<2.5$ region where also the Inner Detector is active. This
high-precision part of the calorimeter
can be used to separate e.g.\ pions, electrons, and photons while
the coarser granularity outside means that only measurements of $E_T$ and
reconstruction of jets is undertaken in that region. The rather crude
parametrization of this structure included in \atlfast\ will be described below.

\subsubsection{Hadronic Calorimeters:}
The ATLAS hadronic calorimetry consists of three subdetectors: the Hadronic Tile
Calorimeter in the central region (barrel and extended barrel, $|\eta|<1.7$), 
the End-Cap (Liquid Argon) Calorimter in the intermediate region
($1.5<|\eta|<3.2$), and the Forward (Liquid Argon)
Calorimeter covering pseudorapidity down to $|\eta| < 4.9$ \cite{atlas_calo}. 
\begin{figure}[b!]
\begin{center}
\includegraphics*[scale=1]{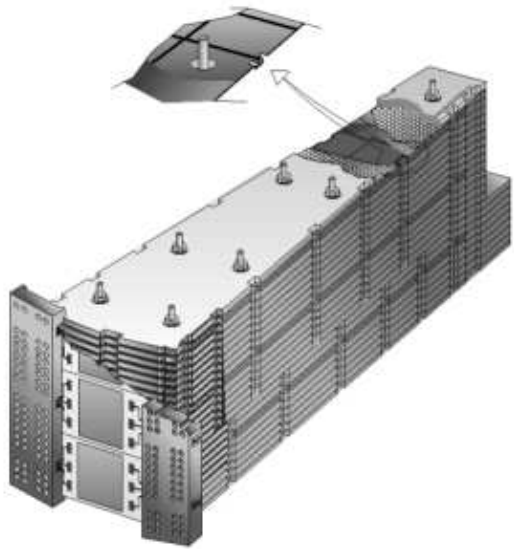}
\caption[\small The ATLAS Hadronic Tile Calorimeter]{Scematic of a module of
the Hadronic Tile Calorimer. The center of the detector lies towards
upper right.\label{fig:tilecal}}
\end{center}
\end{figure}

The Tile Calorimeter has an outer diameter of 8.5\!\ m and is about 13\!\ m
long. A division into a barrel ($|\eta|<1$) and two extended barrel parts
($0.8<|\eta|<1.7$) \cite{atlas_tile} with a small gap between is
necessary simply because the electronics of the inner
parts of ATLAS have to feed out somewhere.
Both use steel as the absorbing material, interspersed with
scintillating plates. The granularity lies at a planned
$\Delta\eta\times\Delta\phi=0.1\times 0.1$, the default for the toy
calorimeter used in \atlfast\ for the central regions of the detector.

Each of the LAr end-caps consists of two copper wheels (2.5 cm and 5cm
thickness, respectively) about 2 metres in radius. Up to $|\eta|=2.5$, the
granularity is $0.1\times0.1$ whereafter it becomes $0.2\times0.2$ up to
$|\eta|=3.1$ \cite{atlas_lar}. Lastly, the most challenging of the
calorimeters are the Forward Calorimeters. Situated a mere 4.7 metres from
the interaction point at high pseudorapidities, the radiation levels here are
considerable. The advantages are equally considerable, however, in terms of
providing uniform coverage up to $|\eta|=4.9$. For technical reasons, the
calorimeter cannot be very long and so must be very dense instead. Three
sections, one of copper and two of tungsten, are included in the design
\cite{atlas_lar}. As a side remark, Liquid Argon detectors aren't cheap. 
All in all, the cost of the Liquid Argon Calorimeters
 is estimated at slightly less than 600
million DKR, roughly a fourth of the total cost for ATLAS, 2.2 billion DKR
 \cite{atlas_tp}.

\subsection{Calorimetry in ATLFAST:}
The complex structure described above would be much too cumbersome to fully
model in a fast simulation. Instead, \atlfast\ uses a toy calorimeter where
the size of the simulated calorimeter cells in $\eta\times\phi$ coordinates is 
 $0.1\times0.1$ in the central region ($|\eta|<3$ \cite{atlfast2.0}) and
$0.2\times 0.2$ outside (down to $|\eta|=5$). Based on the energy
contents of these cells, so-called \emph{clusters} are searched for -- 
groups of cells close together having larger than nominal energy, 
i.e.\ suspected remnants of jets, electrons, or photons. Technically, this is
done using the ``Snowmass Accord'' \cite{stirling96}: 
Begin by defining a cone of a certain radius in $\eta-\phi$ space, 
$\Delta R=\sqrt{\Delta^2\eta +
\Delta^2\phi}$ (\atlfast\ default: $\Delta R=0.4$), sum up the $p_T$ inside it,
and calculate the position of the $E_T$-weighted cluster axis by:
\begin{eqnarray}
E_T^{cluster} & = & \sum_{i\in cone} E_T^i\\
\phi^{cluster} & = & \frac{1}{E_T^{cluster}}\sum_{i\in cone}E_T^i\phi^i \\
\eta^{cluster} & = & \frac{2}{E_T^{cluster}}\sum_{i\in cone}E_T^i\eta^i
\end{eqnarray}
and shift the cone position around 
until the cone and cluster axes line up. Here, $\phi^i,\eta^i$ is the
position of the $i$'th calorimeter cell and we now use $E_T$
instead of $p_T$ since the calorimeters mesaure energies rather than
momenta. This
is a slightly pedantic custom since 
these two quantities are equal for masses that are small compared to the
energy, as is the case for hard leptons and jets. 
The Snowmass definition, however, does not specify how
to deal with overlapping clusters, and so a more refined procedure
is applied in practice. If two clusters overlap,
then merge them into one if there is a lot of $E_T$ in the overlap region,
else split the energy between if there is only little $E_T$. 

In \atlfast, only clusters
with $E_T^{cluster}>10\GeV$ are 
included in the jet/electron/photon reconstruction.  
This reconstruction is fairly accurate with errors around 0.04rad
\cite{atlfast2.0} for the difference between the initiating parton direction and
the center of the reconstructed cluster. The reconstruction of $E_T$ 
is not quite as good, 
due to energy depositions outside the defined cluster cone. Denoting the
difference between the estimated and original parton transverse energies 
by $\Delta E_T = E_T^{est.} - E_T^{par.}$, 
a large tail towards negative
$\Delta E_T/E_T^{par.}$ 
appears with an rms deviation of around 0.2 for the $WH,\ H\to
u\bar{u}$ sample process ($m_H=100\GeV$) 
studied in \cite{atlfast2.0}. Since these jets can be loosely categorized as 
``quark jets from a heavy object'', it is not expected that this error will 
change greatly for the case of hadronic decays of SUSY particles. It is,
however, believed that this error can be corrected for on
averrage when the jet energies are recalibrated. In such procedures, one
estimates the true parton energy from that estimated by the cone algorithm by
\cite{abott01}:
\begin{equation}
E^{par.} = \frac{E^{est.}_{jet}-E_O}{R_{jet} S}
\end{equation}
where $E_O$ is an offset correcting for noise in the detector and energy
depositions not associated with the parton jet itself (i.e.\ the underlying
event and pile-up), $R_{jet}$ is a correction for the calorimeter jet
energy response and energy lost in cracks, and $S$ is the fraction of energy
associated with the jet but not contained within the reconstruction cone.  
See \cite{abott01} for an excellent and more detailed
review of the techniques used in jet energy determinations.

Finally, electrons, photons, and jets are reconstructed from the identified
clusters (muons generally leave so little energy in the calorimeters that the
cluster criteria are impossible to meet). Electrons, muons, and photons which
sit relatively alone in the detector are exceedingly important to measure since they will often be 
the direct decay products of the hard interaction products. A particle
produced in the decay of a fast-moving low-mass particle will generally
tend to follow its mother, whereas a particle produced by a high-mass
particle which is more or less at rest, will generally sit without much
surrounding activity in the detector. Such particles are given the name
\emph{isolated}, and since we shall require all leptons in the analysis to be
isolated, we now present the isolation criteria used in \atlfast\ for
electrons and muons.

Electron candidates from the Inner Detector (i.e.\ with
$|\eta|<2.5$) with $P_T>5\GeV$ are connected to
clusters in the calorimers with a maximum distance between the cluster and
the electron of $\Delta R=0.1$. The isolation criteria used as defaults in
\atlfast\ are a separation from any other clusters by at least $\Delta
R=0.4$ and an energy deposition of maximally $E_T=10\GeV$ in a cone of
$\Delta R=0.2$ around the electron. For the sample process $H\to
e^+e^-e^+e^-$ studied in \cite{atlfast2.0}, the efficiency of these criteria
was 95.3\%, in good agreement with results from full simulation.

For muons, the inner detector is used in
combination with the muon system by default. 
A muon with $|\eta|<2.5$ and $p_T>6\GeV$ is a candidate
for isolation. By default, the same isolation criteria in terms of energy
depositions and cone sizes as for electrons are
applied. For the sample process $H\to\mu^+\mu^-\mu^+\mu^-$ studied in
\cite{atlfast2.0} the efficiency was 97.8\% for \atlfast, significantly
better than what was obtained by full simulation \cite[table 8-1]{atlastdr}:
approximately 85\% (yielding an efficiency for each muon of roughly
$(0.85)^{1/4}=0.96$). 
In this work, an attempt at obtaining more believable numbers has been made by
including by hand estimated electron and muon reconstruction
efficiencies. Using the same electron reconstruction efficiency 
for all energy ranges would be
much too crude. Instead, two numbers, depending on the electron 
$p_T$, are used. The assumed electron and muon reconstruction efficiencies, 
representing educated
guesses based on the ATLAS Physics TDR \cite[table 7-1 and table 8-1]{atlastdr}, are given in table
\ref{tab:receff}. 
\begin{table}[t]
\begin{center}\setlength{\extrarowheight}{0.5pt}
\begin{tabular}{lrr}\toprule
Particle & Eff($p_T<50\GeV$) & Eff($p_T>50\GeV$) \\\cmidrule{1-3}
Electron & 70 \% & 80 \% \\
Muon     & 95 \% & 95 \%\\ \bottomrule
\end{tabular}
\caption[\small Assumed reconstruction efficiencies for muons and
electrons]{Assumed reconstruction efficiencies for muons and electrons.
\label{tab:receff}}
\end{center}
\end{table}

In the end, any clusters which have not been associated with electrons, muons, or
photons, are labelled as reconstructed jets if their $E_T$ is larger than
15\GeV. Furthermore, when the detector is running at high luminosity, some
pile-up is expected in the calorimeters, i.e.\ events begin to overlap,
worsening the energy resolution. For precision physics, i.e.\ measurement of
the Higgs mass and any exclusive measurements, 
the degraded resolution can be a serious
problem. For the study at hand, note that we wish only to determine if there
is and, if so, what type of supersymmetry there is using strictly inclusive
quantities, and so a larger smearing of the energy should not significantly
affect the result (yet this remains to be verified).

\subsection{Muon System}
\begin{figure}[th!]
\begin{center}
\includegraphics*[scale=0.8]{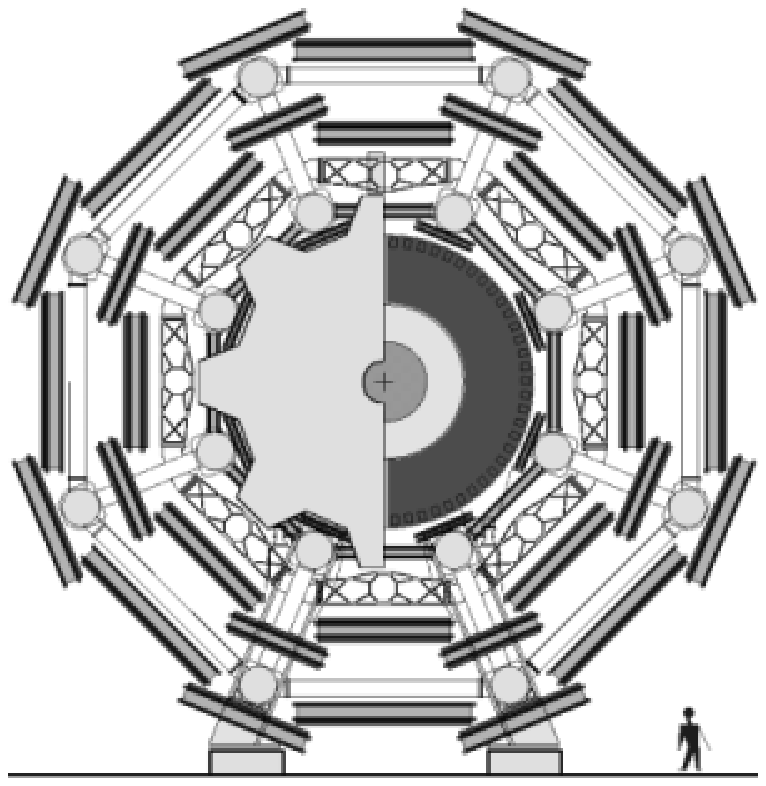}\hspace*{1.5cm}\includegraphics*[scale=2]{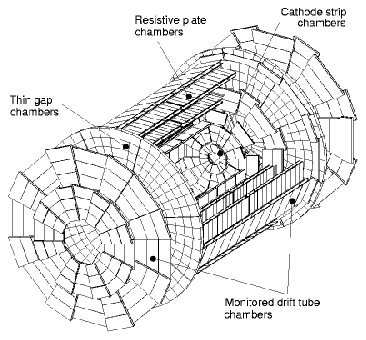}
\caption[\small The ATLAS muon system]{End view and perspective view of the
ATLAS muon chambers. \label{fig:muonchambers}}
\end{center}
\end{figure}

\noindent The ATLAS muon spectrometer 
is composed of three layers of Cathode Strip
Chambers (CSC) close to the interaction point and close to the beam axis and
Monitored Drift Tube (MDT) chambers over the rest of the coverage (up to
$|\eta|=2.7$). The chambers are
arranged symmetrically around the beam axis in the barrel region and
vertically in the end-cap region. Schematic drawings of the arrangement of
the chambers on the sides and on the ends of the detector are shown in figure
\ref{fig:muonchambers}. 
The system is immersed in a magnetic field provided
by superconducting air-core toroid magnets with the resulting bending of the
muon tracks yielding more precise momentum determinations than could be
obtained using the Inner Detector alone. Since one of the
main purposes of the muon spectrometer is to enable a mass measurement for
very narrow Higgs signals, it is quite important that a mass resolution of
about 1\% can be obtained this way. The foreseen resolution, as 
given in the Technical Design Report for the Muon Spectrometer lies at
$\Delta p_T/p_T \approx 2\%$ for muons of $p_T$ between 20 and 100\GeV. For
1000 \GeV muons, this resolution worsens by a factor of roughly 5
\cite{atlas_muon}. 

In addition to the performance at high $p_T$, 
low-$p_T$ muons from $B$ decays require a system containing as
little ``dead'' material as possible. This is among the main reasons that an 
air core design for the magnets was chosen, resulting in
muons down to 3\GeV surviving to be measured by the muon system, the
hadron calorimeter being the largest absorber on the way. 

Also included in the muon spectrometer is a
muon trigger system, extending to $|\eta|<2.4$, increasing the
precision of the muon trigger and the identification of which of the pulsed
beam segments, or ``bunches'', of the LHC (spaced by 25ns \cite{atlastdr}) 
the event belonged to. In addition, the trigger information, though less
precise, can be used to supplement the measurements made in the precision
chambers with an extra coordinate. 

\clearpage
\section{Analysis of ATLAS \LV-SUSY Discovery Potential}
%
\label{sec:analysis}
There is at present no experimental evidence either for
or against supersymmetry in nature, 
although some significant exclusion limits have been acheived at LEP. 
This picture is expected to change drastically over
the next few decades, hopefully already within the next ten years. 
It has been argued above that supersymmetry can only
solve the hierarchy problem if the sparticle masses are not
significantly larger than the \TeV\ scale. This range is within reach of
second generation hadron machines such as the LHC, 
and so we expect to see direct production and decay of
supersymmetric particles relatively soon if supersymmetry exists. 
On the other hand, the
\emph{non-observation} of supersymmetric processes will disfavour
low-energy supersymmetry almost to the point of exclusion, giving us a
powerful clue that we must look to some alternate mechanism for solving the
hierarchy problem. Whatever the case, it is not unreasonable 
to suppose that the
resolution of the hierarchy problem will give \emph{some} observable effects at
the \TeV\ scale, and so we find ourselves almost guaranteed that interesting 
experimental results will be obtained in the not too distant future. 
This section deals with the possible
observable consequences of lepton number violating supersymmetry at the ATLAS
detector, one of four detectors being built for the LHC and scheduled to go
online in 2006. 

The organization of the section is as follows: In \ref{sec:points}, we define
the points in mSUGRA space and the scenarios for the $L$-violating couplings
we will be using in the analysis. Next, in \ref{sec:trigger} we propose
trigger menus dedicated to searches for \LV-SUSY at
mid-luminosity running of the LHC ($L=3\times10\ttn{33}\scm$). At this
luminosity, pile-up is expected, meaning that several events are recorded
simultaneously by the detector, degrading the energy resolution. Since there
are no tools presently available to parametrize this for mid-luminosity
running, we account for this effect in a very crude manner by simply scaling the
simulated event rates by a common factor. 

In \ref{sec:cuts} and \ref{sec:net}, the main part of the analysis is
presented, concentrating on what can be achieved with an amount of data corresponding to an integrated
luminosity of 30\fb$^{-1}$. It is divided into
two parts, one based on cuts and one based on neural networks. The purpose of
the first part is to choose cuts on several kinematical and inclusive
variables which isolate a fairly broad event sample enriched in
supersymmetric events with no emphasis on any particular scenario, except of
course that lepton number is assumed violated. The purpose of the second part
is to process this event sample with neural networks trained to recognize
particular scenarios. Due to the large luminosity at the LHC, it has
not been possible to generate an event sample of comparable magnitude to
30\fb$^{-1}$ of data taking for the highest cross section backgrounds. 
For $Z/W$ production and moderate $p_T$ QCD processes, each
generated event thus corresponds to hundreds of events expected in
data. For these event samples, 
the large rejection factors reached will eventually cause only a very
few or zero events to remain after cuts. At this point, we estimate the event
numbers by 95\% confidence upper limits, fitting to the event distribution
below the cut or by using Poisson statistics on the number remaining after
the cut (see \cite{europhys,feldman98}). 
Rejection factors for these events are in principle unknown but can
be pessimistically estimated using the rejection 
factors for the high $p_T$ QCD sample
and  the double gauge events. 
Lastly, in \ref{sec:results} the results of the cut-based analysis
combined with the nerual network classification are presented.
\subsection{Points of Analysis\label{sec:points}}
\begin{table}[t]
\begin{center}
\setlength{\extrarowheight}{0pt}
{\small
\begin{tabular}{crrrrr}\toprule
 & \hspace*{1cm}$\mathbf{P_2}$ & \hspace*{1cm}$\mathbf{P_7}$ & \hspace*{1cm}$\mathbf{P_9}$ & \hspace*{1cm}$\mathbf{P_{12}}$ & \hspace*{1cm}$\mathbf{F_2}$ \\ \cmidrule{1-6}
\multicolumn{6}{c}{\textbf{GUT Parameters}}\\ 
$\tan\beta$     &   5 &  10 &  20 &  35 &   10 \\
$m_0$           & 170 & 335 & 100 &1000 & 2100 \\
$m_\frac12$     & 780 &1300 & 300 & 700 &  600 \\
sign$(\mu)$     &   + &   + &   + & $-$ &    + \\
$A_0$           &   0 &   0 &   0 &   0 &    0 \\
\cmidrule{1-6}\multicolumn{6}{c}{\textbf{Mass Spectrum}}\\
$h^0$           & 118 & 123 & 115 & 120 & 119 \\
$A^0, H^\pm, H^0$&110 &1663 & 416 & 944 &2125 \\\cmidrule{1-6}
$\neut_1$       & 325 & 554 & 118 & 293 & 239 \\
$\neut_2,\charg_1$&604&1025 & 217 & 543 & 331 \\
$\neut_3$       & 947 &1416 & 399 & 754 & 348 \\
$\neut_4,\charg_2$&960&1425 & 416 & 767 & 502 \\\cmidrule{1-6}
$\ti{g}$        &1706 &2752 & 707 &1592 &1442 \\\cmidrule{1-6}
$\ti{e}_R,\ti{\mu}_R$&336&584&156 &1031 &2108 \\
$\ti{\tau}_1$   & 334 & 574 & 126 & 916 &2090 \\
$\ti{e}_L,\ti{\mu}_L$&546&917&231 &1098 &2126 \\
$\ti{\tau}_2$   & 546 & 915 & 240 &1051 &2118 \\
$\ti{\nu}$      & 541 & 913 & 217 &1095 &2125 \\\cmidrule{1-6}
$\ti{q}_R$      &1453 &2333 & 612 &1612 &2328 \\
$\ti{b}_1$      &1403 &2262 & 566 &1412 &2010 \\
$\ti{t}_1$      &1189 &1948 & 471 &1241 &1592 \\
$\ti{q}_L$      &1514 &2425 & 633 &1663 &2343 \\
$\ti{b}_2$      &1445 &2312 & 615 &1482 &2310 \\
$\ti{t}_2$      &1443 &2286 & 648 &1451 &2018 \\ 
\bottomrule
\end{tabular}}
\caption[\small Selected points in the mSUGRA space]{Selected points of
analysis in the mSUGRA parameter space.\label{tab:sugrapoints}}
\end{center}
\vspace*{-\tfcapsep}\end{table}
\begin{table}[tb]
\center\small \begin{tabular}{cccc}\toprule
& $a$ & $b$ & $n$
\\ \cmidrule{1-4} \multicolumn{4}{c}{\textbf{LLE : Purely Leptonic Lepton Number
        Violation}} 
\\
        $\begin{array}{c}
        \lambda_{ijk} \\ 
        \lambda'_{ijk}
        \end{array}$ &
        $\begin{array}{c}
        10^{-2} \\ 0 
        \end{array}$ &
        $\begin{array}{c}
        10^{-4} \\  0 
        \end{array}$ &
        $\begin{array}{c}
        \sqrt{\hat{m}_{e_i}\hat{m}_{e_j}\hat{m}_{e_k}} \\ 
        0
        \end{array}$
\\ \cmidrule{1-4} \multicolumn{4}{c}{\textbf{LQD : Minimally Leptonic Lepton Number
        Violation}} 
\\
        $\begin{array}{c}
        \lambda_{ijk} \\ 
        \lambda'_{ijk}
        \end{array}$ &
        $\begin{array}{c}
         0 \\ 10^{-2} 
        \end{array}$ &
        $\begin{array}{c}
        0 \\ 10^{-4} 
        \end{array}$ &
        $\begin{array}{c}
        0 \\ 
        \sqrt{\hat{m}_{e_i}\hat{m}_{q_j}\hat{m}_{d_k}}
        \end{array}$
\\ \cmidrule{1-4} \multicolumn{4}{c}{\textbf{LLE + LQD : Mixed Lepton Number
 Violation}}
\\
        $\begin{array}{c}
        \lambda_{ijk} \\ 
        \lambda'_{ijk}
        \end{array}$ &
        $\begin{array}{c}
        10^{-2} \\ 10^{-2} 
        \end{array}$ &
        $\begin{array}{c}
         10^{-4} \\  10^{-4} 
        \end{array}$ &
        $\begin{array}{c}
        \sqrt{\hat{m}_{e_i}\hat{m}_{e_j}\hat{m}_{e_k}} \\ 
        \sqrt{\hat{m}_{e_i}\hat{m}_{q_j}\hat{m}_{d_k}}
        \end{array}$
\\ \bottomrule
\end{tabular}
\caption[\small Selected points in the $\lambda-\lambda'$ space]{Selected
        points of analysis in the $\lambda-\lambda'$ parameter space. The
models with natural couplings (coloumn $n$) 
will be referred to as ``nLLE'', ``nLQD'', and ``nLLE + nLQD'' 
in following sections. If the
couplings get significantly smaller than $10^{-4}$, the LSP lifetime can
become so large that it decays outside the detector, mimicking the $R$-conserving
scenarios. For example, for the mSUGRA point $F_2$, 
setting all $\lambda$ couplings to
$10^{-6}$ and all $\lambda'$ couplings to zero 
results in a decay length for the LSP of $\tau c = 40\mbox{m}$. In the
intermediate range, one may see the effects of neutralino decay
either as a secondary vertex or as a decay inside the fiducial
volume of the detector. 
Naturally, such spectacular signatures 
would greatly lessen the effort required to discover
SUSY, yet we abstain from studying such scenarios here so as not to be overly
optimistic in our results.
\label{tab:lambdapoints}}
\vspace*{-\tfcapsep}\end{table}
For reasons of simplicity, we concentrate on a few points in the mSUGRA
parameter space. These points have not been chosen among the ones 
initially suggested in the ATLAS Physics TDR. This is partly due to the
exclusion of all these points by LEP (essentially from bounds on the Higgs
mass), and partly since it is interesting to enable a direct comparison
between the capabilities of the LHC and other, future experiments such as
CLIC (Compact LInear Collider), a 3\TeV\
electron-positron collider currently under study at CERN. On
these grounds, the analysis is performed on a selection of points defined by
the CLIC physics study group. A general drawback to any selection of points
defined within the MSSM framework is of course that they are constructed to
make the lightest neutralino the LSP. In an $R$-Violating scenario, this
condition does not apply, and so one should bear in mind that the full
parameter space can never be analyzed using just MSSM points, though an
attempt at defining new points for $R$-Violating scenarios would go beyond
the scope of this work. The 5 mSUGRA points shown in table \ref{tab:sugrapoints}
have been selected among 14 points defined by the CLIC
physics study group on the basis that they represent the broadest possible
range within those points. Let it be emphasized that I am not aiming to do
precision physics in this work, but rather to estimate the sensitivity for
\LV-SUSY for various mass hierarchies. It is therefore not interesting
whether we are sensitive to one or another particle or which exact numerical
values the masses have. What is interesting is how sensitive we are for
light/medium/heavy masses for the lightest neutralino in particular and for
various possibilities for the other masses. In this light, there is a great
redundancy in the 14 CLIC points, and thus the 5 points given in table
\ref{tab:sugrapoints} have been chosen for their mutual differences.

In addition, values for the $R$-Violating couplings
must be specified. The experimental upper bounds lie around $10^{-1}-10^{-2}$ 
for most couplings \cite{rvbounds}, 
depending on the masses of the squarks and sleptons, with
heavier masses allowing larger couplings. For the cases of purely leptonic (LLE),
mixed (LLE+LQD), and minimally leptonic (LQD) Lepton number violation, 
three models are
investigated beyond the MSSM without modification. 
Firstly, two points
with common values for all couplings, and secondly a model with
generation-hierarchical couplings defined by \cite{hinchliffe93}: 
\setlength{\extrarowheight}{4pt}
\begin{equation}
\begin{array}{rcl}
|\lambda_{ijk}|^2 & = & \hat{m}_{e_i}\hat{m}_{e_j}\hat{m}_{e_k}\\
|\lambda'_{ijk}|^2 & = & \hat{m}_{e_i} \hat{m}_{q_j} \hat{m}_{d_k}\end{array}
\hspace*{1cm} ; \hspace*{0.7cm} \hat{m}\equiv \frac{m}{v} = \frac{m}{126\GeV} 
\label{eq:natval}
\end{equation}
where 
$m_{q_j}$ represents the mass of ``a quark of generation $j$''. Due to the
mass splittings of the quarks, this definition is inherently ambiguous. A way
to resolve the ambiguity suggested by \cite{hinchliffe_private} is to set $m_j$
equal to the arithmetic mean of $m_{u_j}$ and $m_{d_j}$ where $u$ and $d$ stand
for up-type and down-type respectively:
\begin{equation}
m_{q_j} = {\textstyle\frac12}(m_{u_j}+m_{d_j})
\end{equation} 
This procedure is the one implemented in \pythia. 
The resulting coupling scenarios are given in table 
\ref{tab:lambdapoints}. 

An additional noteworthy remark is that while small
\RV\ couplings have little or no impact on the masses and couplings at the
electroweak scale, large
\RV\ couplings (as compared to the other couplings in the theory) can have
significant effects 
through loops when evolving the Renormalization Group Equations (RGE's)
from the input scale to the \TeV\ scale. This is currently not included in
\pythia. The \RV\
couplings are simply switched on at the low scale and the interplay between
the \RV\ couplings and other physics is neglected. 
Order-of-magnitude-wise, this assumption breaks down
for $\lambda^{(')}>10^{-2}$, and so some effort should be devoted to
including the full RGE's in simulations of \RV-SUSY. Also, 
the CLIC points assume a vanisgin trilinear coupling at the GUT scale, i.e.\ 
$A_0=0$. 
In connection with this work, a small study of the direct consequences of
that assumption upon the results presented here was performed (see
section \ref{sec:assumptions}) with  
the conclusion that the 
semi-inclusive branching ratios (e.g.\ $BR(\neut_1\to qq\nu)$) depend only very weakly on this parameter
($\mathcal{O}(5\%)$), and so the main signatures (number of leptons, number
of jets, etc.) should be only mildly affected by changes to this parameter.

\subsection{Trigger Selection\label{sec:trigger}}
The cross sections for SUSY production at the LHC for the points studied in
this work  range from $10^{-8}$--$10^{-12}$mb. 
For comparison, the cross section for e.g.\ $Z/W$ production is
$10^{-3}$mb, and the total interaction cross section 
(excepting elastic and diffractive
processes which do not give rise to hard jets, leptons, or neutrinos) 
is around 70mb, mostly consisting of small-angle QCD interactions. 
Background rejection is thus extremely essential. In an ideal world, all
data recorded by the experiment would be written to disk for later
scrutiny. With the extreme event rates and sizes (approximately 1
billion events per second and approximately 1MB per event\cite{atlastdr}) 
at the LHC,
such a strategy is entirely out of the question, moreover it is completely
uncalled for. The extreme rates are only
necessary to enable us to \emph{reach} the very small Higgs and New Physics
production cross sections. As already mentioned, the vast majority of events
are QCD processes with small momentum transfers. Relative to how much
statistics we need to study \emph{those} processes, we get a huge
over-abundance. We can then accept to limit the data flow by cutting away a
significant fraction of those events and writing only a very small subset of
them to disk, along with events which contain possible traces of low cross
section physics. To accomplish this reduction, an extensive trigger system is
being developed for ATLAS which 
will identify the (possibly) interesting
events and reject uninteresting ones. For SUSY,
typical signatures that can be triggered on 
include a lot of missing transverse energy (\ET) and
large (hard) jet and lepton multiplicities\footnote{Large
multiplicity = many particles.}.
The usefulness of the ``classical'' SUSY \ET\ signature 
will in general be decreased in \RV\ scenarios due to LSP decay, but
it remains of some discriminating power. On the other hand, the LSP decays to
 jets, leptons, and/or neutrinos, and so the signature is merely transformed,
 not lost. We can use this to define a set of triggers that will reject the
QCD background (which typically has small \ET\ and jet/lepton
multiplicities) but keep the majority of those events that might have been
caused by the production and decay of supersymmetric particles. 

\subsubsection{\LV-SUSY Triggers at Mid-Luminosity}
The acceptable rate of events that can be written to disk is about
100Hz. The sum rate of \emph{all} the triggers which will be implemented in
ATLAS must therefore stay below this number. This includes triggers for Higgs
physics, $B$ physics, $Z/W$ physics, and all kinds of New Physics triggers,
of which \RV-SUSY is only one. A reasonable aim for the trigger rate for the
\RV-SUSY triggers is therefore about 1Hz. Since no detailed trigger rate
studies for mid-luminosity running are yet available, I here present some
crude estimates for the trigger rates for various trigger possibilities.
It should
be understood that there are quite significant uncertainties associated with 
these, as one might well imagine when already uncertain quantities are
extrapolated over orders of magnitudes in their domain. The jet rates,
especially, are not well under control, although interesting 
 work is in progress \cite{webber00,forshaw99,seymour00}.
It has not been possible here to perform a systematic study
of the effects of varying QCD parameters on the trigger rates, 
yet it is useful in the following to keep in mind that the multi-jet rates
can be uncertain by factors of 2 or more \cite{atlfast2.0}.

The only processes which have rates above the 1Hz
domain are QCD $2\to2$ processes and $Z/W$ production, the former with a
cross section around 70mb (but strongly peaked distributions), the latter
with a cross section around $10^{-3}$mb with more broad
distributions
and so more likely to be triggered on (By ``$Z/W$'' production is meant the
sum of $Z$, including photon interference, and $W$ production). 
In this crude study, we shall not focus on
where in the trigger system which triggers are placed (i.e.\ LVL1, LVL2, or
LVL3). This is a technical issue which requires more detailed knowledge of
the detector performance at mid-luminosity than is currently available in
\atlfast. Specifically, running at mid- and high luminosity means that there
can be several interactions recorded simultaneously by the detector (pile-up),
leading to some smearing of event distributions. Since the whole point is
that most events should lie \emph{outside} the trigger domain than inside it, 
more events will be smeared \emph{into} the trigger domain than out of it,
increasing the trigger rates when pile-up is at play. This is
currently only parametrized in \atlfast\ for high luminosity. We here adopt a
crude buest-guess approach, performing the simulation for low luminosity, i.e.\
without pile-up, and 
multiplying the resulting trigger rates by 5 instead of 3 to get from
$L=\tn{33}\scm$ to $L=3\ttn{33}\scm$. This number is based on 
the scaling exhibited
by the trigger rates presented in \cite[chp.11]{atlastdr} from
$L=\tn{33}\scm$ to $L=\tn{34}\scm$ where 
factors between 1 and 5 are found between triggers at equal thresholds 
after dividing out the factor 10 caused by the 
luminosity difference. 
The triggers for which this direct comparison was possible were
the inclusive electron, electron/photon, and \ET+2jets triggers.
Since we are here interested in going from $L=\tn{33}\scm$ to
$L=3\ttn{33}\scm$, a factor of 5/3 was deemed suitable.
These rates should then
to be understood as corresponding to LVL3 trigger rates. Suitable 
intermediate trigger levels will presumably be studied in more detail and
with full detector simulation by the ATLAS trigger/DAQ community in the near
future. 

Returning to the issue at hand, we now propose 11 
trigger menus for SUSY searches in \LV\ scenarios (bearing in mind the many
possible signatures in these scenarios).  
 Granting that the LSP might be the neutralino even in \LV\ scenarios, it is
tempting to define triggers based on the neutralino decay channels, $qq\nu$,
$qq\ell$, and $\ell\ell\nu$. This would work well at all of the SUSY points
analyzed here since all of them have a neutralino LSP, but one should keep in
mind that the lightest neutralino need not be the LSP when $R$ is
violated. Let us therefore go back to the terms in the superpotential that we
are after: $LLE$ and $LQD$, i.e.\ a coupling between three (s)leptons and a
coupling between two (s)quarks and one (s)lepton, where only one of the
particles coupling in each case will be a sparticle. 
From this very general standpoint, 
obvious trigger strategies suggest themselves. 

In the case of the first
coupling (LLE), 
we are searching for a number of leptons (depending on whether one
or two supersymmetric particles were produced and whether they decayed into
two- or three-body channels), possibly accompanied by
\ET. 
We do not here consider an all-neutrino signature since this would be
equivalent to the $R$-conserving SUSY \ET\ signature. 
We require: at least two isolated electrons and/or
muons or one isolated $e$ or $\mu$ accompanied by \ET. 

For the $LQD$ term, we expect jets accompanied by leptons or
\ET. 
The triggers proposed are: at least 3 jets accompanied by either \ET\ or
at least one isolated $\mu$ or $e$, or at least 1 jet, accompanied by
the same with higher thresholds. Each trigger was studied for several
possibilities and combinations of trigger thresholds. The complete 
scans around the selected threshold values are 
shown and commented in Appendix \ref{app:trigger}. 
\begin{table}[t]
\setlength{\extrarowheight}{2pt}
\begin{center}
\begin{tabular}{lcrrr}
\toprule
Trigger & \begin{tabular}{c}Background\\Rate\end{tabular} & 
\begin{tabular}{c}MSSM\\Efficiency\end{tabular} &
\begin{tabular}{c}LLE\\Efficiency\end{tabular}&\begin{tabular}{c}LQD\\Efficiency\end{tabular}\\
\cmidrule{1-5}
mu45I + mu45I   & 0.2 Hz & 1 --- 5 \% & 10 --- 40 \% & 1 --- 10 \% \\ 
e45I  + e45I    & 0.1 Hz & 1 --- 5 \% & 1 --- 35 \% & 1 --- 10 \% \\ 
mu15I + e15I    & 0.1 Hz & 2 --- 5 \% & 20 --- 60 \% & 2 --- 15 \% \\ 
mu40I + me75    & 0.3 Hz & 10 --- 25 \% & 40 --- 75 \% & 10 --- 35 \% \\ 
e40I  + me75    & 0.2 Hz & 10 --- 20 \% & 15 --- 70 \% & 10 --- 35 \% \\ 
j100  + mu40I   & 0.5 Hz & 10 --- 20 \% & 45 --- 70 \% & 10 --- 40 \% \\ 
j100  + e40I    & 0.5 Hz & 5 --- 15 \%  & 15 --- 65 \% & 10 --- 35 \% \\ 
j100  + me175   & 0.3 Hz & 50 --- 80 \% & 35 --- 80 \% & 25 --- 80 \% \\ 
3j50  + mu20I   & 0.1 Hz & 5 --- 15 \% & 45 --- 60 \% & 12 --- 40 \% \\ 
3j50  + e30I    & 0.1 Hz & 5 --- 10 \% & 15 --- 55 \% & 10 --- 35 \% \\ 
3j75  + me125   & 0.1 Hz & 30 --- 65 \% & 30 --- 70 \% & 40 --- 90 \% \\ 
\cmidrule{1-5}
\begin{tabular}{l}Total Rate\\/Combined Efficiency\end{tabular}&   2.1 Hz &
60 --- 90 \% & 90 --- 99.9 \% & 60 --- 96 \%  
\\
\bottomrule
\end{tabular}
\caption[\small Trigger rates and efficiencies for $L=3\times 10^{33}\mathrm{cm}^{-2}\mathrm{s}^{-1}$.]{Estimated trigger rates for background
processes and trigger efficiency ranges for the various MSSM points and \LV-SUSY
  scenarios studied for $L=3\times 10^{33}\mathrm{cm}^{-2}\mathrm{s}^{-1}$
  (summary of plots in appendix \ref{app:trigger}). 
The trigger symbols are described in the text. The total
  trigger rate is slightly smaller than the sum of the individual rates, 
since one event can fulfil several trigger criteria. One sees how much
  cleaner the signatures of the purely leptonic (LLE) coupling are 
  compared to the signatures involving quarks (LQD) where higher thresholds,
  due to the hadronic environment, mean smaller efficiencies.
\label{tab:triggerrates}}
\end{center}
\vspace*{-\tfcapsep}\end{table}
Here, event rates for the selected values of the trigger cuts are given in
table \ref{tab:triggerrates} where also the efficiencies with which these
triggers allow signal events to pass are listed. 
The nomenclature for each trigger item
is such that e.g.\ `mu20I' and `e40I' mean `a muon with $p_T>20\GeV$ which is
isolated', and `an isolated electron with $p_T > 40\GeV$', 
respectively. `3j100' and `me80' mean three reconstructed 
jets of (each) more than 100\GeV\
$p_T$ and more than 80\GeV missing transverse energy, respectively. 
To save space, the signal trigger efficiencies are given as ranges
(min---max), 
obtained by dividing the scenarios studied 
into three categories and scanning for the minimum
and maximum efficiencies within each category. These categories are: the 
MSSM (for reference), purely leptonic \LV\ (i.e.\ decays proceeding
via the $\lambda$ couplings (LLE)), 
and mixed quark-lepton $L$-violation (the $\lambda'$
couplings (LQD)). 
Scenarios where both types of couplings are non-zero generally
lie inbetween the two extremes, depending on the relative coupling
strengths. 
The efficiency ranges reached by combining all the triggers are shown in
the bottom of the table.
\subsubsection{Final Remarks}
Though these trigger proposals are designed explicitly with
\RV-SUSY in mind, they show a certain overlap with triggers proposed for more
conventional physics. The di-muon and 3 jets + lepton triggers, 
for example, have also been
proposed for various Higgs searches. The di-electron
trigger as well as 3 jets + electron are proposed to catch $t\bar{t}$
decays. Finally, the conventional SUSY searches also make use of
multi-lepton, jets + \ET, and multi-jet signatures \cite{bystricky96}. It is
therefore not unthinkable that the triggers here proposed can be incorporated
to some extent into the conventional trigger programme, possibly enabling a
relaxation of the trigger thresholds for the remaining objects.   

Finally, though no study of the mutual redundancy among these trigger objects
has been performed, one sees from table \ref{tab:triggerrates} 
that e.g.\ the di-electron trigger and the 3-jet
+ electron trigger have low efficiencies, this mainly due to the low electron
reconstruction efficiency which is required to be sure one is not mistakenly
believing a jet to be an electron (see \cite[table 7-1]{atlastdr} for
estimated jet rejection factor as function of electron reconstruction
efficiency). As a suggestion for future analyses, it
might be worthwile to investigate what happens if one dispenses 
with these two triggers, replacing them with 
e.g.\ two-jet + leptons or two-jet + \ET\ triggers instead, yet caution in
such undertakings is advisable. The presence of 
electron triggers is, for example,
essential in catching scenarios where the first generation \LV\ couplings are
dominant. This is the reason why as little dependence on lepton flavour 
as possible has been strived for in the menus here proposed.

\subsection{Discriminating Variables \label{sec:cuts}}
We now come to the cut-based analysis. The purpose here 
is to define a set of discriminating variables, i.e.\ variables
capable of distinguishing in a statistical sense 
between the SM and the various SUSY scenarios. Once defined, the idea is to use
these variables and our knowledge of their distributions in SM and non-SM
scenarios to isolate a sample of events with maximal possible enrichment of SUSY
events and minimally contaminated by SM events. At this point, we do not seek
to distinguish between or aim the analysis towards any particular SUSY
scenarios, except that we assume the LSP to decay through violation of lepton
number. 

In the following pages, a number of kinematical and inclusive variables are
presented. The distributions of each variable in the SM and in
the mSUGRA models determine at which values of the variables 
cuts should be placed. In a conventional analysis of
this type, one would seek to maximize the statistical significance with which
a signal can be extracted by adapting the analysis to
maximize quantities like $S/B$, $S/(S+B)$, or $S/\sqrt{S+B}$, where $S$ and
$B$ are the number of signal events and the number of background events,
respectively, remaining after the analysis. Obviously, this requires
knowledge about the shapes of both signal and background distributions 
in each of the discriminating variables. In the present case, we wish to
study \emph{a class} of models rather than individual models, and so no
unique shape can be assigned to the signal we are looking for, 
other than general qualities such as, for example, 
an excess of leptons in the purely leptonic \LV-SUSY scenarios. We are
therefore not in a position to optimize the analysis with respect to exactly
quantifiable estimators. Of course, an analysis could be performed and 
optimized point by point, yet such a strategy would have to be 
carried out on a more general set of mSUGRA points, so as not to
over-tune the analysis to exactly the points considered, risking
to loose sensitivity to points not studied. Moreover, 
more is perhaps learned by generalizing and looking for common discriminators
than spending a large effort studying closely hundreds of scenarios which
may have nothing to do with what the experiment finds. 
This is the real motivation why the last part of the analysis is done using
neural networks. They serve to approximate dedicated search strategies for
specific scenarios.

Let us begin by considering
which backgrounds are most important. Firstly, QCD $2\to2$ processes and $Z/W$
production have the highest cross sections. As before, we mean by $Z/W$ the
sum of $Z$ and $W$ production.  
Secondly, one must expect that the more mass there is in an SM event, the
more  dangerous it is when trying to look for heavy physics.  
$t \bar{t}$ production, $ZZ/ZW/WW$ production, and Higgs
production are examples of heavy SM backgrounds. Again, 
by $ZZ/ZW/WW$ production is
meant the sum of $ZZ$, $ZW$, and $WW$ production.
The Higgs is not expected to be extremely
dangerous, since its mass will presumably be known before LHC SUSY searches
begin, and since the cross sections are low (e.g.\
$\sigma(Z/W+h^0)=2.5\ttn{-9}\mb$ for $m_H=115\GeV$). With respect to the
large lepton and jet multiplicity as well as large \ET, 
one could well ask how significant triple
gauge production could be as background. Unfortunately, these production
cross sections are not yet implemented in \pythia, and so we are forced to
give a rough estimate based on the $\alpha_{EW}^2$ suppression, i.e.\ a
factor $10^{-4}$ relative to $ZZ/ZW/WW$ production. This brings the cross
section down to about $10^{-11}$mb. Special cuts designed to catch pairs of
jets or leptons with $Z$ or $W$ invariant masses and requiring that the
$p_T$ of a jet or a lepton be larger than $\approx$50 \GeV\ (see below) 
would most likely bring this contamination down by at least a factor ten
more. Furthermore, we shall see that even the double gauge events give a 
neglegible contribution to the background at the end of the analysis,
and so I conclude it safe to disregard this background when the SUSY cross
section is larger than $10^{-12}\mb$. For lower cross sections than
this, it would be advisable to conduct a dedicated study of how the triple
gauge background can be dealt with, but in such regions, the total number of
SUSY processes recorded by the ATLAS detector will also be so small that the
LHC is at the limit of its capabilities (see point 7 in the table below), and
 so for such studies to be meaningful, full detector simulation as well as
much more dedicated (i.e.\ specialized) search strategies are necessary.
{\setlength{\extrarowheight}{0pt}
\begin{table}[thb!]
\begin{center}
\begin{tabular}{lrrrrr}\toprule
\bf SM Process &\boldmath$\sigma$ \bf[mb]&\boldmath $N_{\mathrm{gen}}$&\boldmath
$N_{\mathrm{trig}}$ &
\boldmath$N(30\fb^{-1})$&\boldmath$N_{\mathrm{trig}}(30\fb^{-1})$
\\\cmidrule{1-6} 
QCD $2\to 2$\\
\begin{tabular}{l}\scriptsize
$p_T<100\GeV$\end{tabular} 
        & 70 & --- & --- & 2\ttn{15} & $<\tn{5}$\\ 
\begin{tabular}{l}\scriptsize
$p_T=100-150\GeV$\end{tabular} &
        1.42\ttn{-3} & 1.0\ttn{8} & 320 & 4.26\ttn{10} & $(1.4\pm0.1)$\ttn{5}\\
\begin{tabular}{l}\scriptsize
$p_T>150\GeV$\end{tabular} &
        2.88\ttn{-4} & 3.1\ttn{7} & 5.6\ttn{3} & 8.64\ttn{9} & 1.7\ttn{6}\\
$Z/W$      &
        1.19\ttn{-3} & 1.8\ttn{8} & 4.1\ttn{4} & 3.57\ttn{10} & 8.1\ttn{6}\\
$t\bar{t}$ &
        6.08\ttn{-7} & 5.9\ttn{6} & 8.1\ttn{5} & 1.82\ttn{7} & 2.6\ttn{6} \\
$ZZ/ZW/WW$ &
        1.16\ttn{-7} & 5.9\ttn{6} & 1.5\ttn{5} & 3.48\ttn{6} & 8.9\ttn{4} 
\\
\bottomrule
\end{tabular}
\caption[\small Numbers of generated SM events for the analysis]{Cross sections
and event numbers for SM processes. $N_{\mathrm{gen}}$ is the total number of
generated events, and $N_{\mathrm{trig}}$ is the number passing trigger
thresholds. The fourth column simply contains the cross section multiplied by
$30\fb^{-1}$ and the last is the number of generated events passing trigger
thresholds scaled to 30$\fb^{-1}$ of data taking:
$N_{\mathrm{trig}}(30\fb^{-1}) = N_{\mathrm{trig}}\times
N(30\fb^{-1})/N_{\mathrm{gen}}$. For the lowest $p_T$ QCD sample, none of the
generated events passed triggers. The estimate of less than $\tn{5}$ events is
discussed in the text.
For the intermediate $p_T$ QCD sample, the number of generated events passing
triggers was so low that the associated statistical
uncertainty is shown with the scaled number. No correction for
pile-up is included in these numbers.
\label{tab:ananum_sm}}
\end{center}
\vspace*{-\tfcapsep}\end{table}}
\begin{table}[h!]
\setlength{\extrarowheight}{-0.0pt}
\begin{center}{
\begin{tabular}{lrrrrr}\toprule
\bf mSUGRA Point&\boldmath$\sigma$ \bf[mb]&\boldmath $N_{\mathrm{gen}}$&\boldmath
$N_{\mathrm{trig}}$ &
\boldmath$N(30\fb^{-1})$&\boldmath$N_{\mathrm{trig}}(30\fb^{-1})$\\\cmidrule{1-6}
$P_2$   &
        1.3\ttn{-10} & & & 3900 &\\ 
\hspace*{2mm}\scriptsize MSSM
        & & \tn{5} & 8.9\ttn{4} & & 3500\\
\hspace*{2mm}\scriptsize LLE
        & & \tn{5} & 9.9\ttn{4} & & 3900\\
\hspace*{2mm}\scriptsize nLLE
        & & \tn{5} & 9.8\ttn{4} & & 3800\\
\hspace*{2mm}\scriptsize LQD
        & & \tn{5} & 9.4\ttn{4} & & 3700\\
\hspace*{2mm}\scriptsize nLQD
        & & \tn{5} & 9.4\ttn{4} & & 3700\\
$P_7$  & 3.9\ttn{-12} & & & 114 \\
\hspace*{2mm}\scriptsize MSSM
        & & \tn{5} & 8.2\ttn{4} & & 94 \\
\hspace*{2mm}\scriptsize LLE
        & & \tn{5} & 9.9\ttn{4} & & 113\\
\hspace*{2mm}\scriptsize nLLE
        & & \tn{5} & 9.9\ttn{4} & & 113\\
\hspace*{2mm}\scriptsize LQD
        & & \tn{5} & 9.8\ttn{4} & & 111\\
\hspace*{2mm}\scriptsize nLQD
        & & \tn{5} & 9.7\ttn{4} & & 111\\
$P_9$  & 2.4\ttn{-8} &  & & 720000 \\
\hspace*{2mm}\scriptsize MSSM
        & & 8.2\ttn{5} & 6.7\ttn{5} & & 590000\\
\hspace*{2mm}\scriptsize LLE
        & & 8.4\ttn{5} & 8.0\ttn{5} & & 690000\\
\hspace*{2mm}\scriptsize nLLE
        & & 8.2\ttn{5} & 7.3\ttn{5} & & 640000\\
\hspace*{2mm}\scriptsize LQD
        & & 8.4\ttn{5} & 5.9\ttn{5} & & 510000\\
\hspace*{2mm}\scriptsize nLQD
        & & 8.4\ttn{5} & 5.7\ttn{5} & & 490000\\
$P_{12}$ & 1.1\ttn{-10} & & & 3300\\
\hspace*{2mm}\scriptsize MSSM
        & & \tn{5} & 8.8\ttn{4} & & 2900\\
\hspace*{2mm}\scriptsize LLE
        & & \tn{5} & 9.9\ttn{4} & & 3300\\
\hspace*{2mm}\scriptsize nLLE
        & & \tn{5} & 9.7\ttn{4} & & 3200\\
\hspace*{2mm}\scriptsize LQD
        & & \tn{5} & 8.8\ttn{4} & & 2900\\
\hspace*{2mm}\scriptsize nLQD
        & & \tn{5} & 8.6\ttn{4} & & 2800\\
$F_2$  & 1.1\ttn{-10} & & & 3300\\
\hspace*{2mm}\scriptsize MSSM
        & & \tn{5} & 6.1\ttn{4} & & 2000\\
\hspace*{2mm}\scriptsize LLE
        & & \tn{5} & 9.7\ttn{4} & & 3200\\
\hspace*{2mm}\scriptsize nLLE
        & & \tn{5} & 8.8\ttn{4} & & 2900\\
\hspace*{2mm}\scriptsize LQD
        & & \tn{5} & 7.1\ttn{4} & & 2300\\
\hspace*{2mm}\scriptsize nLQD
        & & \tn{5} & 6.0\ttn{4} & & 2000\\
\bottomrule
\end{tabular}}
\caption[\small Numbers of generated SUSY events for the analysis]{Cross sections
and event numbers for mSUGRA. Since single sparticle production is not
included, the choice of a scenario with or without \LV\ 
does not affect the production cross sections. 
The LLE/LQD models
correspond to the models in column $a$ of table \ref{tab:lambdapoints} and
the nLLE/nLQD to column $c$. The models in column $b$ are
generally so similar to the ones in column $a$ that they have been left out
of this table. The same goes for the LLE+LQD models which simply interpolate
between the LLE and LQD scenarios. No correction for
pile-up is included in these numbers.
\label{tab:ananum_susy}}
\end{center}
\vspace*{-\tfcapsep}\end{table}

The numbers of events used in the
analysis are shown in tables \ref{tab:ananum_sm} and \ref{tab:ananum_susy} 
together with cross sections,
the number of generated events passing trigger thresholds, and the number of
events expected after an integrated luminosity of 30\fb$^{-1}$ has been
collected. As can be seen, the number of events expected for $P_7$ is only
around 100. This is most likely too low for ATLAS to see anything, yet the
mass hierarchy in the model is interesting, and so we include it in the
analysis to give an exampel of the performance of this type of model under the
tested cuts.

QCD processes with the $p_T$ of the hard interaction in its rest frame below
100\GeV\ were not possible to include in the analysis, since none of the
events generated for the trigger studies passed the selected 
trigger thresholds. Though the statistical uncertainty on 0 events  
clearly does not make sense to define, 
a conservative estimate on the maximal number of events passing triggers 
can be obtained by noting that a Poisson distribution with mean $\mu=2.99$ 
has less than 5\% probability of resulting in 0 events, and so 
we estimate the event numbers by scaling 3 events out of
each of the generated $p_T<100\GeV$ 
samples to $30\fb^{-1}$ of data taking.
This brings us to conclude that at most 1.9\ttn{7} events in the region
$1\GeV<p_T<10\GeV$, 5.0\ttn{6} events in the region $10\GeV<p_T<75\GeV$,
and at most 4.9\ttn{4} events in the region $75\GeV<p_T<100\GeV$ could pass
trigger thresholds with a 5\% chance that we would not have seen them in
the trigger analysis. Though these numbers are statistically sound, the
estimates for the two lowest $p_T$ samples are likely to be gross
overestimates. That particles 
from the first class of events should be able to gain enough $p_T$ through
parton showering, hadronization, and detector resolution alone
to pass any of the triggers here used borders on the impossible.
With respect to events from the second class ($10-75\GeV$), we note  
that the \ET\ triggers used \emph{begin} 
at 75\GeV\ with the further requirement of hard, isolated leptons and that
the jet triggers require either 3
jets of each 50\GeV\ $p_T$ or 1 jet with $p_T=100\GeV$, 
and so it is also here excessively unlikely to see events passing into the
active trigger range. Towards the high end of the $p_T$ region,
though, and for the third class of events, there are clearly some events in
the far extremes of the distributions
which will pass trigger thresholds. Estimating the upper bound on this 
number to be ten times that estimated for the third class of events alone 
seems the best guess possible at the moment, and so we arrive at an estimated
maximum of \tn{6} $p_T<100\GeV$ QCD events which 
are not included in the analysis. Assuming, pessimistically, that these
events will have the same rejection factors under the cuts applied below as
the $p_T=100-150\GeV$ QCD sample, we will see that an additional 
\tn{6} low-$p_T$ QCD events will not significantly 
affect the conclusions of the analysis. 

No attempt at including the effects of pile-up has been
implemented in the analysis. Generally speaking, 
pile-up results in smearing of the measured calorimeter
energies. Acknowledging that this would shift more events into the active
trigger range than out of it, we made a guess at an overall factor of 5/3 for
the trigger study, i.e.\ the rates obtained for $L=3\ttn{33}\scm$ without
smearing were multiplied by this number to obtain a more realistic
estimate. This was done noting that:
\begin{equation}
R(thres) \propto \int_{thres}^{\infty}\!\!\difd p_T\!\ \phi_{smeared}(p_T) = C(thres)
\int_{thres}^{\infty}\!\!\difd p_T\!\ \phi(p_T)
\end{equation}
i.e.\ the rate passing a certain threshold is proportional to 
the integral from that
threshold to infinity of the smeared distribution of events, 
which can be written as a
(threshold-dependent) constant times the integral of the unsmeared
distribution, or in simpler words: a number can always be written as a
constant times another number. 
In the last section, we assumed $C(thres)=5/3$. In the
analysis, however, we are looking at the distributions themselves rather than
their integrals. Simply pre-weighting each background event by 5/3 rather
than 1, independent of $p_T$, does not give a reliable estimate, since most
of the events smeared into the trigger range will lie just above the $p_T$
thresholds. Also, smearing will cause some signal events to look like they
have lower $p_T$ and some to look like they have higher $p_T$. The net effect
of smearing on the signal ditributions is currently not clear. Therefore,
rather than attempting some best-guess strategy which would in any case end up
rather poor, we do not attempt to include smearing at all in the present
analysis. This will cause the number of background events to be
underestimated, most notably at low $p_T$. This is not deemed a serious
effect since the rejection factors from
the cuts placed are close to 100\% for low $p_T$ events. The cause for worry
lies at higher $p_T$ where smearing will cause 
background and signal events to look more alike, making the purities of the
signals extracted in the analysis too optimistic. As we shall see, however,
we will not be close to the $5\sigma$ discovery border in any of the scenarios,
meaning that the effects of pile-up will not significantly alter the
conclusions reached.  

One further note: Due to limited space, it is of course impossible to show
detailed plots and 
results for every variant of the $\lambda$ couplings for every point in
mSUGRA space here, but it helps to notice that we are trying to discriminate
between particles maximally as heavy as the top and particles which are
typically heavier. The greatest degree of confusion therefore must arise when
the sparticles are relatively light, and so we show detailed results 
only for $P_9$ in the
following subsections. Since the $P_9$ production cross
section is relatively higher than the cross sections for the other points,
this means that the absolute number of events passing cuts is
something of a maximum, yet keep in mind that $P_9$ is typically 
the point where any cut takes away the largest fraction of SUSY events. In
order to still 
give an impression of the spread between the various other SUSY scenarios,
less detailed plots are shown with the full range of models included.

\subsubsection{Missing Transverse Energy}
\begin{figure}[t!]
\begin{center}
\begin{tabular}{cc}
\includegraphics*[scale=0.36]{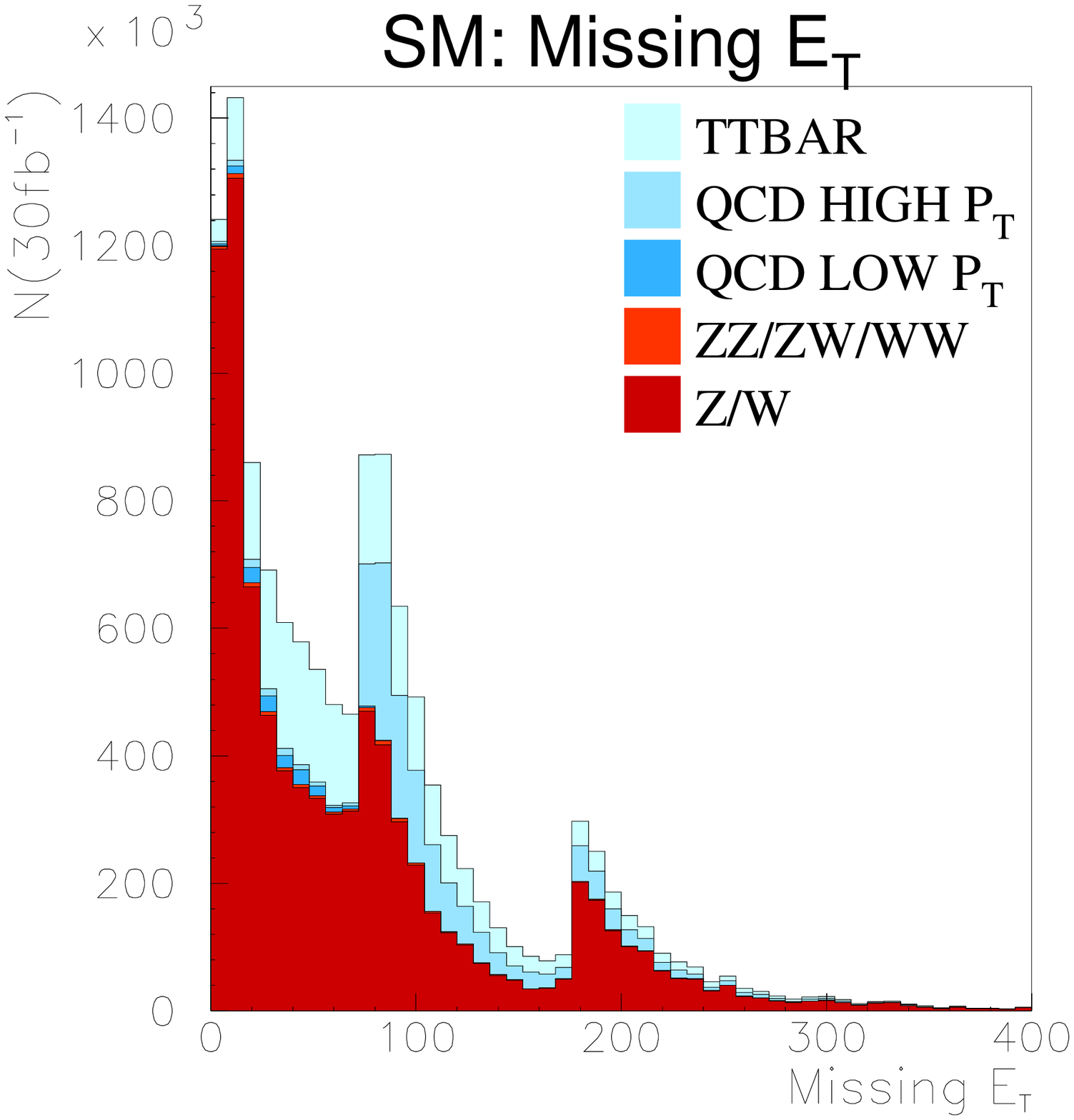}&\hspace*{-.7cm}
\includegraphics*[scale=0.36]{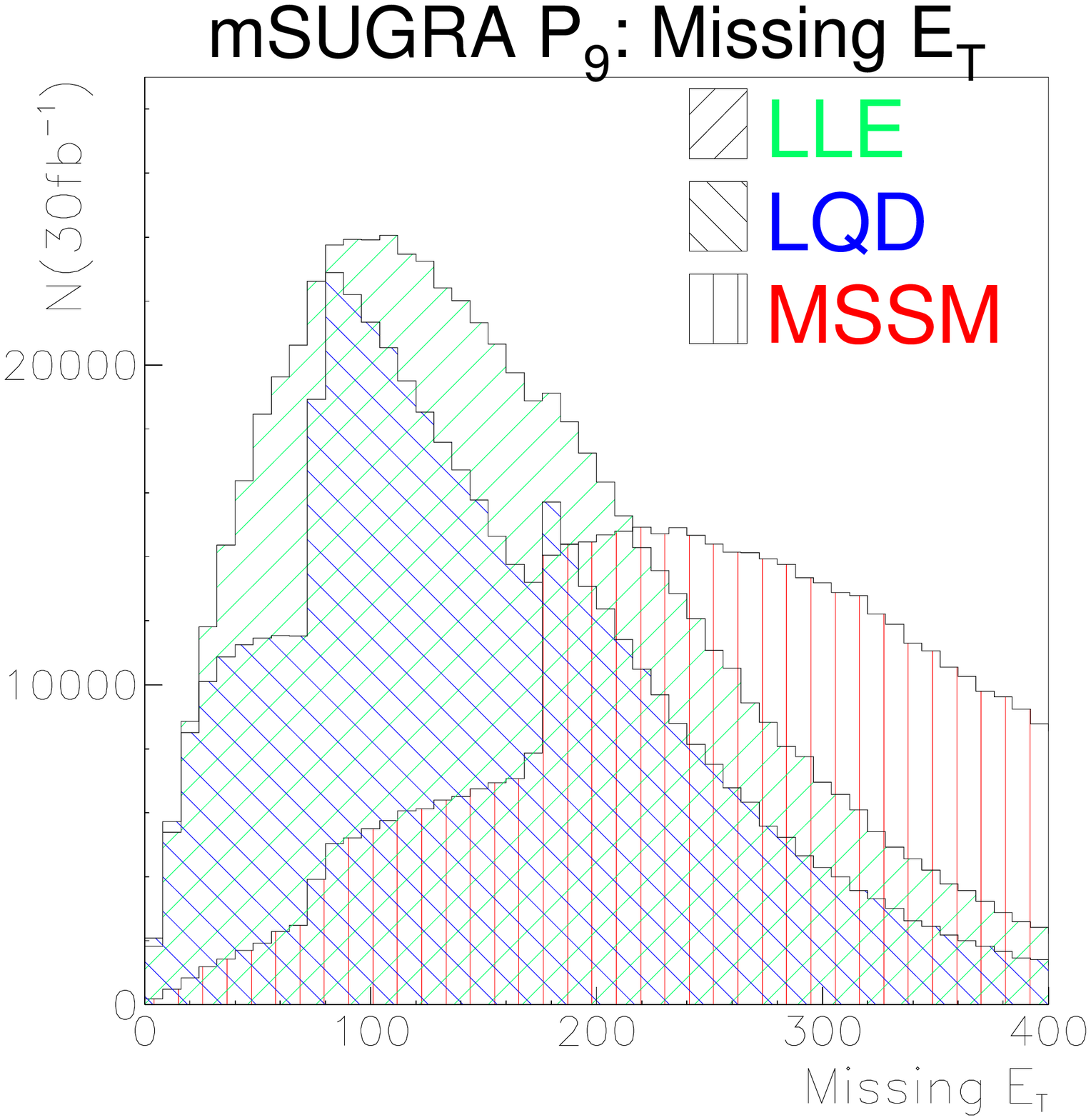}\vspace*{-0.5cm}\\
a) & b) \end{tabular}\vspace*{-8mm}\\
\includegraphics*[scale=0.7]{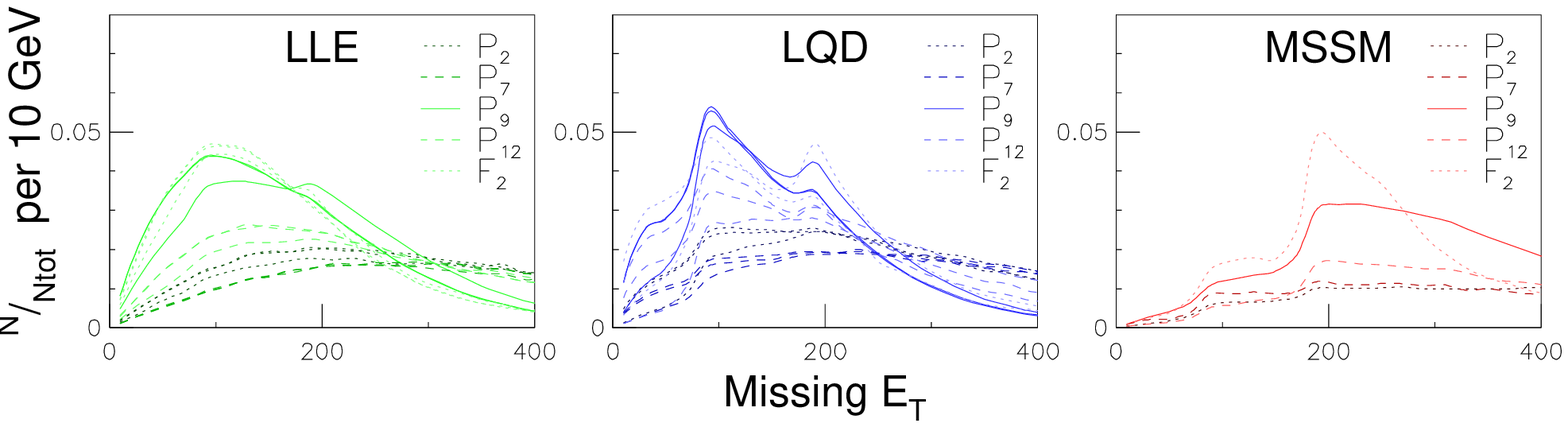}\vspace*{-.7cm}\\
c)\vspace*{-4mm}
\end{center}
\caption[\small Missing $E_T$ for SM and SUSY processes.]{a) and b): \ET\ signatures for
SM and SUSY processes normalized to
30fb$^{-1}$ of data taking. ``QCD LOW $p_T$'' means events from the
$100\GeV<p_T<150\GeV$ sample and ``QCD HIGH $p_T$'' events from the
$p_T>150\GeV$ sample.
Note the large difference in vertical scale
between the two SM plots.  
c): Event
distributions normalized to unit area for LLE, LQD, and the MSSM for all
mSUGRA and \LV\ coupling points studied (commented further in the text).
\label{fig:ET}}
\end{figure}

In $R$-conserving scenarios, LSP's escaping detection give rise to a powerful
signature in \ET. Even when $R$ is violated, escaping neutrinos can give an
enhancement relative to the SM processes. The total background 
distribution after 30\fb$^{-1}$ and its composition
is illustrated for $\ET<400$ in figure
\ref{fig:ET}a. Note that the double gauge events are so few in number that
they are not even visible in the plot. This is a feature which carries
through to the end of the analysis. 
The mSUGRA $P_9$ \ET-distributions for MSSM,
LLE, and LQD scenarios are shown in \ref{fig:ET}b. 
As can clearly be seen, the background with the highest cross sections, 
single gauge production, has an \ET\ distribution which
is sharply peaked at 0 while the $t\bar{t}$ and high $p_T$ QCD proceses 
have more broad distributions. The sharp rises at $\ET=75\GeV$ and $\ET=175$
are due to the me75 and me175 triggers becoming active. 

In figure \ref{fig:ET}c, the full range
of supersymmetric models investigated are plotted for LLE, LQD, and the MSSM,
respectively. As mentioned above, 
these plots are not intended to give detailed information, only
to illustrate the spread between the various scenarios. Smoothed curves 
have been used rather than histograms since it would
otherwise be impossible to disentangle the various models.
In the LLE and LQD scenarios,
three curves are drawn for each mSUGRA scenario corresponding to the three
different \LV\ coupling strength scenarios. Due to the very different
cross sections of the mSUGRA points, the plots have been normalized to equal
areas so that it is the fractions of events per 10\GeV\ which are
plotted. In the LLE scenarios, the sharp rises at 75 and 175 GeV just
mentioned are absent, since these scenarios do not rely as heavily on the
\ET\ triggers as the two others.

Cuts on $\ET>50\GeV$, $\ET>100\GeV$, $\ET>150$, and $\ET>200\GeV$ were 
investigated with results shown in table \ref{tab:ET}. The models are: MSSM, LLE
(all $\lambda$ couplings at \tn{-2}), LQD (all $\lambda'$ couplings at
\tn{-2}), nLLE (natural $\lambda$ couplings, defined by
eq.~(\ref{eq:natval})), and nLQD (natural $\lambda'$ couplings). Note that
the number of trigged SM events given in table \ref{tab:ET}, 
13 million, is not equal to 
$10^7\mbox{s}\times 2.1\mbox{Hz}$ since we are not including the factor of
5/3 from pile-up here. Note also that the estimated number of SM events
\emph{do not} include $p_T<100\GeV$ events. 
As an explicatory note to what one sees in this table, 
it is not surprising that the MSSM does best
under these cuts since escaping neutralinos give extra \ET. 
What really is
noteworthy is that the $R$-Violating models, in spite of neutralino decays,
do so well. With regard to what happens in the other mSUGRA models, 
only $F_2$ has lower efficiencies. $F_2$ is a special
case in the sense that it represents a class of models where 
one really has only a very few species of particle with
low mass available, the rest having very high masses so that the \ET\ spectrum
becomes peaked at low values despite the large GUT parameters. 
\begin{table}[tb!]
\setlength{\extrarowheight}{0pt}
\begin{center}
\textsf{EVENTS PASSING CUTS ON \ET.\vspace*{2mm}}\\
{\footnotesize
\begin{tabular}{lcccccc}\toprule
& SM & $P_9$ MSSM & $P_9$ LLE & $P_9$ nLLE & $P_9$ LQD & $P_9$ nLQD\\ 
\cmidrule{1-7}
$N_{trig}(30\fb^{-1})$&$(13\pm0.1)$ M&590 k&685 k&640 k&500 k&490 k\\\cmidrule{1-7}
$\ET>50\GeV$    & $(7.3\pm0.1)$ M &580 k&450 k&620 k&600 k&465 k\\ \boldmath
$\ET>100\GeV$   & $(3.4\pm0.1)$ M &560 k&350 k&480 k&510 k&380 k\\ 
$\ET>150\GeV$   & $(1.9\pm0.05)$ M&520 k&240 k&340 k&390 k&270 k\\
$\ET>200\GeV$   & $(910\pm30)$ k  &450 k&280 k&220 k&280 k&175 k\\
\cmidrule{1-7} $\frac{N_{\mbox{\tiny post}}}{N_{\mbox{\tiny pre}}}$ 
& 0.26 & 0.95 & 0.70 & 0.70 & 0.80 & 0.79 \\
\cmidrule{1-7}
\end{tabular}}\\ 
\end{center}
\vspace*{-5mm}
\caption[\small Event numbers passing cuts on $\ET$.]{Event numbers
 passing cuts on
$\ET$ for several cut values with associated statistical errors, 
normalized to 30$\fb^{-1}$ of data taking. ``M'' and ``k'' are the standard
abbreviations for \ttn{6} and \ttn{3}, respectively.
Due to the
comparatively large event samples generated for SUSY, the statistical
uncertainties on the SUSY numbers are below the percent level,
and so no uncertainties are shown. 
The selected cut of 100\GeV\ is marked in bold, and 
the ratio of events surviving after this cut to events generated 
is shown for each model.\label{tab:ET}} 
\vspace*{-\tfcapsep}\end{table}

Taking a look at the first row of table \ref{tab:ET}, one notices that a
5$\sigma$ discovery is immediately possible for all the $P_9$ points using just
 the event numbers passing triggers, before any attempt at purifying the
sample is made. By ``a 5$\sigma$ discovery'', we mean exactly the following:
\begin{equation}
\frac{S}{\sqrt{S+B}} > 5  
\label{eq:5sigma}
\end{equation}
where $S$ and $B$ are the number of signal and background events expected,
respectively. In section
\ref{sec:results} we discuss this estimate more closely and how
the uncertainties from the limited numbers of generated events can be taken
into account using 95\% confidence limits. Furthermore, we will
 seek to include, 
albeit in a very crude manner, the effects of QCD uncertainties and pile-up
on the discovery potential as well. Setting this aside for the moment and 
doing the arithmetic
yields that with 500.000 signal events, we 
could get a $5\sigma$ discovery even with background levels a few hundred
times higher. However, since we are not guaranteed to be in quite this
fortuitous situation in the real world, it is worthwhile to pursue the
analysis further. 

\begin{figure}[t!]
\begin{center}
\includegraphics*[scale=0.25]{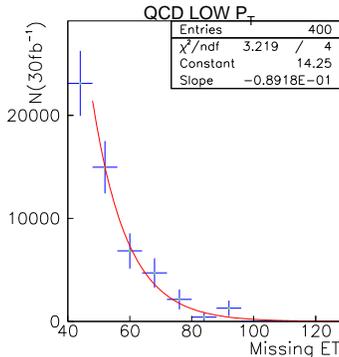} 
\vspace*{-5mm}
\caption[\small Fit to the \ET\ tail of low $p_T$ QCD events.]{Exponential 
fit to the \ET\ tail of the low $p_T$ QCD event sample. 
\label{fig:ETfit}}
\end{center}
\end{figure}

After the cut on \ET, no events remained in the intermediate-$p_T$ QCD
sample ($100\GeV<p_T<150\GeV)$. To estimate
the number of events in the tail beyond \ET=100\GeV, the last 6 bins which 
contained events were fitted to a falling exponential, fig.~\ref{fig:ETfit}. 
Assuming uncorrelated gaussian errors on the fit and dividing the integral by
the bin width gives an estimated $290
\pm 450$ in the tail above 100\GeV, meaning that less than \tn{3} events
pass the cut at 95\% confidence level, yielding a rejection factor for these
events of at least 130 by the \ET\ cut.
Applying this rejection factor to the estimated
$\tn{6}$ events in the $p_T<100\GeV$ QCD sample 
and adding up yields a maximum of 
$(1 + 7.5)\ttn{3} < \tn{4}$ QCD 
events with $p_T<150\GeV$ remaining after the cut. Henceforth we refer to
these \tn{4} events, combined, as the low-$p_T$ (QCD) sample.

\subsubsection{Hard Leptons and Jets \label{sec:lepjets}}
\begin{figure}[t!]
\begin{center}
\begin{tabular}{cc}
\includegraphics*[scale=0.3]{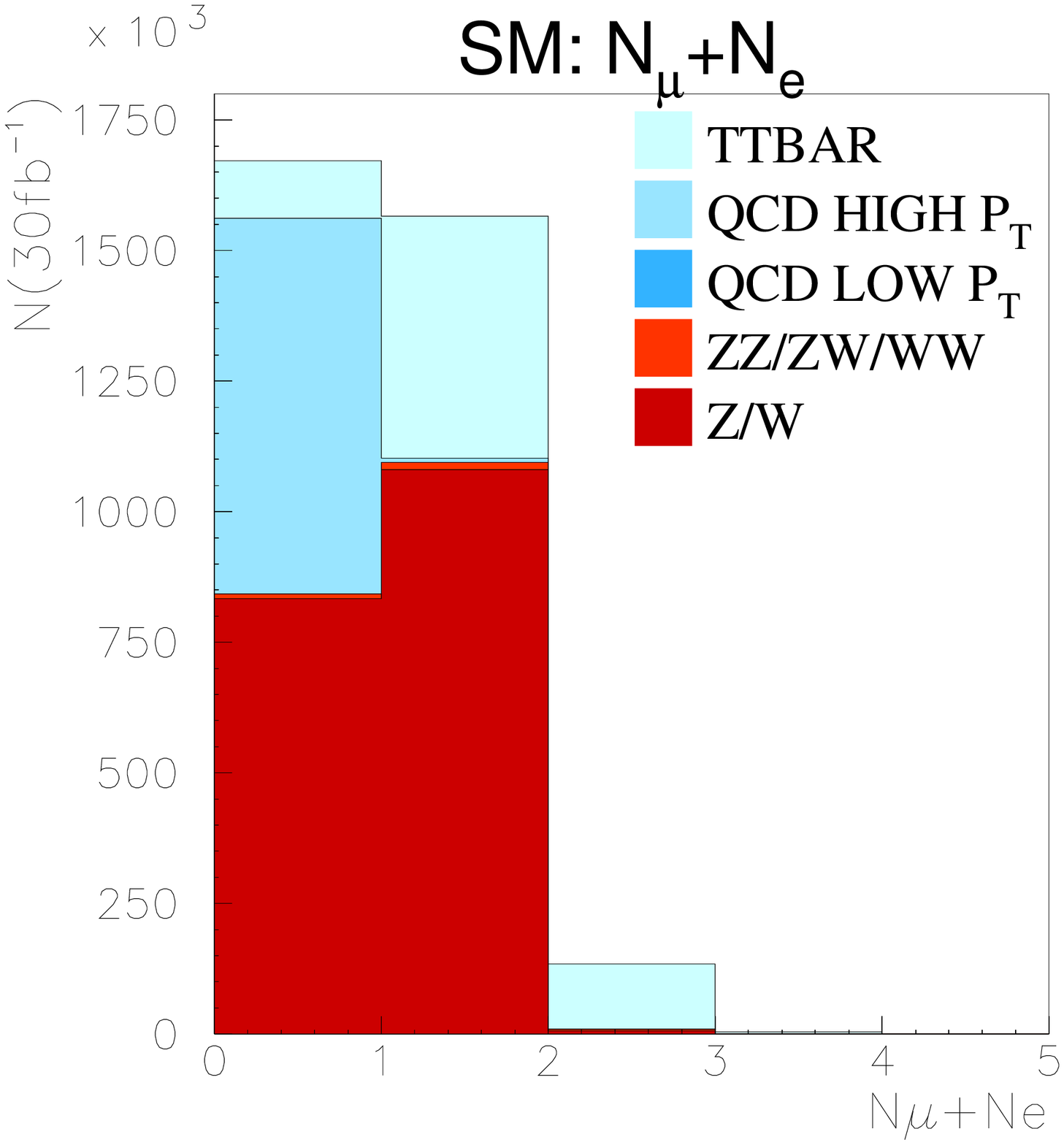}&\hspace*{-.7cm}
\includegraphics*[scale=0.3]{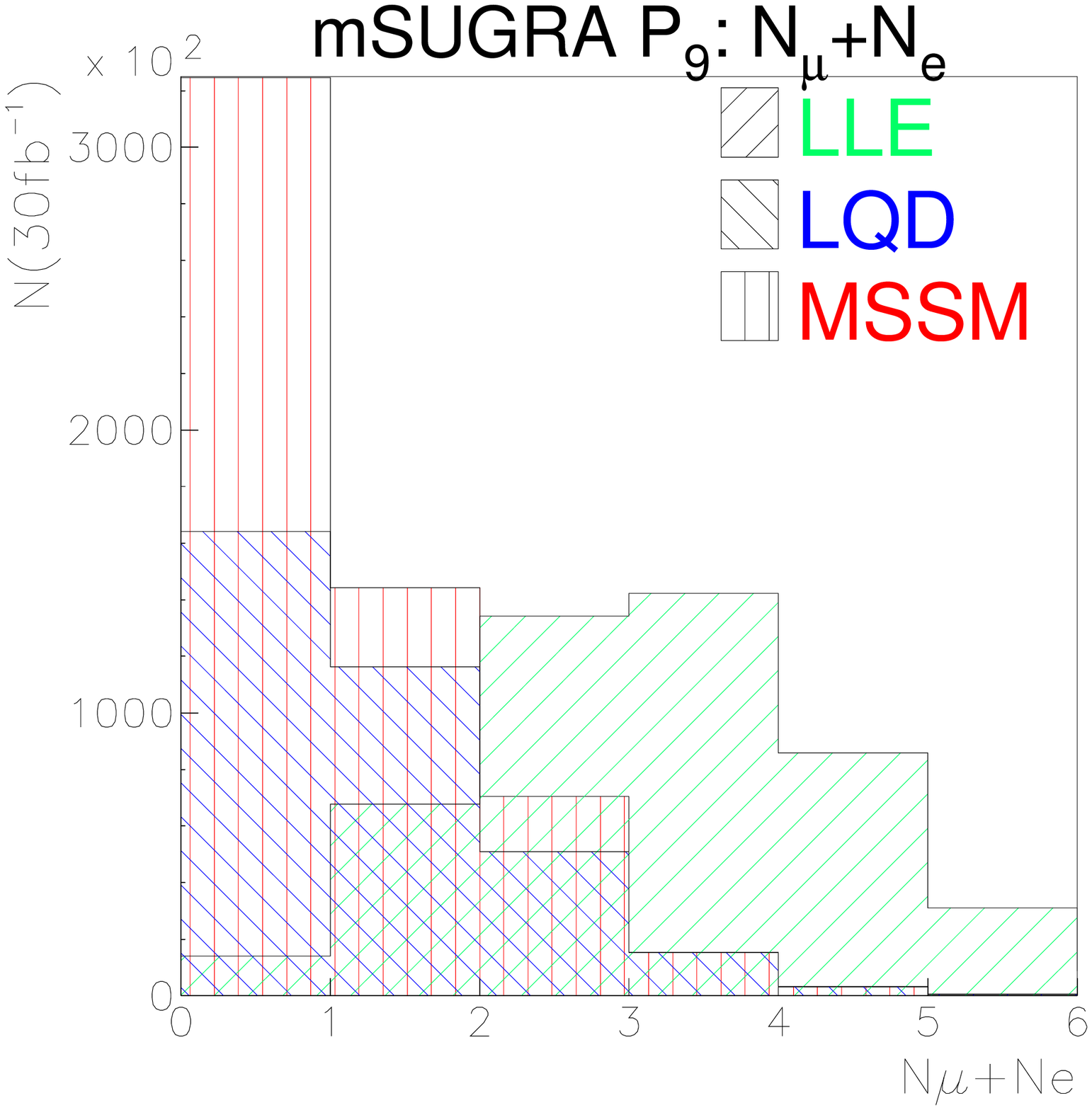}\vspace*{-7mm}\\
a) & b) \end{tabular}\vspace*{-12mm}\\
\includegraphics*[scale=0.66]{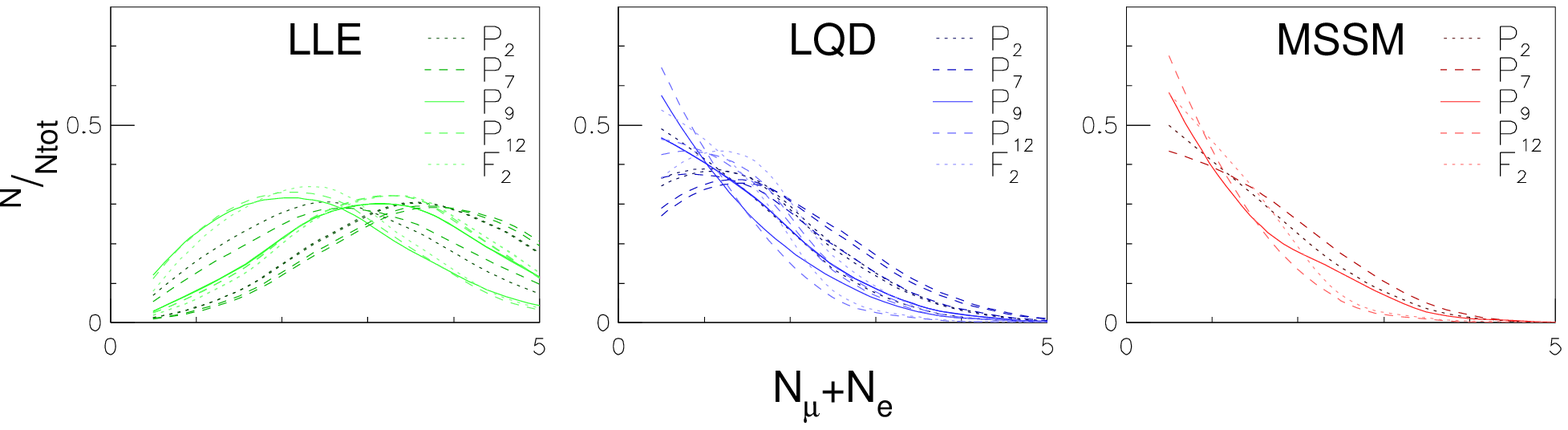}\vspace*{-1.2cm}\\
\hspace*{-1cm}c)\vspace*{-6mm}
\end{center}
\caption[\small Lepton multiplicities in the SM and SUSY.]{a) and b):
Lepton multiplicities in the SM and $P_9$ of events surviving the cut on
\ET. The numbering of the bins is such that the events with 0 leptons are in the
bin to the right of the number 0. c): Fractional distributions of events in
all scenarios investigated.
\label{fig:leptons}}
\begin{center}
\begin{tabular}{cc}
\includegraphics*[scale=0.3]{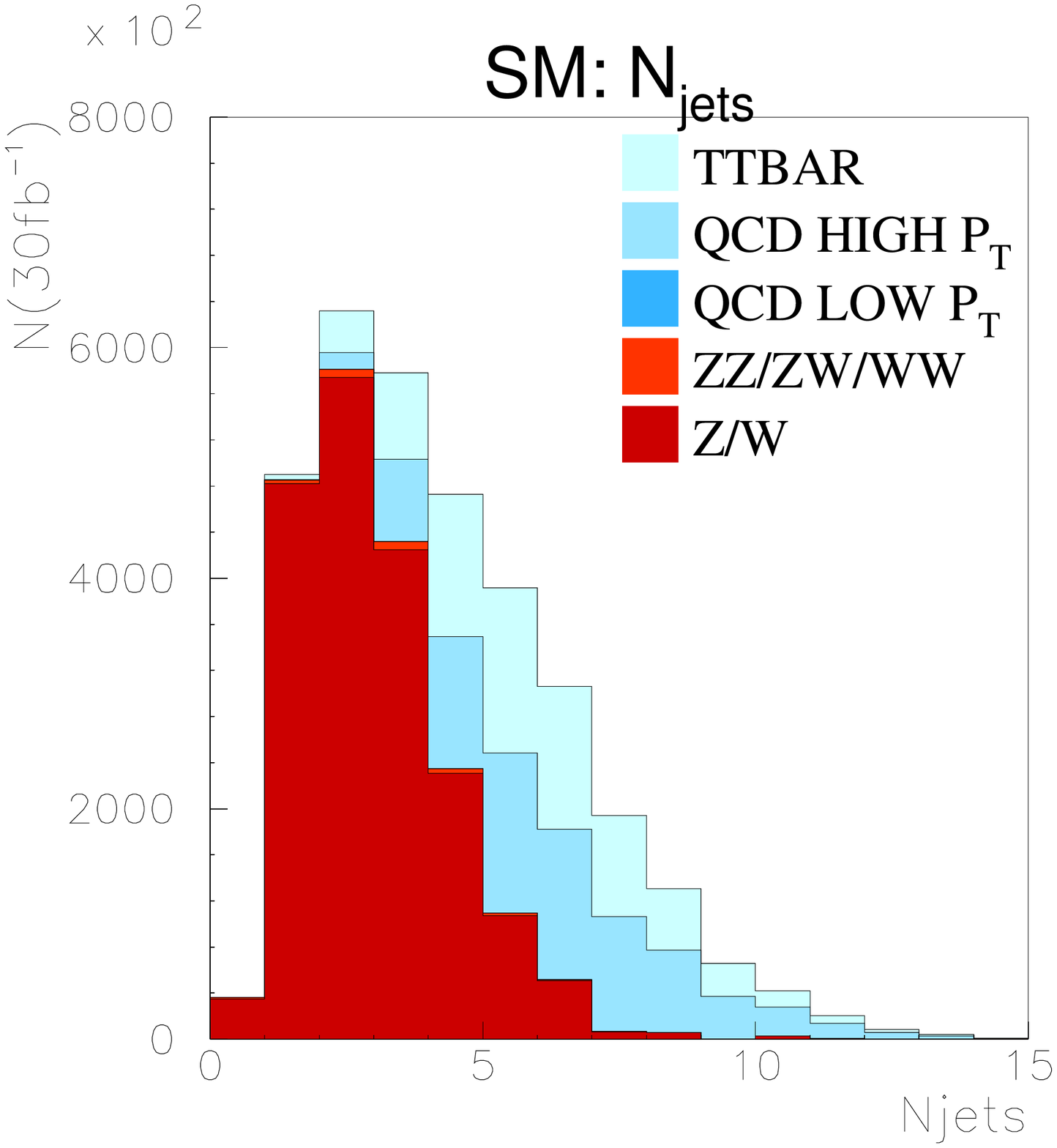}&\hspace*{-.7cm}
\includegraphics*[scale=0.3]{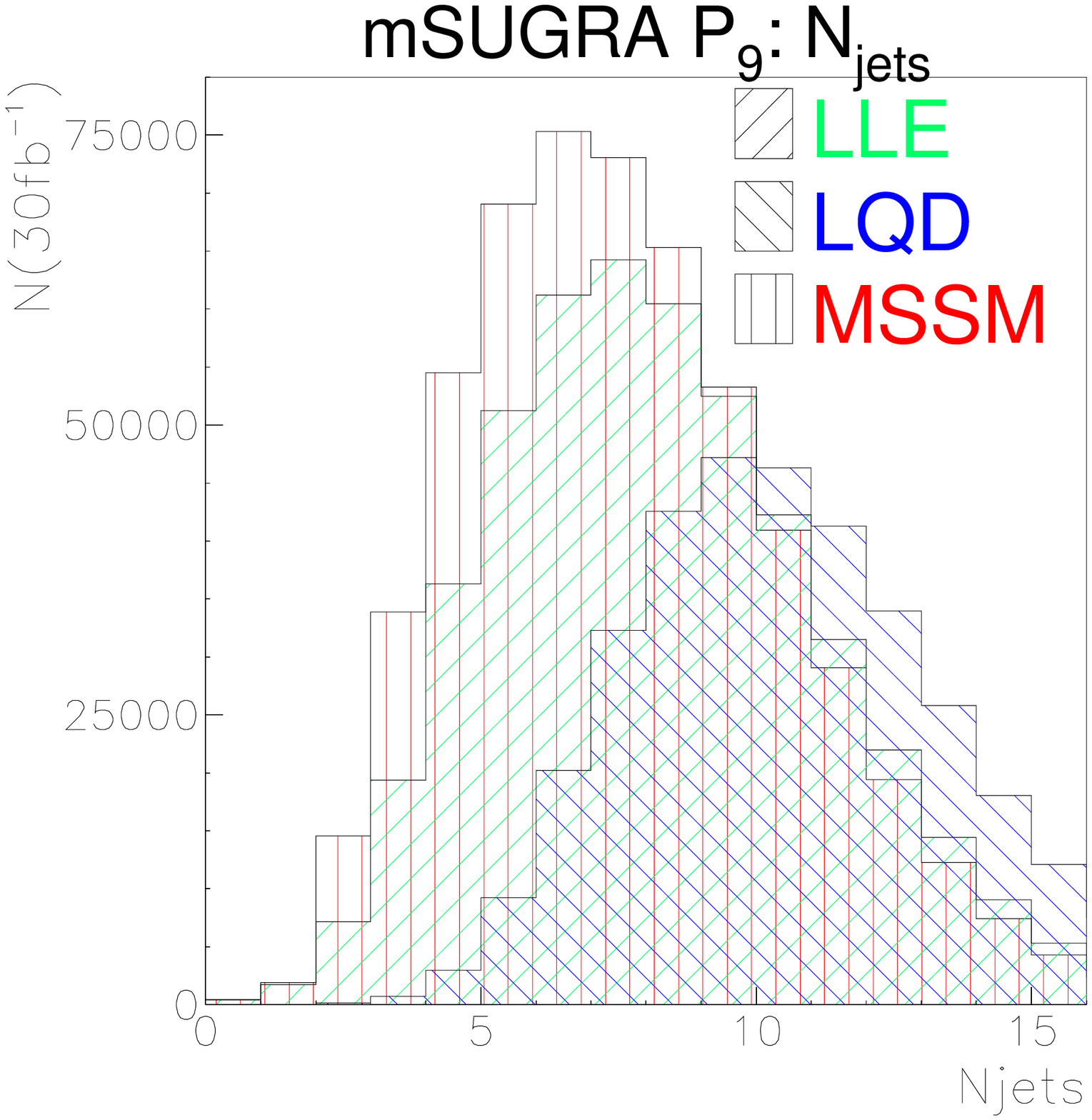}\vspace*{-8mm}\\
a) & b) \end{tabular}\vspace*{-12mm}\\
\includegraphics*[scale=0.66]{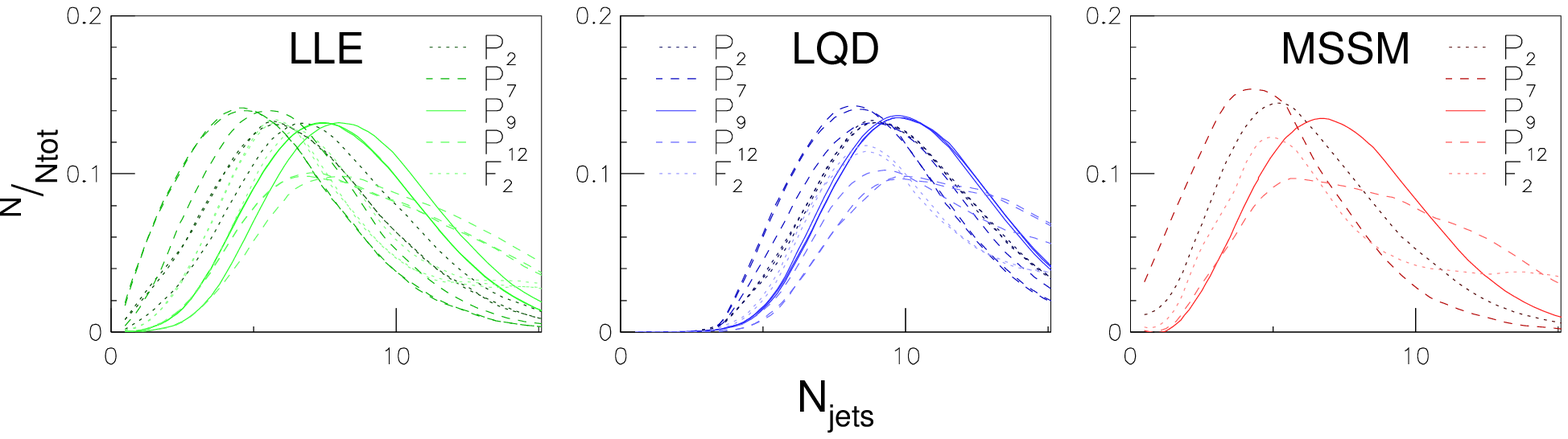}\vspace*{-1.2cm}\\
\hspace*{-8mm}c)\vspace*{-6mm}
\end{center}
\caption[\small Jet multiplicities in the SM and SUSY.]{a) and b):
Jet multiplicities in the SM and $P_9$ of events surviving the cut on
\ET. c): Fractional distributions of events in
all scenarios investigated.
\label{fig:jets}}
\end{figure}

\begin{figure}[t!]
\begin{center}
\includegraphics*[scale=0.2]{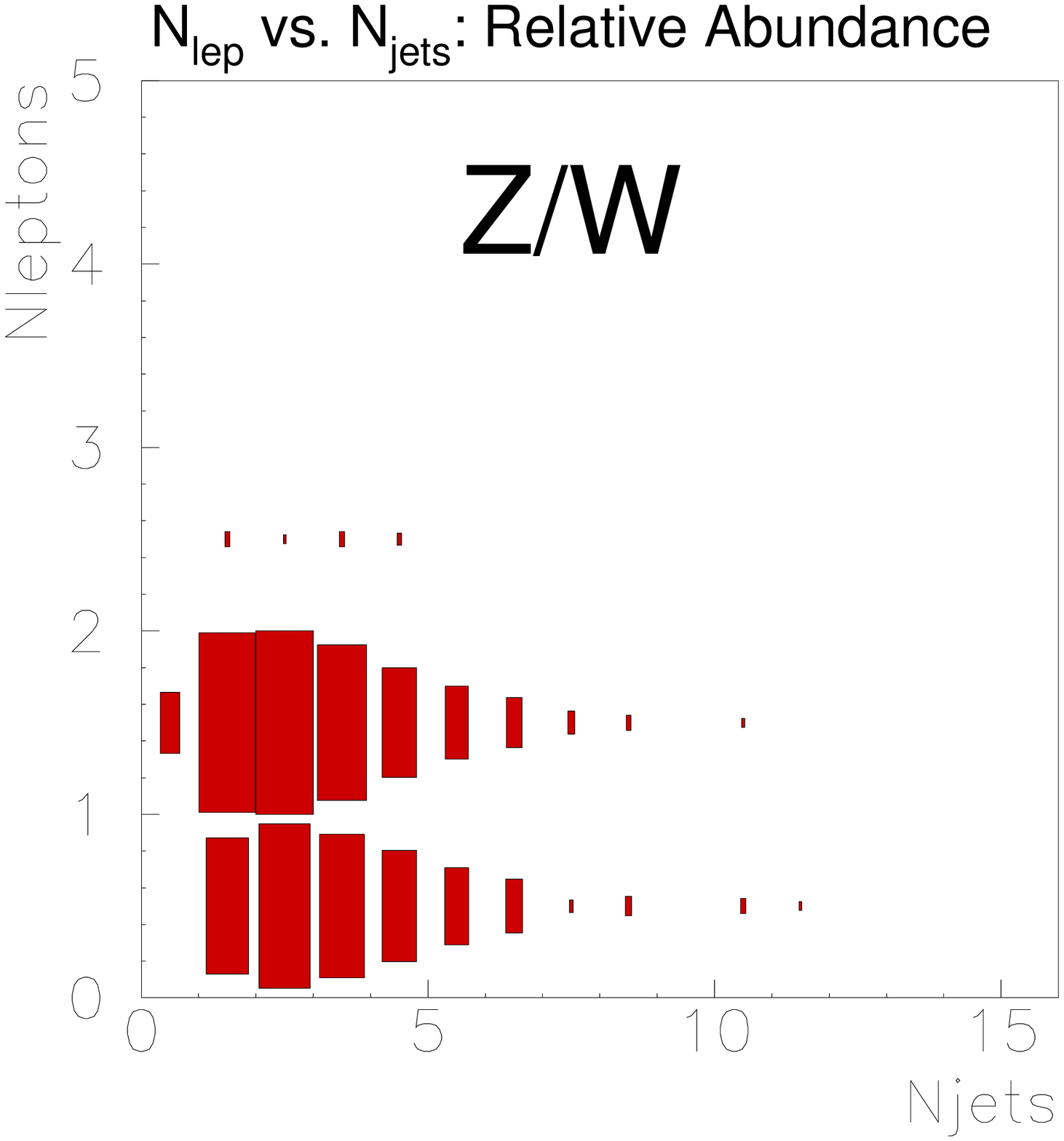}\hspace*{-3mm}
\includegraphics*[scale=0.2]{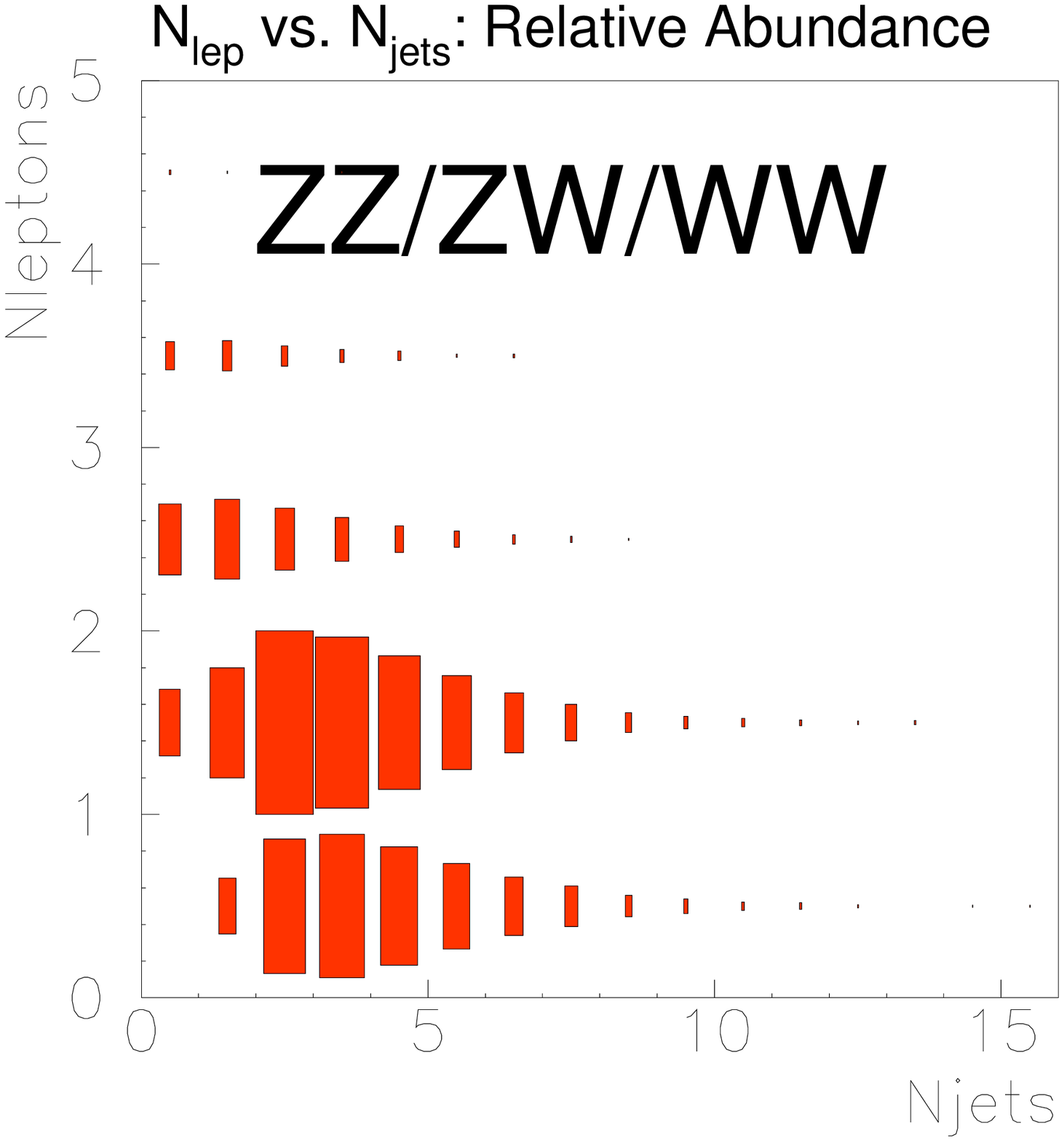}\hspace*{-3mm}
\includegraphics*[scale=0.2]{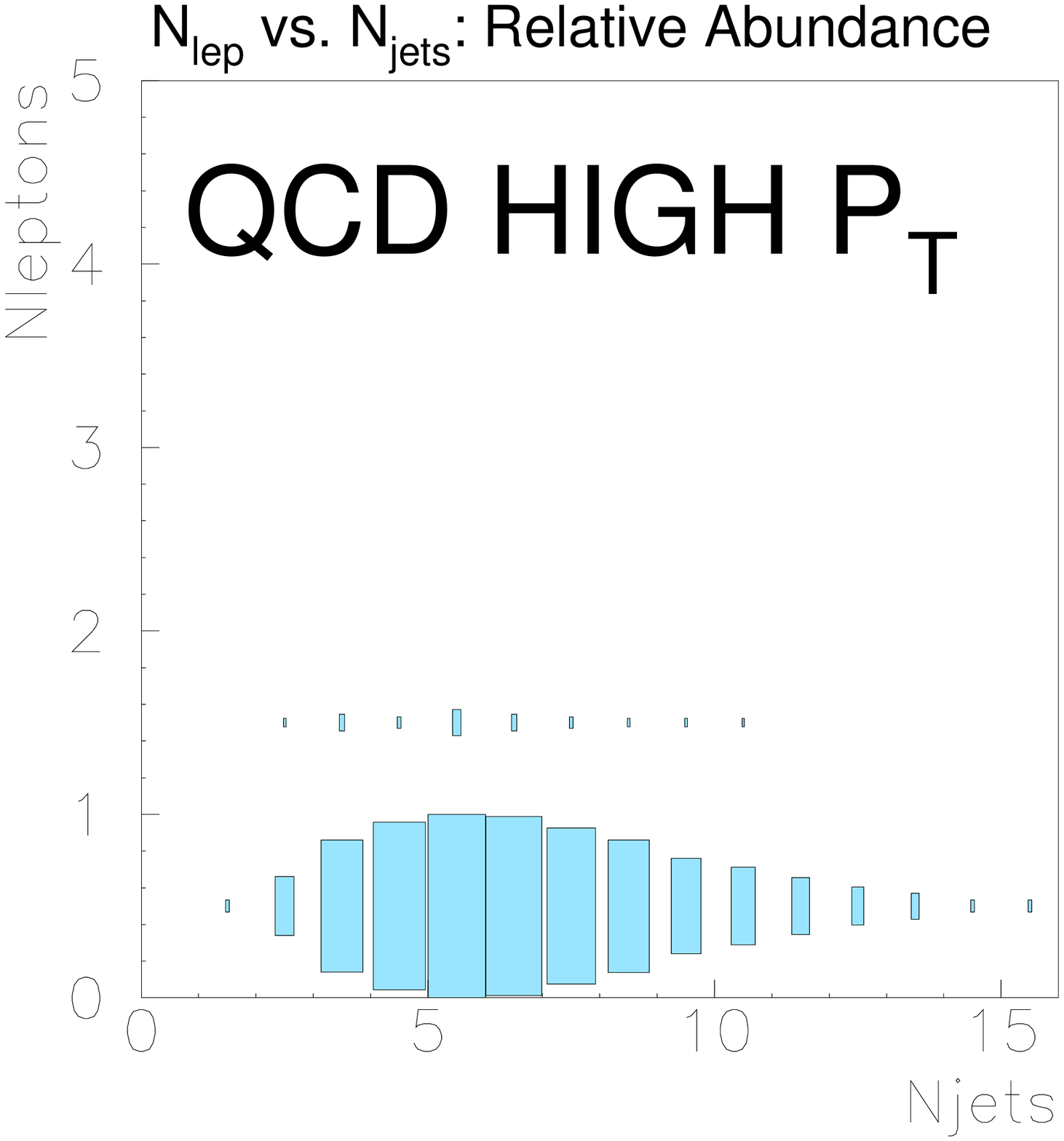}\hspace*{-3mm}
\includegraphics*[scale=0.2]{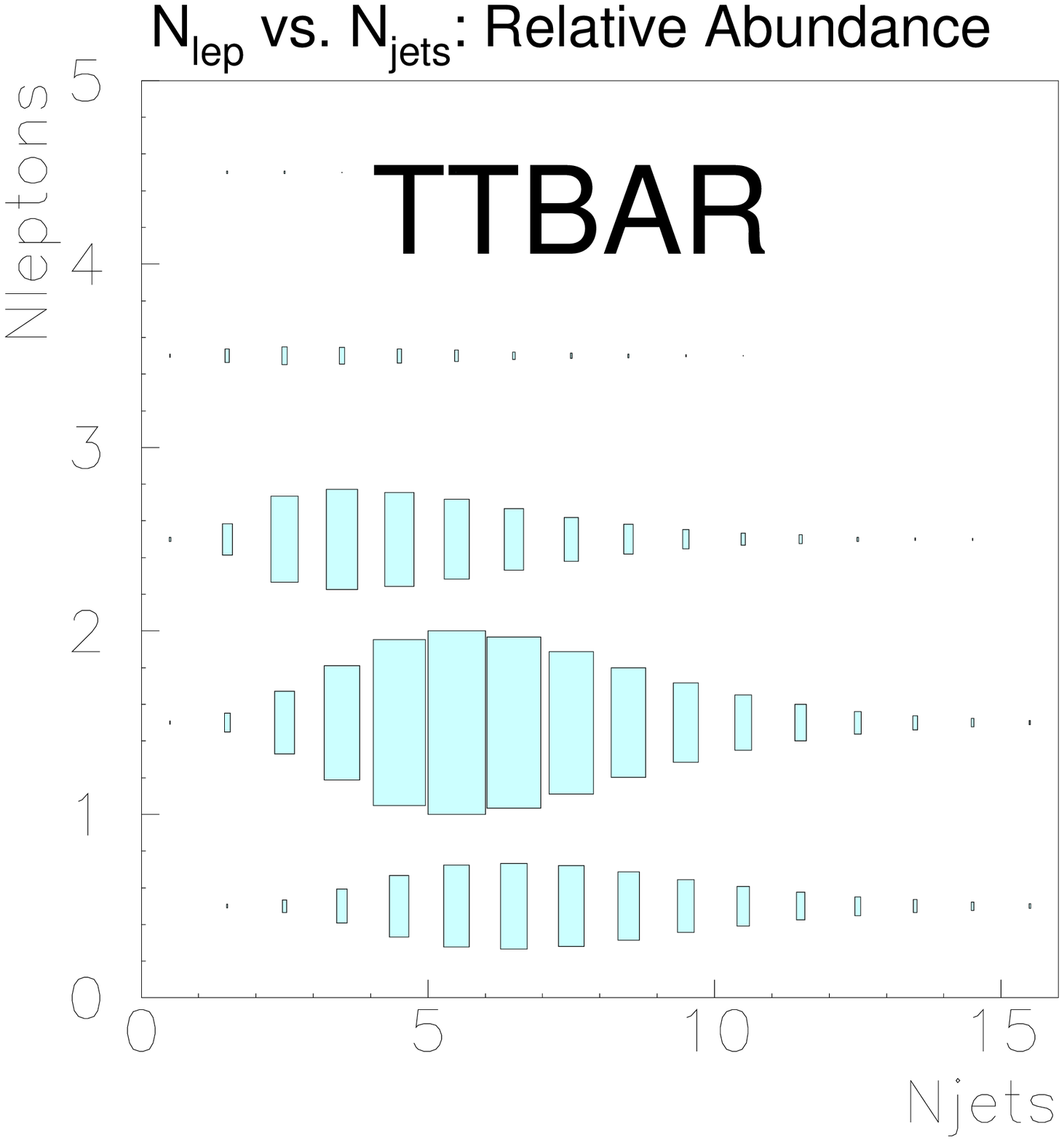}\vspace*{0.5mm}\\
\includegraphics*[scale=0.2]{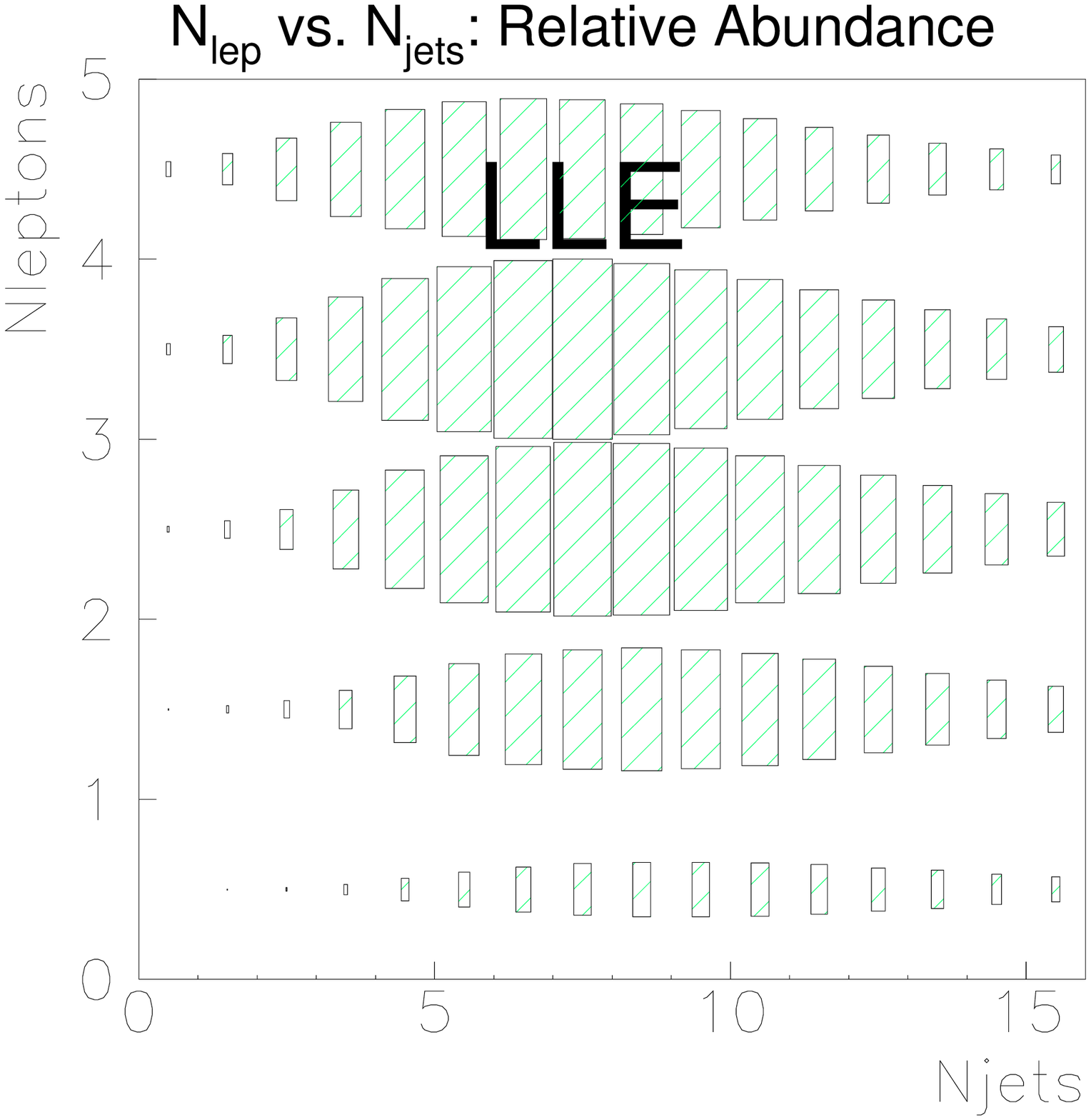}\hspace*{-3mm}
\includegraphics*[scale=0.2]{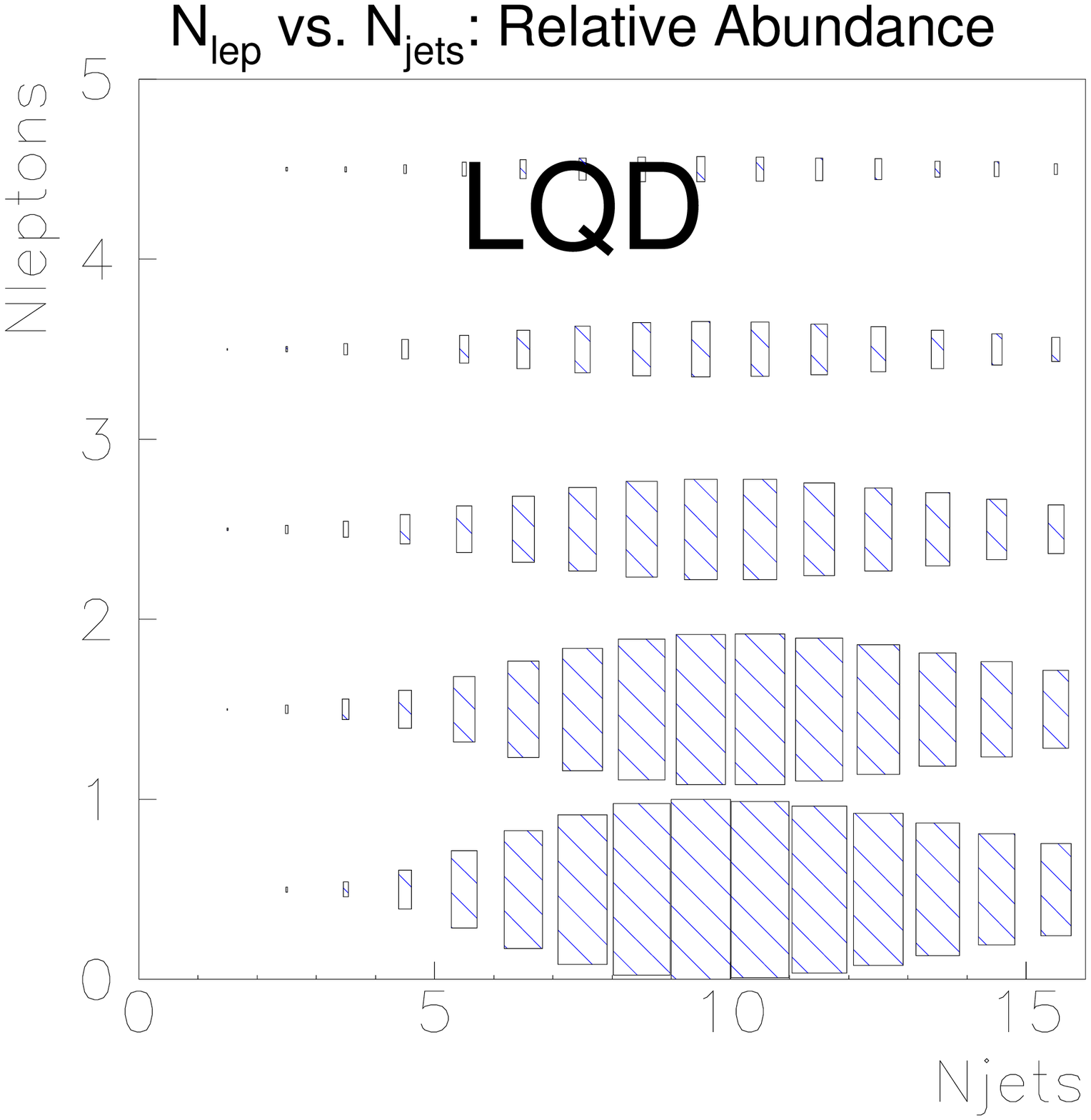}\hspace*{-3mm}
\includegraphics*[scale=0.2]{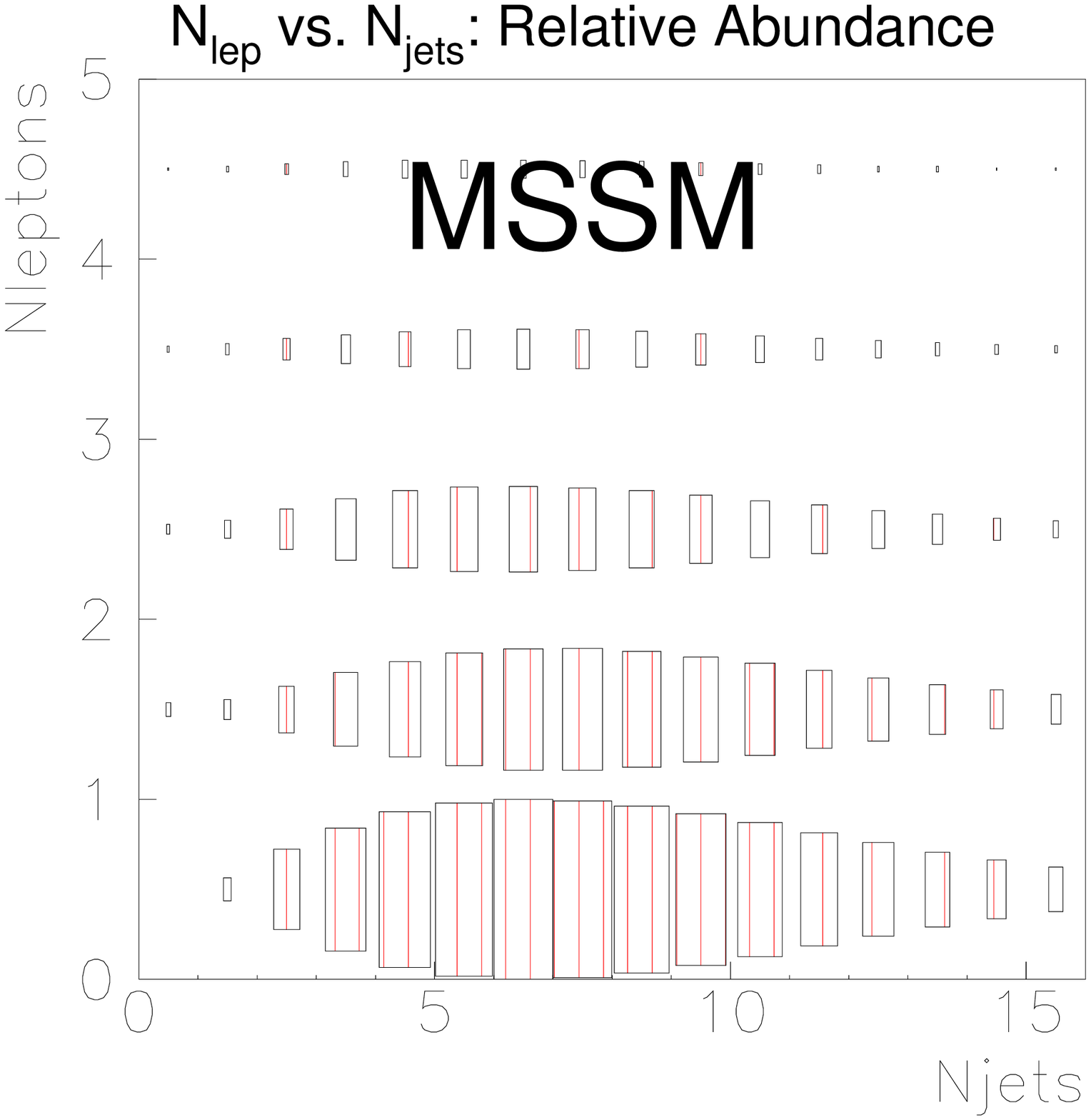}\vspace*{-8mm}
\end{center}
\caption[\small Lepton versus jet multiplicity in the SM and mSUGRA $P_9$.]{
Lepton versus jet multiplicity (see text) 
in the SM and mSUGRA $P_9$ 
of events surviving the cut on \ET. The numbering of the bins is such that the events with 0 leptons are in the
bin to the right of the number 0. The cuts investigated
correspond to cutting out the lower left corner of each plot.
\label{fig:lj}}
\end{figure}

As mentioned, a typical signature for SUSY is the large
number of jets obtained. This, when combined with (possible) 
violation of Lepton Number, may well be
accompanied by a large lepton multiplicity, and so it makes good sense to
combine the analysis here. The lepton multiplicities 
(iso.\ muons + iso.\ electrons) in events 
with $\ET >100\GeV$ are shown in figures \ref{fig:leptons}a (SM) and
\ref{fig:leptons}b (mSUGRA $P_9$ MSSM, LLE, and LQD), and an overwiev of the
distributions in the other scenarios investigated are shown 
in \ref{fig:leptons}c. 
Jet multiplicities are shown in figure \ref{fig:jets}. One sees the larger
relative lepton multiplicity in LLE scenarios and the larger relative jet
multiplicity in LQD scenarios.
Finally, the ``box'' plots in figure
\ref{fig:lj} show the correlations
between the number of leptons and the number of jets with large boxes meaning
that a large fraction of events have the corresponding combination of
$N_{lep}$ and $N_{jets}$ and small boxes meaning that a small fraction of
events have the corresponding combination. 

Based on these distributions, it seems reasonable to require at least $N_J$
jets, at least $(N_J-1)$ jets and at least 1 lepton, at least $(N_J-2)$ jets and
at least 2 leptons, or at least 3 leptons. Values for $N_J=8-11$ have been
investigated. Results are presented in table \ref{tab:leptonjets}. For the
$p_T<100\GeV$ QCD sample, applying the rejection factor of 8.5 found for the
high $p_T$ sample gives an estimated $900$ 
events at most remaining after the cut. 

\begin{table}[t!]
\setlength{\extrarowheight}{0pt}
\begin{center}
\textsf{EVENTS PASSING CUTS ON N$_{\mbox{\scriptsize jets}}$ AND
N$_{\mbox{\scriptsize leptons}}$.\vspace*{2mm}}
{\footnotesize
\begin{tabular}{lcccccc}\toprule
& SM & $P_9$ MSSM & $P_9$ LLE & $P_9$ nLLE & $P_9$ LQD & $P_9$ nLQD\\ 
\cmidrule{1-7}\boldmath
$N_J=8$  &($200\pm10$) k&220 k& 270 k&410 k&380 k&290 k\\ 
$N_J=9$  & ($105\pm4$) k&160 k& 230 k&380 k&340 k&240 k\\ 
$N_J=10$ & ($51\pm 3$) k&120 k& 190 k&360 k&290 k&190 k\\
$N_J=11$ & ($23\pm 2$) k& 80 k& 150 k&330 k&250 k&150 k\\ 
\cmidrule{1-7}
$\frac{N_{\mbox{\tiny post}}}{N_{\mbox{\tiny pre}}}$ 
& 0.06 & 0.39 & 0.85 & 0.78 & 0.76 & 0.75\\\bottomrule
\end{tabular}}\\ 
\vspace*{-5mm}
\end{center}
\caption[\small Event numbers passing cuts on $N_{\mbox{\scriptsize jets}}$
and $N_{\mbox{\scriptsize leptons}}$.]{Events passing cuts on
$N_{\mbox{\scriptsize jets}}$ and $N_{\mbox{\scriptsize leptons}}$ for
several choices of jet cutoff, $N_J$ (see text). In addition, the requirement
that each event pass the \ET\ cut was imposed. The selected cut is marked
in bold, and the ratio of events surviving 
after this cut to events surviving before this cut is shown for each
model. A very good rejection factor for SM events is obtained. The reason for
this is plain to see in figure \ref{fig:lj}. Cutting out the lower left
corner in those plots takes away the majority of $Z/W$ events, and also a
significant reduction in QCD and $t\bar{t}$ events is obtained. 
For the SM, the (gaussian) uncertainty due to the limited event sample is
also shown.  
\label{tab:leptonjets}} 
\vspace*{-\tfcapsep}\end{table}
As for the \ET\ cuts, it is not at all surprising that the MSSM is here 
the model which does the worst. 
After all, the MSSM has a lot of \ET\ exactly because the LSP \emph{escapes}
and does \emph{not} give rise to extra jets and/or leptons. 
Note also that we have power to discriminate between MSSM, $\RV$
with dominant LQD terms, or $\RV$ with dominant LLE terms in these
variables. This, however, is saved for the neural network analysis below.

\begin{figure}[t!]
\begin{center}
\begin{tabular}{cc}
\includegraphics*[scale=0.36]{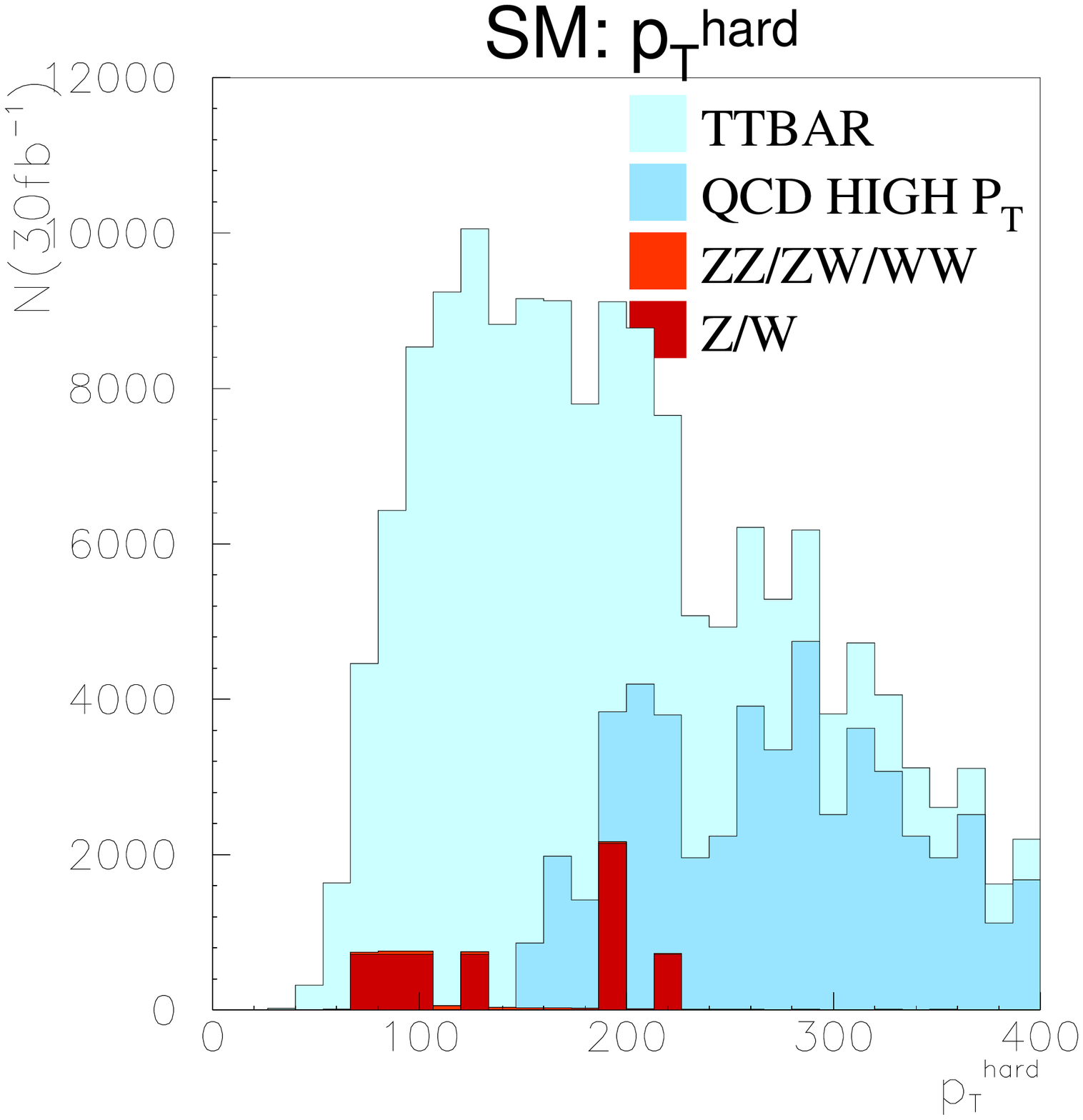}&\hspace*{-.7cm}
\includegraphics*[scale=0.36]{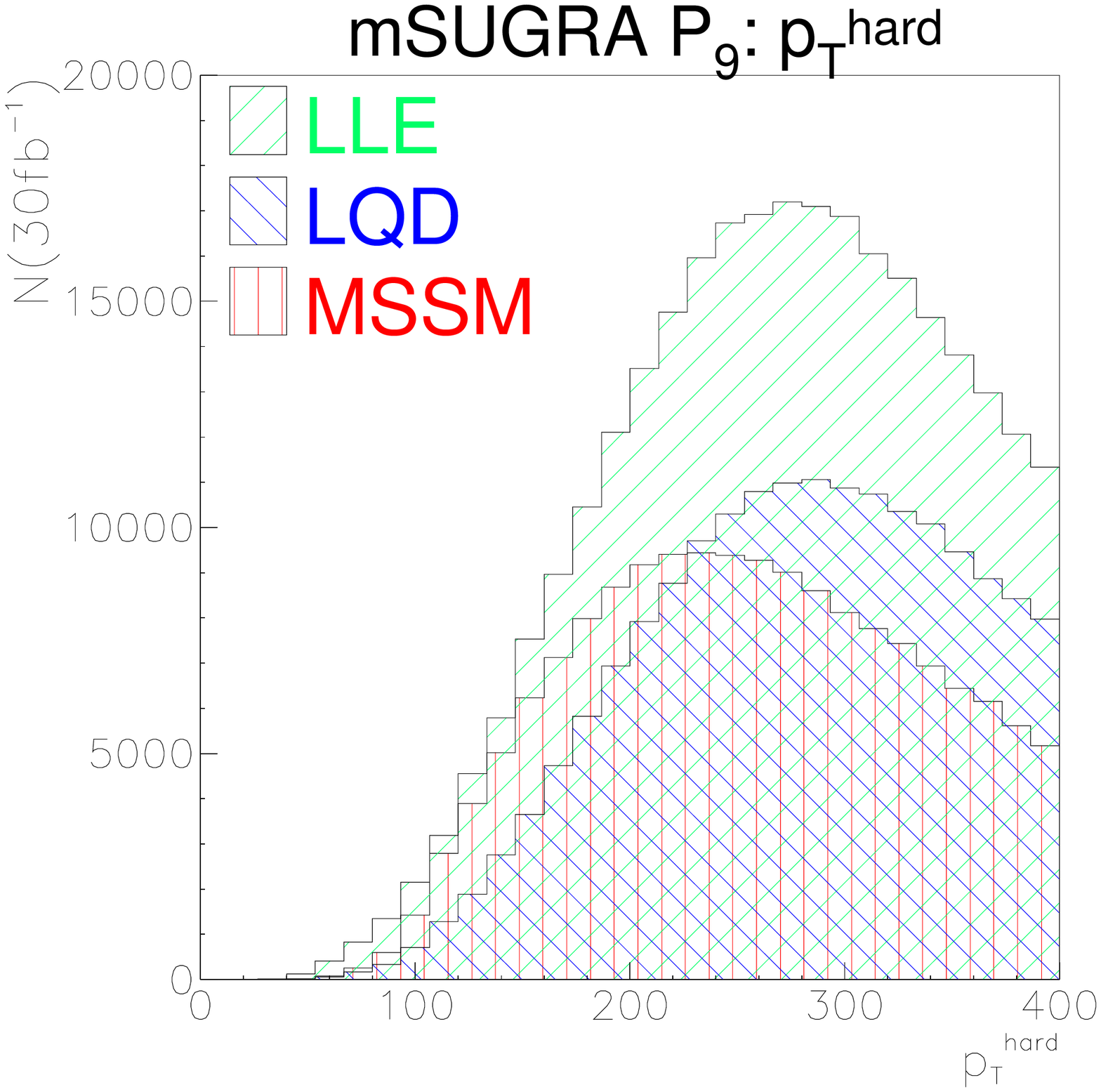}\vspace*{-7mm}\\
a) & b) \end{tabular}\vspace*{-8mm}\\
\includegraphics*[scale=0.7]{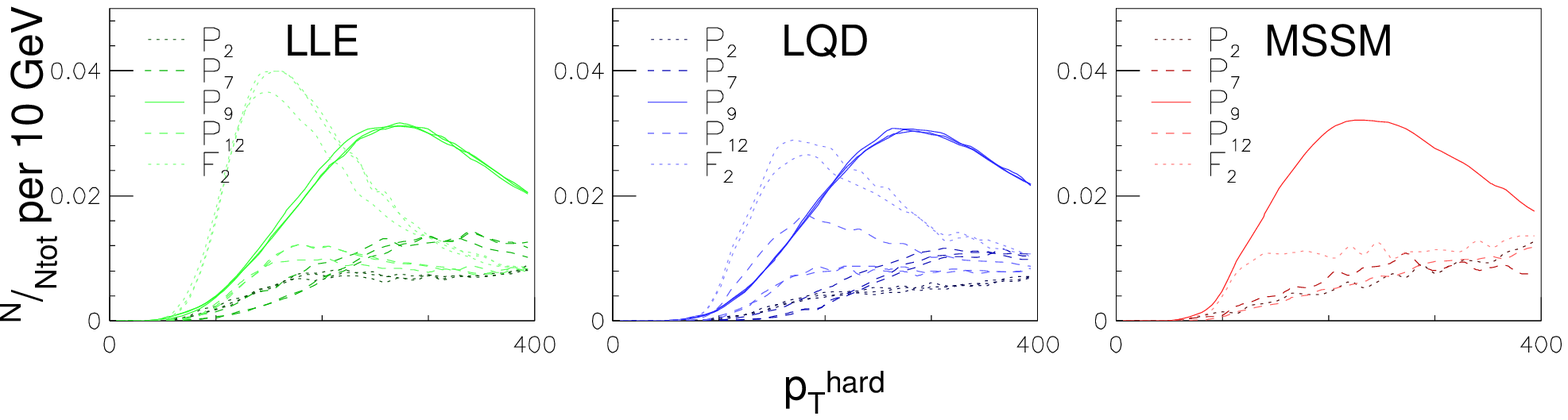}\vspace*{-.7cm}\\
c)\vspace*{-7mm}
\end{center}
\caption[\small $p_T$ distribution for hardest object in SM and SUSY.]{
$p_T$ distribution for hardest object in SM and SUSY, normalized to
30\fb$^{-1}$ of data taking. 
Events shown are the ones surviving previous cuts.
\label{fig:hardobj}}
\end{figure}

Additional variables which are obvious as discriminators when studying the
decays of heavy particles are (transverse) momenta
of the hardest jets and leptons in the event. The transverse momenta of the 
four hardest jets and the two hardest leptons are therefore also used as
inputs in the neural net analysis. 
For events with less than 4 jets and/or less than
2 leptons, the value 0 is assigned to the ``missing'' jet and lepton
$p_T$'s. In the cut-based analysis, 
we simply use the $p_T$ of the hardest object in the
event, $p_T^{hard}$. The SM and $P_9$ distributions for this variable 
are shown in figure
\ref{fig:hardobj}. Cuts at $p_T^{hard}=100,150,200$, and $250$ were
investigated with results as shown in table \ref{tab:hardobj}.
\begin{table}[htb!]
\setlength{\extrarowheight}{0pt}
\begin{center}
\textsf{EVENTS PASSING CUTS ON $p_T^{\mbox{\scriptsize hard}}$.\vspace*{2mm}}
{\footnotesize
\begin{tabular}{lcccccc}\toprule
                      &SM&$P_9$ MSSM&$P_9$ LLE&$P_9$ nLLE&$P_9$ LQD&$P_9$ nLQD\\ 
\cmidrule{1-7}
$p_T^{hard}=100$&$(180\pm10)$ k&220 k&  410 k &   380 k &  270 k   &290 k\\ 
$p_T^{hard}=150$&$(140\pm10)$ k&200 k&  390 k &   370 k &  270 k   &280 k\\ 
\boldmath
$p_T^{hard}=200$&$(110\pm 5)$ k&180 k&  360 k &   330 k &  250 k   &260 k\\
$p_T^{hard}=250$& $(85\pm 4)$ k&140 k&  300 k &   270 k &  210 k   &220 k\\
\cmidrule{1-7}
$\frac{N_{\mbox{\tiny post}}}{N_{\mbox{\tiny pre}}}$ 
& 0.57 & 0.80 & 0.86 & 0.85 & 0.90 & 0.90\\\bottomrule
\end{tabular}} 
\vspace*{-5mm}
\end{center}
\caption[\small Event numbers passing cuts on $p_T^{\mbox{\scriptsize hard}}$.]
{Events passing cuts on
$p_T^{\mbox{\scriptsize hard}}$ for
several choices of cut value. The selected cut is marked
in bold, and the ratio of events surviving 
after this cut to events surviving before this cut is shown for each
model. 
\label{tab:hardobj}}
\vspace*{-\tfcapsep}\end{table}
Due to the higher resonance masses, the $P_9$ scenarios escape these cuts almost
with impunity. The ratio of events surviving before and after the 
cut is even better for the rest of the mSUGRA points. 

After the cut on hardest object, only 1 $Z/W$ event remained in the
sample (out of $5\ttn{7}$ generated). As is apparent from figure
\ref{fig:hardobj}a, it would clearly be nonsense to fit 
the $P^{hard}_{T}$ distribution of $Z/W$ events to thereby obtain an
estimate. Rather, we use the procedure recommended by
\cite{europhys} that a conservative upper bound on the number, $N$,
passing the cut is given by the mean of that Poisson distribution 
which yields a 5\% chance of giving only one event or less surviving the
cut. This gives an estimate for $N<4.75$ at 95\% confidence level,   
translating to $3400$ events after 30\fb$^{-1}$ of data
taking. With respect to
the rejection factor expected for these events under subsequent cuts, we
adopt a slightly pessimistic assumption, reducing the number of events by the
same factor as the double gauge events ($ZZ/ZW/WW$). 
Before leaving the $Z/W$
events, we take one more look at figure \ref{fig:hardobj}a. It is here
evident that there are 3 events \emph{just} below the cut, and one might
 argue that our estimate of 3400 events is therefore likely to be too
optimistic. This argument is incorrect since by using that knowledge we would
invalidate the statistical approach just used. As long as the cut was not
tuned to lie exactly above these events (and 200 was chosen only for its
being a nice round number), the Poisson approach is statistically sound.

For the low-$p_T$ QCD events, the low rejection factor, 1.07, 
found for the high $p_T$ sample gives an estimated 
negligible reduction of event numbers by the $p_T^{hard}$ cut. One must
recall, however, that the high $p_T$ QCD sample consists entirely 
of events where the
hard scattering gave rise to $p_T>150\GeV$ in the CM of the scattering. It is
therefore quite natural that almost 
no reduction is accomplished for these events by
demanding that the $p_T$ of the hardest object in the event be larger than
200\GeV. For the low-$p_T$ QCD sample, 
we expect the reduction from this cut to be
significantly greater, yet to be conservative, we use the same rejection
as for the high $p_T$ sample, yielding maximally 420 events remaining.

\subsubsection{LSP Decay Signature\label{sec:lspdecsig}}
The \RV\ scenarios give us one extra possible signature for SUSY events which
is not present in the $R$-conserving cases, LSP decay. Out of necessity, we
shall here focus on the case of a neutralino LSP. This means that we are
looking for 3-body decays, a more difficult situation than for 2-body
decays. It is, of course,
impossible to say whether the lightest neutralino 
decays into $qq\nu$, $qq\ell$, or
$\ell\ell\nu$ without making assumptions about the relative coupling strengths,
something which is obviously not acceptable when one is interested in
defining as general a search strategy as possible. Whatever coupling
is dominant, there are maximally two neutrinos in an event with
double neutralino decay (see section \ref{sec:lspdecays}). One would therefore expect to see at least 4 hard
jets/leptons with energies not greatly differing from each other. We
therefore introduce the following measure for this ``4-object energy 
correlation'':
\begin{equation}
E_{4C} \equiv \frac13(\frac{E_4}{E_3}+\frac{E_3}{E_2} + \frac{E_2}{E_1})
\end{equation}
where $E_{1-4}$ are the energies of the 4 hardest objects (leptons or jets)
in the event ordered in energy, the hardest being $E_1$. Events with
$N_{lep}+N_{jet}< 4$ are assigned the value zero.
Following the above
line of reasoning we would expect the SUSY events to have 4-object correlations 
close to 1. In contrast, there is no reason to expect this kind of
correlation in e.g.\ $Z/W$ or QCD events. Also, a large number of double
gauge events will have low or zero values since two gauge bosons decaying
leptonically \emph{at most} produce 4 hard objects in the detector. 
The $t\bar{t}$ events, however, are
quite indistinguishable from many of the SUSY scenarios in this
variable. Noting that the more massive a particle is, the larger
the momentum kicks given to its decay products will be,  
one would expect that particles coming from the decays of objects
heavier than the top would, on average, have larger $p_T$ than particles
coming from top decays. We can use this to give the variable just defined
some extra discriminating power against $t\bar{t}$ events. However, it comes
at a cost. A look in table \ref{tab:sugrapoints} reveals that $P_9$, for
instance, has the LSP and several other 
SUSY particles \emph{lighter} than the top. By giving the 4-object energy
correlation some $p_T$
dependence, we will not only get rid of top events, we will also be throwing
away signal events for SUSY scenarios with low-mass particles. 
This is not a serious drawback,
since we can afford to loose a certain amount of signal
in the low-mass 
scenarios due to the relatively high production cross sections. In
return, we get a more pure signal for the heavier scenarios where we don't have
so many signal events and so require a better background rejection.
 
\begin{figure}[t!]
\begin{center}
\begin{tabular}{cc}
\includegraphics*[scale=0.36]{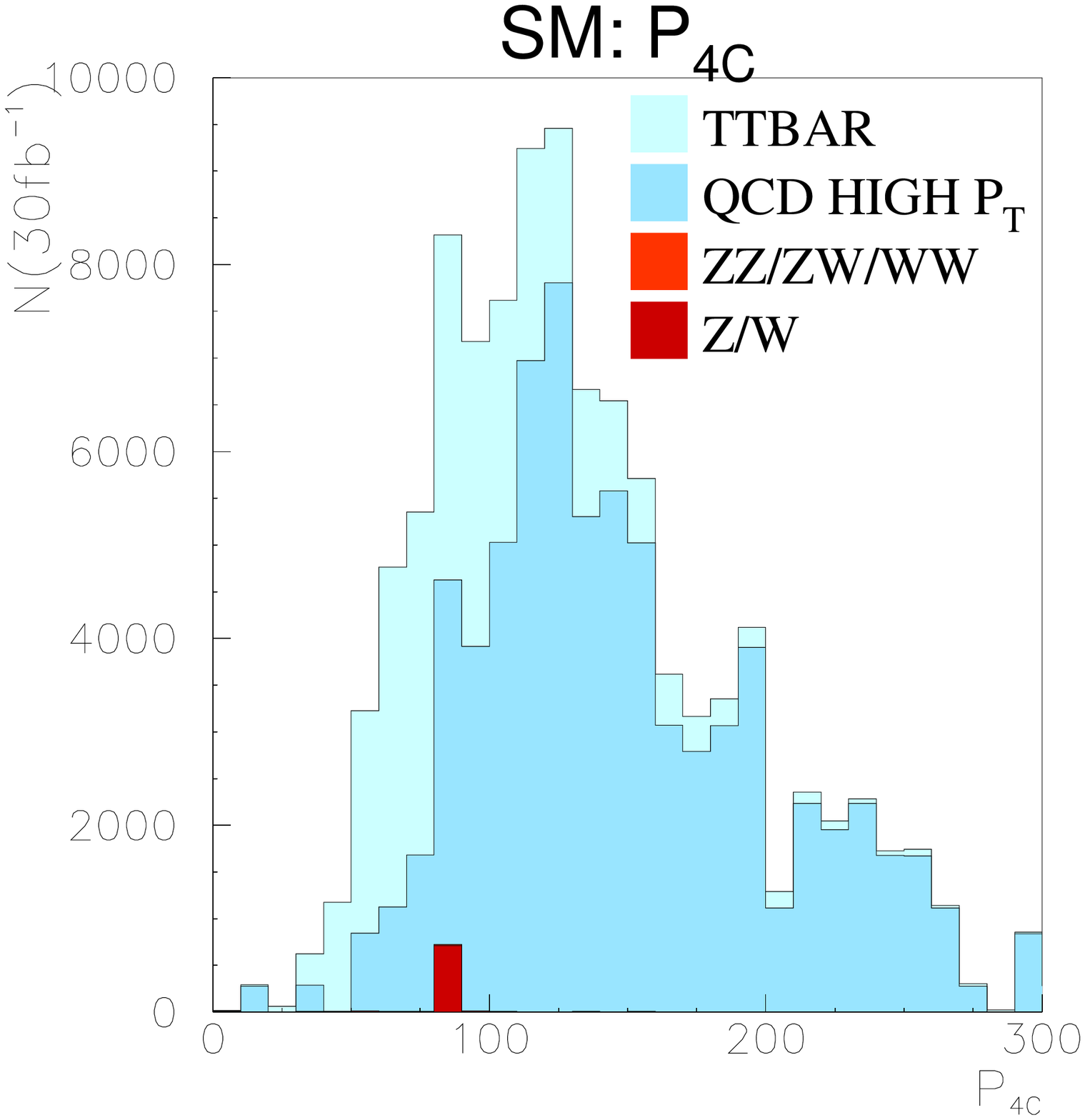}&\hspace*{-.7cm}
\includegraphics*[scale=0.36]{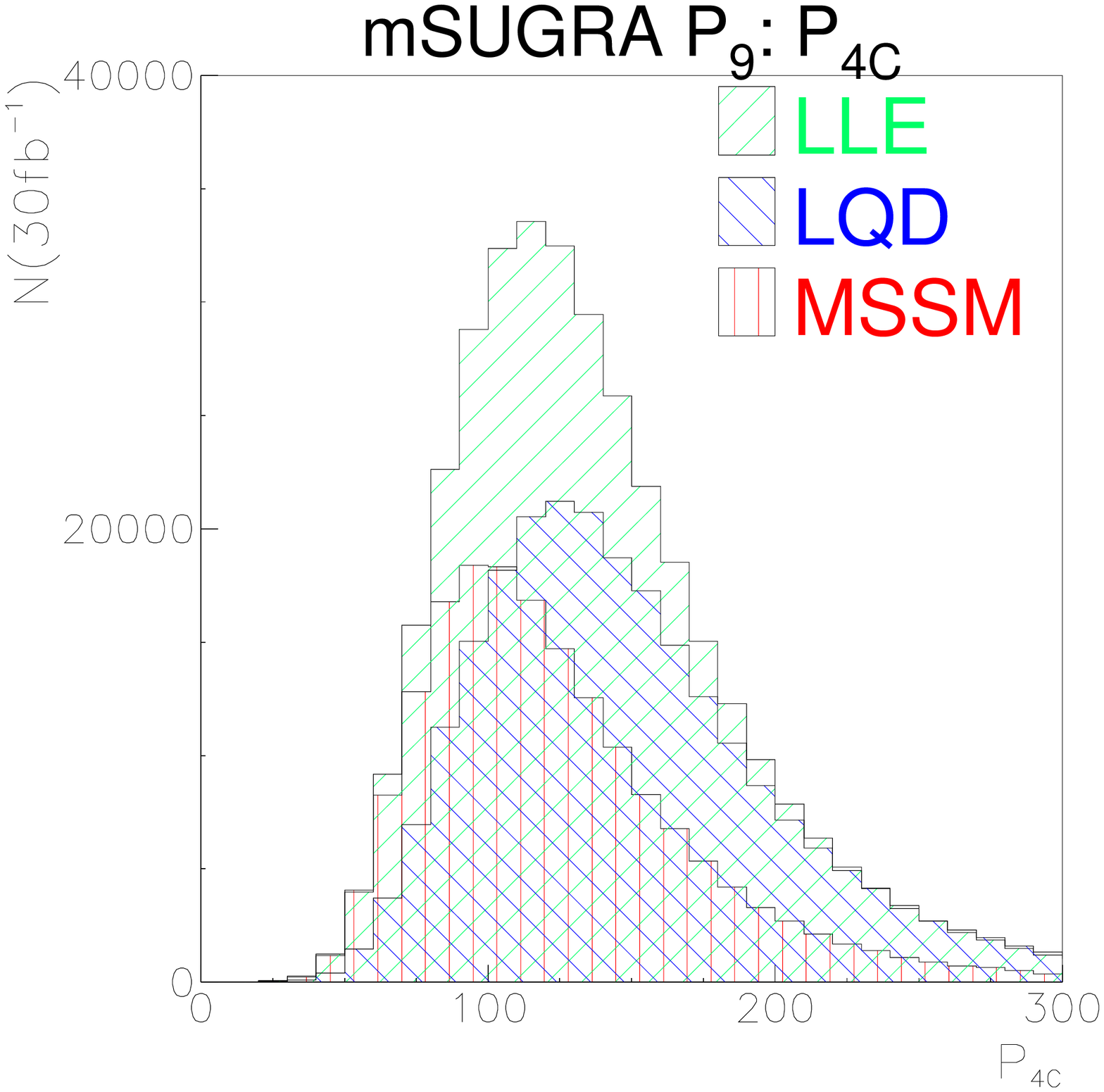}\vspace*{-7mm}\\
a) & b) \end{tabular}\vspace*{-8mm}\\
\includegraphics*[scale=0.7]{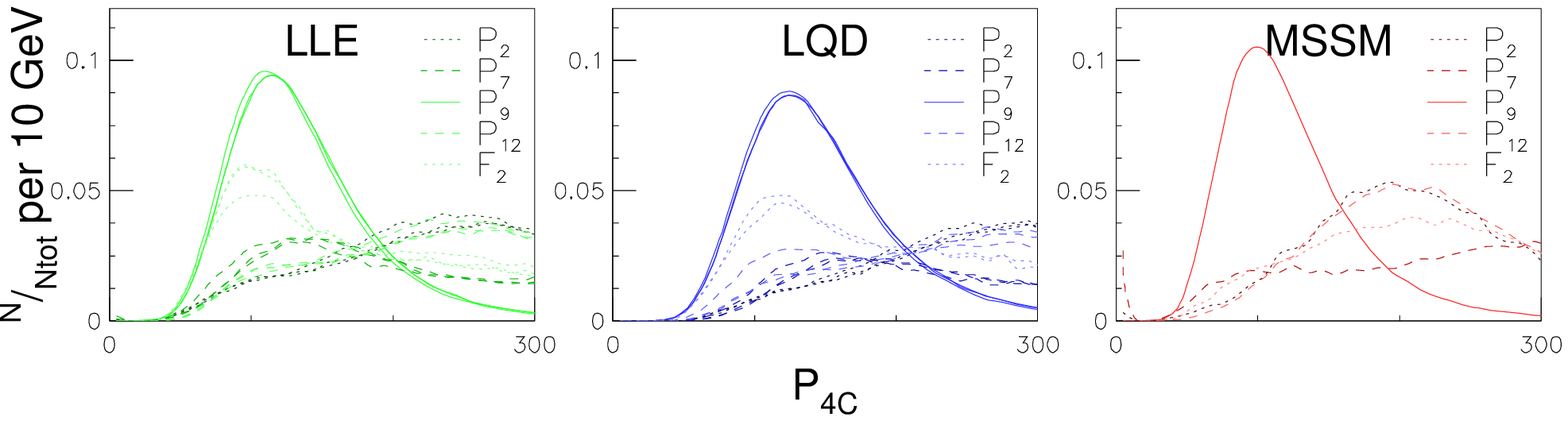}\vspace*{-.6cm}\\
c)\vspace*{-6mm}
\end{center}
\caption[\small LSP decay signature for SM and SUSY.]{Distributions of
$P_{4C}$ in the SM (a and b) and for mSUGRA $P_9$. 
All events used survive previous cuts. 
\label{fig:p4c}}
\end{figure}

This is the basis for using the 4-object energy 
correlation multiplied 
by the average $p_T$ of the four hardest objects rather than the 4-object
energy correlation alone, and so we introduce the $p_T$-weighted 4-object
energy correlation:
\begin{equation}
P_{4C} \equiv \frac{1}{12}\left(\frac{E_4}{E_3}+\frac{E_3}{E_2} + \frac{E_2}{E_1}\right)(
p_{T1}+p_{T2}+p_{T3}+p_{T4})
\end{equation}
The suspicion that the low-mass scenarios will not do well in this variable is
quickly verified by taking a look at figure \ref{fig:p4c}c where peaks around
100\GeV\ are seen for both $P_9$ and $F_2$ whereas the heavier scenarios show
more flat distributions.
\begin{table}[htb!]
\setlength{\extrarowheight}{0pt}
\begin{center}
\textsf{EVENTS PASSING CUTS ON P$_{4C}$.\vspace*{2mm}}
{\footnotesize
\begin{tabular}{lcccccc}\toprule
& SM & $P_9$ MSSM & $P_9$ LLE & $P_9$ nLLE & $P_9$ LQD & $P_9$ nLQD\\ 
\cmidrule{1-7}
$P_{4C}>50$ &$(108\pm5)$ k&170 k&350 k&330 k&250 k&260 k\\
$P_{4C}>75$ & $(97\pm5)$ k&160 k&330 k&310 k&240 k&250 k\\ \boldmath
$P_{4C}>100$ &$(79\pm4)$ k&110 k&270 k&240 k&210 k&220 k\\
$P_{4C}>125$ &$(57\pm4)$ k& 72 k&190 k&170 k&160 k&160 k\\
\cmidrule{1-7}
$\frac{N_{\mbox{\tiny post}}}{N_{\mbox{\tiny pre}}}$ 
& 0.72 & 0.65 & 0.77 & 0.84 & 0.75 & 0.83 \\\bottomrule
\end{tabular}} 
\vspace*{-5mm}
\end{center}
\caption[\small Event numbers passing cuts on $P_{4C}$.]
{Events passing cuts on $P_{4C}$ for
SM and $P_9$ events surviving previous cuts, scaled to correspond to
30\fb$^{-1}$ of data taking. 
The selected cut is marked
in bold, and the ratio of events surviving 
after this cut to events surviving before this cut is shown for each
model. One sees that the MSSM for 
 does even worse than the SM for $P_9$, due to the absence of LSP
decay. The rest of the
mSUGRA points generally have acceptances close to 100\% even for the
MSSM. The only exception is $F_2$ which nonetheless lies above 80\% in all cases.
\label{tab:p4c}} 
\vspace*{-\tfcapsep}\end{table}

For SUSY events where double LSP decay 
does not occur, note that there must be either
one or more heavier particles decaying directly to SM particles or
$R$-parity is conserved. In the first case, the $E_i/E_j$ will generally be
lower, since jet/lepton energies are presumably not equal to so great an
extent, but since heavier particles are decaying, 
the average $p_T$ will be larger, evening out the score. The
$R$-conserving scenarios will look more like the SM in this variable, since
no LSP decay occurs. 

For the high-$p_T$ QCD and the double gauge events, rejection factors of 1.2
and 2.7, respectively, were found, yielding an estimated maximum of 350
low-$p_T$ QCD events and 1300 $Z/W$ remaining.

\subsubsection{Thrust}
\begin{figure}[t!]
\setlength{\extrarowheight}{0pt}
\begin{center}
\begin{tabular}{cc}
\includegraphics*[scale=0.36]{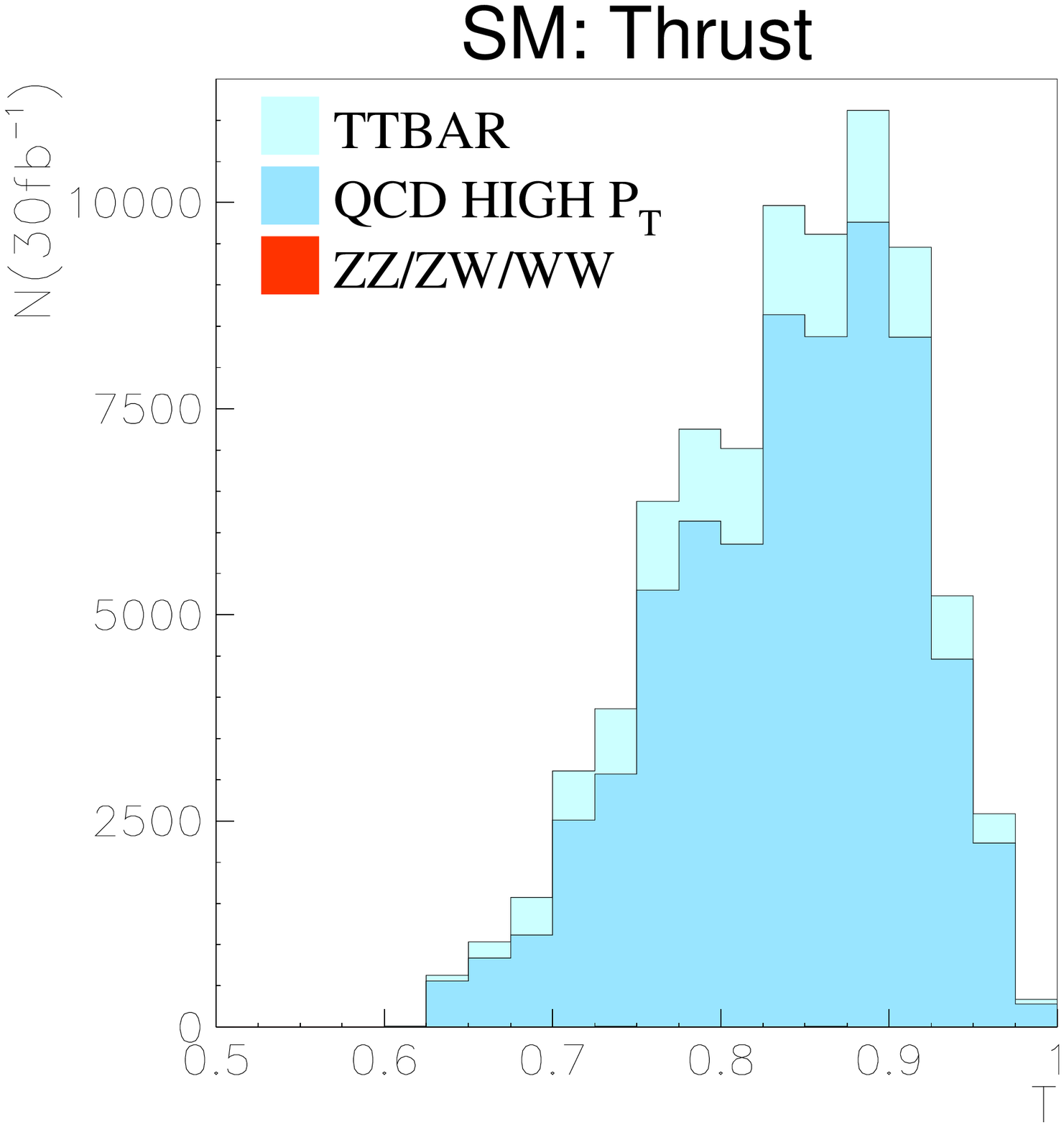}&\hspace*{-.7cm}
\includegraphics*[scale=0.36]{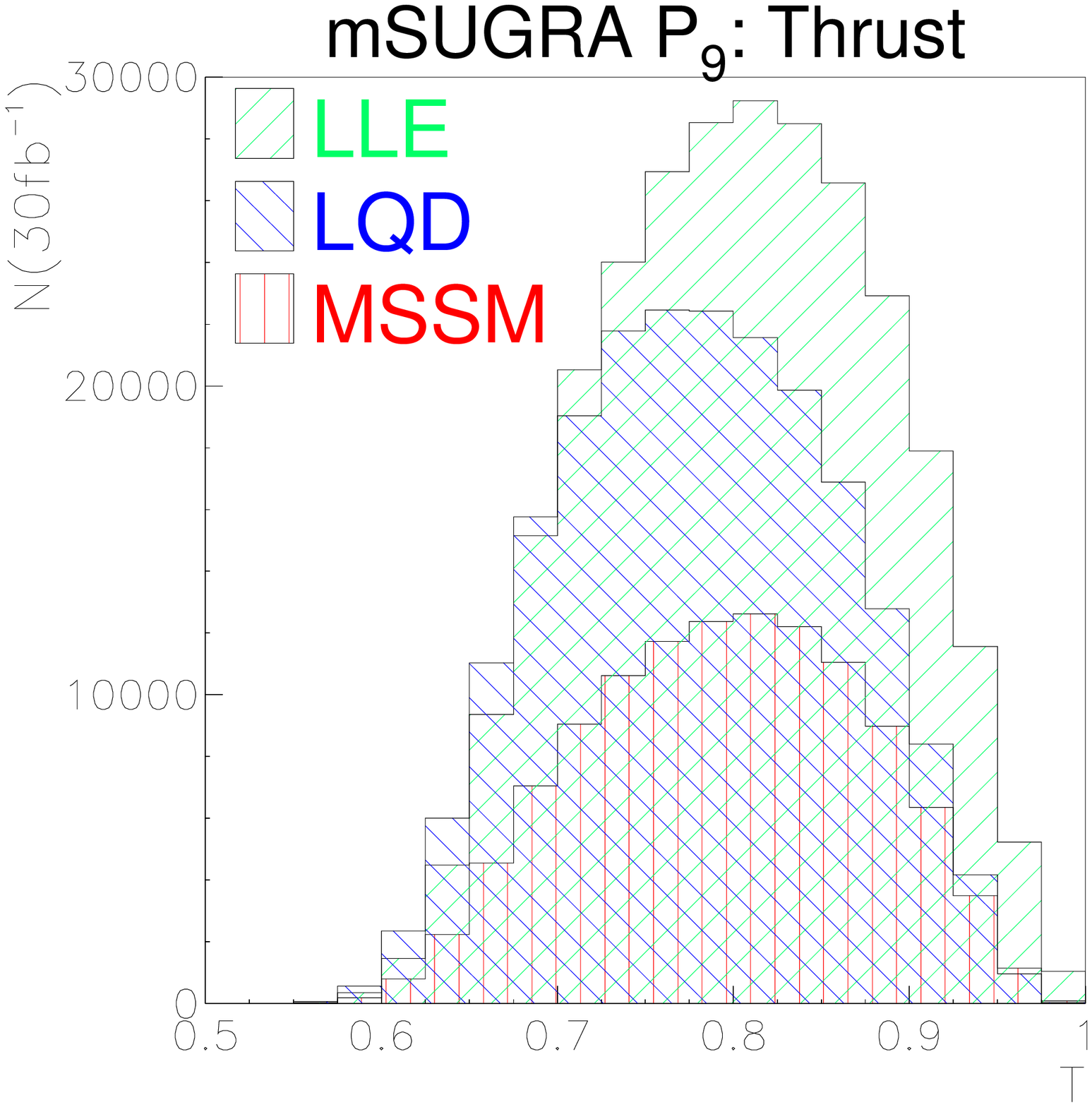}\vspace*{-6mm}\\
a) & b) \end{tabular}\vspace*{-8mm}\\
\includegraphics*[scale=0.7]{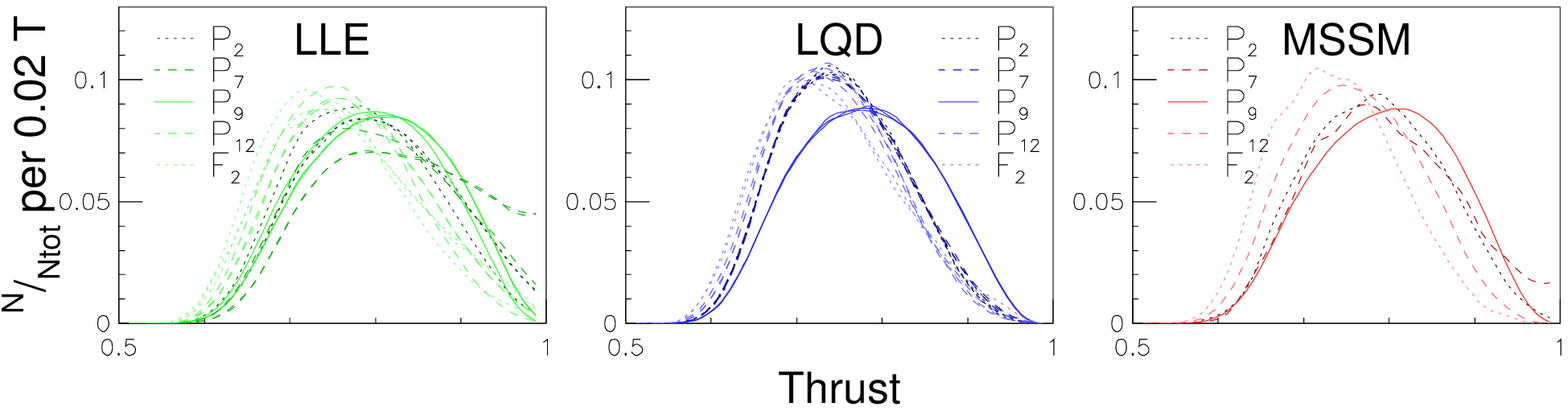}\vspace*{-.7cm}\\
c)\vspace*{-6mm}
\end{center}
\caption[\small Thrust distributions for SM and SUSY.]{Thrust distributions for 
SM processes and mSUGRA $P_9$. Observe that the QCD processes peak around 0.9
whereas $P_9$ peaks around 0.8. All events used survive previous cuts.
\label{fig:thrust}}
\end{figure}

One may intuitively understand that pair production involving less massive
particles is liable to produce decay products which lie more ``on a line''
than processes involving more massive particles (where the decay products get
larger momentum kicks). It is based on this that thrust is expected to be of use
as a discriminating variable. Its value is defined as:
\begin{equation}
T = \mathrm{max}\left\{\frac{\sum_i|p_i\cdot \vec{n}|}{\sum_i|p_i|}\right\}
\end{equation}
where $i$ runs over all particles in the event and $\vec{n}$ is a unit vector
along an arbitrary axis. The axis which maximizes the expression is defined
as the thrust axis. For 
a completely pencil-shaped event, the thrust value is 1, going down to 0.5
for an event where the momenta are evenly distributed throughout the detector.
The thrust distributions for events
satisfying the previous cuts are shown in figure \ref{fig:thrust}.
Cuts on $T<0.75$, $T<0.8$, $T<0.85$, and $T<0.90$ 
were investigated. Results are shown
in table \ref{tab:thrust}. The selected cut, $T<0.85$, is marked in bold. 
\begin{table}[b!]
\setlength{\extrarowheight}{0pt}
\begin{center}
\textsf{EVENTS PASSING CUTS ON THRUST.\vspace*{2mm}}
{\footnotesize
\begin{tabular}{lcccccc}\toprule
& SM            & $P_9$ MSSM & $P_9$ LLE & $P_9$ nLLE & $P_9$ LQD & $P_9$ nLQD\\ 
\cmidrule{1-7}
$T<0.75$ &$(10\pm2)$ k& 35 k &     75 k  &     77 k   & 76 k    &80 k\\
$T<0.8$  &$(24\pm2)$ k& 59 k &    130 k  &    130 k   & 120 k   &130 k\\ 
\boldmath
$T<0.85$ &$(40\pm3)$ k& 83 k &    190 k  &    180 k   & 160 k   &170 k\\
$T<0.90$ &$(62\pm4)$ k&100 k &    240 k  &    220 k   & 190 k   &200 k\\
\cmidrule{1-7}
$\frac{N_{\mbox{\tiny post}}}{N_{\mbox{\tiny pre}}}$ 
& 0.51 & 0.73& 0.69&0.74& \\
\bottomrule
\end{tabular}}
\vspace*{-5mm}
\end{center}
\caption[\small Event numbers passing cuts on thrust.]{Events passing cuts on
thrust for three values of thrust cutoff. The selected cut is marked in bold, 
and the ratio of events surviving after this cut to events surviving before
this cut is shown for each model.\label{tab:thrust}} 
\vspace*{-\tfcapsep}\end{table}
Rejection factors of 2.0 and 2.1 were obtained for the high $p_T$ QCD events
and the double gauge sample, respectively, yielding 180 low-$p_T$ and 620
$Z/W$ events maximally remaining. Note that the $Z/W$ events would most
likely have higher rejections under this cut than their double gauge
counterparts, yet finding it difficult to quantify it, we do not include any
additional suppression in the estimated number.

\subsubsection{Oblateness and Circularity}
Having defined the thrust axis for each event, one can take one step further
and define Major and Minor axes in the plane perpendicular to the thrust axis
in exactly the same way as thrust was defined:
\begin{equation}
\mbox{Major} = \mbox{max}\left\{\frac{\sum_i|p_i\cdot
\vec{n}|}{\sum_i|p_i|}\right\} 
\end{equation}
where the maximum is now to be found perpendicular to the thrust axis. The
Minor axis is then fixed as the axis perpendicular to both the thrust and
Major axes, yet its value is computed exactly like thrust and Major
values. These definitions of the Major and Minor axes, 
used at $e^+e^-$ colliders, get into trouble at a
$pp$ machine like the LHC where variables sensitive
to boosts along the $z$-axis are of very limited use. At $pp$ machines, one
therefore defines the Major and Minor axes to lie orthogonal to the
$z$-direction. This means that the Major axis will most often just be the
projection of the thrust axis onto the $(x,y)$ plane.
A measure for how
``spread out'' the event is in the $(x,y)$ plane is then given by
subtracting the Minor from the Major, yielding the
\emph{oblateness}, $O$. 

\begin{figure}[t!]
\setlength{\extrarowheight}{0pt}
\begin{center}
\begin{tabular}{cc}
\includegraphics*[scale=0.36]{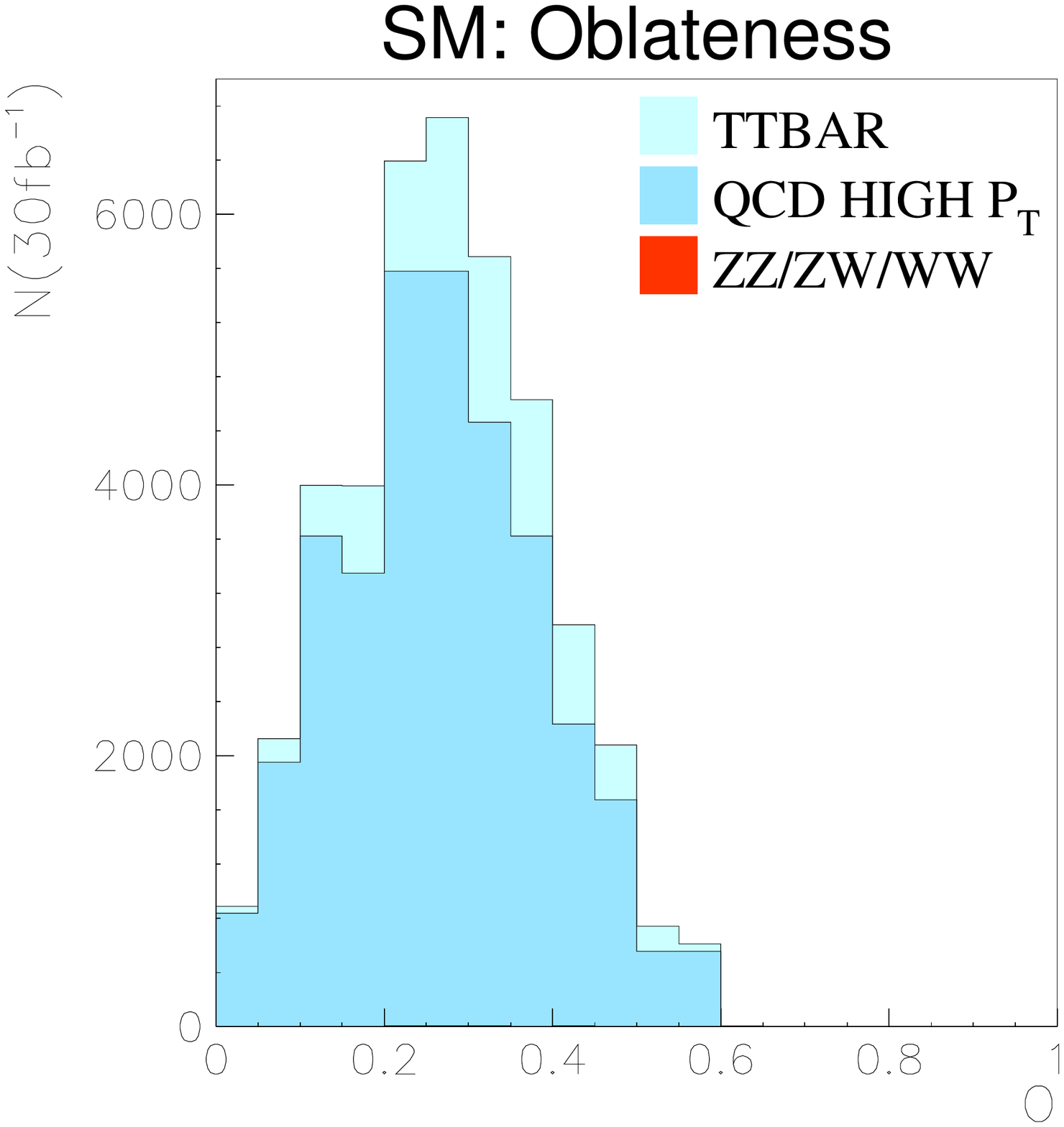}&\hspace*{-.7cm}
\includegraphics*[scale=0.36]{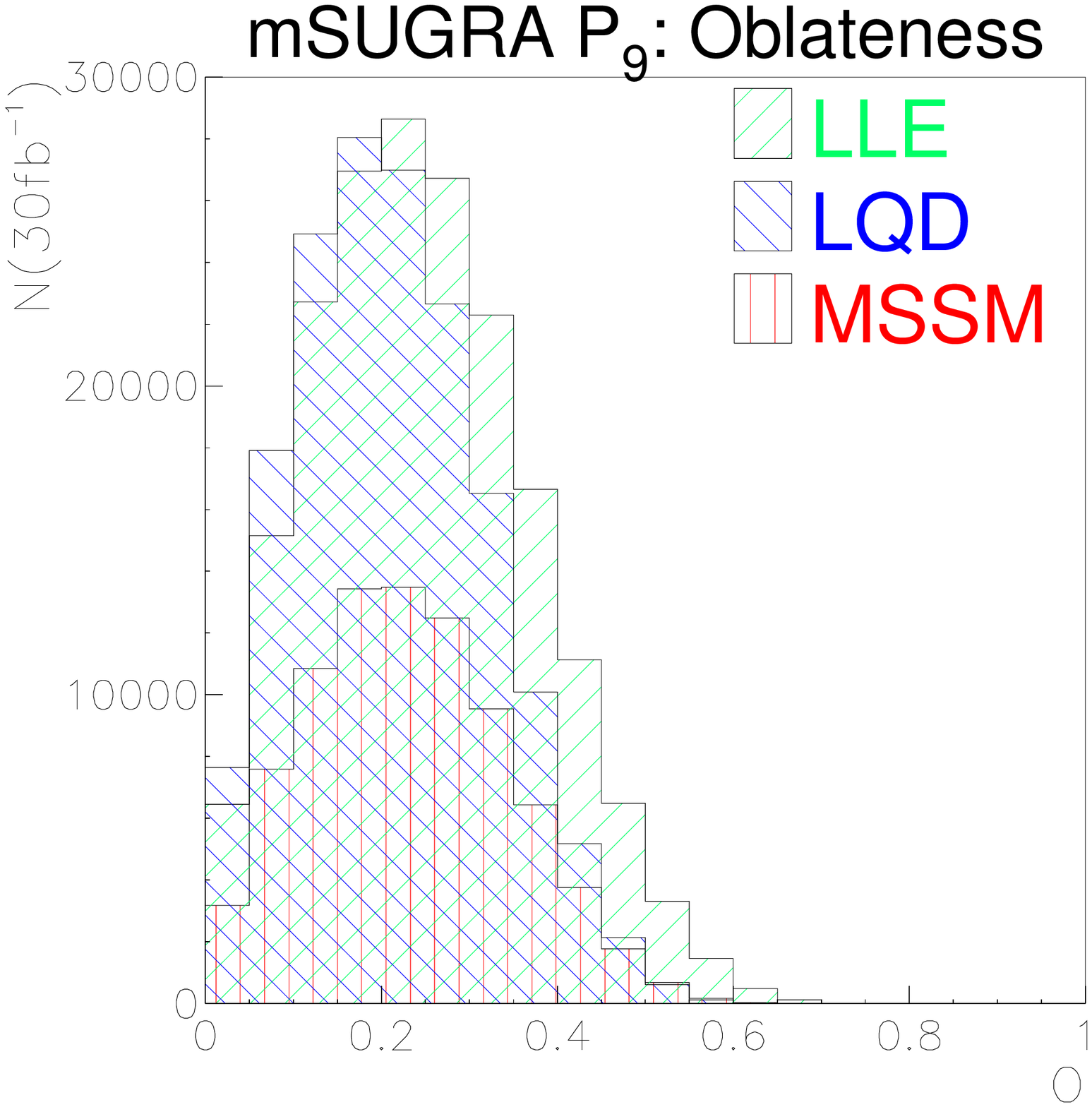}\vspace*{-6mm}\\
a) & b) \end{tabular}\vspace*{-8mm}\\
\includegraphics*[scale=0.7]{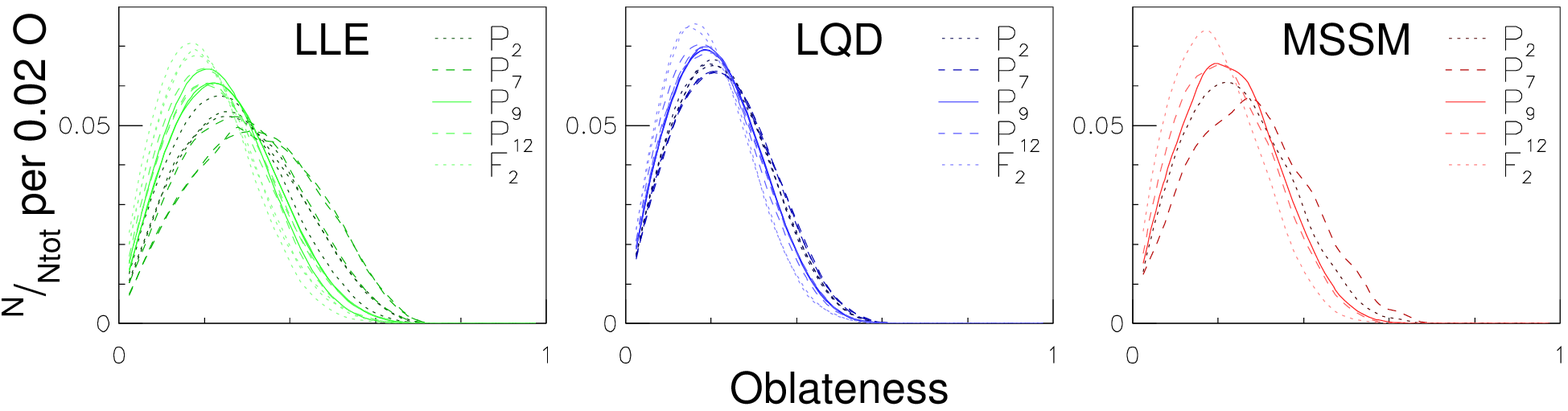}\vspace*{-.7cm}\\
c)\vspace*{-6mm}
\end{center}
\caption[\small Oblateness distributions for SM and SUSY.]{Oblateness distributions for 
SM processes and mSUGRA $P_9$. Observe that the QCD processes peak around 0.3
whereas $P_9$ peaks around 0.2. All events used survive previous cuts.
\label{fig:obl}}
\vspace*{-5mm}
\end{figure}

Essentially, the oblateness compares the fraction of $p_T$ in the direction
where most $p_T$ is going and the fraction in the 
direction where least $p_T$ is going.  
An event with low oblateness is an event in which the $p_T$ is evenly
distributed in the $(x,y)$ plane, and an event with high oblateness has its
$p_T$ concentrated around the direction of the Major axis.
Based on the same arguments that led us to introduce a $p_T$
dependence on the 4-object energy correlation, we expect SUSY events to have
lower oblateness values than SM events, again with a tradeoff involved
between signal loss at low SUSY masses and background rejection for higher
SUSY masses. Distributions are shown in figure \ref{fig:obl}.
The effect of 
requiring the oblateness to be lower than 0.25, 0.3, 0.4, and 0.5 were
investigated with results as given in table \ref{tab:oblat}.
\begin{table}[t!]
\setlength{\extrarowheight}{0pt}
\begin{center}
\textsf{EVENTS PASSING CUTS ON OBLATENESS.\vspace*{2mm}}
{\footnotesize
\begin{tabular}{lcccccc}\toprule
& SM & $P_9$ MSSM & $P_9$ LLE & $P_9$ nLLE & $P_9$ LQD & $P_9$ nLQD\\ 
\cmidrule{1-7}
$O<0.25$&$(17\pm2)$ k&49 k&100 k&100 k&110 k&110 k\\
$O<0.3$ &$(24\pm2)$ k&61 k&130 k&130 k&130 k&140 k\\\boldmath
$O<0.4$ &$(34\pm3)$ k&77 k&170 k&160 k&160 k&160 k\\
$O<0.5$ &$(40\pm3)$ k&83 k&180 k&180 k&160 k&170 k\\
\cmidrule{1-7}
$\frac{N_{\mbox{\tiny post}}}{N_{\mbox{\tiny pre}}}$ 
&0.84 &0.92&0.88&0.95&0.91&0.95\\
\bottomrule
\end{tabular}}
\vspace*{-5mm}
\end{center}
\caption[\small Event numbers passing cuts on oblateness.]{Events passing cuts on
oblateness, normalized to 30\fb$^{-1}$ of data taking. 
The selected cut is marked in bold, 
and the ratio of events surviving after this cut to events surviving before
this cut is shown for each model.\label{tab:oblat}} 
\vspace*{-\tfcapsep}\end{table}

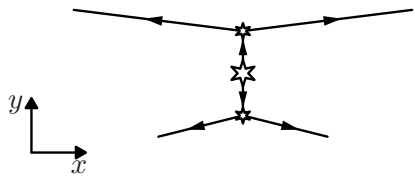
\begin{figure}[b!]
\begin{center}
\begin{fmffile}{p7dec}
\begin{fmfgraph*}(160,60)
\fmfset{arrow_len}{2.5mm}
\fmfforce{0.w,0.h}{origo}
\fmfforce{18.,0.}{xax}
\fmfforce{0.,18.}{yax}
\fmfforce{0.5w,0.5h}{c}
\fmfforce{0.1w,0.9h}{l1}
\fmfforce{0.3w,0.1h}{l3}
\fmfforce{0.9w,0.9h}{r1}
\fmfforce{0.7w,0.1h}{r3}
\fmf{plain}{origo,xax}
\fmf{plain}{origo,yax}
\fmf{fermion}{v1,l1}
\fmf{fermion}{v2,l3}
\fmf{fermion}{v1,r1}
\fmf{fermion}{v2,r3}
\fmf{fermion}{c,v1}
\fmf{fermion}{c,v2}
\fmfv{label=$x$,label.dist=3,label.ang=-90,d.sh=tri,d.siz=5.,d.ang=-90}{xax}
\fmfv{label=$y$,label.dist=3,label.ang=180,d.sh=tri,d.siz=5.}{yax}
\fmfv{d.sh=hexagram,d.f=empty,d.siz=10.}{c}
\fmfv{d.sh=hexagram,d.f=empty,d.siz=6.}{v1}
\fmfv{d.sh=hexagram,d.f=empty,d.siz=6.}{v2}
\end{fmfgraph*}
\end{fmffile}
\end{center}
\caption[\small Decay of a very massive resonance]{Decay of a pair of
extremely massive resonances, projected onto the $(x,y)$ momentum plane.
Here, the decay products of the upper resonance have their 
momenta oriented predominantly along the $x$ axis, whereas the lower
have larger fractions in the $z$ direction\label{fig:p7oblat}. In cases like
this, the $x$
axis is close to being the Major with only little $p_y$ to generate a sizeable
Minor.}
\end{figure}

A remarkable feature of the plots in figure \ref{fig:obl}c is that $P_7$,
by far the heaviest mSUGRA point, has oblateness distributions which are
\emph{broader} than the lower-mass points. To understand the reason for this,
a closer look was taken at the event histories for high-oblateness events in
an nLLE $P_7$ scenario. 
A plausible explanation is that the
extremely massive resonances in $P_7$ are
produced more often than not with $p_T$ less than their masses. 
When we argued above that we would expect 
low oblateness for events containing more massive resonances, 
we implicitly 
assumed that the original direction of flight of the resonances would 
define the Major axis around which a non-zero Minor value would be generated
by the decay of the resonances in proportion to their masses. 
But for resonances with masses as heavy as those
found in $P_7$ (around 2\TeV) this argument is turned upside down. 
At $P_7$, the (two-body) 
decays of the resonances produce momentum kicks of such large magnitudes that
the original direction of flight can be completely erased, resulting in
events looking roughly as depicted in figure \ref{fig:p7oblat} in the $(x,y)$
plane. 

\begin{figure}[t!]
\setlength{\extrarowheight}{0pt}
\begin{center}
\begin{tabular}{cc}
\includegraphics*[scale=0.36]{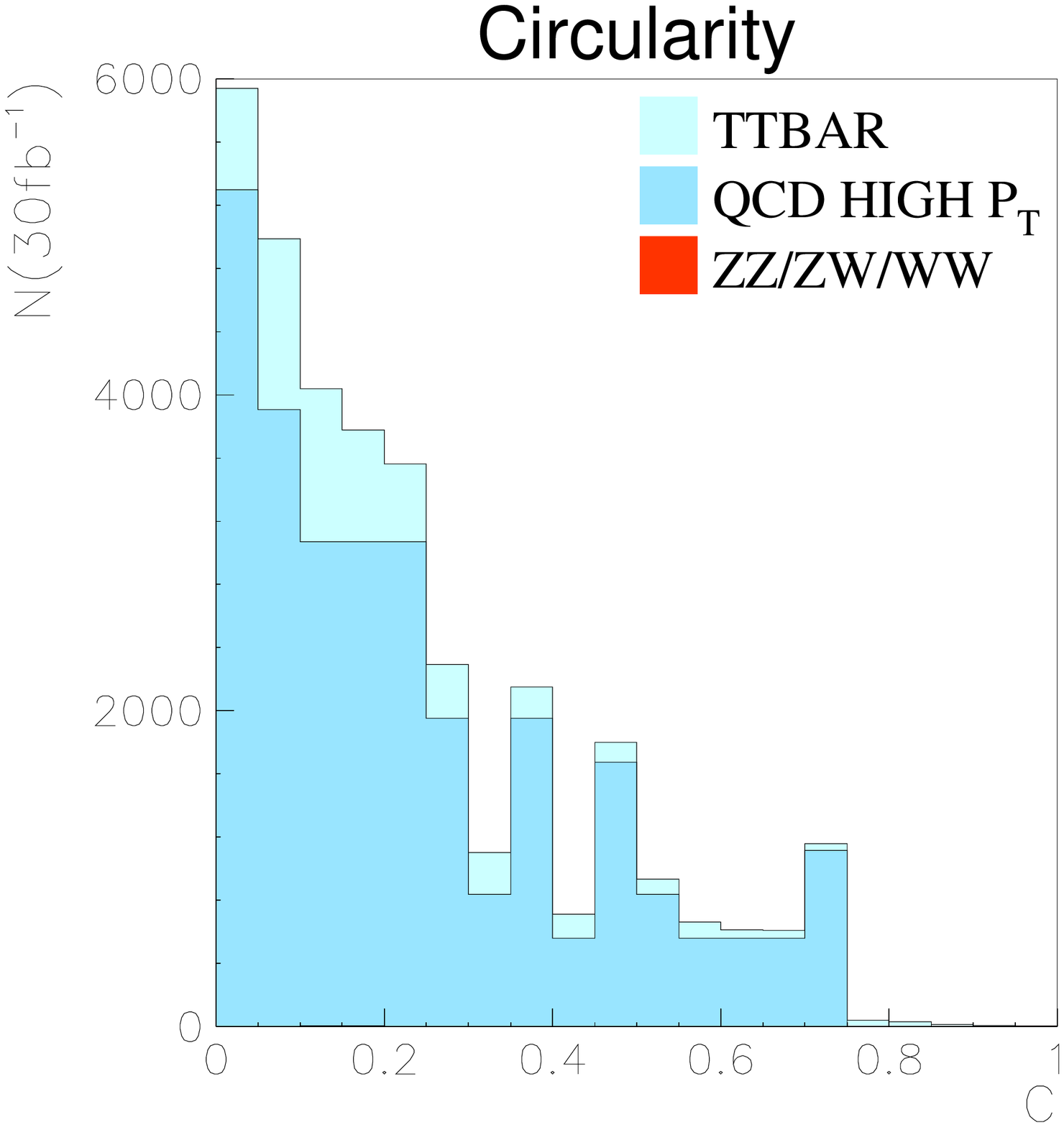}&\hspace*{-.7cm}
\includegraphics*[scale=0.36]{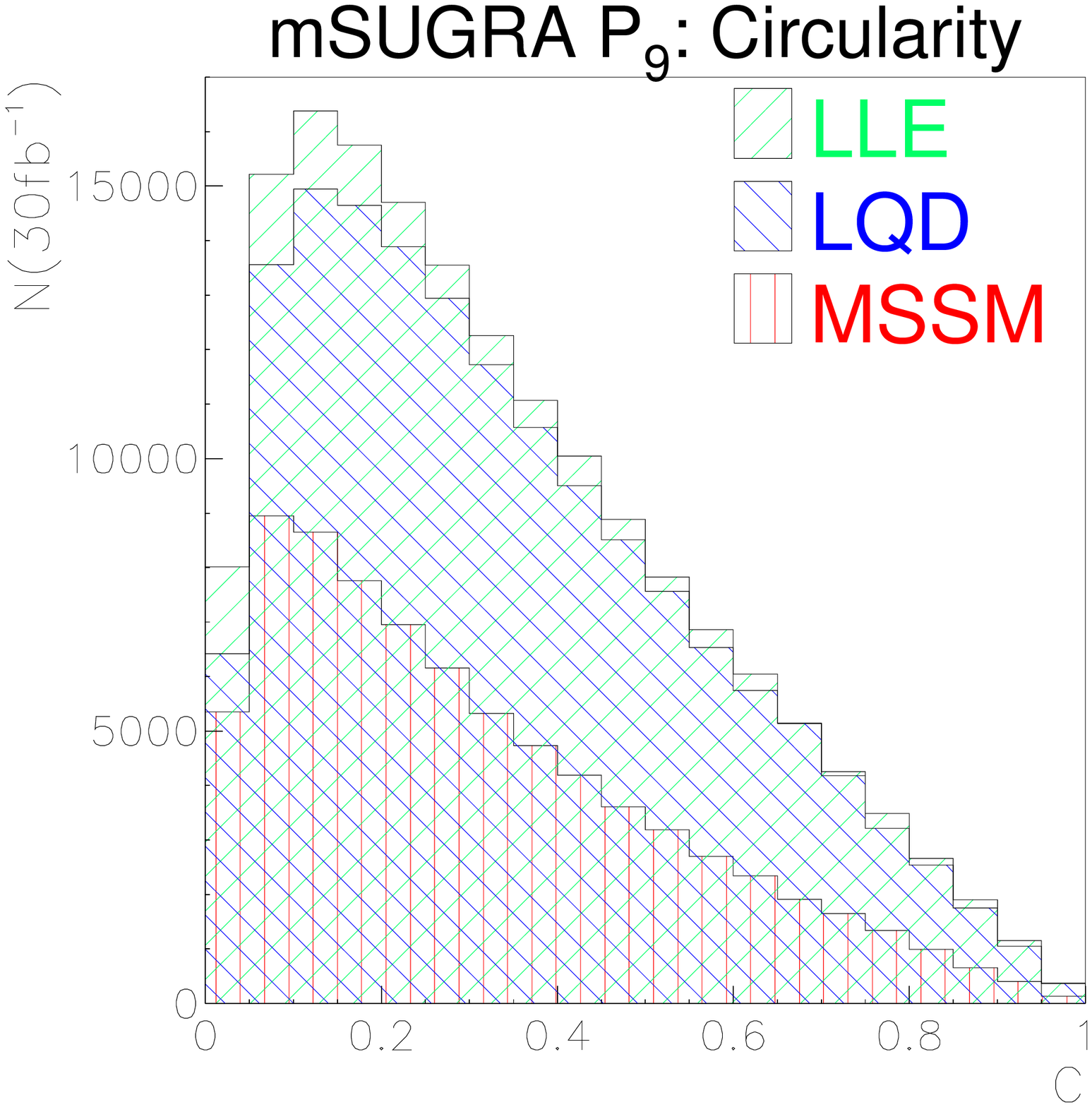}\vspace*{-6mm}\\
a) & b) \end{tabular}\vspace*{-8mm}\\
\includegraphics*[scale=0.7]{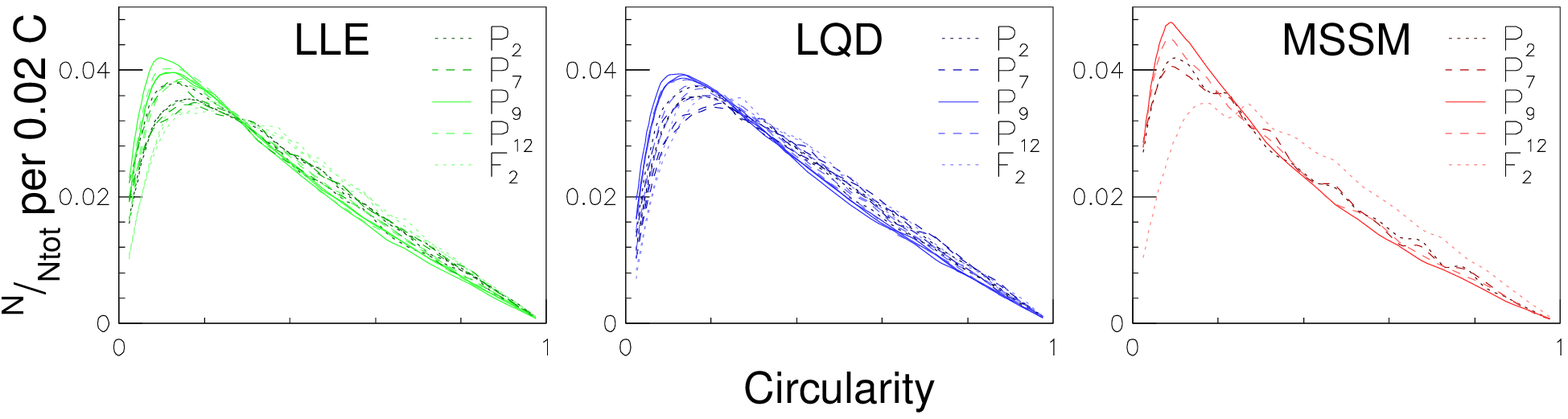}\vspace*{-.7cm}\\
c)\vspace*{-6mm}
\end{center}
\caption[\small Circularity distributions for SM and SUSY.]{Circularity
distributions for  
SM and mSUGRA scenarios. Observe that the number of QCD events rises almost
linearly with $C$ as one goes to smaller $C$ wheras the 
mSUGRA distributions peak around $0.1-0.2$. All events used survive previous
cuts. 
\label{fig:circ}}
\end{figure}

The last event shape variable used in this work is the circularity, again a
transverse version of a variable commonly used in $e^+e^-$ colliders, the
sphericity \cite{stirling96}. It is similar, but not identical, to
1 minus oblateness, in that events which have their $p_T$ evenly distributed
in the transverse plane have high circularities and events with more uneven
distributions low circularities. 
Having cut away the highest oblateness
events, we try to catch a few more fish (background events) by cutting away
the lowest circularity events as well. It is defined through the eigenvalues
of the circularity matrix:
\begin{equation}
C = \frac{1}{\sum_i (p_T^i)^2}\left[ \begin{array}{cc}
\displaystyle\sum_i (p_x^i)^2 & \displaystyle\sum_i p_x^ip_y^i\vspace*{2mm}\\
\displaystyle\sum_i p_y^ip_x^i & \displaystyle\sum_i (p_y^i)^2
\end{array}\right]  
\end{equation}
where $i$ runs over the reconstructed particles/jets in the event. The
Circularity value is defined as twice the smallest eigenvalue of this
matrix, making it a measure of the momentum fraction along the smaller of the
principal axes.
The distributions of events in the SM and SUSY surviving all previous cuts
can be seen in figure \ref{fig:circ}. Taking a look at the very first few
bins of the plots, one sees that circularity does indeed have
some discriminating power beyond what was contained in the oblateness though
it has not been possible to identify the cause. Cuts requiring a circularity
greater than 0.05, 0.1, 0.15, and 0.20 were investigated with results shown
in table
\ref{tab:circ}.  
\begin{table}[t!]
\setlength{\extrarowheight}{0pt}
\begin{center}
\textsf{EVENTS PASSING CUTS ON CIRCULARITY.\vspace*{2mm}}
{\footnotesize
\begin{tabular}{lcccccc}\toprule
& SM & $P_9$ MSSM & $P_9$ LLE & $P_9$ nLLE & $P_9$ LQD & $P_9$ nLQD\\ 
\cmidrule{1-7}
$C>0.05$&$(29\pm3)$ k&72 k&160 k&160 k&150 k&150 k\\\boldmath
$C>0.10$ &$(24\pm2)$ k&63 k&140 k&140 k&140 k&140 k\\
$C>0.15$ &$(20\pm2)$ k&54 k&130 k&120 k&120 k&120 k\\
$C>0.20$ &$(16\pm2)$ k&46 k&110 k&110 k&110 k&110 k\\
\cmidrule{1-7}
$\frac{N_{\mbox{\tiny post}}}{N_{\mbox{\tiny pre}}}$ 
&0.68 & 0.81 & 0.86 & 0.84 & 0.87 & 0.86\\ 
\bottomrule
\end{tabular}}
\vspace*{-5mm}
\end{center}
\caption[\small Event numbers passing cuts on circularity.]{
Events passing cuts on
circularity, normalized to 30\fb$^{-1}$ of data taking. 
The selected cut is marked in bold, 
and the ratio of events surviving after this cut to events surviving before
the cut is shown for each model. Note that only 70 high-$p_T$ QCD 
events remain after the cut, and so the
gaussian errors used here will be replaced by Poisson statistics in what
follows.
\label{tab:circ}} 
\vspace*{-\tfcapsep}\end{table}

A last thing worth noticing about the circularity is that whereas the thrust
and the oblateness are \emph{linear} in the momenta, circularity is
quadratic. This means that two events which are identical in every respect
except that one parton in the first event is replaced by two collinear partons
with some sharing of the momentum in the other 
will not have the same circularities, since
$(p_a + p_b)^2\neq p_a^2 + p_b^2$. This, in turn, means that the circularity
is sensitive to uncertainties in the fragmentation model (where such
splittings occur), in contrast to thrust and oblateness. 

The combined rejection factors under the oblateness and circularity cuts were
1.7 and 1.6 for the high-$p_T$ QCD and double gauge events respectively. We
thus estimate a maximum of 110 and 400
low-$p_T$ and single gauge events
remaining, respectively. These numbers 
should be compared to 4000 $t\bar{t}$ and 20000
high-$p_T$ QCD events. Based on the 22 generated double gauge events
remaining, an upper limit of 31.4 events can be set using the conservative 
Poisson estimate discussed in section \ref{sec:lepjets}, translating to a
maximum of 20 events after $30\fb^{-1}$ of data taking. One thus sees that
the double gauge events are completely negligible as background, 
giving some justification
of the earlier made statement that triple gauge events should also be 
negligible.
\subsection{Pattern Recognition with Neural Networks \label{sec:net}}
We now turn to the more specialized part of the analysis where we shall seek
to extract signals for LLE, LQD, and MSSM SUSY scenarios 
separately using three neural networks trained to regocnize the specific
event shapes associated with each scenario. The results of the analysis are
presented in section \ref{sec:results}. Here, we concentrate on the structure
and function of neural networks and their application to the present problem,
essentially one of pattern classification. Does this event ``look'' like a
Standard Model event, or does it look like a SUSY event?

\subsubsection{What Neural Networks Do} 
The first step in any analysis based on
cuts is to find as optimal variables as possible 
to place cuts on, the second step to find the
optimal \emph{placements} of the cuts. What neural networks do is 
to \emph{learn} 
which variables to use and where to place the cuts, based on a teaching 
sample of background and signal events. The simplest type of network consists
of a single neuron which computes a linear combination of the input variables
in the problem, in our case the discriminating variables just discussed. 
It then places a cut on this ``activation level'' and returns
1 if the activation was above the cut value and 0 otherwise. In our case,
these two outputs would  correspond to the event having been classified as
either a signal or a background event. The learning
algorithm then serves to adjust the coefficients in the linear combination
and the placement of the cut according to the average error the network makes
over the learning sample such that next time it goes over the sample it
will classify more events correctly. The way this works is by a procedure
called ``gradient descent'' where the network calculates
the gradient of the error squared, or some other function that one
wishes to minimize, with respect to each network parameter. It then 
adjusts each parameter, taking a small step in parameter space 
in the minimizing direction each time it has processed an event or, to
decrease the effect of insignificant
fluctuations, it sums up the required changes over a number of processed
events before it applies them. This latter approach smoothes out the
otherwise jittering movement of the network across parameter space, often 
allowing faster progress towards the minimum. 

For gradient descent to work, note that the neuron cannot be allowed to
compute a sharp cut on it its activation level, since the step function 
is discontinuous and hence not
differentiable. Instead, one uses so-called sigmoidal functions which look
like smoothed out versions of the step function. We consider these
functions and the gradient descent algorithm in more detail below, yet let us
first extend our network beyond just a single neuron. 

In problems where the classification is
not quite so easy that it can be performed using a cut on just one linear
combination of the inputs, more neurons are needed, each one computing a 
sigmoidal function of its 
activation level, resulting in an output from each of these ``cut neurons'' 
between zero and one. These outputs 
then serve as inputs to the output neuron who sums them up in
a new linear combination, the output neuron activation. In the present case,
this activation is used directly as the network output, alternatively one may
let the output neuron compute a function of its input. It remains that the
computing power of the network lies in the cut neurons. A function may or may
not be handy to apply to the output, but it will not increase the amount of
information there is in the output value. It now also becomes apparent why
the cut neurons are customarily 
referred to as hidden neurons. The world outside the
network interacts with it by giving it input on the input neurons and by
reading the output from the output neuron. The cut neurons 
communicate only with other neurons. Henceforth, we refer to these internal
neurons as hidden neurons. 

The function of a neural network is thus nothing but 
a number of smoothed 
cuts on the same number of linear combinations of the inputs, with the
results of the cuts being used as variables in a last linear combination
defining the output of the network -- similar to what is being done
in an ordinary cut-based analysis. The benefit is that neural networks
automatically pick up correlations and anti-correlations between arbitrarily
many of the input variables. A hypothetical example of high-dimensional
correlations would be if we
imagine that many signal events have high jet multiplicities
when there is little \ET\ in the event, but that they have very few jets when
\ET\ is high. Furthermore, let us suppose that, at high \ET\, a certain
fraction of signal
events with few jets have high lepton multiplicities, but that at low
\ET\ high lepton multiplicities would be a characteristic of background
events, unless there was also a high thrust in the event, or failing that at
least a high oblateness. These
correlations would of course follow from physical arguments related to the
processes involved in shaping the hypothetical 
background and signal processes in this example, and linear combinations of
the variables designed to make use of the correlations could be constructed and
optimized manually, yet this would be an extremely 
time-consuming task considering the more than 50 different scenarios
investigated in this work. Moreover, it is a task which neural networks are
ideally suited for by construction.

\subsubsection{Network Layout and Network Learning:}
\begin{figure}[b]
\begin{center}
\includegraphics*[scale=0.6]{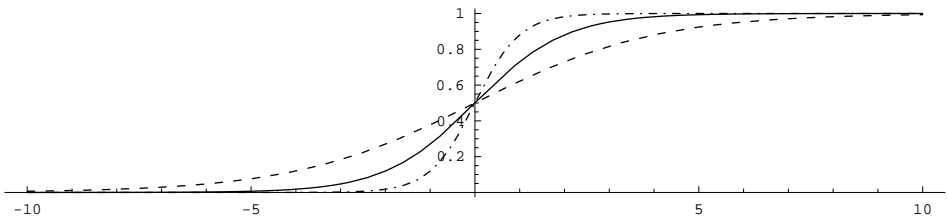}
\caption[\small The logistic function]{The logistic (sigmoid) function for
$T=2$ (dashed), $T=1$ (solid), and $T=0.5$ (dot-dashed). \label{fig:sigmoid}}
\end{center}
\end{figure}

\begin{figure}[h!]
\vspace*{5mm}
\begin{fmffile}{neural}
\begin{fmfgraph*}(320,150)
\fmfset{arrow_len}{2mm}
\fmftop{i1,i2,i3,i4,i5,i6,i7}
\fmfbottom{o}
\fmfforce{0.1w,0.43h}{v1}
\fmfforce{0.25w,0.43h}{v2}
\fmfforce{0.41w,0.43h}{v3}
\fmfforce{0.5w,0.43h}{v4}
\fmfforce{0.61w,0.43h}{v5}
\fmfforce{0.8w,0.43h}{v6}
\fmfforce{0.95w,0.7h}{wij}
\fmfforce{0.93w,0.43h}{tm}
\fmfforce{0.95w,0.1h}{oj}
\fmfforce{0.93w,0h}{tl}
\fmfforce{0.95w,1h}{tr}
\fmfv{label=$I_1$,d.sh=circ,d.siz=0.08w,d.fill=empty,foreground=(0.7,,0,,0),label.dist=0}{i1}
\fmfv{label=$I_2$,d.sh=circ,d.siz=0.08w,d.fill=empty,foreground=(0.8,,0.4,,0),label.dist=0}{i2}
\fmfv{label=\large\boldmath$...$}{i3}
\fmfv{label=$I_i$,d.sh=circ,d.siz=0.08w,d.fill=empty,foreground=(0.7,,0.7,,0),label.dist=0}{i4}
\fmfv{label=\large\boldmath$...$}{i5}
\fmfv{label=$I_{N_{\mathrm{in}}}$,d.sh=circ,d.siz=0.08w,d.fill=empty,foreground=(0.5,,0.7,,0),label.dist=0}{i6}
\fmfv{label=$H_1$,d.sh=square,d.siz=0.075w,d.fill=empty,foreground=(0,,0.75,,0),label.dist=0}{v1}
\fmfv{label=$H_2$,d.sh=square,d.siz=0.075w,d.fill=empty,foreground=(0,,0.7,,0.5),label.dist=0}{v2}
\fmfv{label=\large\boldmath$...$}{v3}
\fmfv{label=$H_j$,d.sh=square,d.siz=0.075w,label.ang=-30,d.fill=empty,foreground=(0,,0.5,,0.8),label.dist=0}{v4}
\fmfv{label=\large\boldmath$...$}{v5}
\fmfv{label=$H_{\!N_{\mathrm{hid}}}$,d.sh=square,d.siz=0.075w,d.fill=empty,foreground=(0,,0,,0.9),label.dist=0}{v6}
\fmfv{label=$\mathcal{O}$,d.sh=circ,d.siz=0.08w,d.fill=empty,foreground=black,label.dist=0}{o}
\fmf{fermion,right=0.3,foreground=(0.7,,0,,0)}{i1,v1}
\fmf{fermion,right=0.1,foreground=(0.7,,0,,0)}{i1,v2}
\fmf{fermion,foreground=(0.7,,0,,0)}{i1,v4}
\fmf{fermion,left=0.08,foreground=(0.7,,0,,0)}{i1,v6}
\fmf{fermion,right=0.3,foreground=(0.8,,0.4,,0)}{i2,v1}
\fmf{fermion,right=0.3,foreground=(0.8,,0.4,,0)}{i2,v2}
\fmf{fermion,left=0.05,foreground=(0.8,,0.4,,0)}{i2,v4}
\fmf{fermion,left=0.15,foreground=(0.8,,0.4,,0)}{i2,v6}
\fmf{fermion,right=0.3,foreground=(0.7,,0.7,,0)}{i4,v1}
\fmf{fermion,right=0.3,foreground=(0.7,,0.7,,0)}{i4,v2}
\fmf{fermion,foreground=(0.7,,0.7,,0)}{i4,v4}
\fmf{fermion,left=0.2,foreground=(0.7,,0.7,,0)}{i4,v6}
\fmf{fermion,right=0.2,foreground=(0.5,,0.7,,0)}{i6,v1}
\fmf{fermion,right=0.1,foreground=(0.5,,0.7,,0)}{i6,v2}
\fmf{fermion,left=0.05,foreground=(0.5,,0.7,,0)}{i6,v4}
\fmf{fermion,left=0.2,foreground=(0.5,,0.7,,0),label=$w_{ij}$,label.side=left}{i6,v6}
\fmf{fermion,right=0.25,foreground=(0,,0.75,,0)}{v1,o}
\fmf{fermion,right=0.1,foreground=(0,,0.7,,0.5)}{v2,o}
\fmf{fermion,foreground=(0,,0.5,,0.8)}{v4,o}
\fmf{fermion,left=0.15,foreground=(0,,0,,0.9),label=$o_j$,label.side=left}{v6,o}
\fmfv{label=$I_i = \mathrm{In}_i - \delta^I_i$}{tr}
\fmfv{label=$H_j = \frac{1}{1+\exp(\sum_i I_i w_{ij}-\delta^H_j)}$}{tm}
\fmfv{label=$\mathcal{O}=\frac{1}{N_{\mathrm{hid}}}\left(\sum_j H_jo_j - \delta^O\right)$}{tl}
\end{fmfgraph*}
\end{fmffile}
\vspace*{6mm}
\caption[\small Layout of the neural networks]{Layout of the neural
networks used in the final part of the analysis (before brain-damage -- see
below).  
\label{fig:netlayout}}
\end{figure}
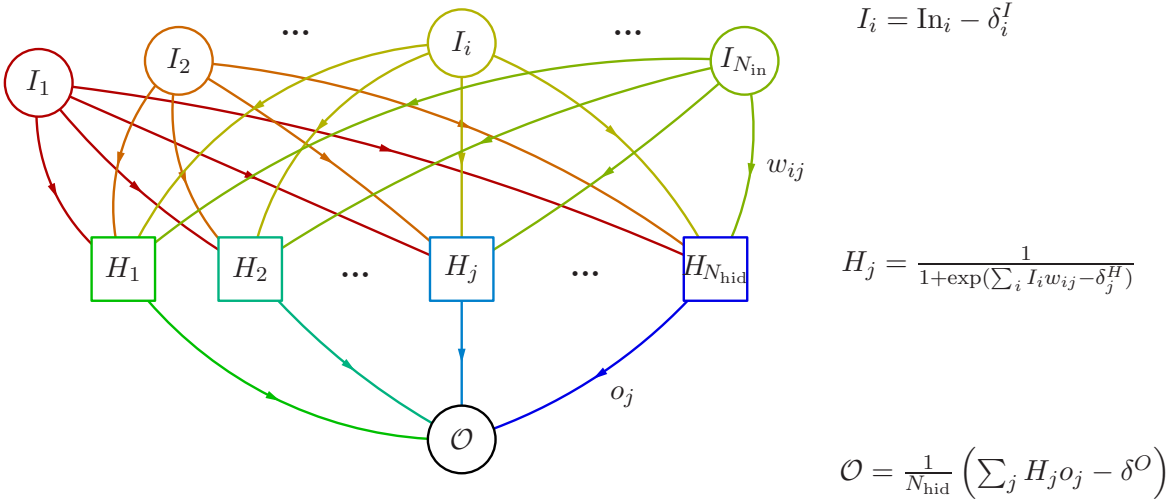

For each event to be processed, 
each of the discriminating variables defined above are presented as inputs to
the network. Since the network is initizalized with random weights between
0 and 1, it is sensible to scale these inputs to typically 
lie in the range $[0,1]$ as well for faster learning. Otherwise, 
the input-to-hidden weights (the coefficients in the
linear combination mentioned above)
have to be corrected, possibly for a long time, until 
the right ball-park is found.
The input normalizations used here are listed in table \ref{tab:inputnorms}. 
\begin{table}[tb]
\begin{center}
\setlength{\extrarowheight}{0pt}
\begin{tabular}{cccccccc}\toprule
\boldmath$i$    & \bf1 & \bf2 & \bf3 & \bf4 & \bf5 & \bf6 & \bf7
\\
\boldmath$\mathrm{In}_{i}$&$\displaystyle\frac{\ETs}{200}$&$\displaystyle\frac{N_{\mathrm{jets}}}{15}$&
$\displaystyle\frac{N_{\mu}^{\mathrm{iso}}}{5}$&$\displaystyle\frac{N_e^{\mathrm{iso}}}{5}$& 
$\displaystyle\frac{P_{4C}}{500}$&\textup{Thrust}&\textup{Circularity}\\
\cmidrule{1-8}
\boldmath$i$&\bf8&\bf9&\bf10&\bf11&\bf12&\bf13&\bf14
\\\boldmath$\mathrm{In}_i$ & Oblateness
&$\displaystyle\frac{p_{T,\mathrm{jet}}^1}{100}$&$\displaystyle\frac{p_{,T\mathrm{jet}}^2}{100}$&$\displaystyle\frac{p_{T,\mathrm{jet}}^3}{100}$&$\displaystyle\frac{p_{T,\mathrm{jet}}^4}{100}$&$\displaystyle\frac{p_{T,\ell}^1}{100}$&$\displaystyle\frac{p_{T,\ell}^2}{100}$\\\bottomrule
\end{tabular}
\caption[\small Inputs to the neural net]{Inputs to the neural network and
their normalizations. In the text, $i$ is used as an index denoting 
input neurons and 
$\mathrm{In}_i$ the value of the input variable as given in this table. Note
that $\mathrm{In}_i$ is 
not necessarily identical to the
output of the input neuron which we denote by $I_i$. 
$P_{4C}$ is defined in section \ref{sec:lspdecsig},
$p_{T\mathrm{jet}}^{1-4}$ are the transverse momenta of the four
hardest jets, and $p_{T,\ell}^{1-2}$ of the two hardest
leptons.\label{tab:inputnorms}} 
\end{center}
\vspace*{-\tfcapsep}\end{table} 
The hidden layer in most applied networks 
normally has fewer neurons than the input layer, 
representing that some generalization can already be made at this stage: it
is not always necessary to form $N$ linear combinations of $N$ variables since
some mutual interdependence can usually be eliminated. In
the present analysis with 14 inputs, 
it is found that a network with 10 hidden neurons
performs with negligible loss of discriminating power compared to networks
with more hidden neurons. As described above, each hidden neuron computes a
sigmoidal of its activation level, the name sigmoidal coming from the tilted
$S$ shape of these functions. The
particular sigmoidal used in this work is the logistic function (the most
commonly used). This function assigns an
output value for the $j$'th neuron in the hidden layer of:
\begin{equation}\vspace*{2mm}
H_j=\frac{1}{1+e^{-\sum_{i=1}^{N_{\mathrm{in}}} 
(I_i w_{ij}) - \delta^H_j }}\label{eq:logistic}\vspace*{1mm}
\end{equation}
where $I_i$ is the output of the $i$'th input neuron, 
$w_{ij}$ is the weight of the synapse
connecting input $i$ to hidden neuron $j$, $\delta^H_j$ is a bias term for the
hidden neuron, and $N_{\mathrm{in}}$ is the number of neurons in the input
layer. Henceforth, we follow the convention that subscript $i$ refers to the
input layer whereas subscript $j$ refers to the hidden layer. The slope of
the sigmoid is sometimes also adjusted by introducing a ``temperature'', $T$:
\begin{equation}
\vspace*{2mm}H_j=\frac{1}{1+e^{-\left(\sum_{i=1}^{N_{\mathrm{in}}} 
(I_i w_{ij}) - \delta^H_j\right)/T_j }}\vspace*{1mm}
\end{equation}
The effect of this modification is shown in figure \ref{fig:sigmoid}.
However, since introducing a temperature different from unity 
simply corresponds to rescaling all the weights connecting to $H_j$ and the
bias by a common factor $1/T_j$, there is nothing gained by introducing such a
parameter. Moreover, the network becomes slower and there is the risk that it
begins to oscillate between changing $T_j$ and rescaling the weights in the
learning procedure, and so we stick with eq.~(\ref{eq:logistic}). 
For the input neurons, only a bias is added to the value of the input variables:
\begin{equation}
I_i=\mathrm{In}_i - \delta^I_i
\end{equation}
Taking one more look at figure \ref{fig:sigmoid} one also sees the reality of
the earlier made comment that these functions look like smoothed out step
functions and so can be regarded as smooth versions of cuts. 
The complete network layout looks as depicted in figure
\ref{fig:netlayout}. 
\noindent 

\subsubsection{Network Training}
The neural networks used in this work all learn by adjusting their
synaptic weights and biases 
in the direction which minimizes the error squared of the network, the error
being defined on an event-by-event basis by:
\begin{equation}
e = t - \mathcal{O}
\end{equation} 
where $t$ is the target output (0 for background, 1 for signal), 
and $\mathcal{O}$ is the output that was actually obtained for the event.

As the name implies, the \emph{gradient descent} learning 
algorithm \cite{mcclelland89}
consists of a learning rule defined by taking a step in weight space in the
direction of the gradient of the (squared) error:
\begin{equation}
p \to p + \Delta p = p - \alpha\frac{\partial e^2}{\partial p} = p + 2\alpha e
\frac{\partial \mathcal{O}}{\partial p}
\end{equation} 
where $p$ represents any of the adjustable parameters in the network, 
$e$ is the error on the output, and $\alpha$ is a learning
rate parameter specifying how large steps the network takes.
This rule does not necessarily have to be applied immediately after each
event in the learning sample has been processed. 
In order to wash out sharp, irregular 
changes called for from
extreme events in the distributions, possibly pulling in opposite directions,
we do better in accumulating the parameter changes over some processing
period before invoking them. 
After the period, the \emph{accumulated} change to each
parameter is invoked, and the accumulated error is reset. 
In order to get as little statistical noise in
the parameter changes as possible, the period was here set to be the number of
events in the learning sample. Denoting event number by $n$,
the learning rule above becomes:
\begin{equation}
p \to p + \Delta p = p + 2\alpha \sum_ne(n)
\frac{\partial \mathcal{O}(n)}{\partial p}
\end{equation} 
One further improvement can be made on the learning rule. Seeing as two
successive changes to the parameters of the network 
often go in roughly the same direction in parameter
space, we add a mixture of the last parameter change to the current change, 
something which often increases the learning speed of the network:
\begin{equation}
p \to p + \Delta p = p + 2\alpha \sum_ne(n)
\frac{\partial \mathcal{O}(n)}{\partial p} + \beta \Delta p^{\mathrm{last}}
\label{eq:learnrule}
\end{equation} 
where $\Delta p^{\mathrm{last}}$ is the change which was made after the previous
period, and $\beta<1$ specifies the ``inertia'' of the system. 
We are now ready to specify the learning rules for each of the parameters of
the network in fig.~\ref{fig:netlayout}. 
These can be easily derived (using the chain rule). 
\begin{equation}
\begin{array}{rclcrcl}\displaystyle
\frac{\partial \mathcal{O}}{\partial \delta^O} 
& = & \displaystyle-1/N_{\mathrm{hid}} & & \displaystyle\frac{\partial \mathcal{O}}{\partial o_j} 
& = & \displaystyle\frac{H_j}{N_{\mathrm{hid}}}
 \vspace*{3mm}\\ 
\displaystyle\frac{\partial \mathcal{O}}{\partial \delta^H_j} & = &
\displaystyle o_jH_j(H_j-1) & & \displaystyle\frac{\partial
  \mathcal{O}}{\partial w_{ij}} & = &
\displaystyle\frac{o_j}{N_{\mathrm{hid}}}  
H_j(1-H_j)I_i \vspace*{3mm}\\ \displaystyle
\frac{\partial \mathcal{O}}{\partial \delta_i^I} & = &\displaystyle
I_i(1-I_i)\sum_j o_jH_j(H_j-1)w_{ij} 
 & \hspace*{5mm} & \\
\end{array}
\end{equation}
Replacing these quantities back into eq.~(\ref{eq:learnrule}) directly 
yields the required learning rules. 
\subsubsection{Optimal Brain Damage}
\begin{figure}[t]
\begin{center}
\input{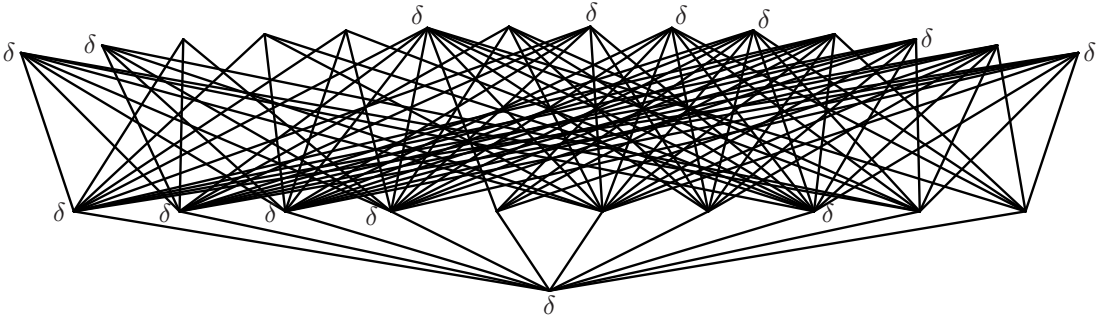}
\caption[\small Sketch of the MSSM network]{Sketch of the MSSM
network after brain damage. 
Neurons with biases
still on are shown with the symbol $\delta$. The numbering of the inputs 
goes from left to right.\label{fig:braindamage}}
\end{center}
\end{figure}

As described above, the first derivates of the squared error with respect to
the network parameters are used in training by gradient descent. 
It was shown by LeCun, Solla, and Denker 
\cite{lecun90} that the \emph{second} derivatives can be used to
trim the network by getting rid of the most redundant parameters. This is
desireable since
redundant parameters are the ones that will eventually cause the network to
overfit the sample space. This can have a severe effect on the
generalizational ability of the network, i.e.\ its performance on data
samples it has not been in contact with during the learning process. 
The idea
of Optimal Brain Damage
is to introduce a measure for how much the squared error will change as a
result of deleting each network parameter. The parameters whose deletion will
have the least effect can then be discarded if over-fitting is a problem. The
measure proposed in \cite{lecun90} is the \emph{saliency}, defined for the
network parameter $p$ as:
\begin{equation}
s_p = \frac{p^2}{2}\frac{\partial^2 e^2}{\partial p^2} 
\end{equation}
This definition is appropriate when the learning process is near its end and
the network is almost in the minimum (otherwise a more general formula would
apply. See \cite{lecun90}). The procedure is quite simple. One begins with a
network that contains many parameters. One then trains it and deletes the
parameters with lowest saliency. 
The resulting, ``brain-damaged'', network is then retrained
until it converges. This procedure is repeated until a satisfactory 
compromise between the mean squared error and the generalizational ability of
the network is found. As a side benefit, the finished 
network contains fewer parameters and is therefore faster to run. What
happens is typically that the smallest coefficients in the linear
combinations forming the hidden neuron activations get thrown away while the
larger coefficients are kept.
The diagonal second
derivates of the parameters for the present network are:
\begin{eqnarray}
\frac{\partial^2 e^2}{\partial (\delta^I_i)^2} & = & \frac{2}{N_{\mathrm{hid}}}
  \left[\left(\sum_{j=1}^{N_{\mathrm{hid}}}\!w_{ij}H_jo_j(1\!-\!H_j)\right)^2\!\! - e
  \left(\sum_{j=1}^{N_{\mathrm{hid}}}\!w_{ij}^2o_jH_j(1\!-\!H_j(3\!-\!2H_j))\right)\right]\\
\frac{\partial^2 e^2}{\partial w_{ij}^2} & = &
\frac{2}{N_{\mathrm{hid}}}I_i^2H_jo_j\left(o_jH_j(1-H_j)^2+e(1-H_j(3-2H_j))\right)\\
\frac{\partial^2 e^2}{\partial (\delta^H_j)^2} & = & \frac{2}{N_{\mathrm{hid}}}H_jo_j\left(o_jH_j(1-H_j)^2+e(1-H_j(3-2H_j)))\right)\\
\frac{\partial^2 e^2}{\partial o_j^2} & = & \frac{2H_j^2}{N_{\mathrm{hid}}}\\
\frac{\partial^2 e^2}{\partial (\delta^O)^2} & = & \frac{2}{N_{\mathrm{hid}}}
\end{eqnarray}
Each of the networks used thus started out with 14 inputs (= 14 biases), 
10 hidden units (= 10 biases), 
140 input-to-hidden synapses, 10 hidden-to-output synapses, and a bias on the
output for a total of 175 parameters. Approximately a quarter of 
the parameters can
be deleted with very little effect on the mean squared error of the networks
over the learning samples, but with an increase in convergence rate and
processing speed. 
An example of a finished, brain-damaged network, the one used for MSSM
recognition, is illustrated in figure \ref{fig:braindamage}. 

To improve the convergence rate, two further improvements were made in the
design. Firstly, a fixed learning rate is not optimal since there may be
regions in ``error space'' which look like endless plateaus to one whose legs
are not long enough. The network was therefore equipped with the ability to
increase its learning rate if the relative change in the squared 
error (averaged over the learning sample) after a learning cycle 
is small. A little experimenting showed an error change of less than 
5\% from cycle to cycle to be a good indicator of when a larger learning rate
was needed. 
Secondly, unlimited growth is not desireable. At the other side of
a plateau, a mountaneous region may yet exist, and so a mechanism to decrease
the rate is also included. If the error 
change is greater than zero, meaning
that the error \emph{increases}, the network ``concludes'' that it has taken
too big a step, unlearns the direct effects of
the last learning step (``direct'' here meaning that the effects of the momentum
term are not unlearned) 
and goes forward again with a smaller learning rate. After some more
experimenting to reach a good balance between increase and decrease, this
technique proved highly efficient, typically improving the convergence rate
by factors of ten.

\subsection{Results\label{sec:results}}
\begin{figure}[t!]
\begin{center}
\begin{tabular}{cc}
\includegraphics*[scale=0.37]{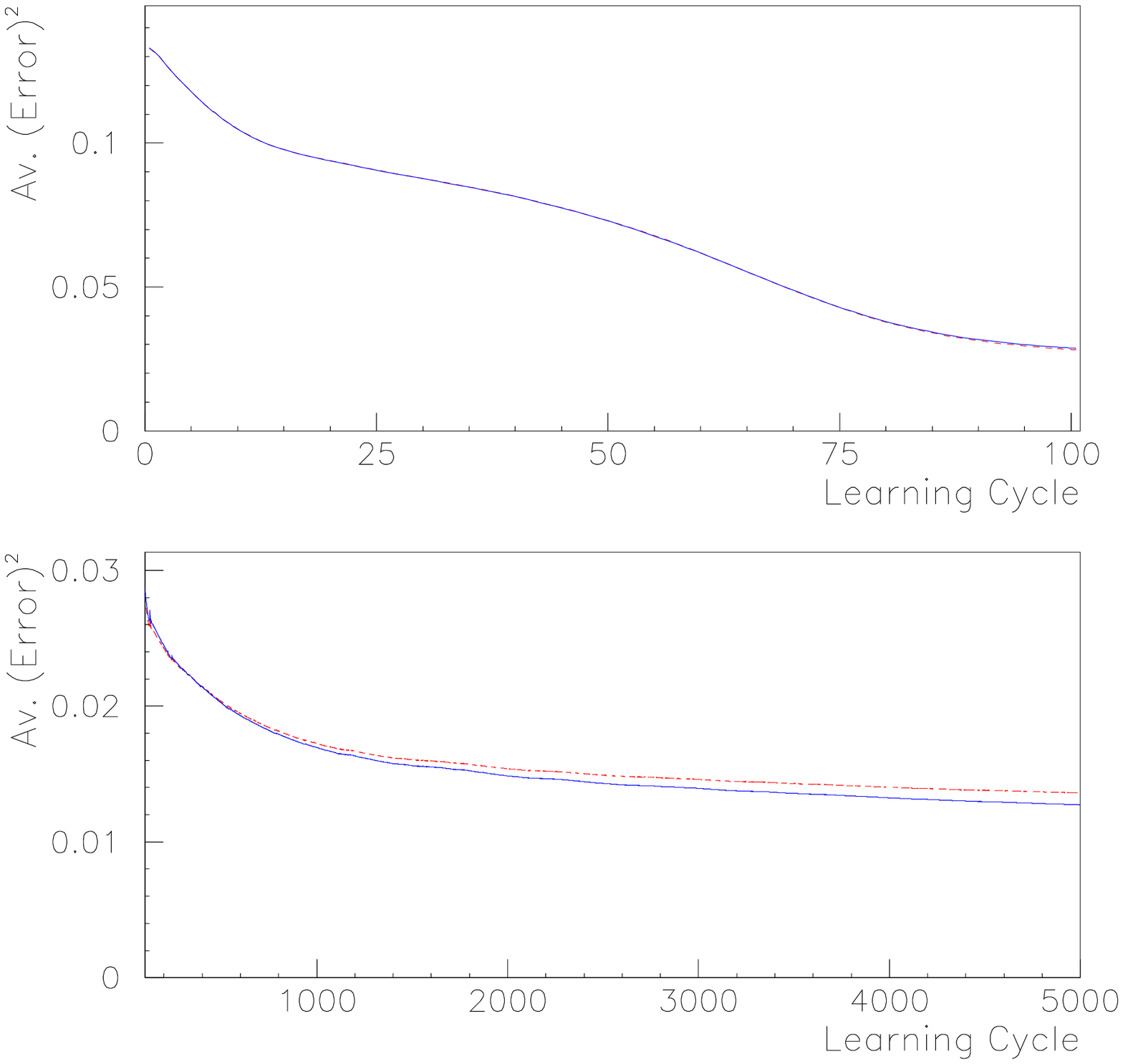} &
\includegraphics*[scale=0.54]{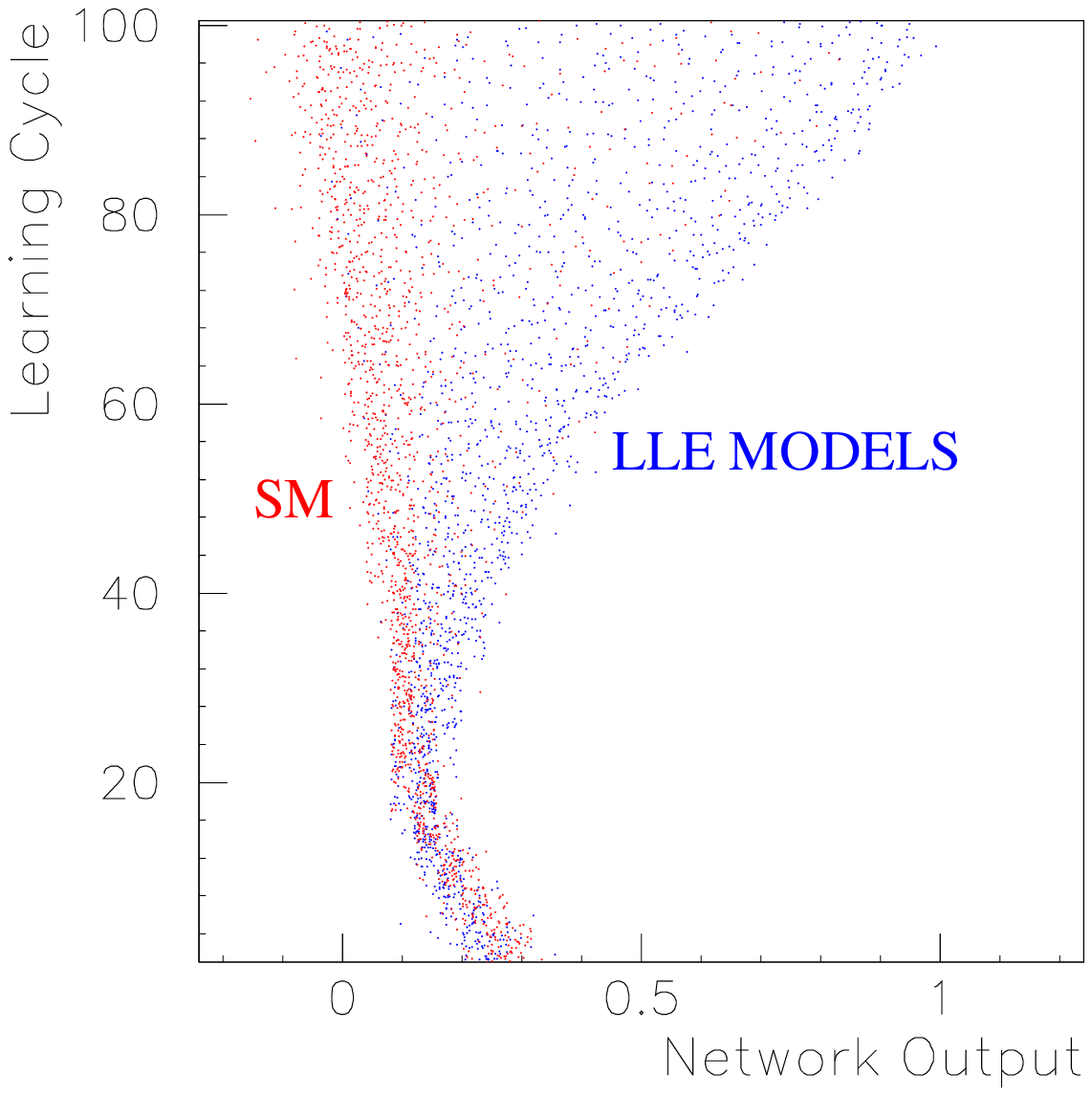}\vspace*{-8mm}\\
a) & b)\end{tabular}
\caption[\small Learning Curves for the LLE network]{a) Learning Curves for the
LLE network, average squared error versus learning cycle. 
The blue curve represents the average squared error on the
learning sample and the red (dashed) 
the performance on the independent sample. b)
learning cycle versus network output. Each red point represents an SM event
and each blue point a SUSY event. In the beginning, the network is not capable
of distinguishing between them, yet it quickly learns to separate the two
distributions. 
\label{fig:learncurve}}
\end{center}
\vspace*{-14pt}
\end{figure}

As the last item on the agenda, results for signal extraction in all \LV-SUSY
scenarios studied are now presented. For this purpose, three networks were
constructed and trained to recognize MSSM, LLE, and LQD signals
respectively. Approximately 8000 background events, scaled to represent 
approximately
$\tn{7}$ events in
the learning algorithm, and 6000 signal events were used in each
training process, all events required only to have passed the triggers. The
relative numbers of signal and background events, as seen by the learning
algorithm, are thus fairly similar to the post-trigger event numbers expected
for SUSY cross sections around \tn{-10}\mb. For the LQD network, 
the background numbers were scaled by twice
as much to obtain a better rejection factor in view of the less clean
signatures of the LQD couplings. 
For each sample, half of the events
were set aside as an
independent sample on which the performance of the network was tested for
over-fitting cycle by cycle. 
A typical learning curve is shown in figure \ref{fig:learncurve},
together with the gradually improving separation of the two distributions
early on in the learning process (only discernible in colour). 

Selecting for each network events with network outputs above 0.9,
we obtain a number of signal and background events for which we define 
the statistical significance of the signal (which we shall call the discovery
potential) by:
\begin{equation}
P = \frac{S}{\sqrt{S+B}}\label{eq:P}
\end{equation}
where $B$ should be understood as the number of background events, $N$,
coming out of the analysis plus 1.64 standard deviations to account for the
statistical uncertainty related to the limited number of generated events,
1.64 being chosen since 95\% of a gaussian distributed event sample will lie
below the mean plus 1.64$\sigma$.
For low numbers of generated events, we use the conservative Poisson estimate
discussed in section \ref{sec:lepjets}
to reach the same 95\% confidence on $B$. 
If $P$ is above 5 for any of the networks, 
we draw the preliminary conclusion that a 5$\sigma$ discovery is
possible. 

In reality, $P$ should be corrected for QCD uncertainties and 
the effects of pile-up, and so we can only be confident that a $5\sigma$
discovery is possible if $P$ is somewhat larger than 5. 
Therefore, aside from working with the definition, eq.~(\ref{eq:P}), 
we attempt to obtain a more
believable estimate by including the effects of pile-up and QCD uncertainties
in a very crude, \emph{ad hoc} manner. To accomplish this,
we now take a look at why eq.~(\ref{eq:P}) is a
reasonable quantity to use for the statistical significance. When we ask for
a 5$\sigma$ discovery, we are really asking that the background hypothesis be
more than 5 standard deviations away from the number of observed events. 
Assuming the event numbers to
be gaussian distributed, we arrive at the requirement: 
\begin{equation}
\begin{array}{lrcl}
 & N_{obs} - 5\sigma_{N_{obs}} & > & B \\
\implies & S + B - 5\sqrt{S + B} & > & B \\
\implies & P = \displaystyle\frac{S}{\sqrt{S+B}} & > & 5 
\end{array}\label{eq:stat1}
\end{equation}
Note that other definitions are also possible. One could equally well ask
that the background plus 5$\sqrt{B}$ should be less than the observed
number, resulting in more optimistic $P$ values
for larger numbers of signal events. 

The problem with 
pile-up lies in that the ratio of signal to background events
passing the analysis is too optimistic. To include an estimate of the
reduction of this ratio, we rewrite eq.~(\ref{eq:stat1}) to:
\begin{equation}
P = \frac{\sqrt{S}}{\sqrt{1+B/S}} > 5
\end{equation}
where we include the effects of pile-up by multiplying $B/S$ by some
factor. That twice as many background events per signal event could be
passing the analysis if pile-up was included seems a reasonably pessimistic
guess. 
Furthermore, assuming that the intrinsic uncertainty on both $S$ and
$B$ coming from uncertainties on QCD parameters will, 
to a first approximation, work in the same direction and
with a comparable magnitude for
both $B$ and $S$, we expect that the denominator in the above formula is not
affected by this uncertainty, and so we 
include the QCD-related uncertainties by reducing the number of
signal events in the numerator by a factor of 1.5. This yields the following
form for the ``corrected discovery potential'':
\begin{equation}
P_{corr} = \frac{S}{\sqrt{1.5S + 3B}}\label{eq:Pcorr}
\end{equation}
Both $P$ and $P_{corr}$ will be listed in the results below.

\begin{figure}[ht]
\begin{center}
\begin{tabular}{ccc}
\includegraphics*[scale=0.23]{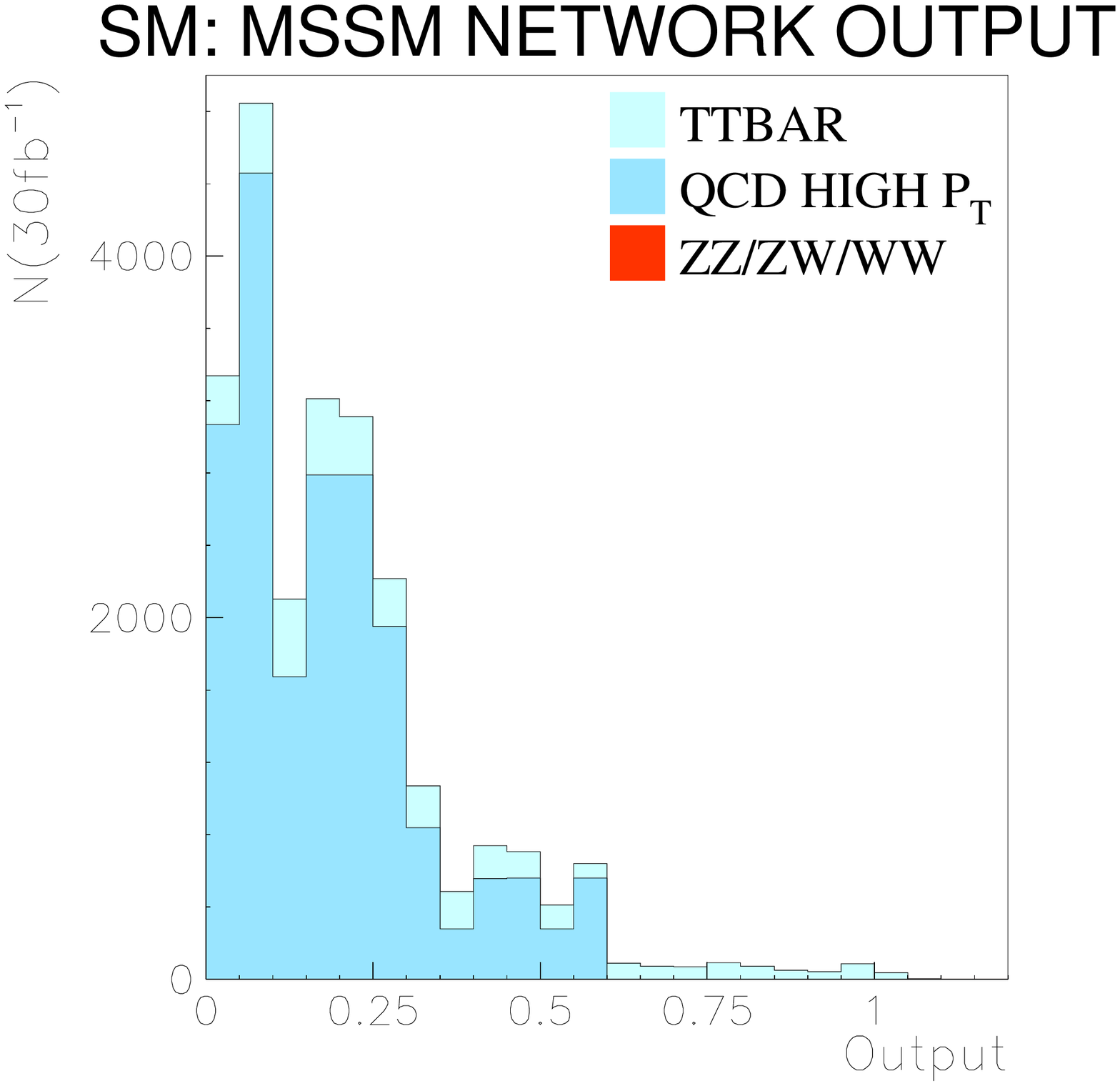} & 
\includegraphics*[scale=0.23]{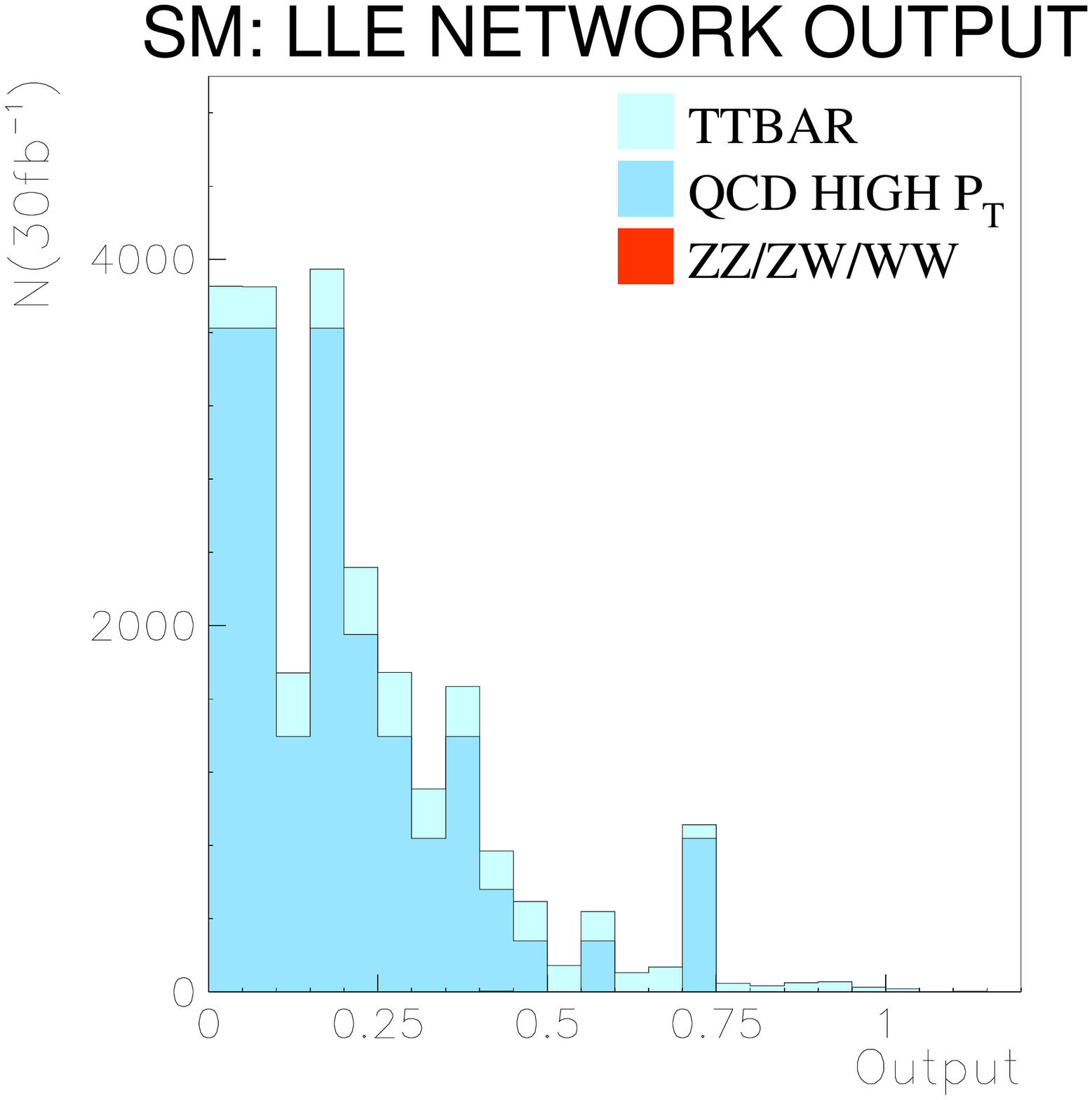} &
\includegraphics*[scale=0.23]{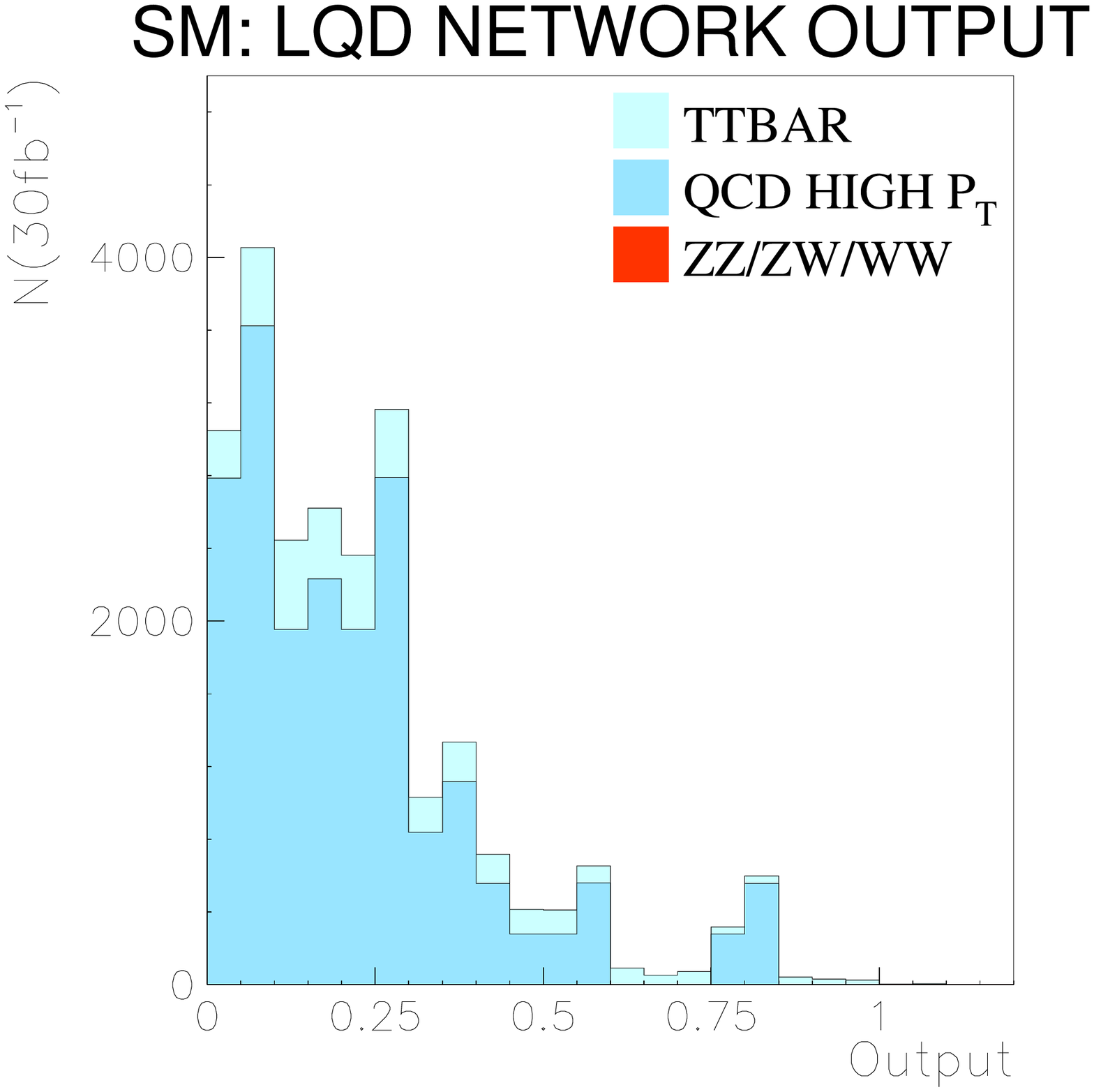} \vspace*{-6mm}
\end{tabular}
\end{center}
\caption[\small Network outputs for background events]{Network outputs for
background events. As for the previous plots, the numbers of double gauge
events are too small to be visible. We include them in the legend merely 
to signify that they have been taken into account. \label{fig:net_bg_out}}
\end{figure}

As was indeed the purpose, the network classifications gave very few background
events with outputs above 0.9. The response to background events surviving
the cut-based analysis over the entire output range is shown in figure
\ref{fig:net_bg_out}.
For all three networks, no QCD events survived the cut at 0.9. 
Knowing that the network is a highly non-trivial function, especially near 0
and 1, we do not attempt to fit the distribution. Rather, we note that  
zero events has less than 5\% probability of coming from a distribution with
mean larger than 2.99. Adopting and scaling 
this estimate yields the numbers given in
table \ref{tab:netevents}.
\begin{table}[t]
\begin{center}
\setlength{\extrarowheight}{0.2pt}
\begin{tabular}{lrrr}\toprule
\bf Process 
&\boldmath  $B_{MSSM-net}$
&\boldmath  $B_{LLE-net}$ 
&\boldmath  $B_{LQD-net}$\\\cmidrule{1-4}
Low-$p_T$ QCD
        & 5   &   5 &   5 \vspace*{2mm}\\
High-$p_T$ QCD
        & 840 & 840 & 840
 \vspace*{2mm}\\
Z/W 
        & 150 & 150 & 150
 \vspace*{2mm}\\
$t\bar{t}$
        & 210 & 140 & 100   
 \vspace*{2mm}\\
ZZ/ZW/WW
        & 5 & 5 & 5
\\ \cmidrule{1-4}
\bf Total SM
        & 1210 & 1140 & 1100  
\\
\bottomrule
\end{tabular}
\caption[\small Event numbers remaining after network cuts]{Estimated maximal
event numbers remaining after network cuts. Maximal here means that the
Poisson estimate discussed in section \ref{sec:lepjets} has been used to
estimate $B$ as the mean of the distribution that would 
result in 5\% probability of getting the number of generated events or less
which remained in the event samples actually used. Since any distribution with a
higher mean would have less than 5\% probability of resulting in the
generated numbers, this is equivalent to saying that we have 95\% confidence
in the numbers here being maximal.
\label{tab:netevents}}
\end{center}
\vspace*{-14pt}\end{table}
The same kind of Poisson
estimate was used for the 
double gauge events, mainly to estimate the rejection factor
which should be applied to the remaining single gauge events. Also 
for the $t\bar{t}$
event sample, the Poisson estimate was used where the number of 
generated events remaining was less than 100,
else the gaussian 95\% confidence estimate, $B < N + 1.64\sqrt{N}$. 

The total
number of background events passing each network, as listed in table
\ref{tab:netevents}, can now be used as upper limits on 
the quantity $B$ entering eqs.\
(\ref{eq:P}) and (\ref{eq:Pcorr}) by which we have 
defined the discovery potential and the corrected discovery potential,
respectively. Typically, around $500-1000$ signal events remain after cuts
for $P_2$, $P_{12}$, and $F_2$. More remain for $P_9$ because of the larger
cross section, but not nearly as large a fraction as for the other
scenarios. $P_7$, of course, was impossible from the start, with only 114
events expected after  $30\fb^{-1}$, yet one should keep in mind that large
values of the \RV\ couplings can lead to a significant increase in the
production cross section for the heavy-mass points, $P_7$ and $F_2$, due to
single sparticle production not included in the present analysis. For a
hadron machine like the 
LHC, this effect will be largest in the LQD scenarios since single squark
production will then be possible.

The discovery potentials, corrected and uncrrected, for all scenarios
are given in table \ref{tab:discoverypotential}. The labelling of the models
in the table 
should be self-explanatory to a large extent. The subscripts $a$, $b$,
and $n$ refer to the column labels in table \ref{tab:lambdapoints} where the
\LV\ scenarios were defined, so that
$a$ and $b$ are the constant coupling scenarios, and $n$ is the scenario with
natural (hierarchical) couplings. This distinction, of course, does not exist
for the MSSM where no lepton number violating couplings exist. The individual
numbers in the table are
not extremely interesting, except for the quite significant fact that, 
except for $P_7$, they are all above 5 for at least one of the network
types. It is also of some interest to note 
that signals can be extracted reliably for the mixed (LLE+LQD)
scenarios which neither entered in the cut optimization nor in the network
training. 
Our estimate of the LHC discovery potential in the case of Lepton
Number Violating SuperGravity scenarios is thus that a $5\sigma$ discovery is
possible
for cross sections down to at least $10^{-10}\!\ \mathrm{mb}$  
with 95\% certainty when the effects of pile-up and uncertainties on QCD
parameters are neglected. Depending on whether our reduced estimate of the
discovery potential does a reasonable job, these uncertainties should not
be able to affect this general conclusion, yet one sees that 
there could be cause for concern 
for models with non-zero LQD couplings in SUSY scenarios of the $F_2$
type. As mentioned above, however, one would expect single sparticle
production to be enhancing the cross section for this model for large values
of the $\lambda'$ couplings whereas low values would mean longer LSP
lifetimes and hence either secondary vertices or an MSSM-like signature if
the LSP escapes detection altogether. In the first case, we would
have an extra discriminating variable, and in the second we note that even in
the present analysis (in which the
cuts were not optimized with the MSSM in mind) we have a reduced discovery
potential of up to $6.5\sigma$ for the $F_2$ MSSM model. 
\clearpage
\begin{table}[th!]
\begin{center}
{\hspace*{-0.4cm}
{\large\sf ATLAS \LV-SUSY DISCOVERY POTENTIAL}\\
{\hspace*{-0.4cm}\setlength{\extrarowheight}{3.pt}\sf
\begin{tabular}[t]{c|c}\toprule
\begin{tabular}[t]{lccc}
& \multicolumn{3}{c}{\sf NETWORK}\\
SUSY & MSSM & LLE & LQD \vspace*{-\extrarowheight}\\
Point &                         $P/P_{corr}$ & $P/P_{corr}$ & $P/P_{corr}$
\\ \cmidrule{1-4}
$ P_{2a}^{\mbox{\tiny LLE    }}$
&
$ 24.3
/
 16.2$
&
$ 25.5
/
 17.1$
&
$ 25.4
/
 17.1$
\\
$ P_{2b}^{\mbox{\tiny LLE    }}$
&
$ 24.5
/
 16.3$
&
$ 25.8
/
 17.3$
&
$ 25.7
/
 17.3$
\\
$ P_{2n}^{\mbox{\tiny LLE    }}$
&
$ 23.2
/
 15.4$
&
$ 24.7
/
 16.5$
&
$ 24.3
/
 16.3$
\\
$ P_{7a}^{\mbox{\tiny LLE    }}$
&
\color{red}
$  0.7
\color{black}
/
\color{red}
  0.4$
\color{black}
&
\color{red}
$  0.7
\color{black}
/
\color{red}
  0.4$
\color{black}
&
\color{red}
$  0.8
\color{black}
/
\color{red}
  0.5$
\color{black}
\\
$ P_{7b}^{\mbox{\tiny LLE    }}$
&
\color{red}
$  0.8
\color{black}
/
\color{red}
  0.4$
\color{black}
&
\color{red}
$  0.8
\color{black}
/
\color{red}
  0.4$
\color{black}
&
\color{red}
$  0.8
\color{black}
/
\color{red}
  0.5$
\color{black}
\\
$ P_{7n}^{\mbox{\tiny LLE    }}$
&
\color{red}
$  0.7
\color{black}
/
\color{red}
  0.4$
\color{black}
&
\color{red}
$  0.7
\color{black}
/
\color{red}
  0.4$
\color{black}
&
\color{red}
$  0.8
\color{black}
/
\color{red}
  0.5$
\color{black}
\\
$ P_{9a}^{\mbox{\tiny LLE    }}$
&
$191
/
153$
&
$315
/
256$
&
$218
/
176$
\\
$ P_{9b}^{\mbox{\tiny LLE    }}$
&
$190
/
153$
&
$316
/
256$
&
$218
/
176$
\\
$ P_{9n}^{\mbox{\tiny LLE    }}$
&
$166
/
133$
&
$257
/
208$
&
$169
/
135$
\\
$P_{12a}^{\mbox{\tiny LLE    }}$
&
$ 23.4
/
 15.5$
&
$ 25.6
/
 17.2$
&
$ 25.5
/
 17.2$
\\
$P_{12b}^{\mbox{\tiny LLE    }}$
&
$ 23.4
/
 15.5$
&
$ 25.5
/
 17.2$
&
$ 25.5
/
 17.2$
\\
$P_{12n}^{\mbox{\tiny LLE    }}$
&
$ 21.8
/
 14.4$
&
$ 24.2
/
 16.2$
&
$ 24.3
/
 16.3$
\\
$ F_{2a}^{\mbox{\tiny LLE    }}$
&
$ 11.3
/
  7.0$
&
$ 14.0
/
  8.8$
&
$ 13.3
/
  8.4$
\\
$ F_{2b}^{\mbox{\tiny LLE    }}$
&
$ 11.2
/
  6.9$
&
$ 13.7
/
  8.7$
&
$ 13.1
/
  8.2$
\\
$ F_{2n}^{\mbox{\tiny LLE    }}$
&
$  9.9
/
  6.1$
&
$ 12.4
/
  7.8$
&
$ 12.3
/
  7.7$
\\
\vspace*{-3ex}\\
\cmidrule{1-4}
$ P_{2a}^{\mbox{\tiny LQD    }}$
&
$ 20.9
/
 13.7$
&
$ 24.3
/
 16.2$
&
$ 24.8
/
 16.6$
\\
$ P_{2b}^{\mbox{\tiny LQD    }}$
&
$ 21.4
/
 14.1$
&
$ 24.7
/
 16.5$
&
$ 25.3
/
 17.0$
\\
$ P_{2n}^{\mbox{\tiny LQD    }}$
&
$ 21.5
/
 14.1$
&
$ 23.3
/
 15.5$
&
$ 24.2
/
 16.2$
\\
$ P_{7a}^{\mbox{\tiny LQD    }}$
&
\color{red}
$  1.0
\color{black}
/
\color{red}
  0.6$
\color{black}
&
\color{red}
$  1.0
\color{black}
/
\color{red}
  0.6$
\color{black}
&
\color{red}
$  1.1
\color{black}
/
\color{red}
  0.6$
\color{black}
\\
$ P_{7b}^{\mbox{\tiny LQD    }}$
&
\color{red}
$  1.0
\color{black}
/
\color{red}
  0.6$
\color{black}
&
\color{red}
$  1.0
\color{black}
/
\color{red}
  0.6$
\color{black}
&
\color{red}
$  1.1
\color{black}
/
\color{red}
  0.7$
\color{black}
\\
$ P_{7n}^{\mbox{\tiny LQD    }}$
&
\color{red}
$  1.0
\color{black}
/
\color{red}
  0.6$
\color{black}
&
\color{red}
$  1.0
\color{black}
/
\color{red}
  0.6$
\color{black}
&
\color{red}
$  1.1
\color{black}
/
\color{red}
  0.6$
\color{black}
\\
$ P_{9a}^{\mbox{\tiny LQD    }}$
&
$116
/
 91.6$
&
$153
/
122$
&
$125
/
 99.0$
\\
$ P_{9b}^{\mbox{\tiny LQD    }}$
&
$113
/
 88.7$
&
$151
/
121$
&
$123
/
 97.6$
\\
$ P_{9n}^{\mbox{\tiny LQD    }}$
&
$113
/
 88.5$
&
$131
/
104$
&
$110
/
 86.6$
\\
$P_{12a}^{\mbox{\tiny LQD    }}$
&
$ 15.7
/
 10.0$
&
$ 19.2
/
 12.5$
&
$ 20.9
/
 13.7$
\\
$P_{12b}^{\mbox{\tiny LQD    }}$
&
$ 15.8
/
 10.1$
&
$ 19.4
/
 12.6$
&
$ 21.1
/
 13.9$
\\
$P_{12n}^{\mbox{\tiny LQD    }}$
&
$ 16.6
/
 10.6$
&
$ 18.6
/
 12.1$
&
$ 20.9
/
 13.8$
\\
$ F_{2a}^{\mbox{\tiny LQD    }}$
&
$  7.0
/
\color{red}
  4.2$
\color{black}
&
$  9.5
/
\color{dred}
  5.9$
\color{black}
&
$ 10.5
/
  6.5$
\\
$ F_{2b}^{\mbox{\tiny LQD    }}$
&
$  7.0
/
\color{red}
  4.2$
\color{black}
&
$  9.4
/
\color{dred}
  5.8$
\color{black}
&
$ 10.5
/
  6.5$
\\
$ F_{2n}^{\mbox{\tiny LQD    }}$
&
$  6.9
/
\color{red}
  4.2$
\color{black}
&
$  8.8
/
\color{dred}
  5.4$
\color{black}
&
$ 10.3
/
  6.4$
\\
\vspace*{-3ex}\\
\end{tabular}
&
\begin{tabular}[t]{lccc}
& \multicolumn{3}{c}{\sf NETWORK}\\
 SUSY & \multicolumn{1}{c}{MSSM} & 
\multicolumn{1}{c}{LLE} &
\multicolumn{1}{c}{LQD}\vspace*{-\extrarowheight}\\
Point                           
&$P/P_{corr}$ 
&$P/P_{corr}$ &
$P/P_{corr}$  
\\ \cmidrule{1-4}
$ P_{2 }^{\mbox{\tiny MSSM   }}$
&
$ 10.4
/
  6.4$
&
$ 10.2
/
  6.3$
&
$ 10.2
/
  6.3$
\\
$ P_{7 }^{\mbox{\tiny MSSM   }}$
&
\color{red}
$  0.2
\color{black}
/
\color{red}
  0.1$
\color{black}
&
\color{red}
$  0.2
\color{black}
/
\color{red}
  0.1$
\color{black}
&
\color{red}
$  0.2
\color{black}
/
\color{red}
  0.1$
\color{black}
\\
$ P_{9 }^{\mbox{\tiny MSSM   }}$
&
$136
/
108$
&
$121
/
 95.9$
&
$ 93.5
/
 72.8$
\\
$P_{12 }^{\mbox{\tiny MSSM   }}$
&
$ 16.1
/
 10.3$
&
$ 15.5
/
  9.9$
&
$ 16.1
/
 10.3$
\\
$ F_{2 }^{\mbox{\tiny MSSM   }}$
&
$  9.4
/
\color{dred}
  5.8$
\color{black}
&
$  9.7
/
\color{dred}
  6.0$
\color{black}
&
$ 10.5
/
  6.5$
\\
\vspace*{-3ex}
\\ \cmidrule{1-4}
$ P_{2a}^{\mbox{\tiny LLE+LQD}}$
&
$ 24.3
/
 16.2$
&
$ 25.9
/
 17.4$
&
$ 25.8
/
 17.4$
\\
$ P_{2b}^{\mbox{\tiny LLE+LQD}}$
&
$ 24.5
/
 16.3$
&
$ 25.9
/
 17.4$
&
$ 25.9
/
 17.4$
\\
$ P_{2n}^{\mbox{\tiny LLE+LQD}}$
&
$ 21.4
/
 14.0$
&
$ 23.2
/
 15.4$
&
$ 24.2
/
 16.2$
\\
$ P_{7a}^{\mbox{\tiny LLE+LQD}}$
&
\color{red}
$  0.8
\color{black}
/
\color{red}
  0.5$
\color{black}
&
\color{red}
$  0.8
\color{black}
/
\color{red}
  0.5$
\color{black}
&
\color{red}
$  0.9
\color{black}
/
\color{red}
  0.5$
\color{black}
\\
$ P_{7b}^{\mbox{\tiny LLE+LQD}}$
&
\color{red}
$  0.8
\color{black}
/
\color{red}
  0.5$
\color{black}
&
\color{red}
$  0.8
\color{black}
/
\color{red}
  0.5$
\color{black}
&
\color{red}
$  0.9
\color{black}
/
\color{red}
  0.5$
\color{black}
\\
$ P_{7n}^{\mbox{\tiny LLE+LQD}}$
&
\color{red}
$  1.0
\color{black}
/
\color{red}
  0.6$
\color{black}
&
\color{red}
$  1.0
\color{black}
/
\color{red}
  0.6$
\color{black}
&
\color{red}
$  1.1
\color{black}
/
\color{red}
  0.6$
\color{black}
\\
$ P_{9a}^{\mbox{\tiny LLE+LQD}}$
&
$179
/
143$
&
$289
/
234$
&
$203
/
164$
\\
$ P_{9b}^{\mbox{\tiny LLE+LQD}}$
&
$178
/
143$
&
$291
/
236$
&
$204
/
164$
\\
$ P_{9n}^{\mbox{\tiny LLE+LQD}}$
&
$114
/
 89.3$
&
$132
/
105$
&
$111
/
 87.0$
\\
$P_{12a}^{\mbox{\tiny LLE+LQD}}$
&
$ 20.7
/
 13.6$
&
$ 23.8
/
 15.8$
&
$ 24.3
/
 16.3$
\\
$P_{12b}^{\mbox{\tiny LLE+LQD}}$
&
$ 20.8
/
 13.6$
&
$ 23.8
/
 15.8$
&
$ 24.4
/
 16.4$
\\
$P_{12n}^{\mbox{\tiny LLE+LQD}}$
&
$ 16.5
/
 10.6$
&
$ 18.5
/
 12.0$
&
$ 20.8
/
 13.7$
\\
$ F_{2a}^{\mbox{\tiny LLE+LQD}}$
&
$  9.0
/
\color{dred}
  5.5$
\color{black}
&
$ 11.9
/
  7.4$
&
$ 11.9
/
  7.5$
\\
$ F_{2b}^{\mbox{\tiny LLE+LQD}}$
&
$  9.0
/
\color{dred}
  5.5$
\color{black}
&
$ 11.7
/
  7.3$
&
$ 11.8
/
  7.4$
\\
$ F_{2n}^{\mbox{\tiny LLE+LQD}}$
&
$  7.1
/
\color{red}
  4.3$
\color{black}
&
$  8.9
/
\color{dred}
  5.5$
\color{black}
&
$ 10.4
/
  6.4$
\\
\vspace*{-3ex}
\end{tabular}\\
\bottomrule
\end{tabular}}}
\caption[\small ATLAS Discovery Potential]{ATLAS discovery potential and
corrected discovery potential (see text) for all
SUSY scenarios investigated using each of the three networks.\label{tab:discoverypotential}}
\end{center}
\end{table}
\clearpage

\clearpage
\section{Outlook and Conclusion\label{sec:conc}}
\subsection{Outlook}
Though some preliminary studies have been performed in the present work, many
important things remain to be done in this field. From my perspective, the
most important theoretical/phenomenological issues which remain are:
\begin{enumerate}
\item The inclusion of Baryon number violating processes in the \pythia\
  generator. In addition, this will require a study of the exceptional colour
  flows that are possible when baryon number is broken. Such a study has
  already been carried out for the \herwig\ generator (see \cite{dreiner00}),
  but it remains to be done in \pythia. 
\item The inclusion of resonant slepton and squark production in the \pythia\
generator. In \RV\ scenarios, the
  production of single sparticle resonances is possible and can extend 
the discovery potential of the LHC towards higher SUSY masses. 
\item The existence of \RV\ couplings of comparable magnitude to the
  gauge couplings would have a significant effect on the renormalization
  group evolution of the masses and couplings from the input scale to the
  electroweak. At present, \pythia\ can be told to call on \isasusy\ to
  perform this evolution, but the $R$-violating couplings are not yet included.
\end{enumerate}
On the experimental side, several studies would be advisable:
\begin{enumerate}
\item The present work has only dealt with lepton
  number violation, the signatures of which are most likely easier to
  identify than baryon number violating signatures (an excess of
  jets). Preliminary studies indicate a lessening of the reach of the LHC in
  these scenarios \cite{baer97}. 
  Therefore, dedicated studies which could push this reach to the limit would
  be advisable. These can be performed either with the present version of the 
  \herwig\ generator or with the \pythia\ generator when \BV\ has been
  included. Since the reach is lessened, it is advisable to wait with such a
  study until single squark production (an enhancing mechanism) has also been
  included. Studying only decays is likely to give too pessimistic
  predictions. Again, single sparticle production \emph{is} included in the
\herwig\ generator in its present form.
\item Having concentrated on signal isolation and discovery potential, no
mass reconstruction has been undertaken in this
work. Taking e.g.\ the events isolated by the current analysis as a basis,
it would be interesting to determine how well the SUSY mass spectrum can be
disentangled in the various scenarios. 
\item A systematic study of the consequences of hierarchical structure in the
\RV\ couplings. E.g.\ one sees that large 1$^{\mathrm{st}}$ generation
couplings will lead to electrons or electron-neutrinos in the final state
etc. It would be of interest to study how well we can expect to ``measure''
the individual \RV\ couplings. 
\item The mSUGRA points studied in this work were all based on the MSSM and
as such had a neutralino LSP. As mentioned, this is not required in
\RV-SUSY. It is therefore of some importance to study the effects of having
non-neutralino LSP's. A special case is if one imagines 
e.g.\ a slepton (or, less likely, a squark)
LSP, for which 
the three-body LSP decays studied here would be replaced by two-body
decays with the associated much simpler kinematics, allowing more precise
invariant mass reconstruction.
\item Re-evaluation of the trigger rates and of the trigger objects proposed
  here for both high and mid luminosities with better detector simulation,
  either a parametrization of the effects of pile-up at mid-luminosity in
  \atlfast\ or full detector simulation.
\item A study of to what extent the trigger
menus here proposed 
can be combined with trigger menus for other kinds of physics.
\end{enumerate}
\subsection{Conclusion}
In the first part of this work, it was seen that the most general space-time
symmetries possible in an interacting 
quantum field theory includes a symmetry between
bosons and fermions 
which is not present in the currently accepted theory,
the so-called Standard Model of Particle Physics. That this extra symmetry,
Supersymmetry, is not forbidden gave us our first motivation to study the
physical consequences of having such a symmetry in nature.
Disjoint from this, it was argued that the discovery of a
fundamental Higgs boson would lead, through the hierarchy problem, 
to a requirement of the existence of physics not contained within the
framework of the Standard Model itself. It was with the
realization that Supersymmetry could cure the hierarchy problem \emph{and} 
give a natural 
explanation for the size of the electroweak scale that we found our
second, more compelling 
motivation. It was then noted that supersymmetry is not without
defects in that it must be broken at low energies and some \emph{additional}
symmetry must exist to assure the experimentally observed high degree of 
proton stability. 

Basically, three choices for this symmetry exist: the
conservation of both lepton and baryon number or the conservation of only one of
them, in the supersymmetric interactions. 
The former is usually cast in the shape of a conserved, multiplicative
quantum number, $R$, and has the additional property of giving a natural dark
matter candidate, since it results in   
the Lightest Supersymmetric Particle (the LSP) being stable. 
The latter two do not have
this property in most of their parameter spaces. Moreover, 
they give rise to more free parameters and more complex phenomenologies, 
i.e.\ many additional production and decay mechanisms for the supersymmetric
particles. On these grounds, suppersymmetrized versions of the Standard Model
are most often found with $R$-parity conservation being implicitly assumed.

In section \ref{sec:lspdecays}, some effort was devoted to explain the
potential fallacies and the dangers of this assumption with the conclusion
that $R$-parity cannot assure proton stability when Supersymmetry is
embedded into more fundamental frameworks containing baryon and lepton number
violation exterior to Supersymmetry, as is the case, for example, 
in a wide range of Grand Unified Theories. The danger in focussing too much
on $R$-parity conserving scenarios in accelerator searches becomes clear when
one considers the ramifications of LSP decay on event topologies
in the detector. Particularly, the reduction of the missing transverse energy
signature associated with escaping LSP's in $R$-parity conserving
scenarios could be greatly reduced if $R$-parity is not
conserved. 

1278 decay modes of Supersymmetric particles into Standard Model
particles through lepton number violating
couplings in the Minimal Supersymmetric Standard Model
were therefore studied and implemented in the \pythia\ event
generator. Combining this augmented version of the generator with a crude
simulation of the ATLAS detector, trigger menus for mid-luminosity running of
the LHC were proposed and seen to have a high acceptance of supersymmetric
events in several $L$-violating SuperGravity scenarios while still giving
event rates in the 1Hz region. 

Taking these trigger menus as basis, the possibility for a 5$\sigma$
discovery after 30\fb$^{-1}$ data taking was estimated for each investigated
model, including also the $R$-conserving MSSM for reference. The analysis
was divided into two parts, the first of which consisted of
a series of cuts on kinematical and inclusive variables, placed so as to have
good background rejection factors
while accepting 
events from as many of the various SUSY models as possible (excepting
the MSSM scenarios used only as reference). The second part consisted of
processing the remaining events through three neural networks trained to
recognize $R$-conserving scenarios and two different variants of lepton
number violating scenarios. For cross sections down to \tn{-10}\mb\ it was
found that a $5\sigma$ discovery was possible for all scenarios with
30\fb$^{-1}$ of data. It is not estimated that uncertainties related to QCD
parameters or pile-up in the detector, both of which have not been taken into
account in the present analysis, could significantly affect this conclusion.
\vfill
\begin{center}
\textbf{Acknowledgements}
\end{center}
I am deeply grateful to my two supervisors, both for their excellent 
guidance and for having allowed me a considerable freedom in what I have
occupied myself with this last year. In many ways, the master thesis is
likely to be the only chance a student gets to
learn about and discuss within so few pages so many aspects of
his/her field, here from quantum gravity effects on
global symmetries through supersymmetric phenomenology, 
Monte Carlo simulations, and detector studies to 
brain-damaged neural networks (though of
course not one of these subjects has been treated in as much detail as it might
deserve). From my perspective, such freedom is a luxury,
and I am very happy to have found it in carrying out this work. On the same
note, I thank all the members of the HEP group at NBI for discussions, an
inspiring atmosphere, and for the $n\to\infty$ CPU hours I spent on the
farm. In addition, I would like to thank Dr.\ Alexander Khodjamirian for the depth of
physical insight he gave to me before I embarked on this work.
Lastly, I would like to thank Paula Eerola and 
the Nordic Academy for Advanced Studies for having made it possible
for me to stay in Lund on a regular basis through the spring of 2001.

%
\clearpage
\appendix
\section{Decays of SUSY particles: Conventions and References \label{app:A}}
\label{app:susydecays}
\label{app:neutralino}
Both
\spythia\ \cite{mrenna97} and \herwig\ \cite{herwig6} 
follow \cite{haber85} (identical to \cite{gunion86}) for
the neutralino and chargino mixing conventions 
whereas \isasusy\ follows \cite{baer87}. 
\herwig\ uses \cite{baer94} for the sfermion mixing conventions,
corresponding to changing the sign of the mixing angles in \cite{haber85}, 
used in \isasusy\ and \spythia\ (as well as the work performed here). 
All matrix elements used in this work have been calculated in
\cite{dreiner00}. Note that it is assumed that there are no 
right-handed neutrinos (and thus no $\ti{\nu}_R$
SUSY particles). The $a$ and $b$ couplings defined in 
\cite{dreiner00} are for particles. The antisparticle couplings are obtained by
$a(\ti{c}^*)=b(\ti{c})$. Furthermore, $a(\ti{c})$ (and thus also $b(\ti{c}^*)$)
change sign for all couplings pertaining to the neutralino $\chi$ if the
diagonalization of the neutralino mass matrix results in a negative
eigenvalue for the $\chi$ mass (this sign change is the effect of changing
the physical field to $\gamma_5 \chi$, thus obtaining a field with positive
mass, but with slightly different couplings). 
With respect to the signs of mixing matrix elements and $a$ and $b$
couplings, one should not be distressed by noticing a number of sign
differences between \spythia\ and \herwig\ for the neutralino mixing. 
These are artifacts of the
diagonalization procedure, and as such of no physical importance. If they
exist, they are always such that all mixing matrix elements pertaining to
e.g.\ $\neut_1$ have their sign flipped, both in
$N_{ij}$ and $N'_{ij}$ (defined in section \ref{sec:mixing}). 
Since the decay rates of neutralino $i$ depend only on products of mixing
matrix elements like $N_{ij}N'_{ik}$ (see \cite{dreiner00}), 
the overall signs of the rows in $N$ and $N'$ have no importance. 

With respect to the scalar mixing, there is indeed a convention
difference, as mentioned above. \herwig\ uses \cite{baer94} while \spythia\ uses
\cite{haber85}. The difference is in the sign of the mixing angle, but the
mixing elements themselves should still be the same. One can quickly see this
by taking a look at the mixing matrices as expressed in the two conventions
with their two different choice for the sign of the mixing angle:
\begin{equation}
\twovec{\ti{s}_L}{\ti{s}_R} = \left( \begin{array}{cc}
	\cos\theta_+ & \sin\theta_+ \\
	-\sin\theta_+ & \cos\theta_+
\end{array} \right) \twovec{\ti{s}_1}{\ti{s}_2} \hspace*{1.5cm}
\twovec{\ti{s}_L}{\ti{s}_R} = \left( \begin{array}{cc}
	\cos\theta_- & -\sin\theta_- \\
	\sin\theta_- & \cos\theta_-
\end{array} \right) \twovec{\ti{s}_1}{\ti{s}_2} 
\end{equation}
in the conventions of \cite{baer94} and \cite{haber85}, respectively, and
noting that $\theta_+ = -\theta_-$.

\section{Phase Space Integrations in Dalitz Variables}
\label{app:phasespace}
\setlength{\extrarowheight}{5pt}
The kinematics of an unpolarized three-body decay $0\to 1,2,3$ can be
parametrized in terms of the Dalitz variables:
\begin{equation}
m_{12}^2\equiv (p_1+p_2)^2 \hspace*{1cm} m_{23}^2\equiv (p_2+p_3)^2
\hspace*{1cm} m_{13}^2 \equiv (p_1+p_3)^2
\end{equation}
\begin{equation}
M_0^2 = m_{12}^2 + m_{23}^2 + m_{13}^2 - m_1^2 -m_2^2 - m_3^2
\label{appeq:dalitzrelation} 
\end{equation}
Since $|\overline{M}|^2$ can contain resonances in any of these variables, we
wish to study how we can parametrize the phase space in terms of \emph{any}
combination of two of these, enabling an optimization of numerical
integration procedures.

Changing variables inside $|\overline{M}|^2$ is an easily accomplished
task. If one desires to chose e.g.\ $m_{12}^2$ and $m_{23}^2$ to parametrize
the degrees of freedom, one simply uses eq.~(\ref{appeq:dalitzrelation}) to
interpret any dependence of $|\overline{M}|^2$ on $m_{13}^2$ as a dependence
on $m_{12}^2$ and $m_{23}^2$. The integration must then be performed over
phase space parametrized in terms of the two chosen variables. The limits for
this integration can be derived as follows:

In all generality, we know that
\begin{eqnarray}
m_{ab}^2 & = &(p_a+p_b)^2 = m_a^2+m_b^2 + 2E_aE_b - 2\vec{p}_a\cdot\vec{p}_b
\label{eq:mab1} = (E_a+E_b)^2 - (\vec{p}_a+\vec{p}_b)^2\\ & = & (p_0-p_c)^2 =
M_0^2 + m_c^2 - 2E_0E_c + 2\vec{p_0}\cdot\vec{p_c} = (E_0-E_c)^2 -
(\vec{p}_0-\vec{p}_c)^2
\label{eq:mab2}
\end{eqnarray}
As the Dalitz variables are Lorentz invariant, we can chose any frame to
evaluate them in. Going to the CM of $a$ and $b$, we find that the minimum
value of $m_{ab}^2$ must occur when the two particles are lying completely
still. On the other hand, the maximum value must occur when particle $c$ is
lying completely still. In this case, the CM of $a$ and $b$ is also the CM of
the decaying particle, and so the decaying particle is also lying completely
still. Using eqs.~(\ref{eq:mab1}) and (\ref{eq:mab2}) we then have:
\begin{equation}
m_{ab,\mbox{\scriptsize min}}^2 = (m_a+m_b)^2 \hspace*{20mm}
m_{ab,\mbox{\scriptsize max}}^2 = (M_0-m_c)^2
\end{equation}
These are the limits for the `outside' integration.

For the `inside' integration, the limits follow almost as easily. $m_{ab}^2$
should now be thought of as having some specific value. Again going to the
rest frame of $a$ and $b$, we have the following picture:
\begin{center}
\vspace*{0.5mm}
\begin{fmffile}{mbclimits}
\begin{fmfgraph*}(75,50)\fmfpen{thin}\fmfset{arrow_ang}{30}\fmfset{arrow_len}{0.04w}
\fmfforce{0.5w,0.5h}{CM}
\fmfforce{0.0w,0.5h}{a}
\fmfforce{0.7w,0.1h}{v0}
\fmfforce{0.3w,0.9h}{c}
\fmfforce{w,0.5h}{b}
\fmfv{d.sh=circle,d.siz=0.05w}{CM}
\fmf{plain}{CM,a}
\fmf{plain}{CM,b}
\fmf{dashes}{v0,v1}
\fmf{dashes}{v1,v2}
\fmf{scalar}{v2,CM}
\fmf{plain}{CM,c}
\fmfv{d.sh=triangle,d.siz=0.04w,d.ang=90,label=$\vec{p}_a$}{a}
\fmfv{d.sh=triangle,d.siz=0.04w,d.ang=35,label=$\vec{p}_c$}{c}
\fmfv{label=$\vec{p}_0$,label.dist=3}{v0}
\fmfv{d.sh=triangle,d.siz=0.04w,d.ang=-90,label=$\vec{p}_b$}{b}
\end{fmfgraph*}
\end{fmffile}

\vspace*{2.5mm}
\end{center}
Taking a look at eq.~(\ref{eq:mab1}), we see that that $m_{ac}^2$
($m_{bc}^2$) attains its minimum when $c$ goes parallel to $a$ (parallel to
$b$), and its maximum when $c$ goes antiparallel to $a$ ($b$).  Thus, using
$^*$ to denote that we are to evaluate energies and momenta in the CM of $a$
and $b$, we have:
\begin{equation}
m_{(a,b)c}^2 = (E^*_{a,b}+E^*_c)^2 - (\vec{p}_{a,b}^* + \vec{p}_c^*)^2 =
(E^*_{a,b}+E^*_c)^2 - (|\vec{p}_{a,b}^*|^2 + |\vec{p}_c^*|^2 + 2
|\vec{p}_{a,b}^*||\vec{p}_c^*|\cos\theta)
\end{equation}\vspace*{-6mm}
\begin{equation}
\implies 
\begin{array}[c]{rcl} m_{(a,b)c,\mbox{\scriptsize min}}^2 & = & (E_{a,b}^* + E_c^*)^2 -
\left(\textstyle{\sqrt{E_{a,b}^{*2}-m_{a,b}^2}}+\sqrt{E_c^{*2}-m_c^2} 
\right)^2 \\
m_{(a,b)c,\mbox{\scriptsize max}}^2 & = & (E_{a,b}^* + E_c^*)^2 -
\left(\textstyle{\sqrt{E_{a,b}^{*2}-m_{a,b}^2}}-\sqrt{E_c^{*2}-m_c^2} 
\right)^2
\end{array}
\end{equation}
with the CM energies \cite[chp.34]{europhys}:
\begin{equation}
E_{a,b}^* = \frac{m_{ab}^2-m_{b,a}^2+m_{a,b}^2}{2m_{ab}} \hspace*{1.5cm}
E_c^* = \frac{M_0^2 - m_{ab}^2-m_c^2}{2m_{ab}}
\end{equation}

\subsection{Numerical Integration Procedure for Virtual Contributions}
\label{app:numint}
Now that the most advantageous integration variables can be chosen without
problems, we need to select the most advantageous integration
technique. In general, the integrations could 
contain fairly narrow peaks (resonances), and so a
simple step-by-step integration would be either too coarse or too
time-consuming. A good solution in many cases is to resort to an optimized
Monte Carlo technique, but when the number of dimensions is small (in our
case $d=1-2$), a better idea is to use a transformation of integration
variables that results in approximately flat integrands (this is
usually referred to as variance reduction)
which can then be integrated using a simple stepping algorithm on a uniform
lattice. This optimizes both CPU time and accuracy simultaneously.

For the case at hand, however, note that the resonances \emph{will never} lie
inside the integration limits. If they did, it would mean that the process
could happen with the intermediate particle on shell, and so we could really
speak of the process as being composed of two distinct parts, namely 1) the
($R$ conserving) decay of a neutralino/chargino to an on-shell sfermion and
an SM fermion, and 2) the ($R$ violating) decay of the sfermion to two SM 
fermions. The branching ratio for the first process is already calculated by the
\spythia\ code, and the branching ratio for the second is calculated by the
sfermion part of the $R$-violation code. Thus we would be doing double
counting if we added such a contribution again. 
The only contribtuions not already
handled by other parts of the code are the off-shell contributions where the
intermediate sfermion is truly virtual and therefore outside the kinematical
limits for the integrations we will perform.\\ \\
\emph{It is therefore exceedingly important to note that the $R$-Violating 
3-body widths and branching ratios for the neutralinos and charginos as output
by \pythia\ only contain the off-shell contributions! For the complete
exclusive branching ratios, the on-shell contributions should be computed and
added separately.}\\ \\
Since the integrands for virtual contributions can be expected to be
reasonably flat, the integrations can be handled using standard gaussian
quadrature without problems. With respect to interference terms, these are
neglected if either of the interfering diagrams have an on-shell propagator,
since they are then suppressed by $(\Gamma_{\mathrm{res}}/M_{\mathrm{res}})^2$. 

\subsection{Numerical Integration Procedure for Total Exclusive
Contributions}
Above, we have only given justification for why the integrands in this
specific implementation can be expected to be smooth and thus integrable by
standard methods. In addition, we shall now give a method to calculate the
full, exclusive width (i.e.\ including both
off-shell and on-shell intermediate states) where gaussian quadrature would 
be a hazardous enterprize since it cannot be relied upon to
catch sharp peaks in the integrand. Rather, we should make use of our
knowledge of the shape of the resonances to integrate efficiently those
contributions that are peaked and reserve gaussian quadrature for the
``flat'' contributions.

In all generality, we know that the matrix elements are expressible as sums
of terms of which the most complicated are the interferences between
different diagrams with different resonances in different Dalitz
variables. As mentioned in section \ref{sec:neutralino}, the interference
terms can be expected to be relatively small and do not exhibit sharp peaks.
We then
concentrate on the pure resonance terms.  A rough idea of the shape of these
terms is given by (with $x$ denoting the resonant Dalitz variable
squared):\vspace*{-1mm}
\begin{equation}
R(x) + C(x)
\end{equation}
where $C$ is a slowly varying function, and $R$ is the Breit-Wigner resonance
function:
\begin{equation}
R(x,M,\Gamma) =
\frac{1}{(x-M^2)^2 + M^2\Gamma^2}
\end{equation}
with $M$ being the mass of the resonance and $\Gamma$ its decay width.
Since we shall only need a rough
approximation, let $C$ be approximated by a constant.  For a specific
resonance term charactarized by the resonance mass, $M_{res}$, and the width
$\Gamma_{res}$, the Phase Space integral then becomes:
\vspace*{-2mm}
\setlength{\extrarowheight}{18pt}
\begin{equation} 
\begin{array}[b]{c}
\displaystyle
\int_{x_{min}}^{x_{max}}\int_{y_{min}}^{y_{max}} \left[ R(x) +
C\right] \difd y\difd x = \Delta y \int_{x_{min}}^{x_{max}}
\frac{1}{(x-M_{res}^2)^2+\Gamma_{res}^2M_{res}^2} \difd x 
 + C\Delta x\Delta y \\
\begin{array}[b]{l}
\displaystyle = \frac{1}{\Gamma_{res}M_{res}}\left[
\arctan\left(\frac{x_{max}-M_{res}^2}{\Gamma_{res}M_{res}}\right)  
- \arctan\left(\frac{x_{min}-M_{res}^2}{\Gamma_{res}M_{res}}\right)
\right] 
+ C \Delta x \Delta y
\end{array}
\end{array}
\end{equation}
\setlength{\extrarowheight}{0pt}
We now introduce the following functions, normalized to integrate to unity,
which will represent the slowly varying and resonant contributions,
respectively:
\begin{equation}
f_1(x) \equiv \frac{1}{\Delta x} \hspace*{0.5cm},\hspace*{0.5cm}
f_2(x)\equiv\frac{R_1(x)}{\int\! R_1(x)\ \!\difd x}
\end{equation}\vspace*{-3.5mm}
\begin{equation}
f(x) \equiv f_1(x) + f_2(x)
\end{equation}
with the primitive functions:
\begin{eqnarray}
F_{1,2}(x) = \int_{x_{min}}^x\!\!\! f_{1,2}(x)\!\ \difd x
\hspace*{0.2cm}&\implies&\hspace*{0.2cm} \difd F_{1,2}(x)=f_{1,2}(x)\difd x
\end{eqnarray}
From the normalization of the $f_i$, it is clear that $F_i(x_{min})$ and
$F_{i}(x_{max})=1$.  The usefulness of these definitions becomes clear when
we rewrite the original integral (using $|\overline{T}|^2$ to denote a
resonant term in $|\overline{M}|^2$ and asuming that $f(x)$ is positive
difinite):
\begin{equation}
\int_{x_{min}}^{x_{max}}\int_{y_{min}}^{y_{max}} |\overline{T}(x,y)|^2 \difd
y\difd x =
\int_{x_{min}}^{x_{max}}\int_{y_{min}}^{y_{max}}
\frac{|\overline{T}(x,y)|^2}{f(x)}
\left(f_1(x)+f_2(x)\right) \difd y \difd x
\end{equation}
The quantity $|\overline{T}|^2/f(x)$ must be reasonably flat. If, for some
$x$, there is suddenly a large deviation from flatness in $|\overline{T}|^2$,
this must be due to the resonance, and so $f(x)$ will also deviate from
flatness there, making the quotient of the two approximately flat over the
entire integration region. We can get rid of the factor multiplying the
quotient by changing integration variables using the primitive functions:
\begin{equation}
\int_{x_{min}}^{x_{max}}\int_{y_{min}}^{y_{max}} |\overline{T}(x,y)|^2 \difd
y\difd x = \int_0^1\int_{y_{min}}^{y_{max}}
\frac{|\overline{T}(x,y)|^2}{f(x)}
 \difd y\difd F_1(x) + \int_0^1\int_{y_{min}}^{y_{max}}
\frac{|\overline{T}(x,y)|^2}{f(x)}
 \difd y\difd F_2(x)\label{eq:MCM2integral}
\end{equation}
Furtermore, the resonance terms in the matrix elements considered in this
work (see appendix \ref{app:susydecays}) only depend on the resonant
variable, here denoted $x$, so
$|\overline{T}(x,y)|^2=|\overline{T}(x)|^2$. The integral over $\difd y$ then
simply becomes $\Delta y$, and there is now no problem in discretizing
eq.~(\ref{eq:MCM2integral}) using a simple grid lattice in $F_i$ (i.e.\ with
constant grid spacing).  The definitions of the primitive functions can then
be inverted to yield $x$ as a function of $F_i$.  Using an $N\times N$
lattice, one obtains:
\begin{eqnarray}
\Delta y\int_{0}^{1}
\frac{|\overline{T}(x)|^2}{f(x)}
 \difd F_{1,2}(x) & \to & \frac{\Delta y}{N}\sum_{i=1}^N
 \frac{|\overline{T}(x_i)|^2}{\frac{1}{\Delta x} +
\frac{R(x_i)}{\int\!\! R\ \! \difd
 x}} \label{eq:discretization}
\end{eqnarray}
It is clear that a uniform stepping in the $F_1$ variable corresponds
directly to a uniform stepping in $x$:\vspace*{-3mm}
\begin{equation}
F_{1}(x) = \frac{1}{\Delta x}
\int_{x_{min}}^{x}\hspace*{-3mm}\difd x
\hspace*{0.5cm}\implies\hspace*{0.5cm} x = x_{min} + F_{1} \Delta x 
\end{equation}
so that a stepping of $F_1$ between 0 and 1 will produce a uniform
distribution of $x$ values between $x_{min}$ and $x_{max}$.  For the second
integral in eq.~(\ref{eq:MCM2integral}), the expressions become a bit more
involved. Translating the simple $F_2$ grid to $x$ values involves the
inversion of a complicated function:
\begin{eqnarray}
F_2(x) = \int_{x_{min}}^{x} \frac{R(x)}{\int\!\! R(x)\ \! \difd x} \difd x =
\frac{
\arctan\left(\frac{x-M_{res}^2}{\Gamma_{res}M_{res}}\right)  
-
\arctan\left(\frac{x_{min}-M_{res}^2}{\Gamma_{res}M_{res}}\right)}{
\arctan\left(\frac{x_{max}-M_{res}^2}{\Gamma_{res}M_{res}}\right)  
-
\arctan\left(\frac{x_{min}-M_{res}^2}{\Gamma_{res}M_{res}}\right)}
\end{eqnarray}
Isolating $x$ in this equation (using
\mathematica) yields 
\begin{equation}
x =
M_{res}^2+\Gamma_{res}M_{res}
\tan\left[
  F_2\arctan\left(\frac{x_{max}-M_{res}^2}{\Gamma_{res}M_{res}}\right)
  -(F_2-1)\arctan\left(\frac{x_{min}-M_{res}^2}{\Gamma_{res}M_{res}}\right)
\right]
\label{eq:x2}
\end{equation}
This is not as bad as it looks. Most of the involved quantities need to be
evaluated only once, since only $F_2$ changes. A quick check is furnished by
letting $F_2$ be zero in which case it is obvious from the above formula that
$x$ becomes simply $x_{min}$ (only the +1 of the last term in (\ref{eq:x2})
contributes, and $\tan(\arctan(f(x)))=f(x)$). For $F_2=1$, it is also quick
to verify that $x=x_{max}$. We have thus arrived at a complete prescription
for how to press the resonances flat, obtaining the sum of a uniform and a
non-uniform grid in $x$ instead. 

\section{\LV-SUSY in PYTHIA. User's Reference.}
\label{sec:pythia}
\subsection{Switches \& Parameters set by the user in the
\pythia\ calling program}
\begin{itemize}
\item IMSS(51): $\lambda$ couplings. Default is off (=0).
\begin{itemize}
\item (=0): $\lambda$ couplings off. 
\item (=1): All allowed couplings set to fixed value given by
$10^{\mathrm{RMSS(51)}}$. Default value for RMSS(51) is 0. 
\item (=2): Couplings set to ``natural values'',
$\lambda_{ijk}\propto\sqrt{m_im_jm_k/(126\GeV)^3}$, as defined in
\cite[eq.~(17)]{hinchliffe93}, 
giving a hierarchical structure in generations. RMSS(51)
is here used to give the constant of proportionality. If the user does not
have other specific wishes, its value should be set to 1 when using 
this scenario.
\item (=3): The program initializes with default values ($\lambda_{ijk}=0$),
but expects the user to enter all non-zero couplings by hand into the
RVLAM(I,J,K) array in the calling program. 
\end{itemize}
\item IMSS(52): $\lambda'$ couplings. Default is off (=0).
\begin{itemize}
 \item (=0): $\lambda'$ couplings off. 
 \item (=1): All couplings set to fixed value given by
   $10^{\mathrm{RMSS(52)}}$. Default value for RMSS(52) is 0. 
 \item (=2): Couplings set to ``natural values'',
$\lambda'_{ijk}\propto\sqrt{m_im_jm_k/(126\GeV)^3}$, as defined in
\cite{hinchliffe93}, with the constant of proportionality given by
 RMSS(52). If the user does not
have other specific wishes, its value should be set to 1 when using 
this scenario.
\item (=3): The program initializes with default values ($\lambda'_{ijk}=0$),
but expects the user to enter all non-zero couplings by hand into the
RVLAMP(I,J,K) array in the calling program. 
\end{itemize}
\item (IMSS(53)): Should be reserved for $\lambda''$ couplings on/off
together with RMSS(53).
\item PYMSRV : This is the commonblock in \pythia\ where the $L$-violating
couplings reside; the $\lambda_{ijk}$ couplings are stored in RVLAM(i,j,k),
and the $\lambda'_{ijk}$ in RVLAMP(i,j,k). Space has been reserved for an
eventual future study of $B$-violation in the RVLAMB(i,j,k) array which is
intended to store the $\lambda''_{ijk}$ couplings.
\end{itemize}
\subsection{Subroutines and functions handling $R$-violation inside \pythia}
\begin{itemize}
\item PYRVST : Subroutine to print information about semi-inclusive branching
ratios and coupling values. Called from PYSTAT if R-Violation has been turned 
on (IMSS(51)=1 or IMSS(52)=1).
\item PYRVSF : Subroutine to calculate \RV\ decay widths of sfermions based
on the selected SUSY scenario and coupling values. If there is no phase space
available for a given mode, the width for that mode is set to zero and
\pythia\ is told to ignore the mode completely. 
\item PYRVN  : Same as above for \RV\ decays of neutralinos.
\item PYRVCH : Same as above for \RV\ decays of charginos.
\item PYRVSB : Calculates sfermion two-body widths.
\item PYRVGW : Routine for calculating three-body 
decay widths of neutralinos and charginos.
\item PYRVI1, PYRVI2, PYRVI3 : Functions to integrate resonant terms, L-R
interference terms, and true interference terms, respectively over phase
space.
\item PYRVG1, PYRVG2, PYRVG3, PYRVG4 : Integrands for the phase space
integrations.
\end{itemize}
In addition, a consistency check has been added to the routine which
diagonalizes the neutralino mass matrix (PYEIG4) so that error messages will
be printed if the off-diagonal elements of the diagonalized matrix
(recomputed from the original matrix and the diagonalizing matrix given by
the routine) are larger than $10^{-6}$ or if the diagonal elements calculated
by brute force deviate by more than $10^{-6}$ from the values in the mass
vector given by the routine. This does not encumber the performance of
\pythia\ in any way, since the matrix needs to be diagonalized only once for
each run. 

\subsubsection{Warnings and error messages:}
\begin{itemize}
\item \texttt{PYRVGW ERROR: Negative Width in
$\ti{a}\to b\ c\ d$}. This error, if encountered, is not as serious as one
should think. Even though
resonant diagrams are not included in the calculations, neither are the
interferences between resonant and non-resonant diagrams (which are
proportional to $\Gamma_{\mathrm{res}}^2/M_{\mathrm{res}}^2$ and so
negligible for narrow resonances). This means that the calculation should
still be positive definite, and so a negative width can only occur as a
result of numerical error. Such errors are most
likely to occur in the presence of strong threshold effects where the diagram
values are small (due to limited phase space) and 
can be greatly differing, quickly building up a sizeable numerical
error. Since the decay rate in these cases is at any rate very small
(compared to rates for modes with bigger phase spaces), the program just
keeps running, setting the offending width to zero. One may want to check,
however, that the process concerned really does lie near the edge of its
phase space. 
\end{itemize}
\subsection{Resonant decay treatment}
\label{sec:resdec}
Each time a physical particle is
generated in \pythia, a random mass is calculated, distributed according to
the Lorentz profile for that particle. As decay widths and therefore branching
fractions depend on the mass of the incoming particle\footnote{This becomes
especially important when some of the channels contain particles with masses
close to  the decaying particle's mass. In that case, there will be a
threshold where only particles above the threshold mass can decay via that
channel.},  
the most detailed method for determining in which channel the particle will
decay is to calculate its widths to all available channels at that mass and
from them determine the branching fractions. This is feasible
for low-mass particles with few and/or simple decay channels, and so every
time a $W$ boson, for example, decays in \pythia, its branching fractions are
evaluated at the specific mass of that $W$. For supersymmetric particles
with hundreds of decay channels and many adjustable SUSY parameters entering
the calculations, this 
method would be very time consuming with \emph{all} decay widths having to be
recalculated each time a particle is generated. Instead,
an approximate method is used. All branching fractions are evaluated once and
for all at the resonance mass during each program execution. This saves CPU
time, yet
it has the potential danger that some of the sparticles produced in the
generator have masses that could be lower than some threshold where some
channel reaches zero phase space. Allowing such a sparticle to decay into
that channel would result in violation of energy conservation. The way that
\pythia\ is told to handle this in the present case is 
to go away from the full Lorentz profile when
generating masses, using instead a limited range (symmetrically, so as not to
skew the center point) where the lower bound is the kinematical threshold
causing the trouble. 


\clearpage
\section{Trigger Shapes\label{app:trigger}}
\subsection{Background Rates}
A comprehensive collection of trigger rates ass functions of thresholds
is here given. 
Note that the vertical scale is not the same in the plots. 
It is implictly
understood that all threshold values are on $p_T$ in \GeV.
With respect to the normalizations of these plots, one
should allow for significant uncertainties since we are extrapolating already
uncertain parton distributions (GRV 94L) over orders of magnitude in
energy. Caution in too strong interpretations of these figures should also be
exercized since the impact of the PDF uncertainties on the \emph{shape} of
the distributions can be non-trivial. 

Plots show the total background rates
passing the trigger $p_T$ 
thresholds as functions of the thresholds. For the two-muon
and two-electron trigger, \emph{both} of the muons/electrons are required to
be above threshold. Since the number of events passing a threshold should
obviously be a monotonically decreasing function of the threshold value, 
all plots shown should be strictly decreasing functions of the
thresholds. In some of the plots, there is some ``noise'' or ``jitter''
superimposed on this decreasing behaviour, due to the way muon and electron
reconstruction efficiencies were taken into account in the analysis. For each
event for each trigger threshold a random number was compared to the trigger
efficiency for each lepton candidate. This creates a natural ``jitter''
since more events could be thrown away for one threshold than another if
generated event numbers are low, as happens on the tails of the $p_T$
distributions. This problem is especially significant for the low-$p_T$ QCD
$2\to 2$ processes where rates are extremely high and the number of events in
the $p_T$-tails surviving cuts are extremely low. Even though comparably
large event samples, $\mathcal{O}(10^8)$ events, have been generated for
the lowest $p_T$ processes, 
still only two make it to the e+mu10I trigger (see
below), creating sudden drops in the trigger rate at the $p_T$'s of the
electrons in those events. With larger samples,
these problems could be avoided, yet considering the uncertainties already
associated with QCD processes in the extreme parts of the $p_T$ spectrum, it
has not been deemed worth the (significant) extra computer resources and time
required to do so here. The overall numbers of generated events used in the
trigger analysis are given in table \ref{tab:ngentrig}. \clearpage
\begin{table}[h!]
\begin{center}
\begin{tabular}{lrrrr}\toprule
\textbf{Process} & \begin{tabular}{c}QCD $2\to
2$\\{\scriptsize$p_T=1-10\GeV$}\end{tabular} & \begin{tabular}{c}
QCD $2\to 2$\\{\scriptsize$p_T=10-75\GeV$}\end{tabular}
&\begin{tabular}{c}QCD $2\to 2$\\{\scriptsize$p_T=75-150\GeV$}\end{tabular}
& \begin{tabular}{c}QCD $2\to 2$\\{\scriptsize$p_T>150\GeV$}\end{tabular}
\\\cmidrule{1-5}
\boldmath\bf$\sigma$ & 55\!\ mb & 12\!\ mb & 5.5\ttn{-3}mb& 2.9\ttn{-4}mb\\
{\bf\boldmath Rate \footnotesize[s$^{-1}$]} 
& 1.6\ttn{8} & 3.7\ttn{7} & 1.7\ttn{4} & 8.7\ttn{2}\\
\boldmath\bf$N_{gen}$    & 2.5\ttn{8} & 2.2\ttn{8} & 1.4\ttn{7} & 1.1\ttn{7}\\\vspace*{1.ex}\\ \cmidrule{1-5}
\bf Process & $t\bar{t}$ & $Z/W$ & $ZZ/ZW/WW$ \\
\cmidrule{1-5} 
\boldmath\bf$\sigma$ & 6.2\ttn{-7}mb & 1.2\ttn{-3}mb & 1.2\ttn{-7}mb\\
{\bf\boldmath Rate \footnotesize[s$^{-1}$]} &1.9 & 3.6\ttn{3} & 0.36\\
\boldmath\bf$N_{gen}$ & 5.9\ttn{6}&1.8\ttn{8}&5.9\ttn{6} \\
\bottomrule
\end{tabular}
\caption[\small Numbers of generated events for the trigger study]{Numbers of
generated events for the trigger study. The rates listed are total rates
before trigger for $L=3\ttn{33}\scm$. To avoid unnecessarily messy plots,
the two lowest $p_T$ QCD $2\to 2$ samples have been combined and 
are shown in the same colour below, as are the two highest $p_T$
samples.\label{tab:ngentrig}} 
\end{center}
\end{table}
\begin{center}
\setlength{\extrarowheight}{0pt}
\includegraphics*[scale=0.25]{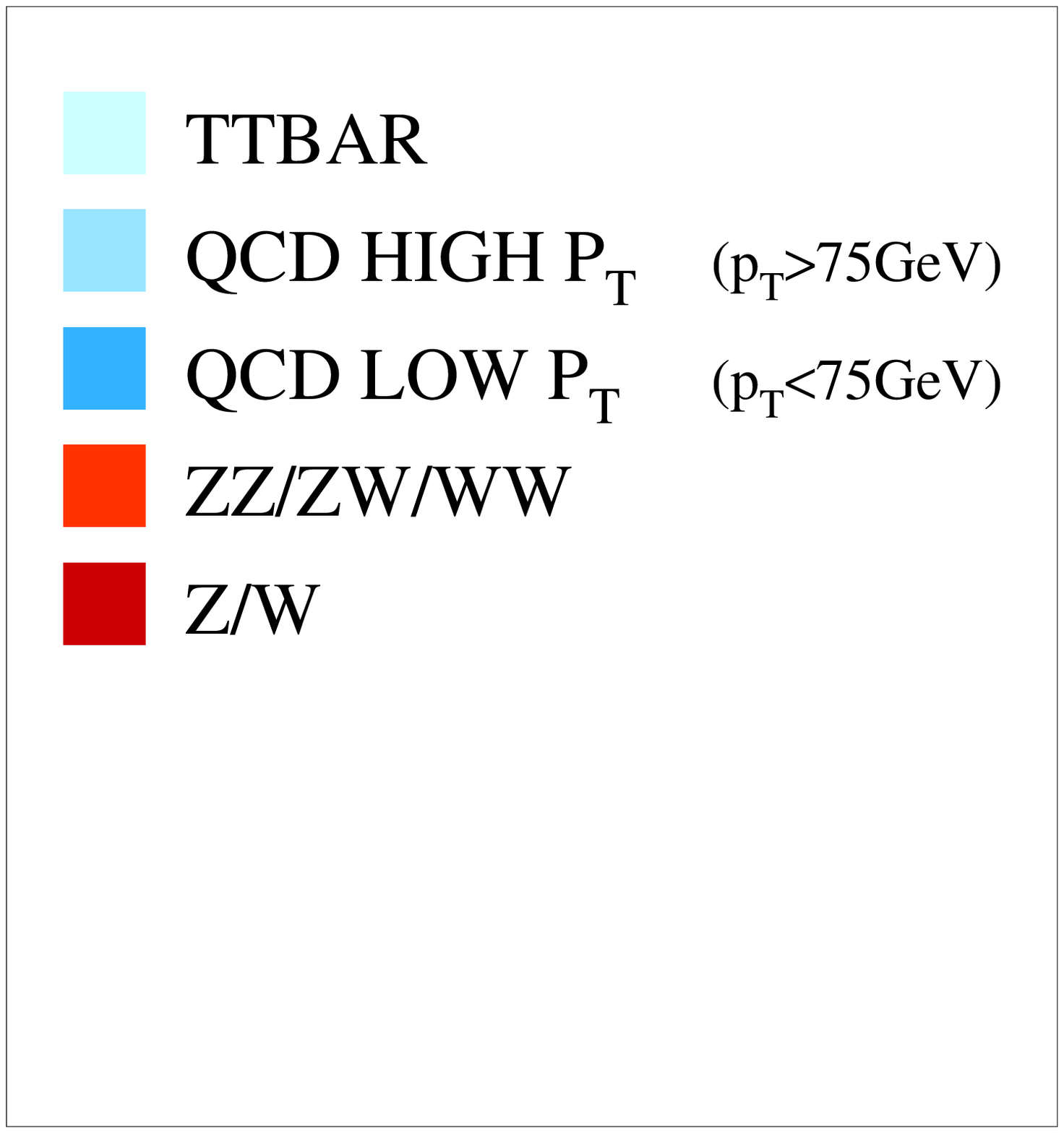} \hspace*{-3mm} 
\includegraphics*[scale=0.25]{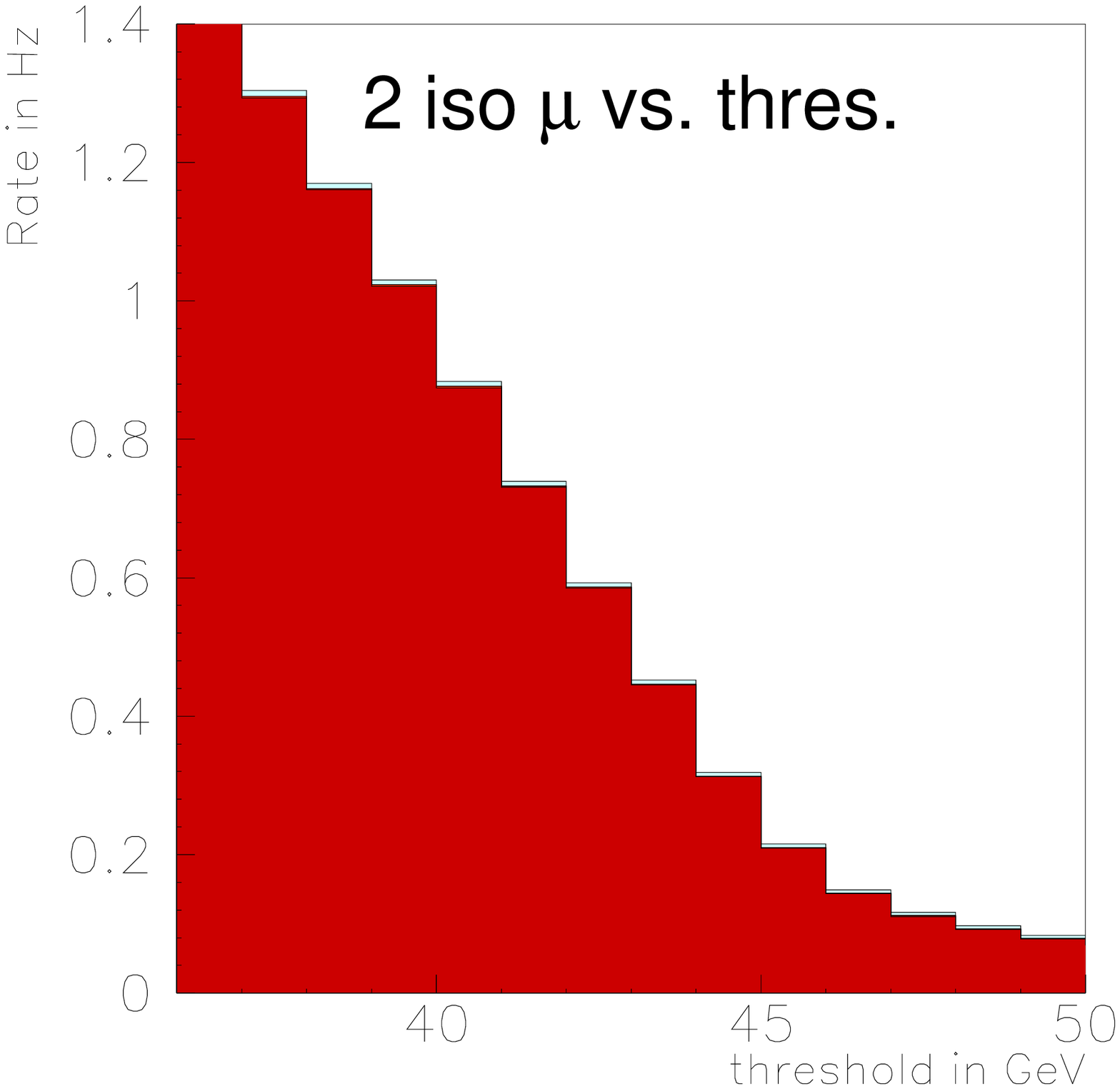} \hspace*{-3mm} 
\includegraphics*[scale=0.25]{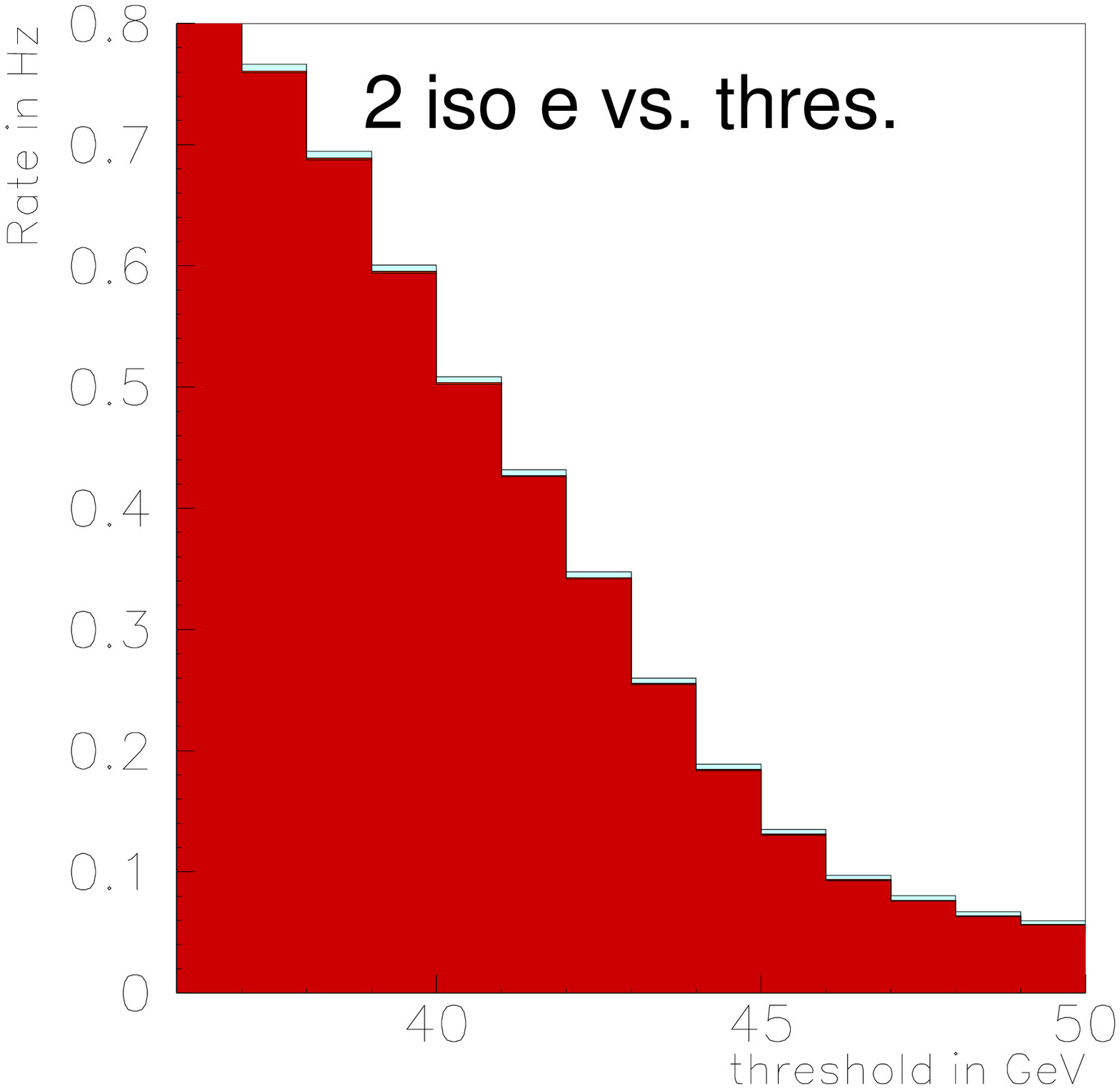} \vspace*{-3mm}\\ 
\includegraphics*[scale=0.25]{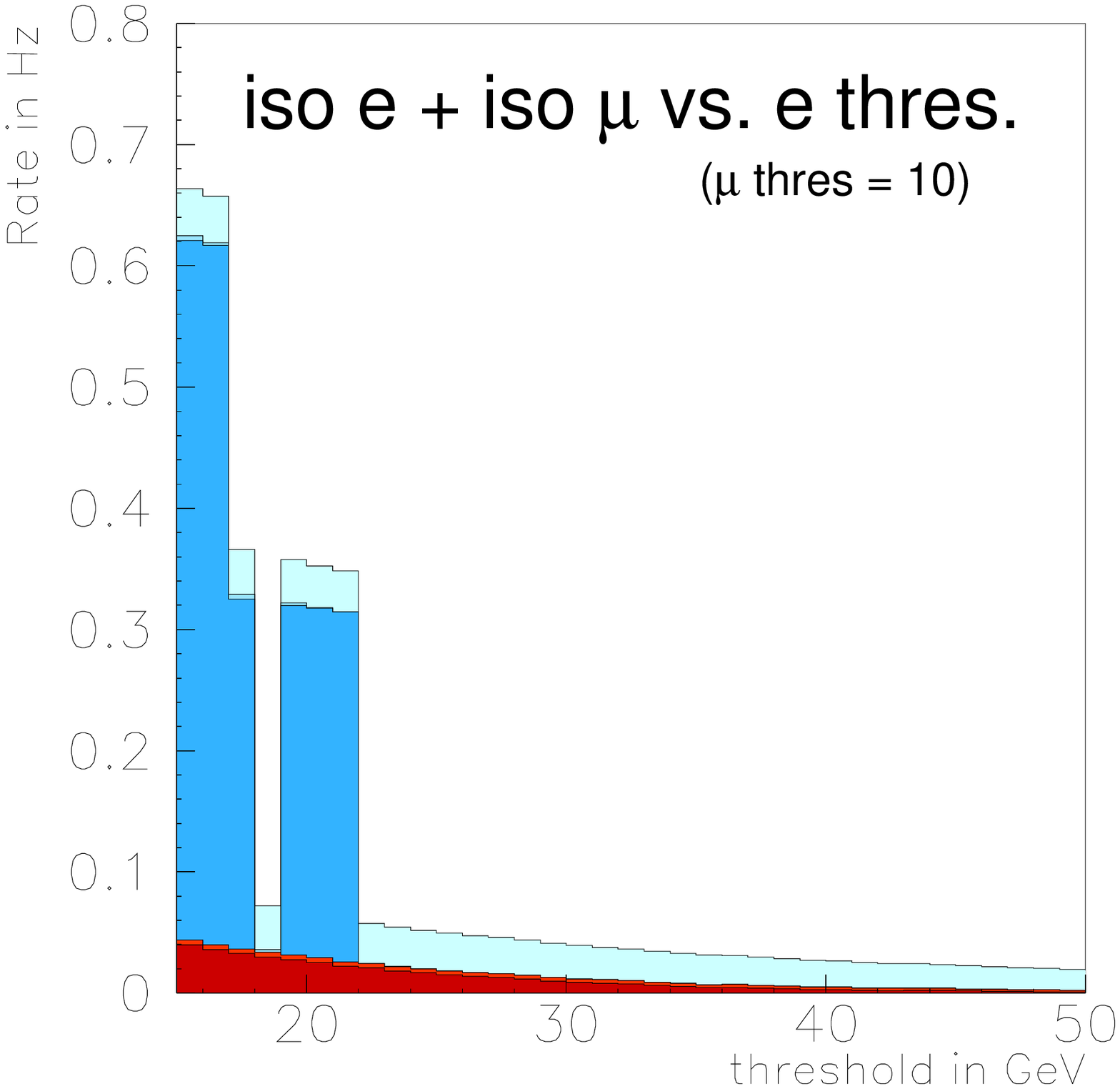} \hspace*{-3mm} 
\includegraphics*[scale=0.25]{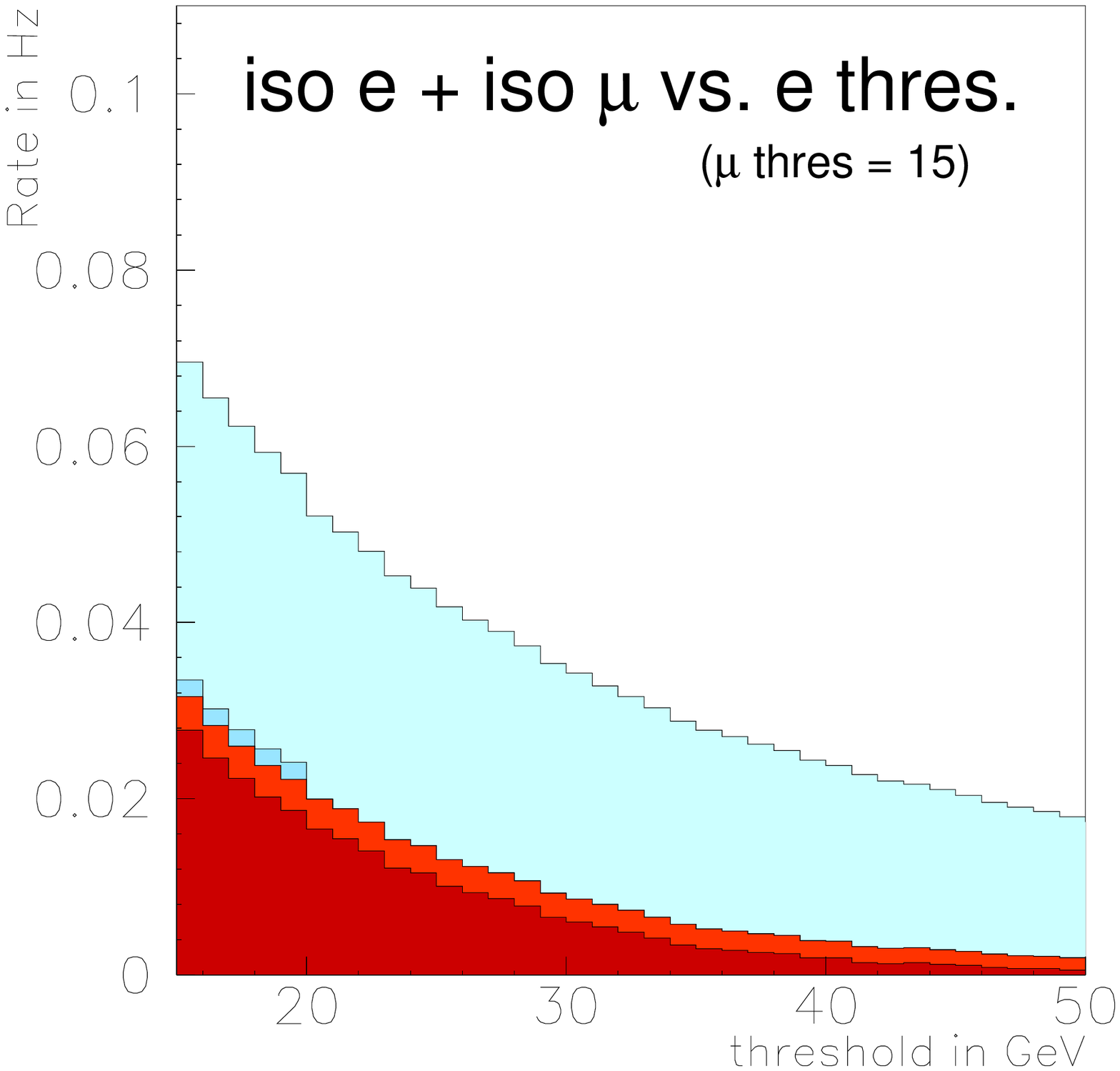} \hspace*{-3mm} 
\includegraphics*[scale=0.25]{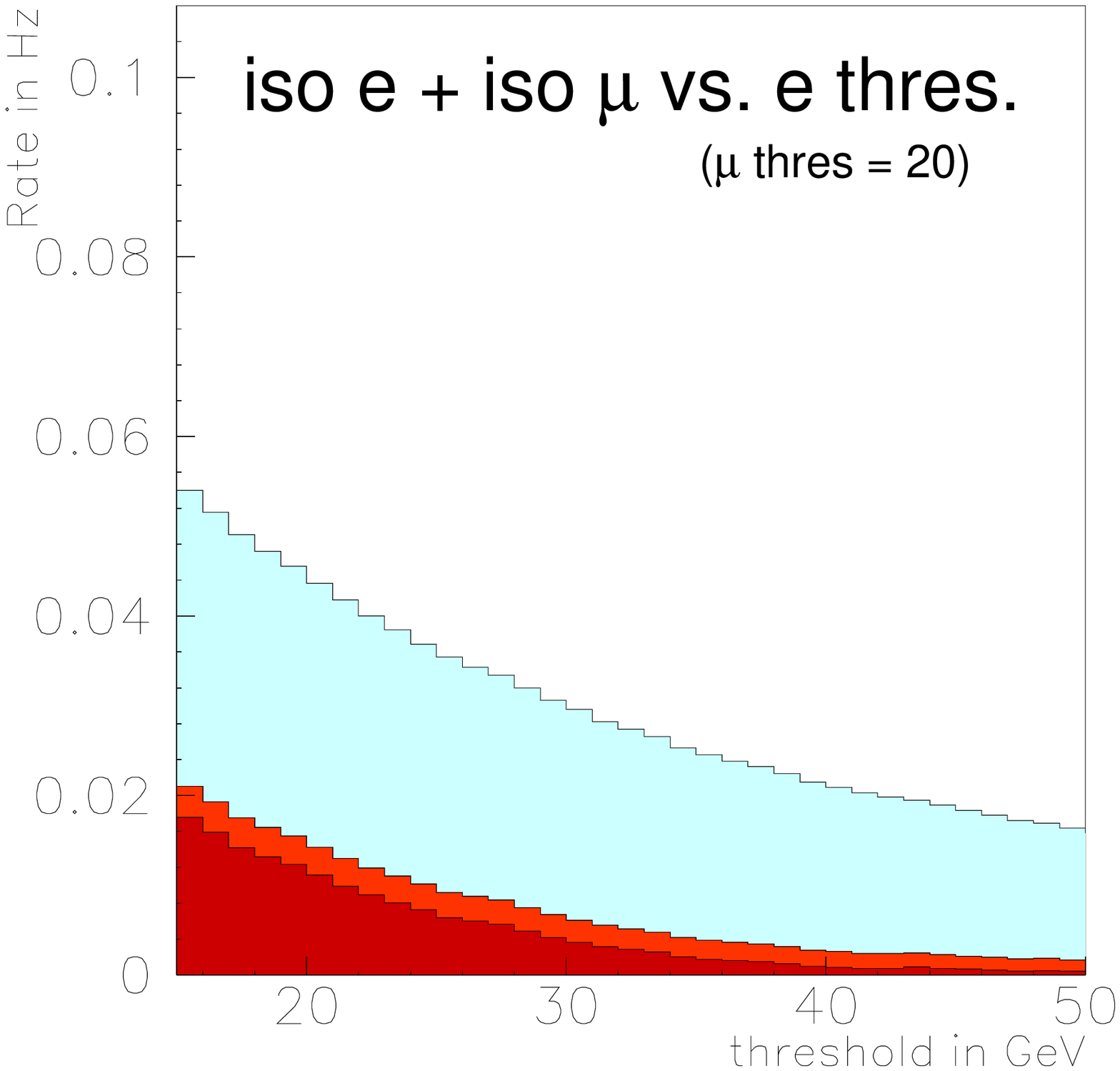} \vspace*{-3mm} \clearpage
\includegraphics*[scale=0.25]{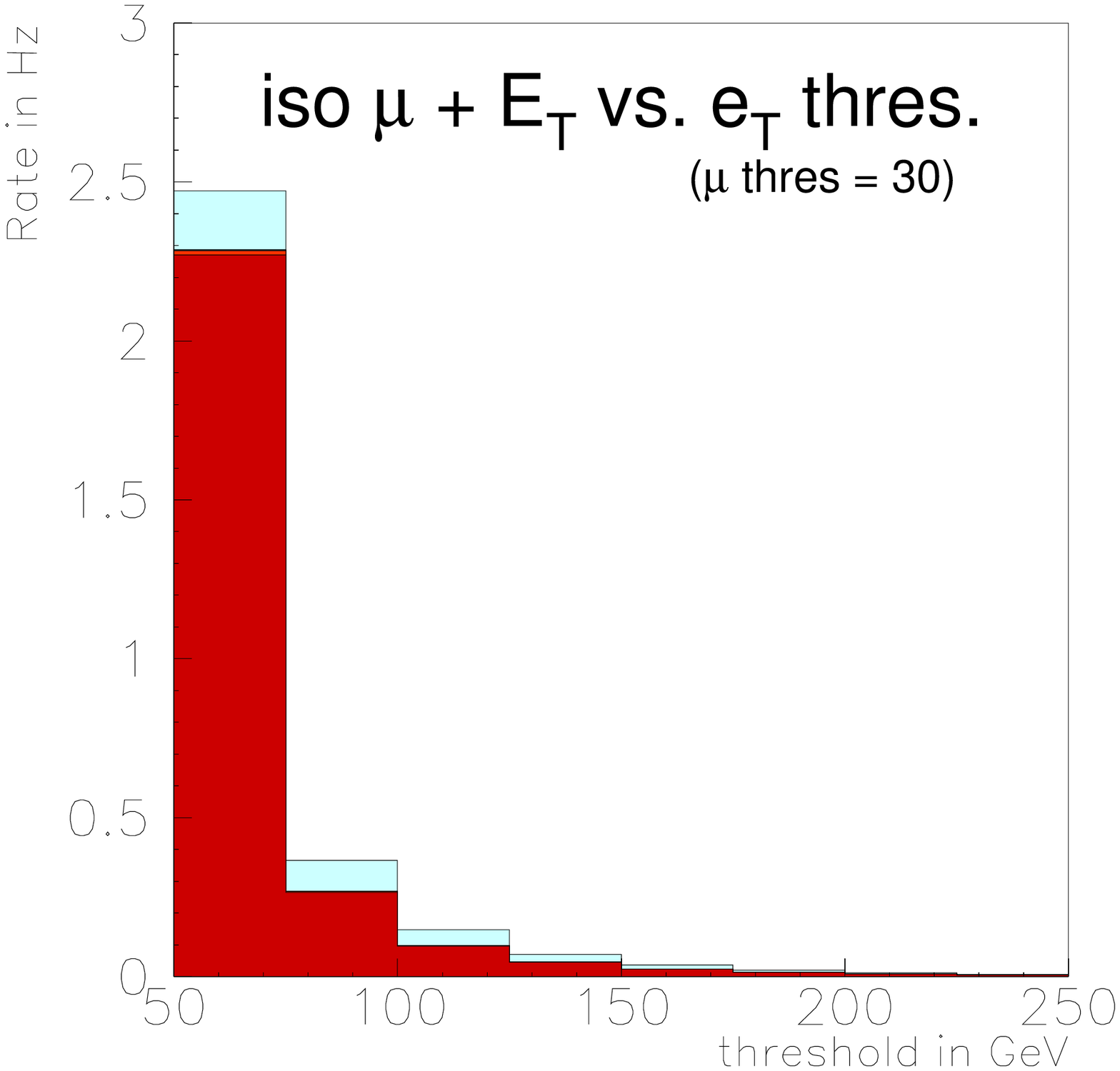} \hspace*{-3mm} 
\includegraphics*[scale=0.25]{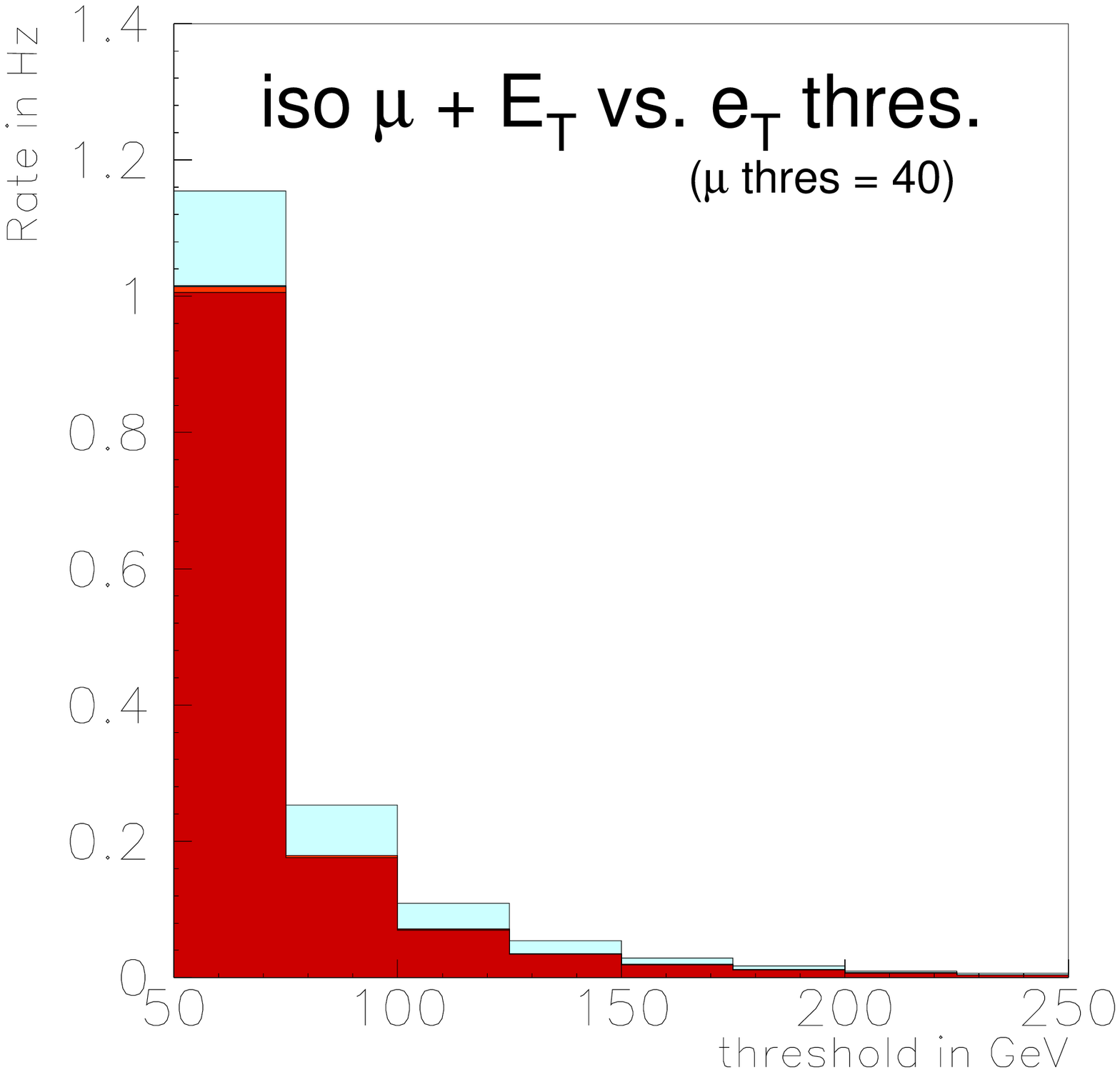} \hspace*{-3mm} 
\includegraphics*[scale=0.25]{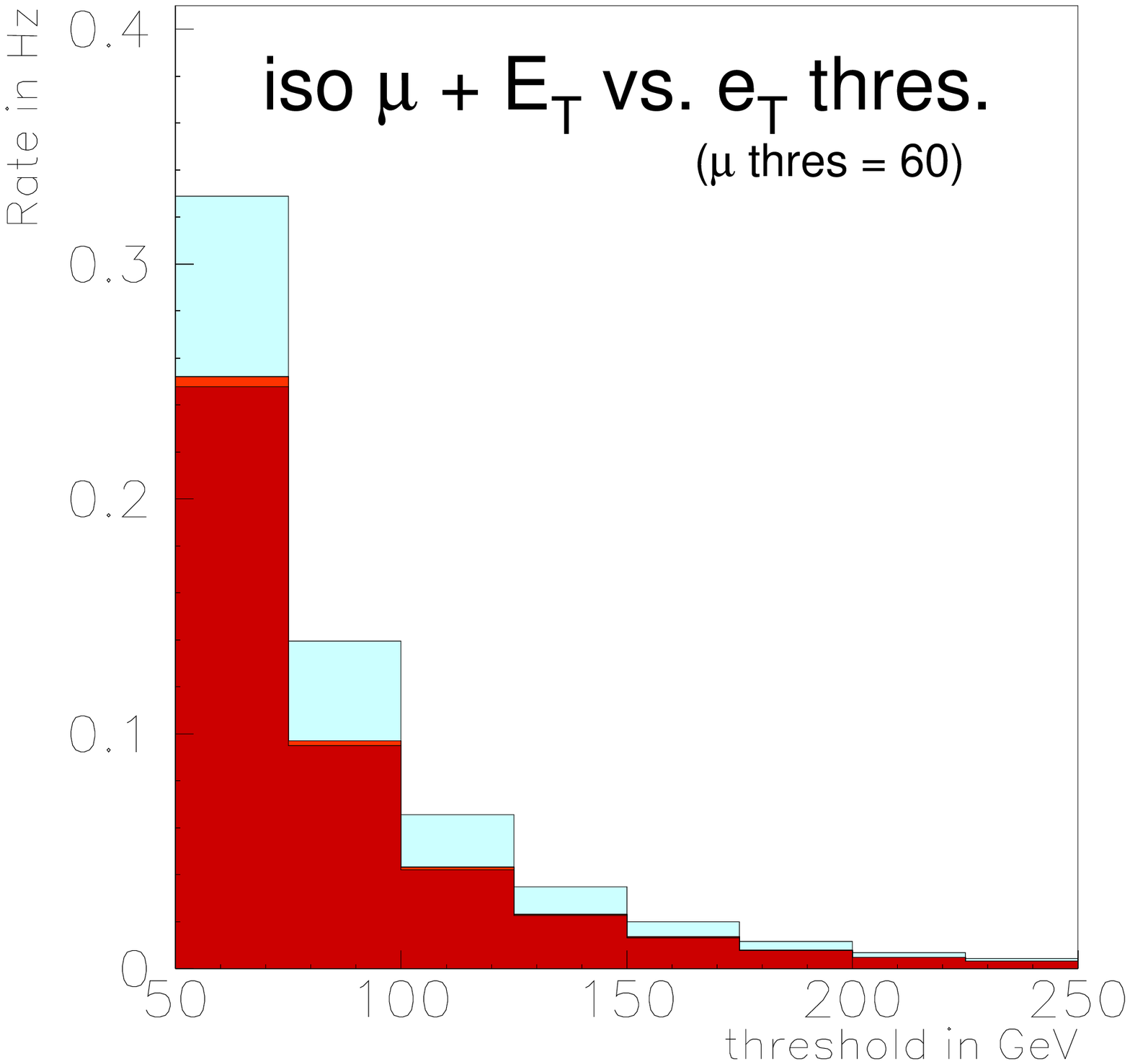} \vspace*{-3mm} \\
\includegraphics*[scale=0.25]{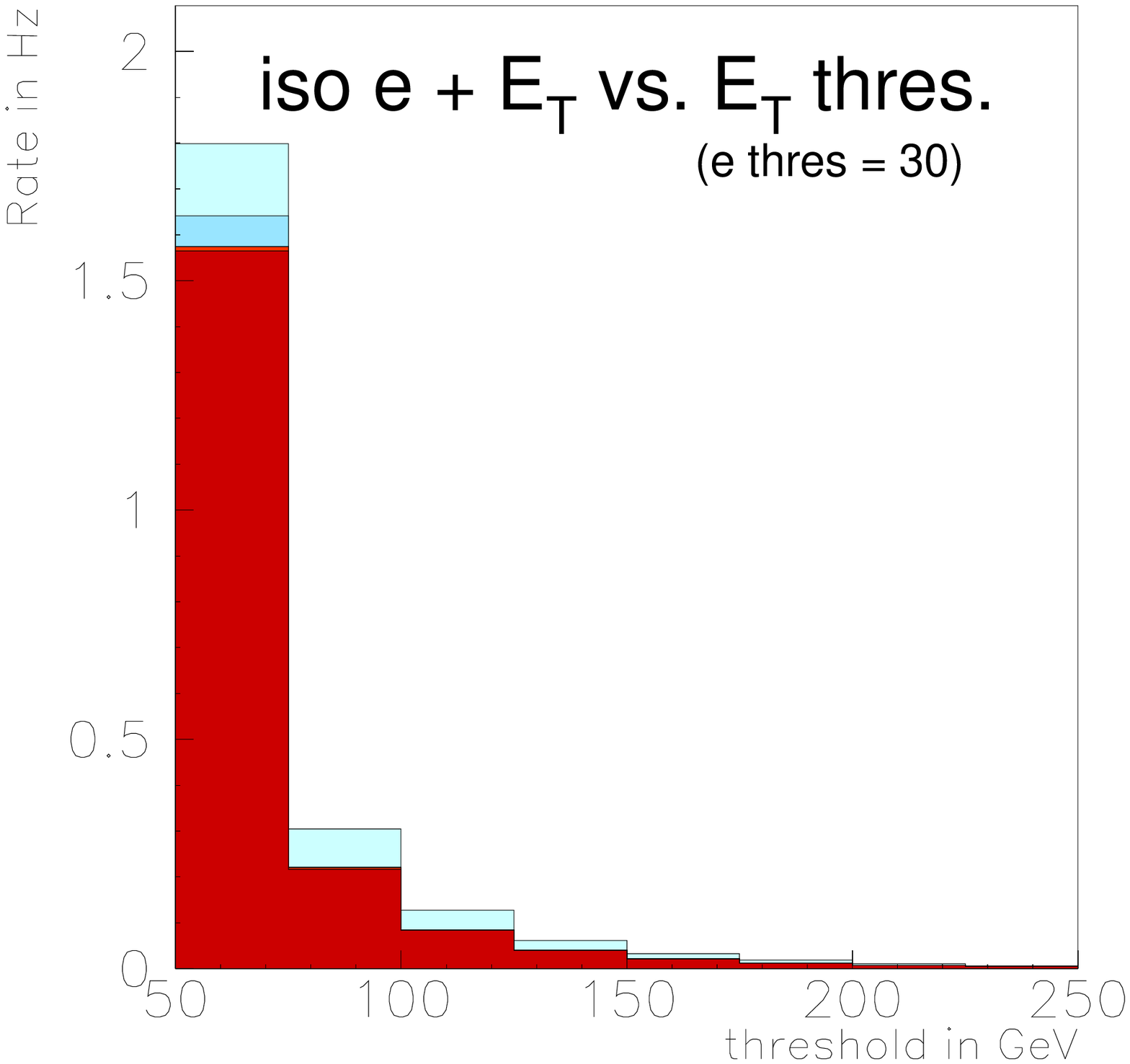} \hspace*{-3mm} 
\includegraphics*[scale=0.25]{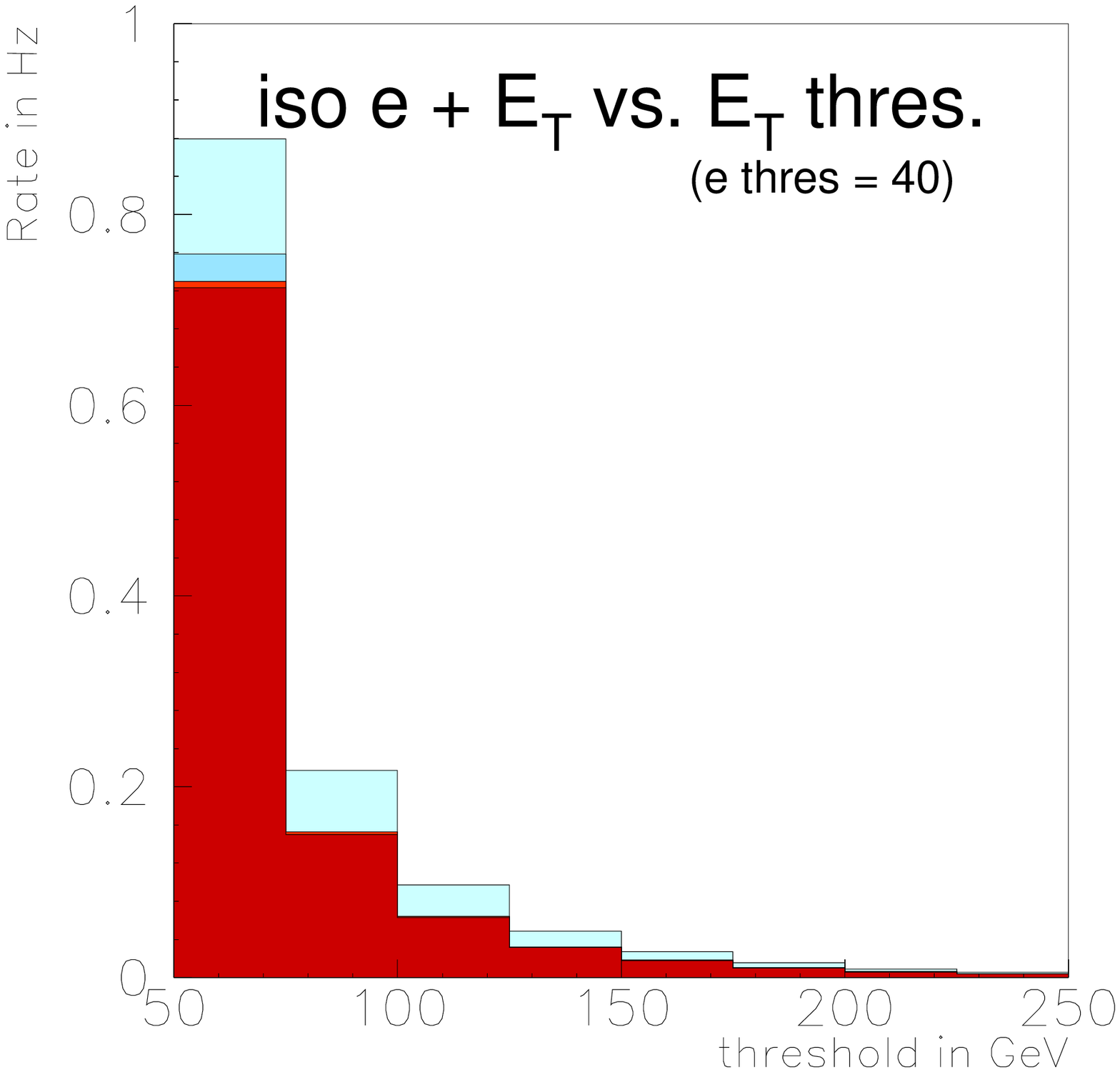} \hspace*{-3mm} 
\includegraphics*[scale=0.25]{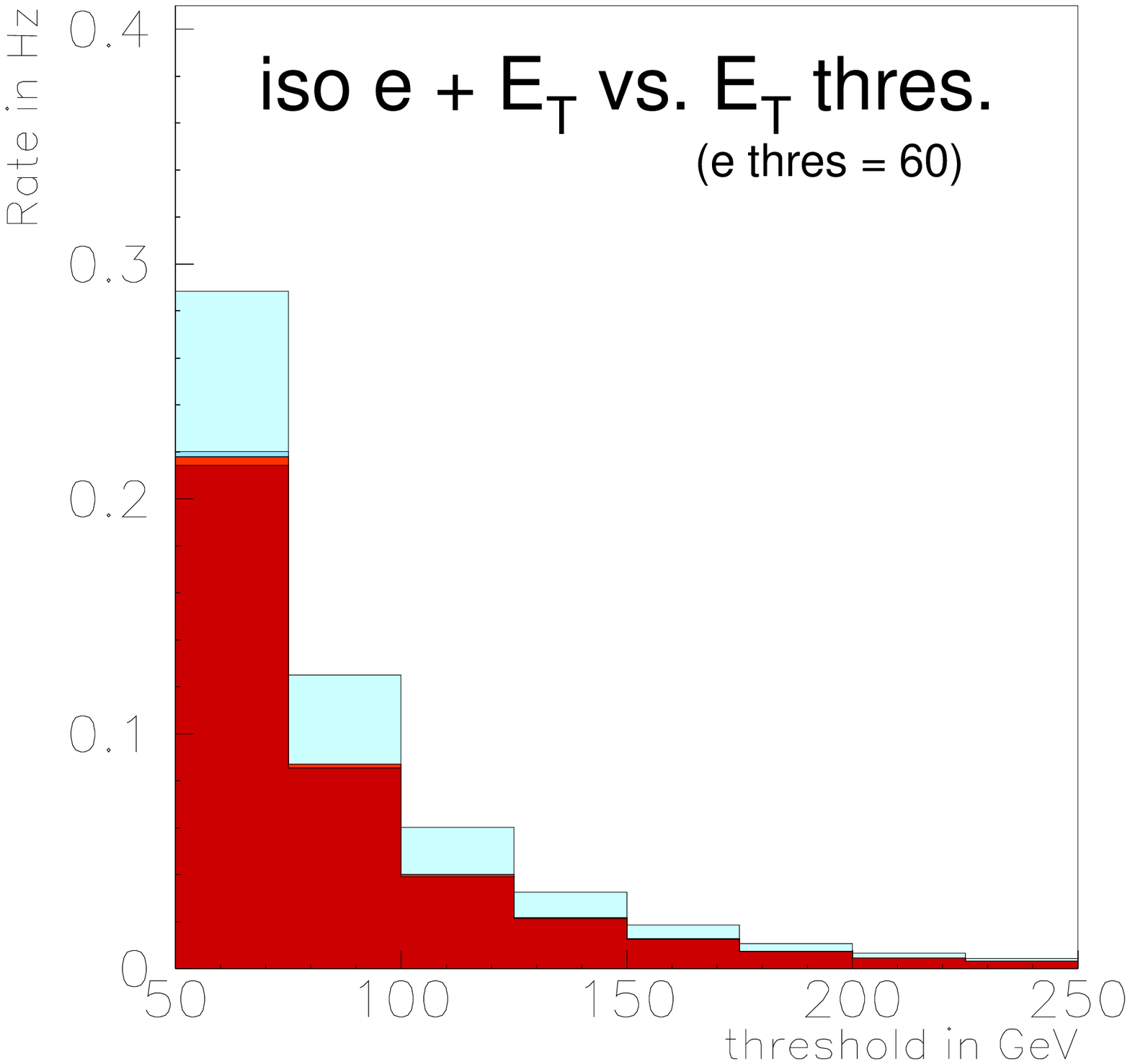} \vspace*{-3mm} \\
\includegraphics*[scale=0.25]{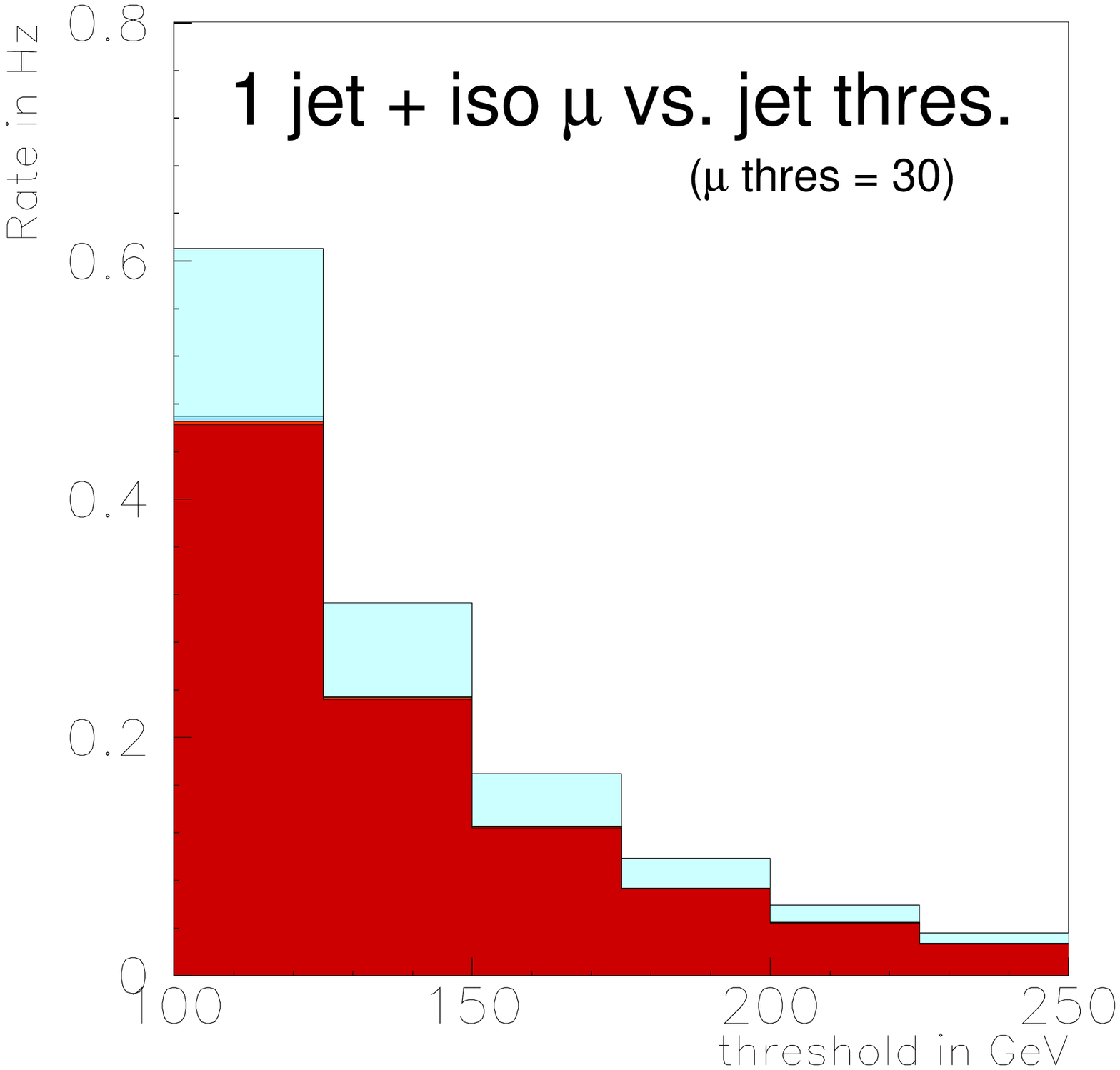} \hspace*{-3mm} 
\includegraphics*[scale=0.25]{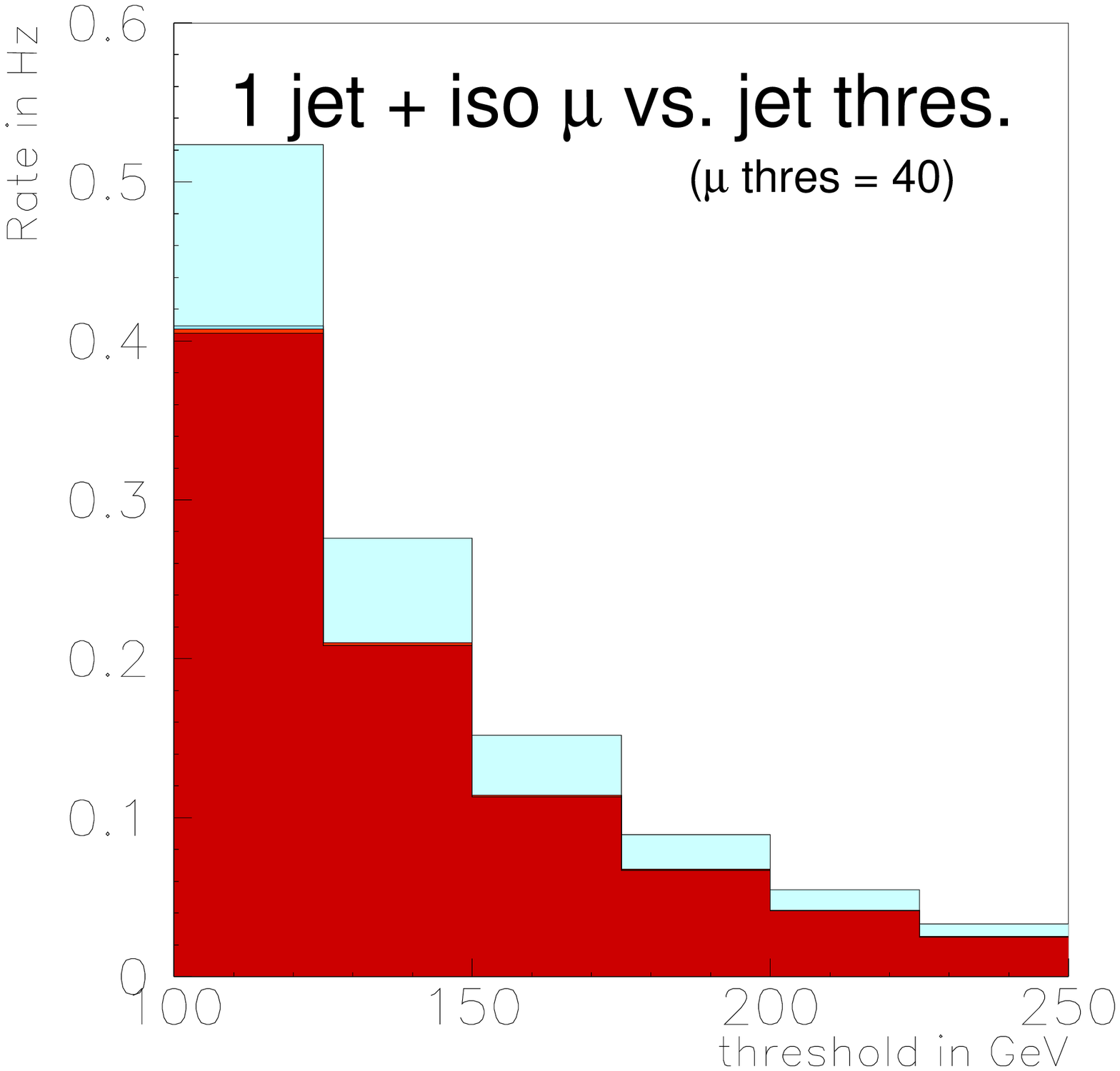} \hspace*{-3mm} 
\includegraphics*[scale=0.25]{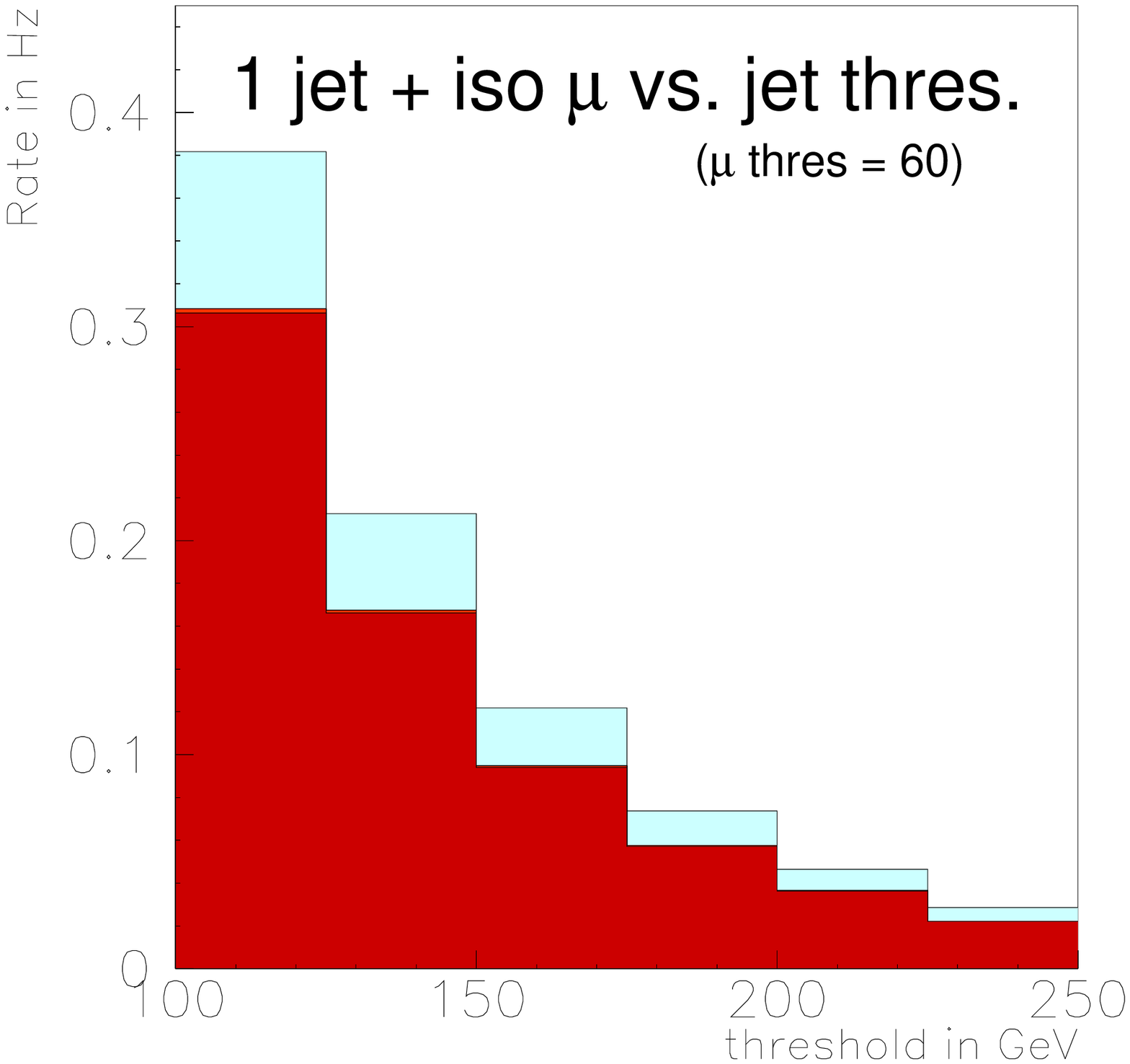} \vspace*{-3mm} \\
\includegraphics*[scale=0.25]{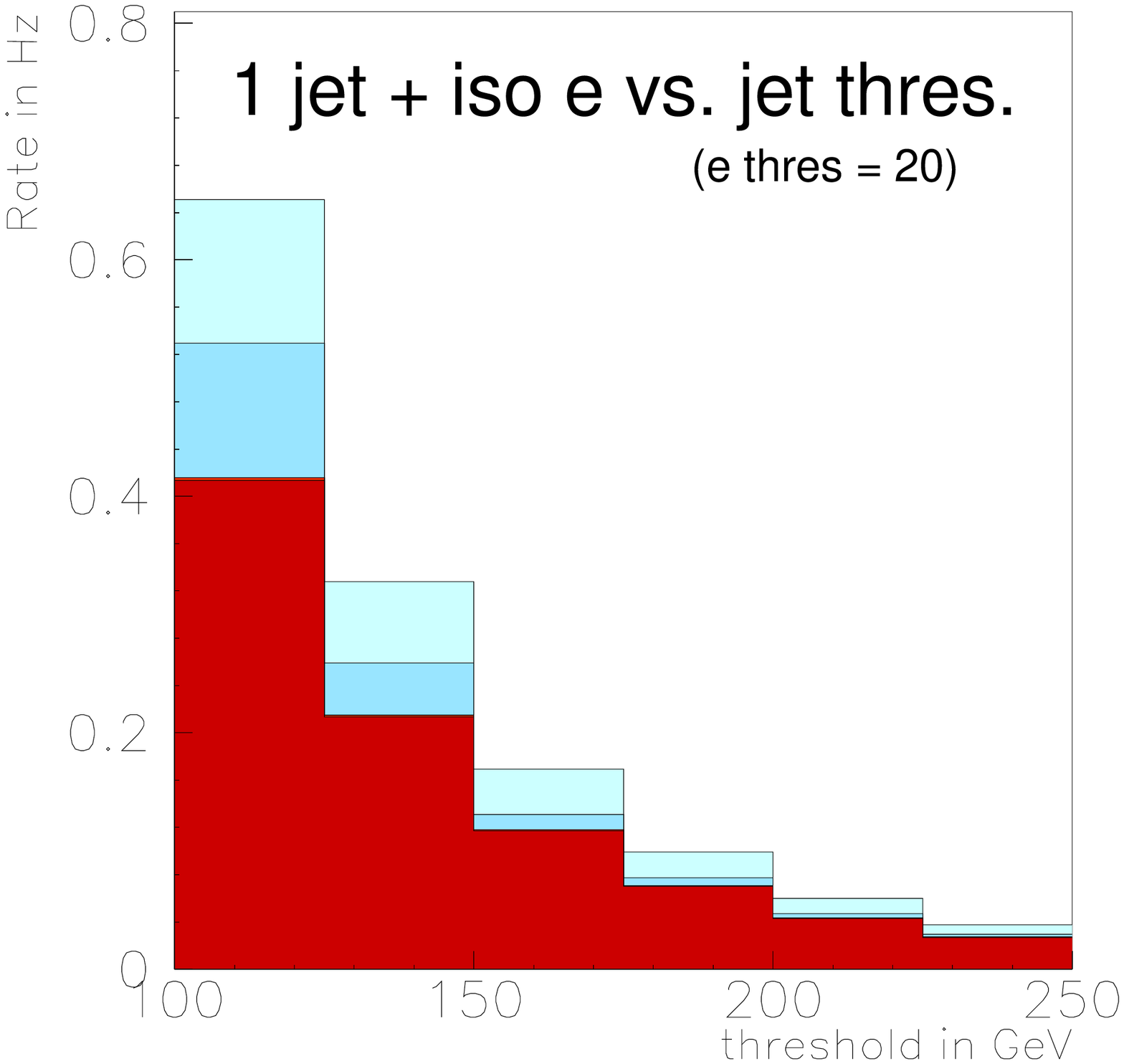} \hspace*{-3mm} 
\includegraphics*[scale=0.25]{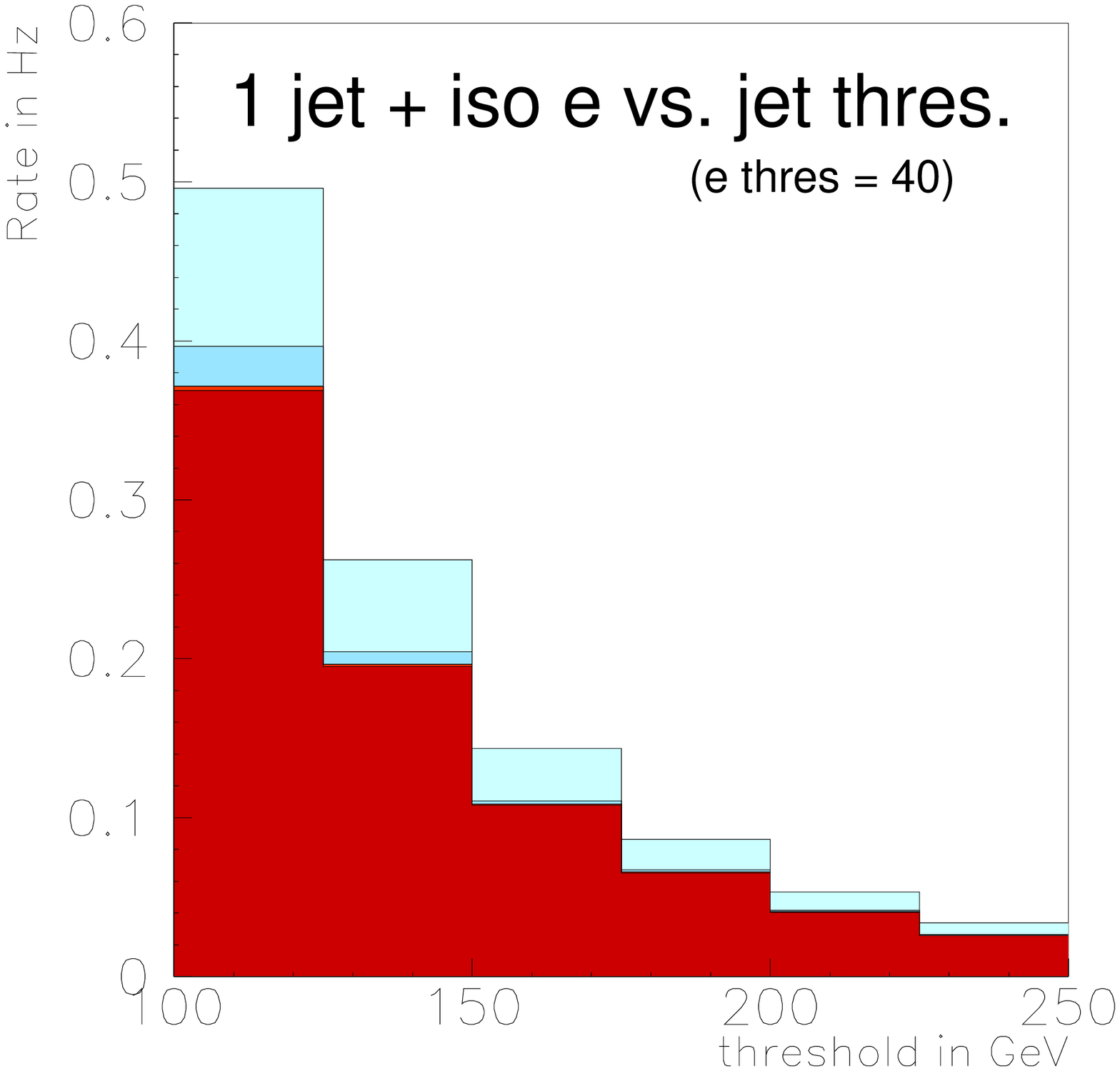} \hspace*{-3mm} 
\includegraphics*[scale=0.25]{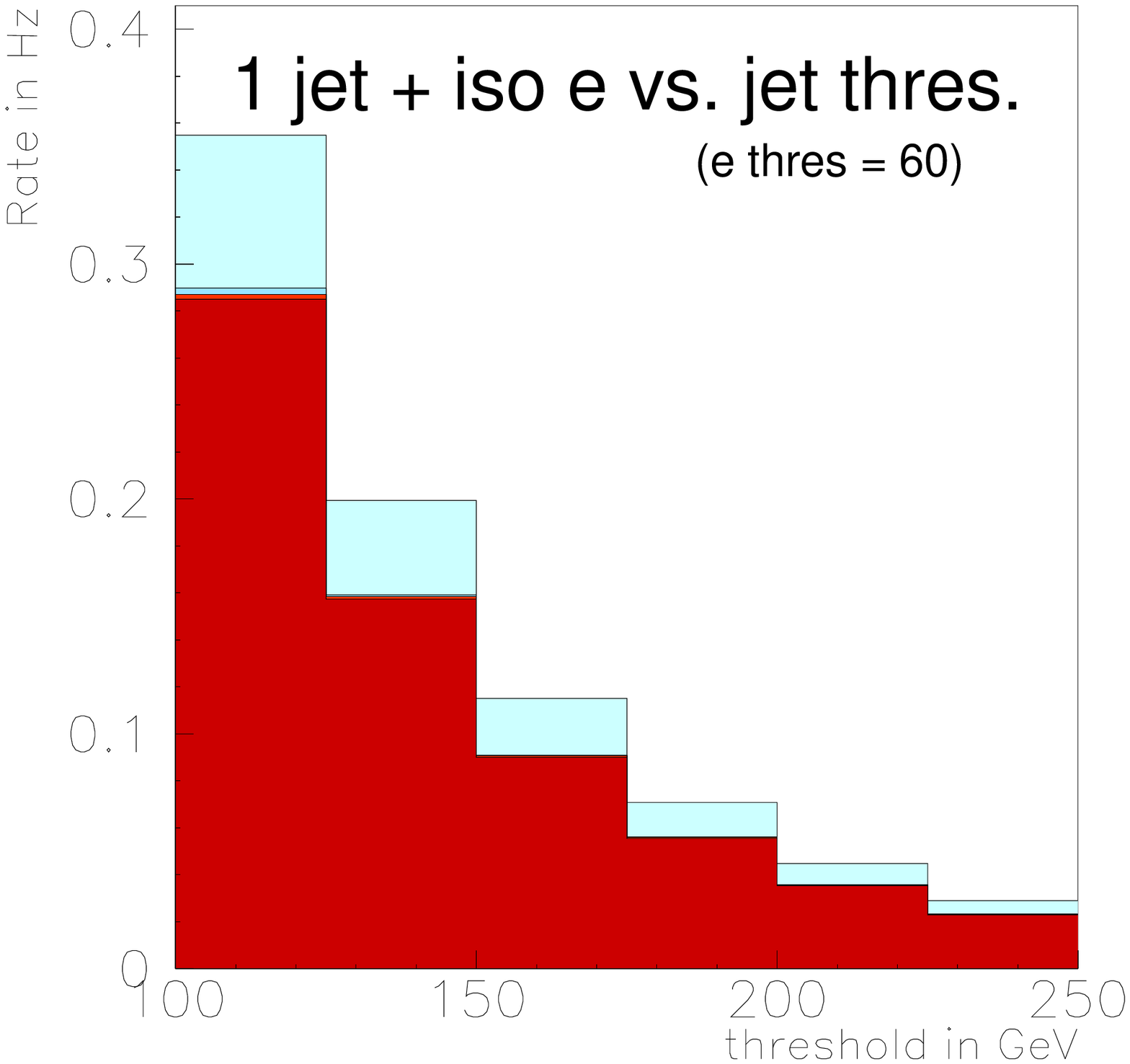} \vspace*{-3mm} \clearpage
\includegraphics*[scale=0.25]{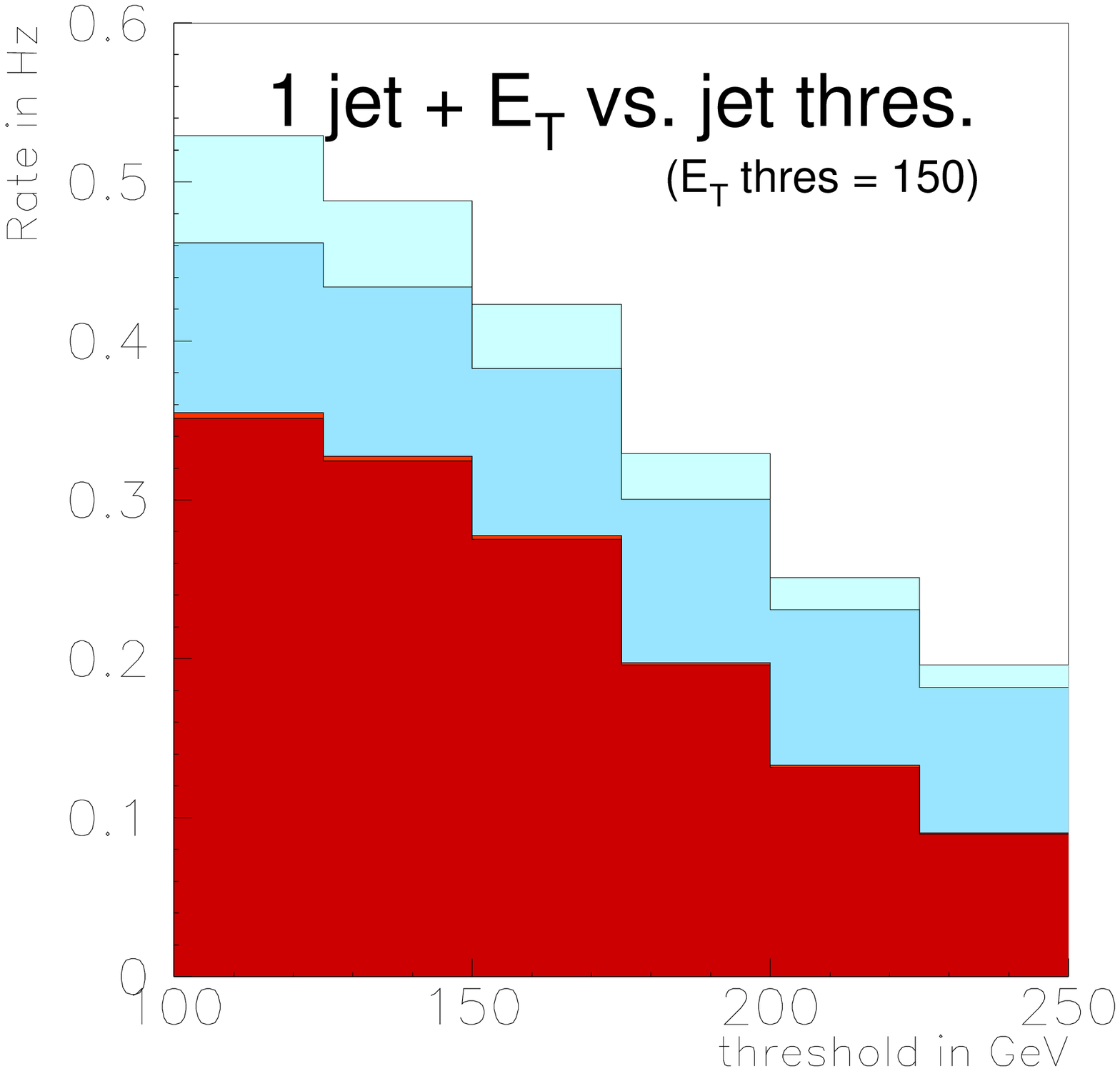} \hspace*{-3mm} 
\includegraphics*[scale=0.25]{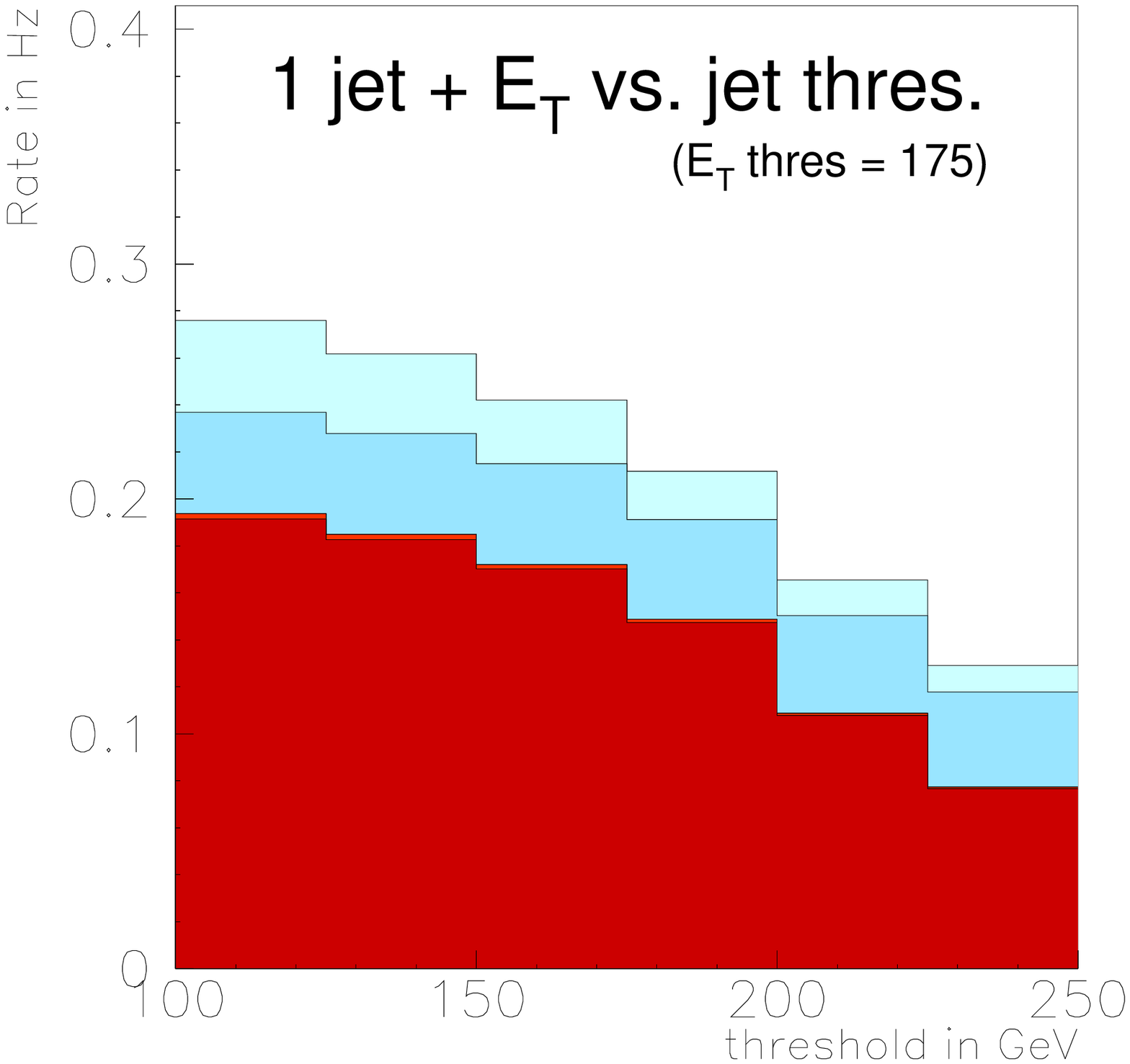} \hspace*{-3mm} 
\includegraphics*[scale=0.25]{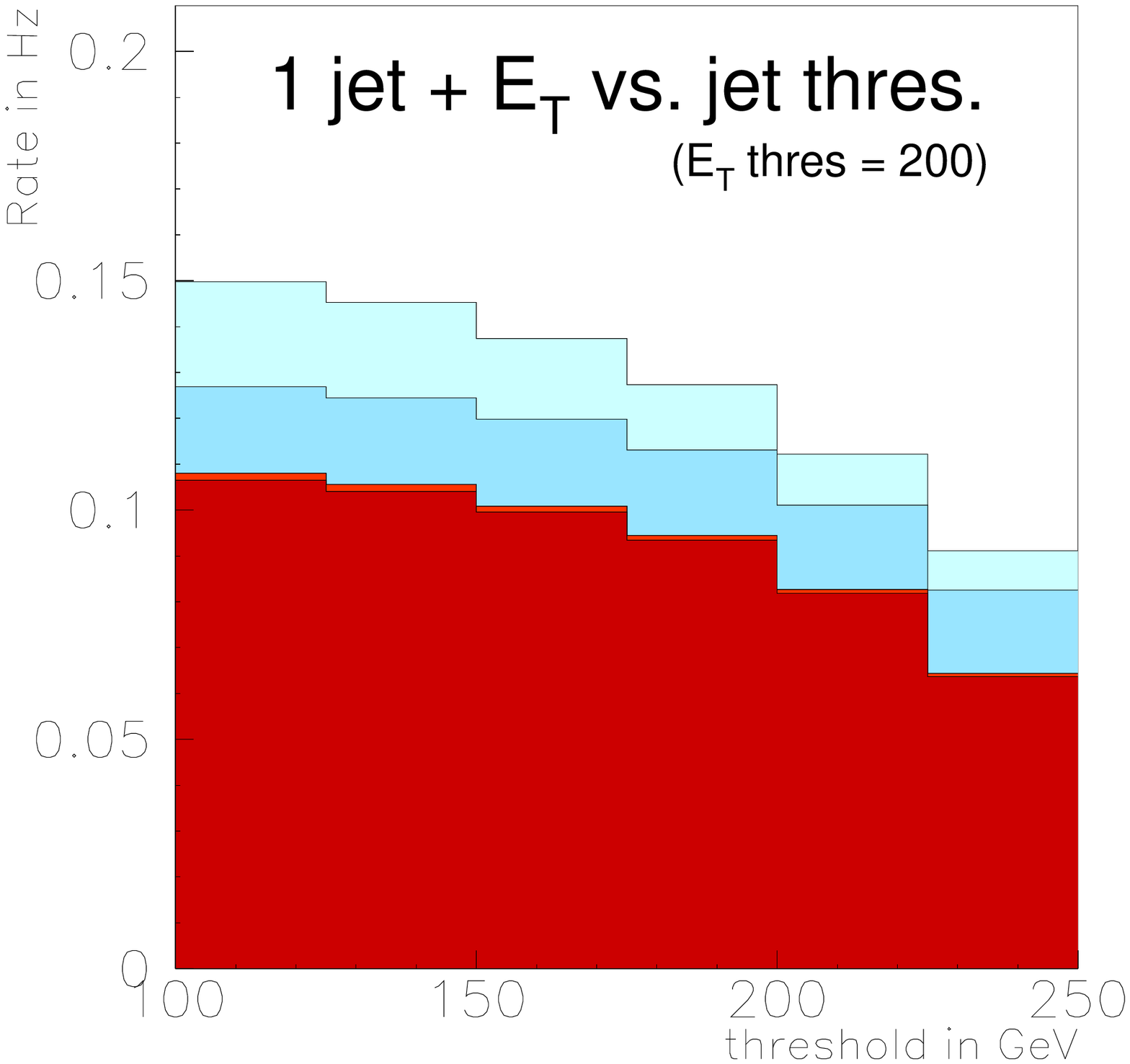} \vspace*{-3mm} \\
\includegraphics*[scale=0.25]{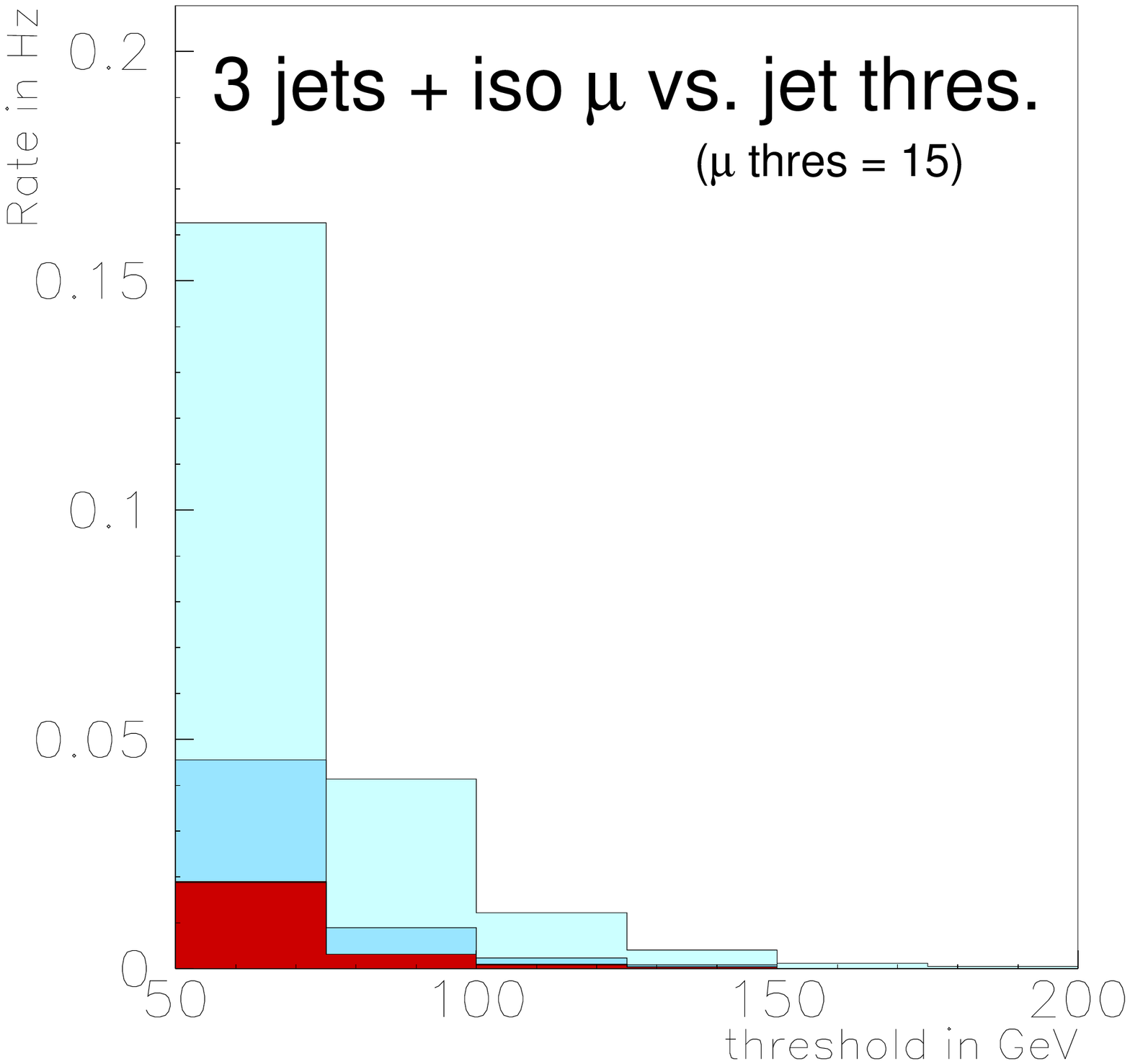} \hspace*{-3mm} 
\includegraphics*[scale=0.25]{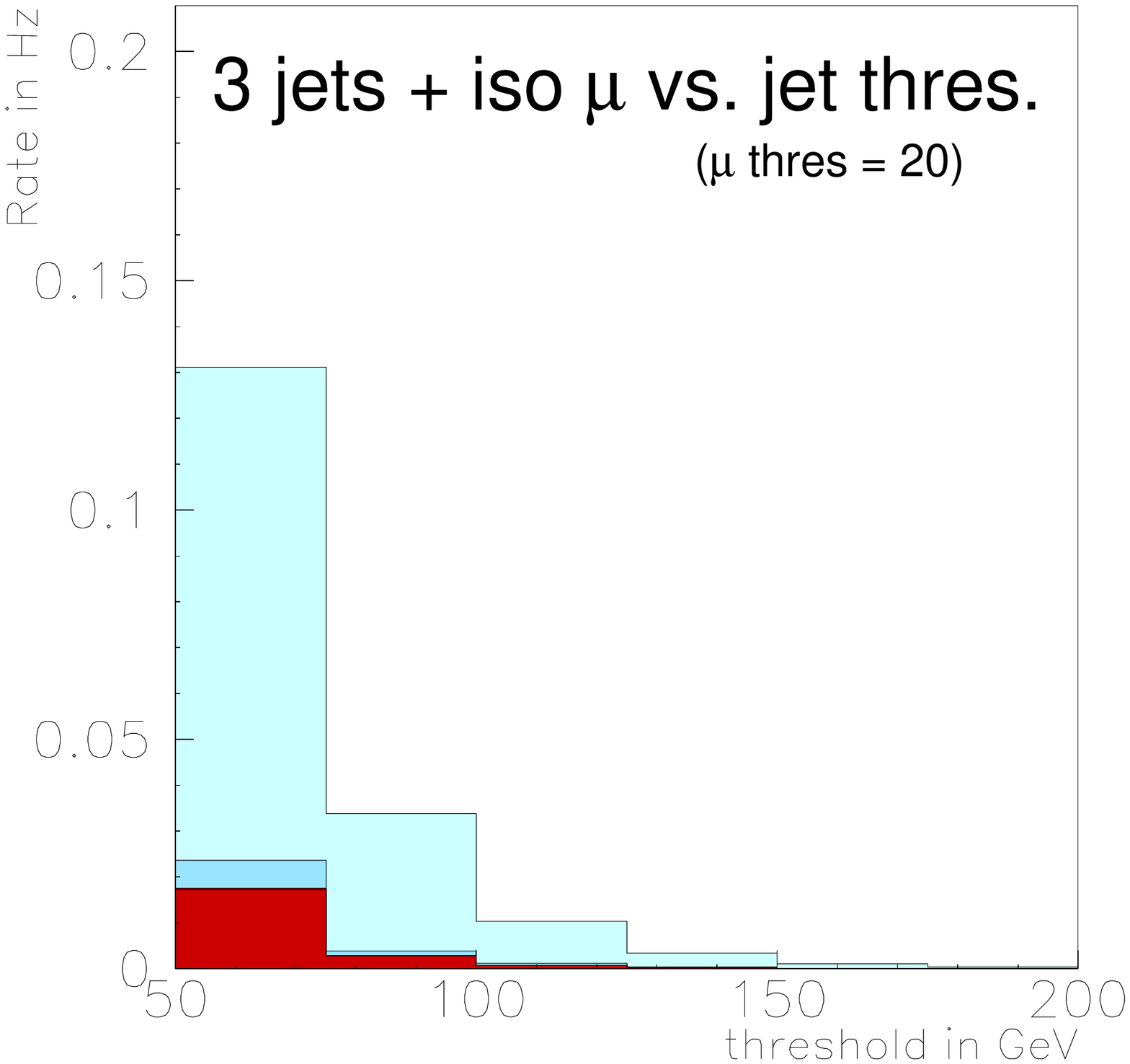} \hspace*{-3mm}  
\includegraphics*[scale=0.25]{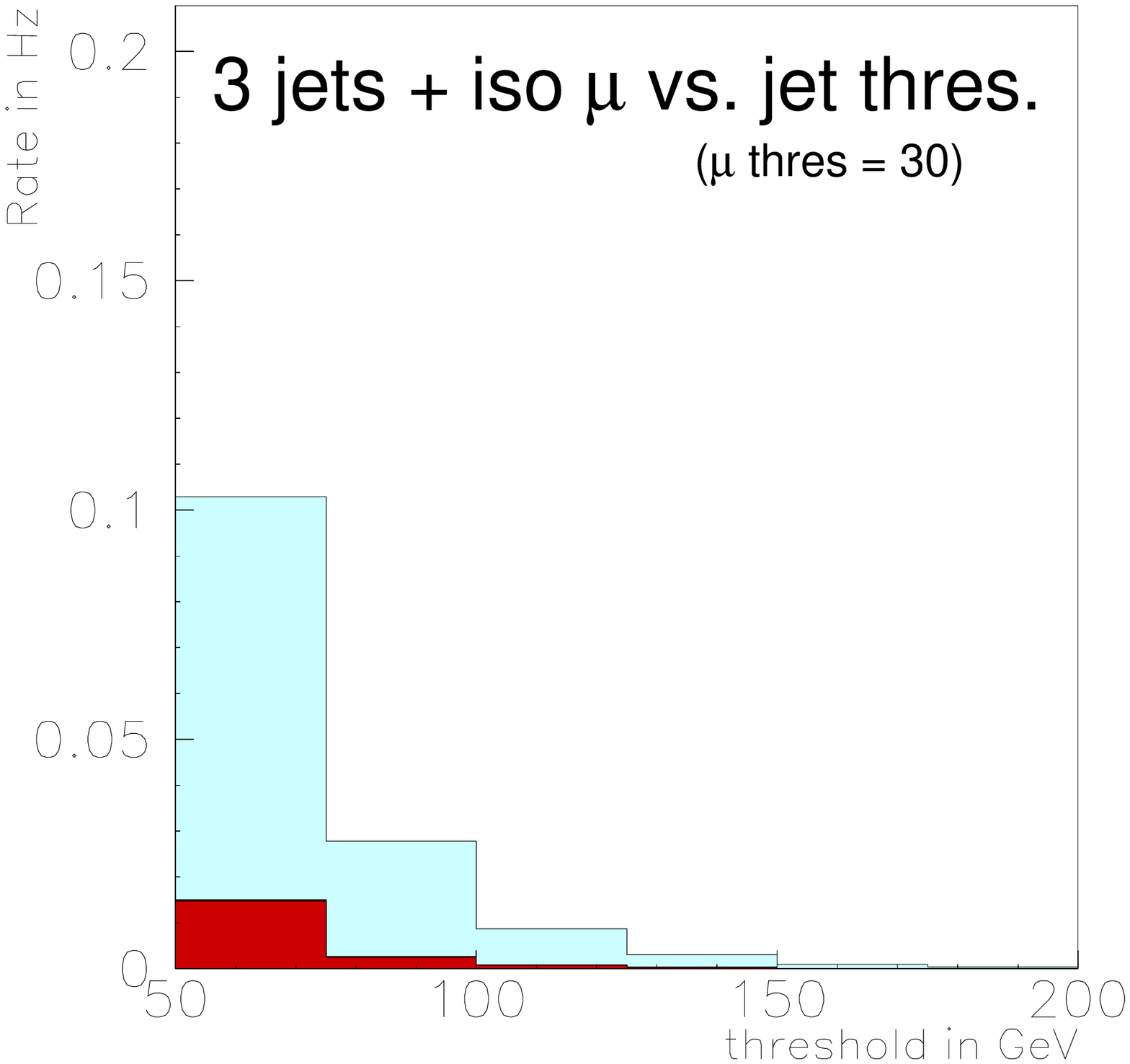} \vspace*{-3mm} \\
\includegraphics*[scale=0.25]{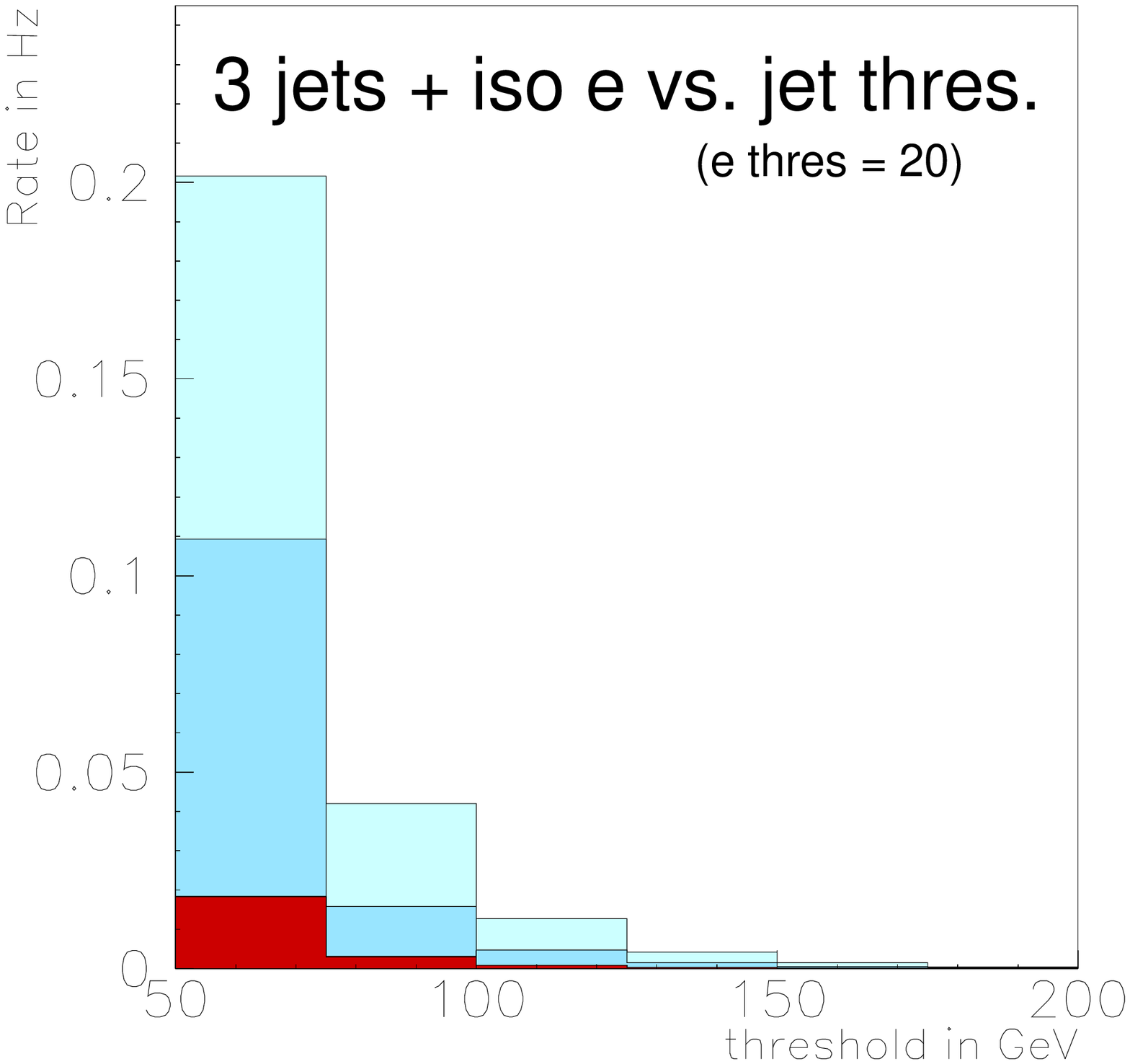} \hspace*{-3mm} 
\includegraphics*[scale=0.25]{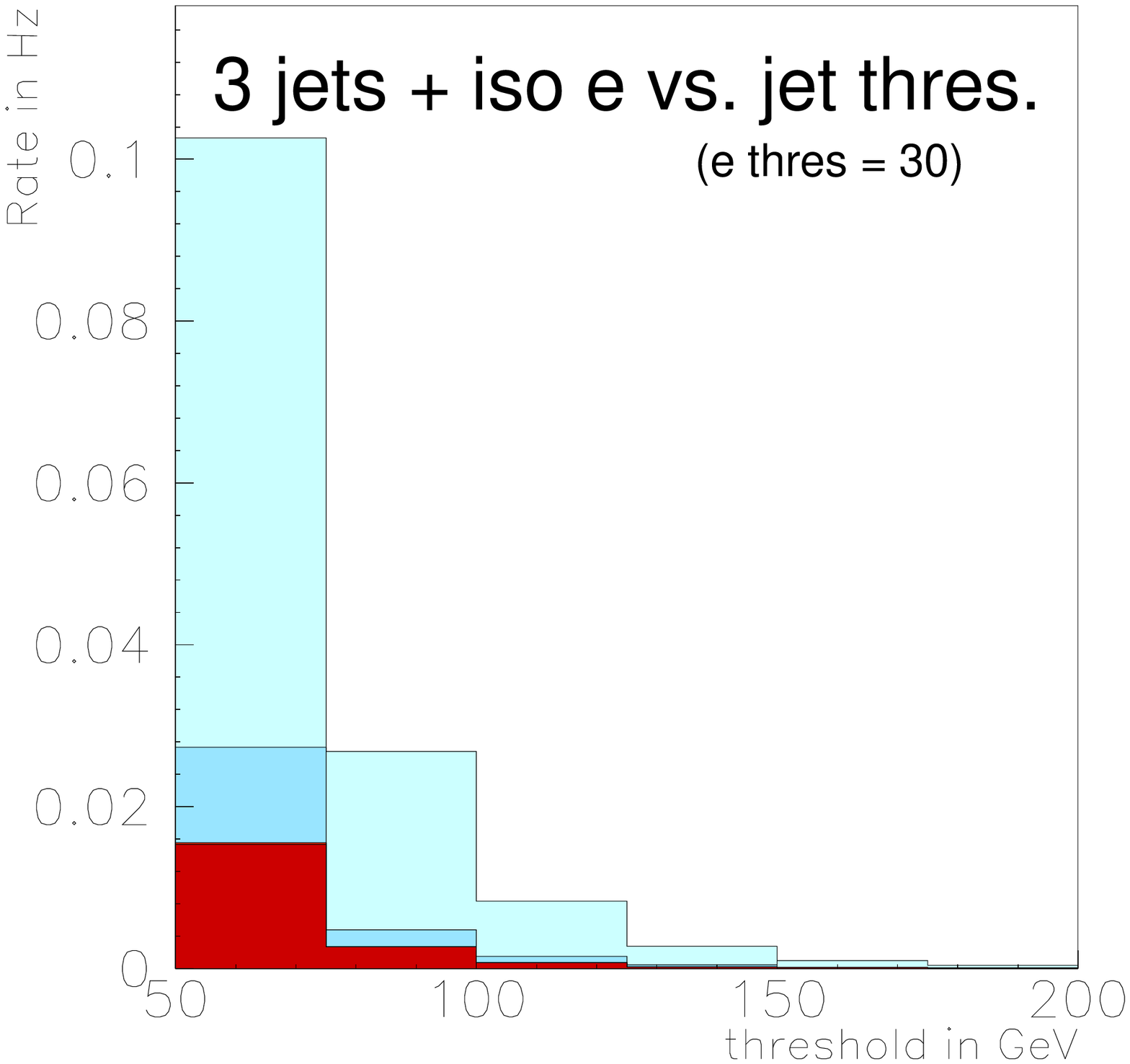} \hspace*{-3mm} 
\includegraphics*[scale=0.25]{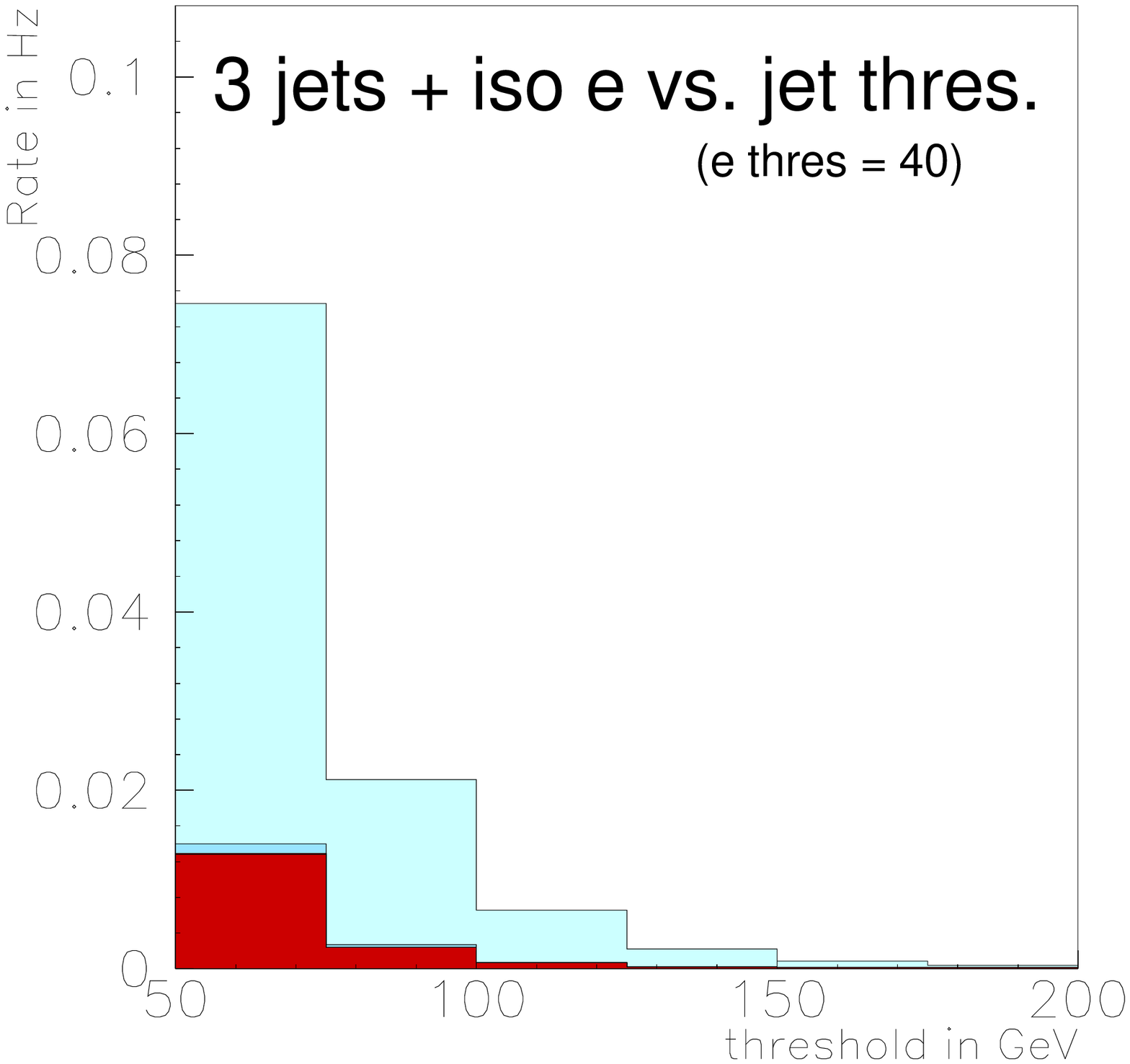} \vspace*{-3mm} \\
\includegraphics*[scale=0.25]{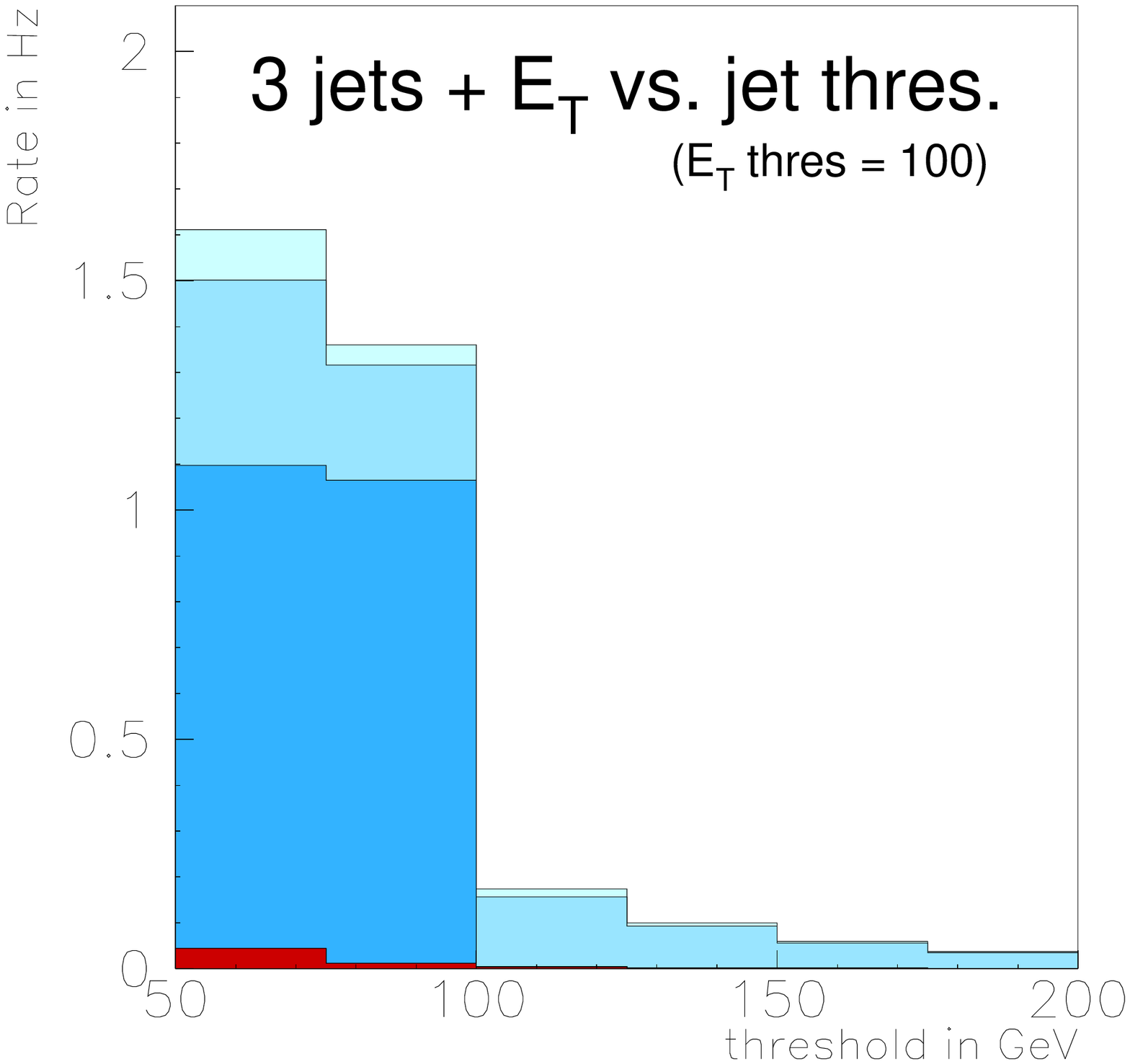} \hspace*{-3mm} 
\includegraphics*[scale=0.25]{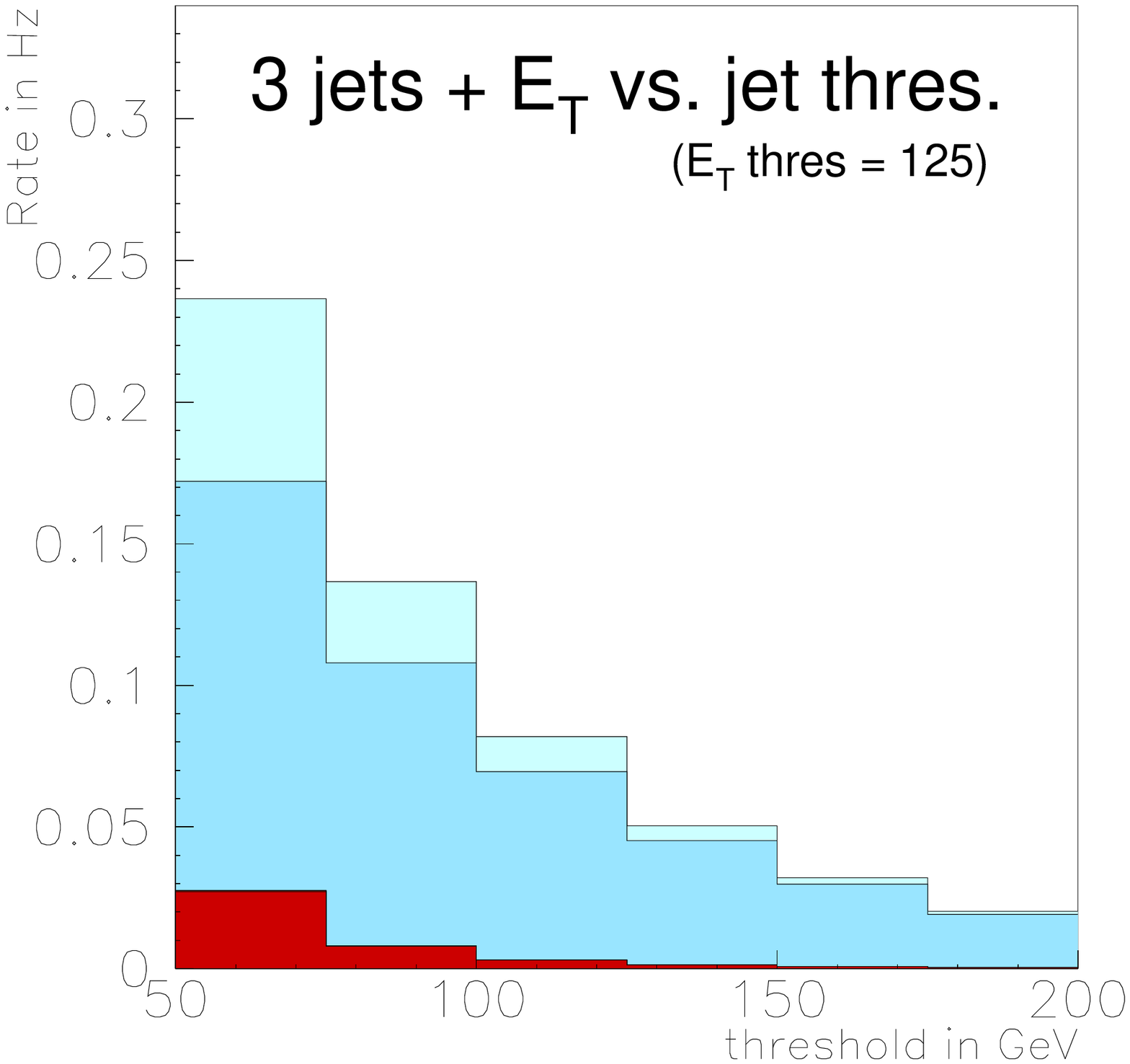} \hspace*{-3mm} 
\includegraphics*[scale=0.25]{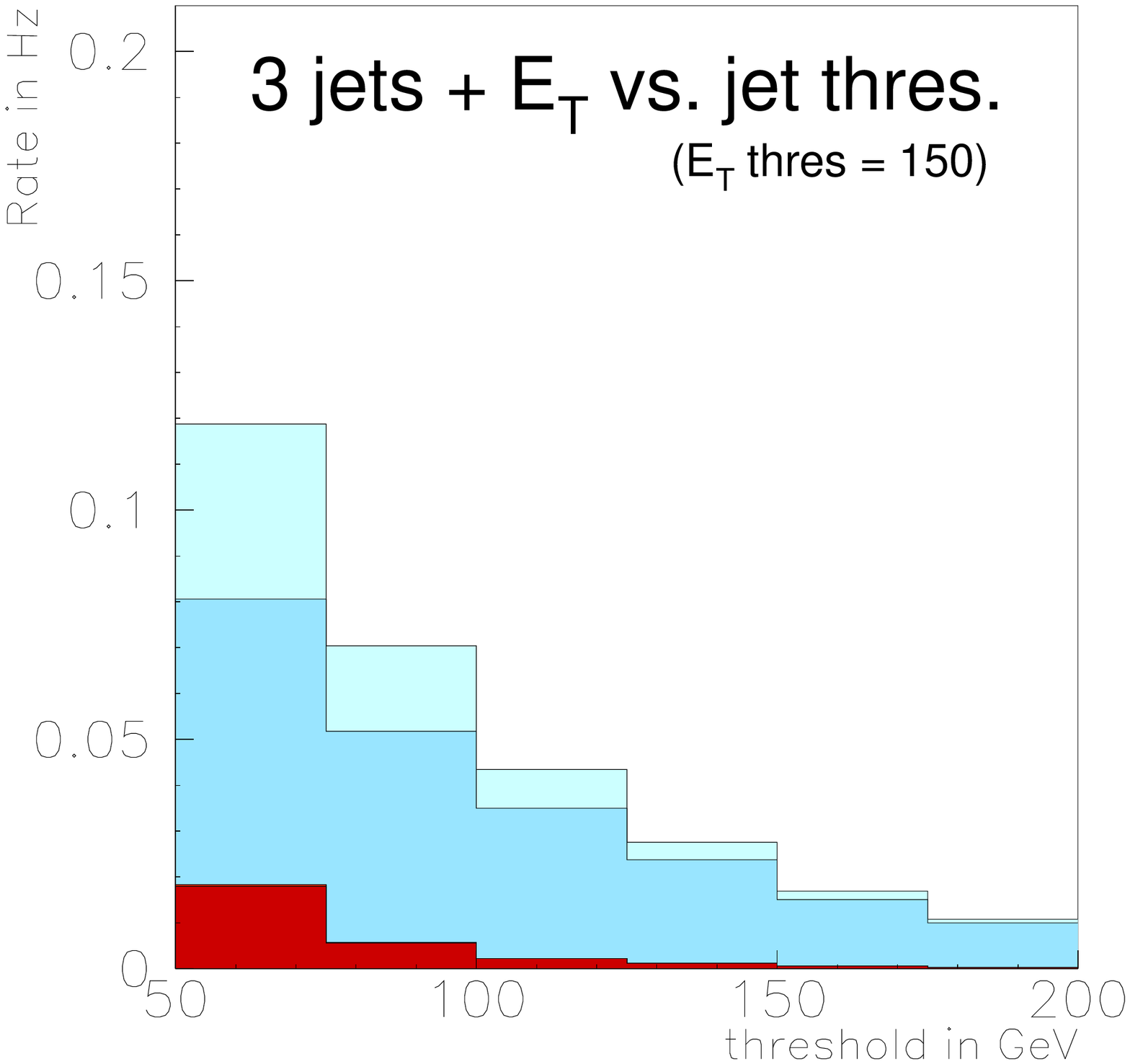} \vspace*{-3mm}\clearpage
\end{center}
\subsection{Signal Efficiencies}
The following plots attempt to combine a large amount of information in as
small a space as possible. For each of the two \LV\ terms in the
superpotential, LLE and LQD, a scan has been made over all five mSUGRA points
and over the three different possibilities for the individual coupling
strengths investigated. 
The range of efficiencies (min to max found in the scan over
mSUGRA and coupling scenarios) for the scenarios with non-zero LLE couplings
is shown in green and the efficiencies for LQD in blue. 
In order not to over-clutter the plots, the LLE+LQD scenarios studied are not
shown. Their efficiencies all lie within the boundaries set by the two
``pure'' scenario types. As one would expect,
the LLE scenarios do well in the leptonic triggers while the LQD terms do not
fare quite as well as one could have hoped in the jet triggers. This is
simply due to the fact that the lightest neutralino has a tendency to decay into
$qq\nu$ in these scenarios, caught mainly by the jets + \ET\ triggers which
must necessarily have high thresholds due to the large background rates. 
In
the generation of each plot, $10^5$ events were generated per scenario. Since
more than 1000 events pass the triggers in all cases, we do not see the same
jitter caused by the electron and muon reconstruction efficiencies 
as in the previous plots. The total efficiency is, of
course, a complicated function of each of these triggers. These plots are
merely presented to give the reader some feeling for how useful each trigger
item is by itself and as a potential starting point for future studies.
\begin{center}
\includegraphics*[scale=0.25]{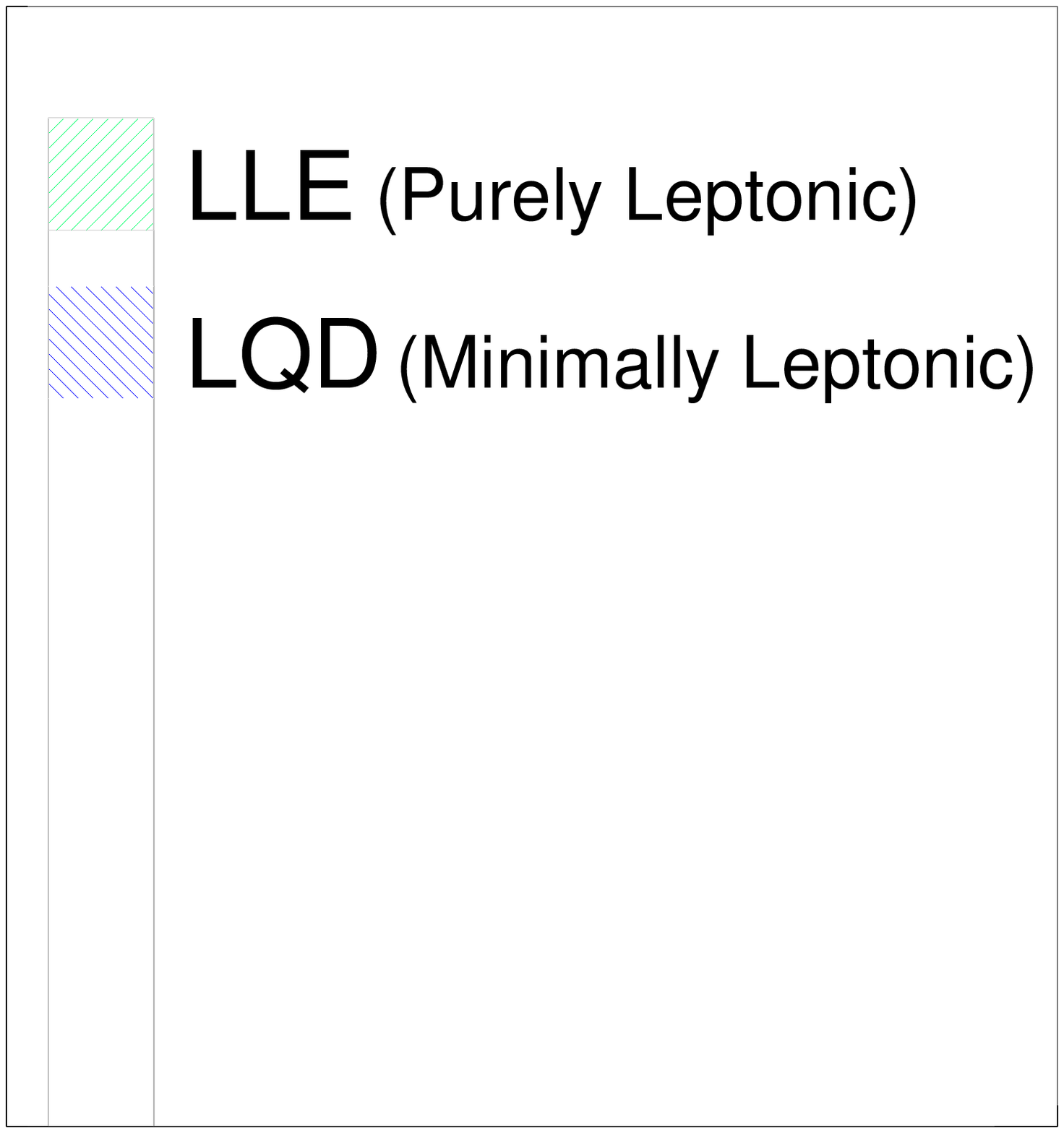} \hspace*{-3mm} 
\includegraphics*[scale=0.25]{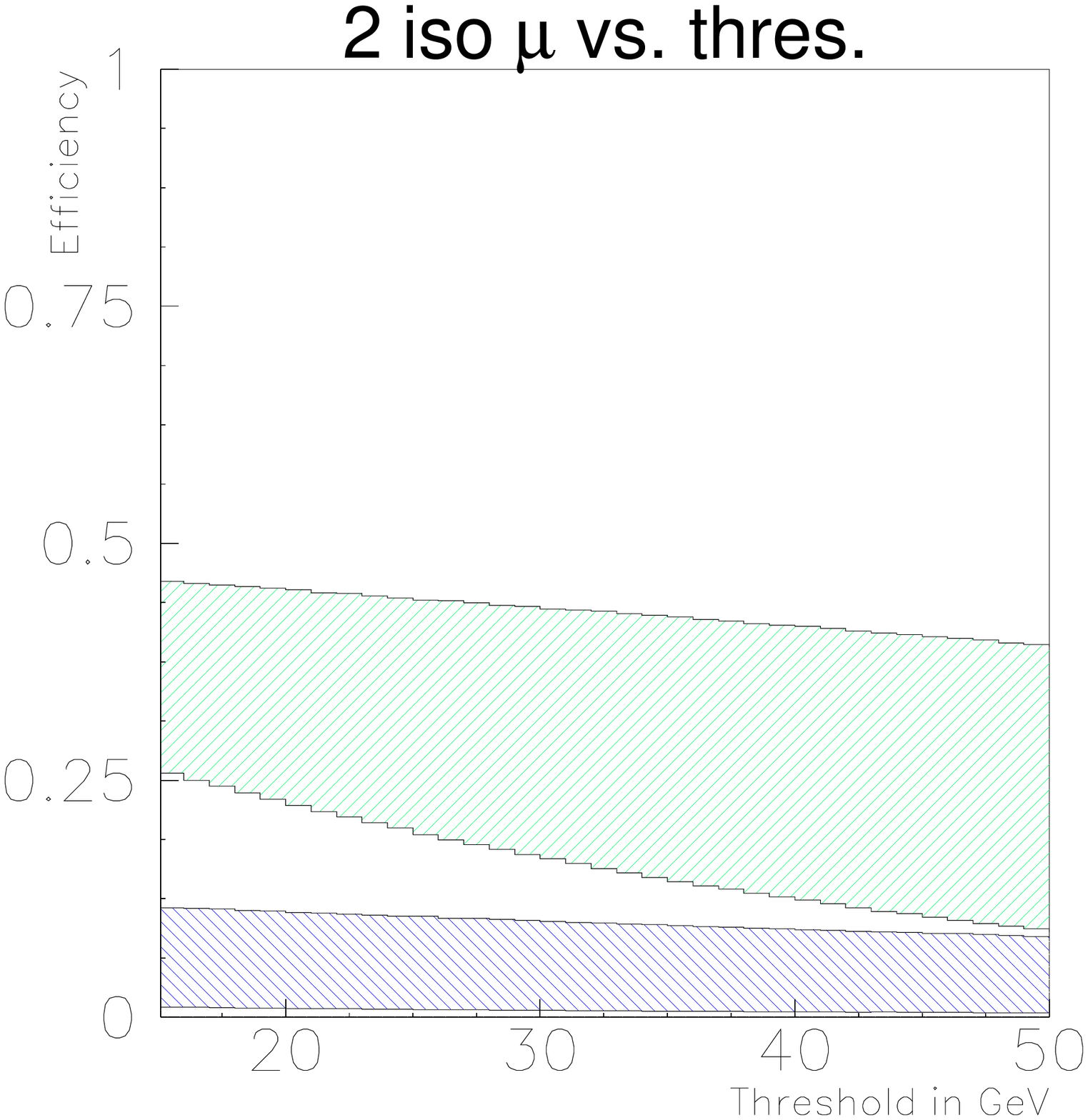} \hspace*{-3mm} 
\includegraphics*[scale=0.25]{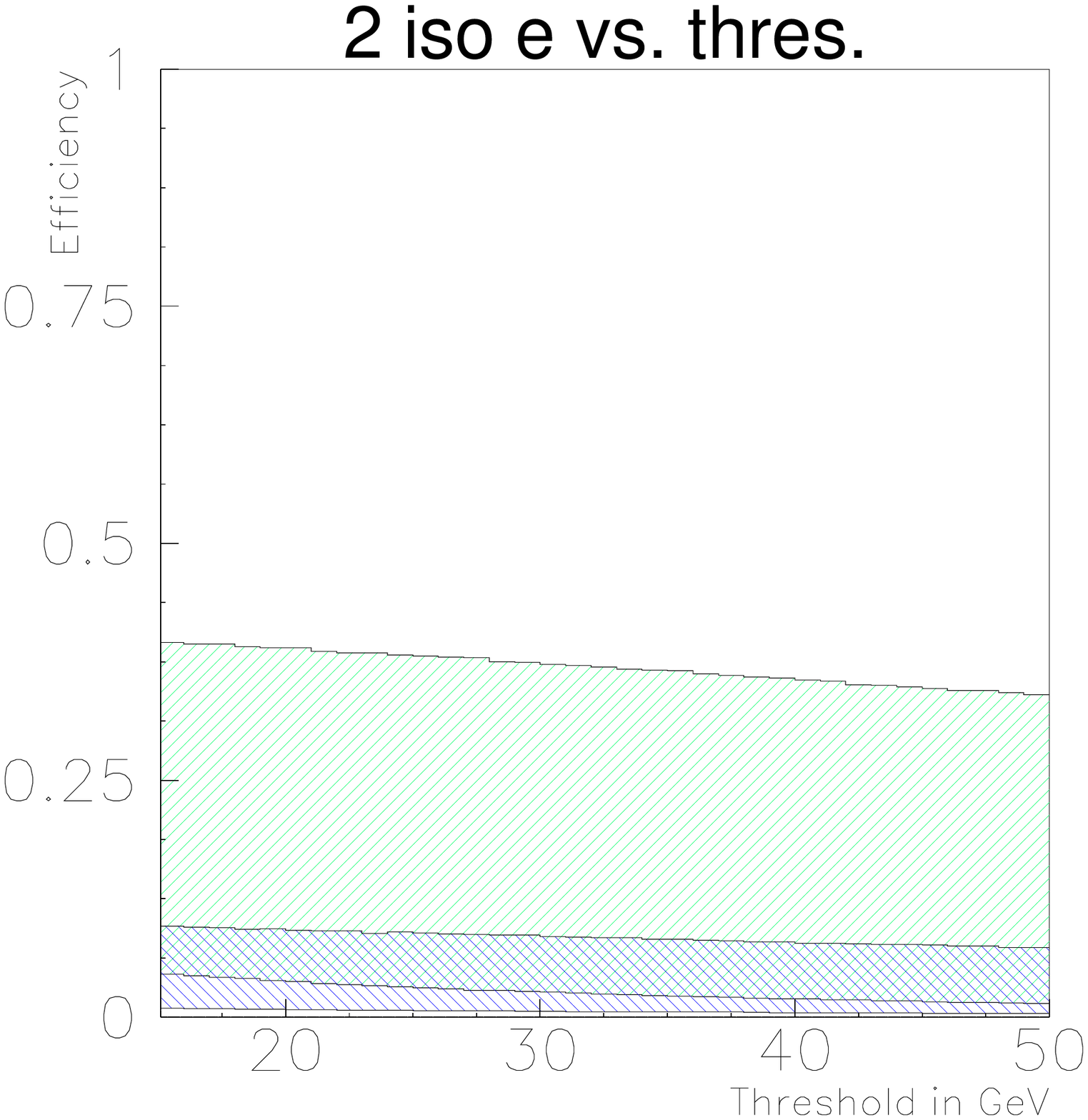} \vspace*{-1mm}\\ 
\includegraphics*[scale=0.25]{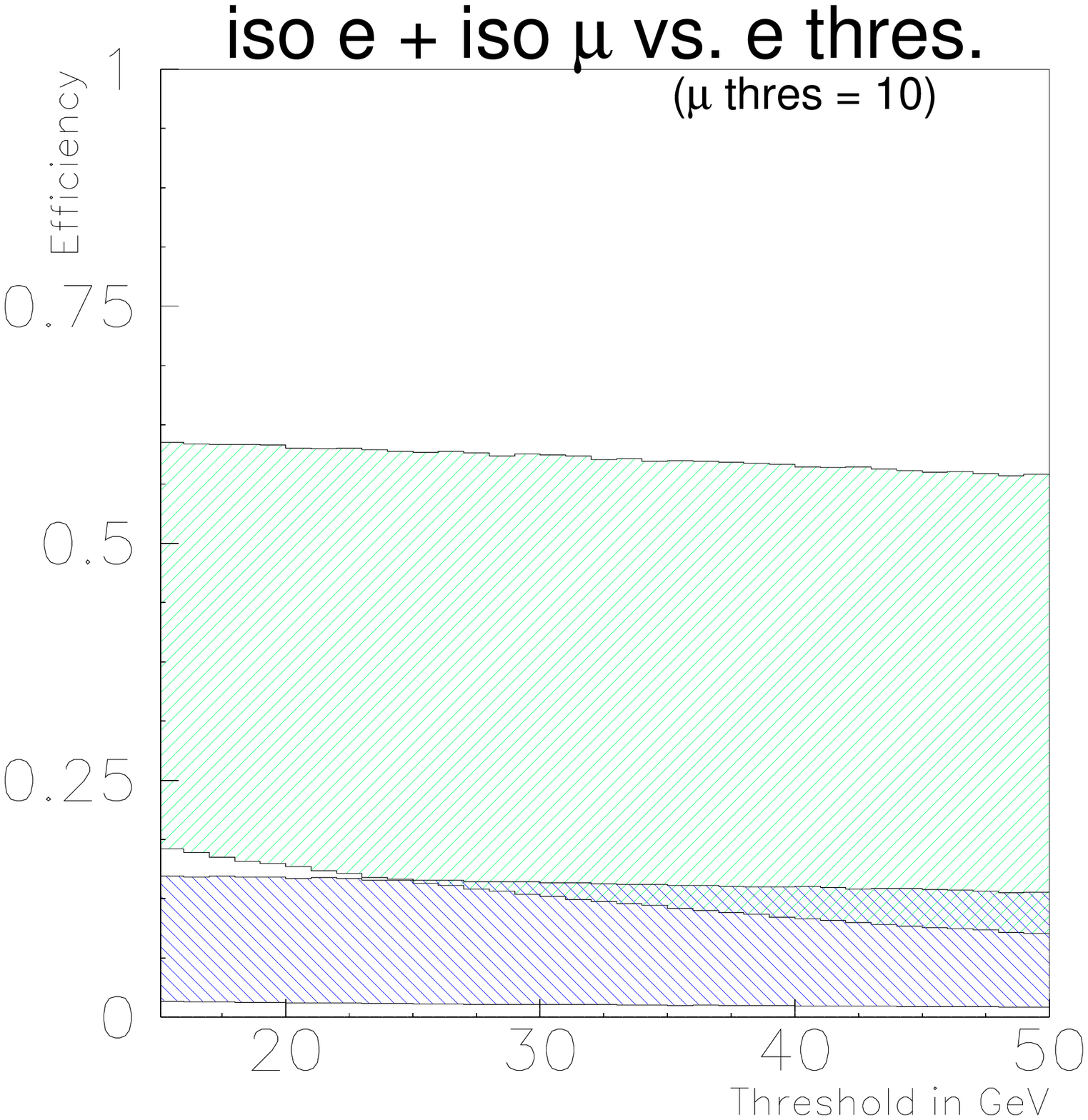} \hspace*{-3mm} 
\includegraphics*[scale=0.25]{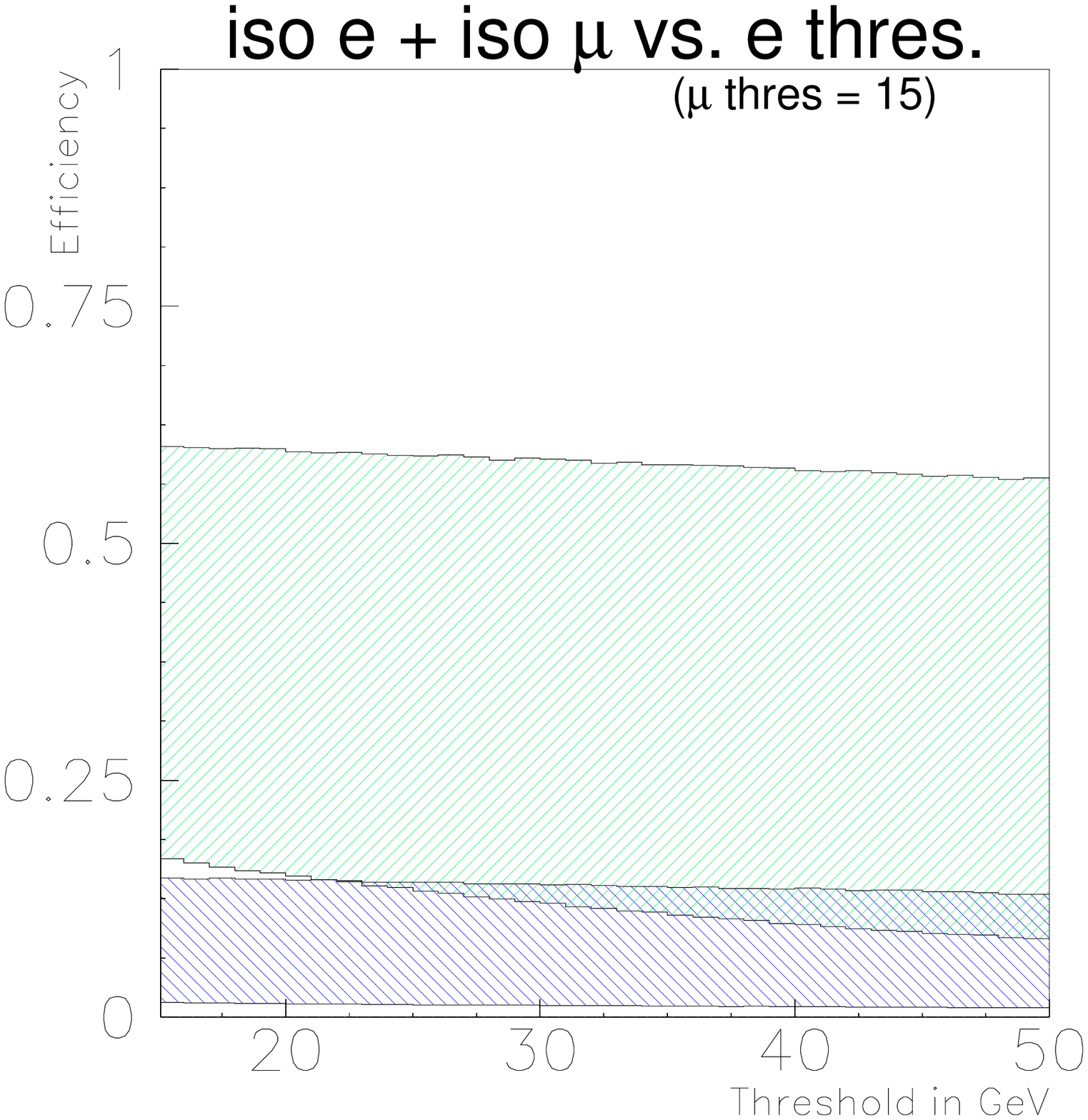} \hspace*{-3mm} 
\includegraphics*[scale=0.25]{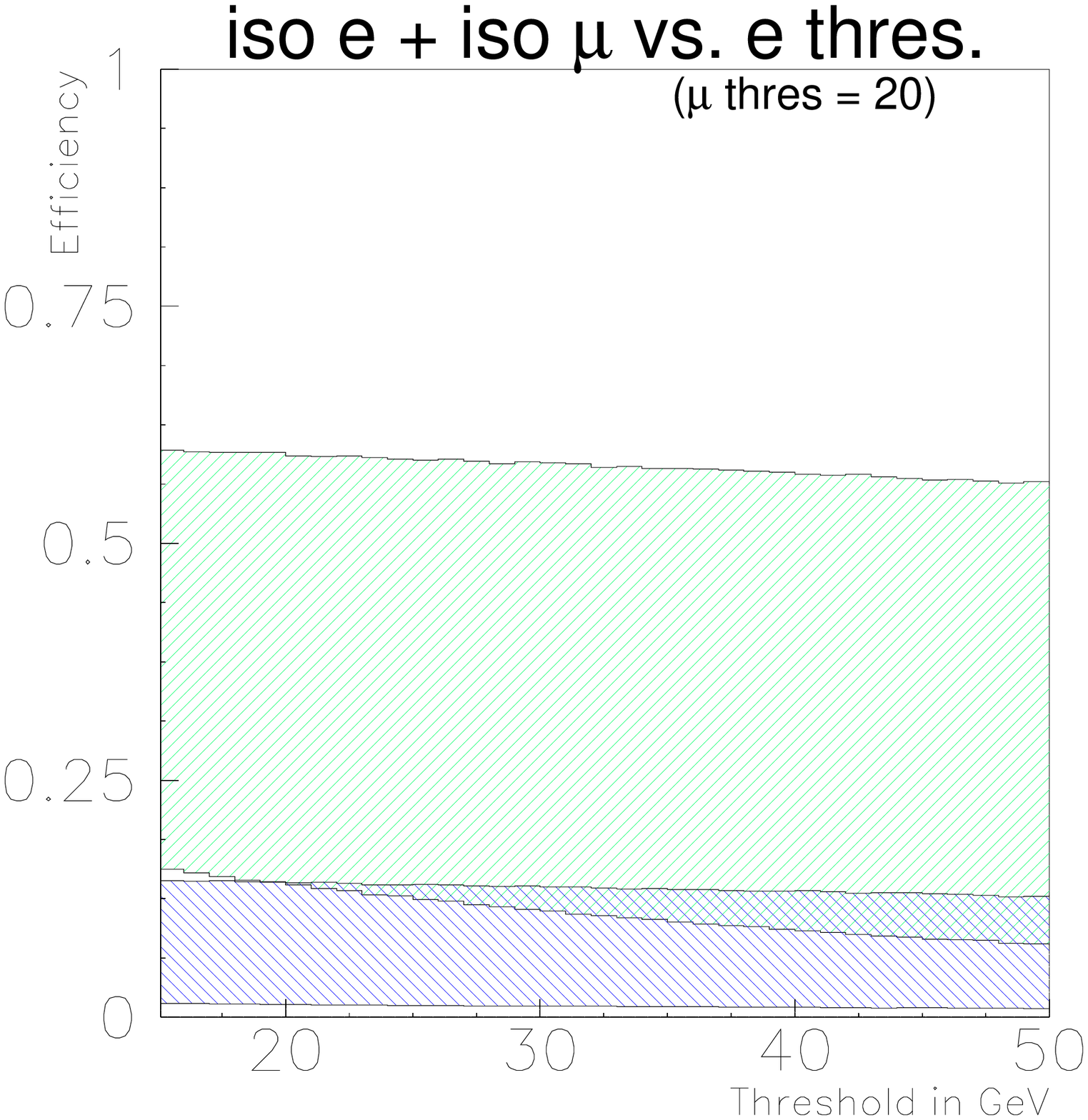} \vspace*{-1mm} \clearpage
\includegraphics*[scale=0.25]{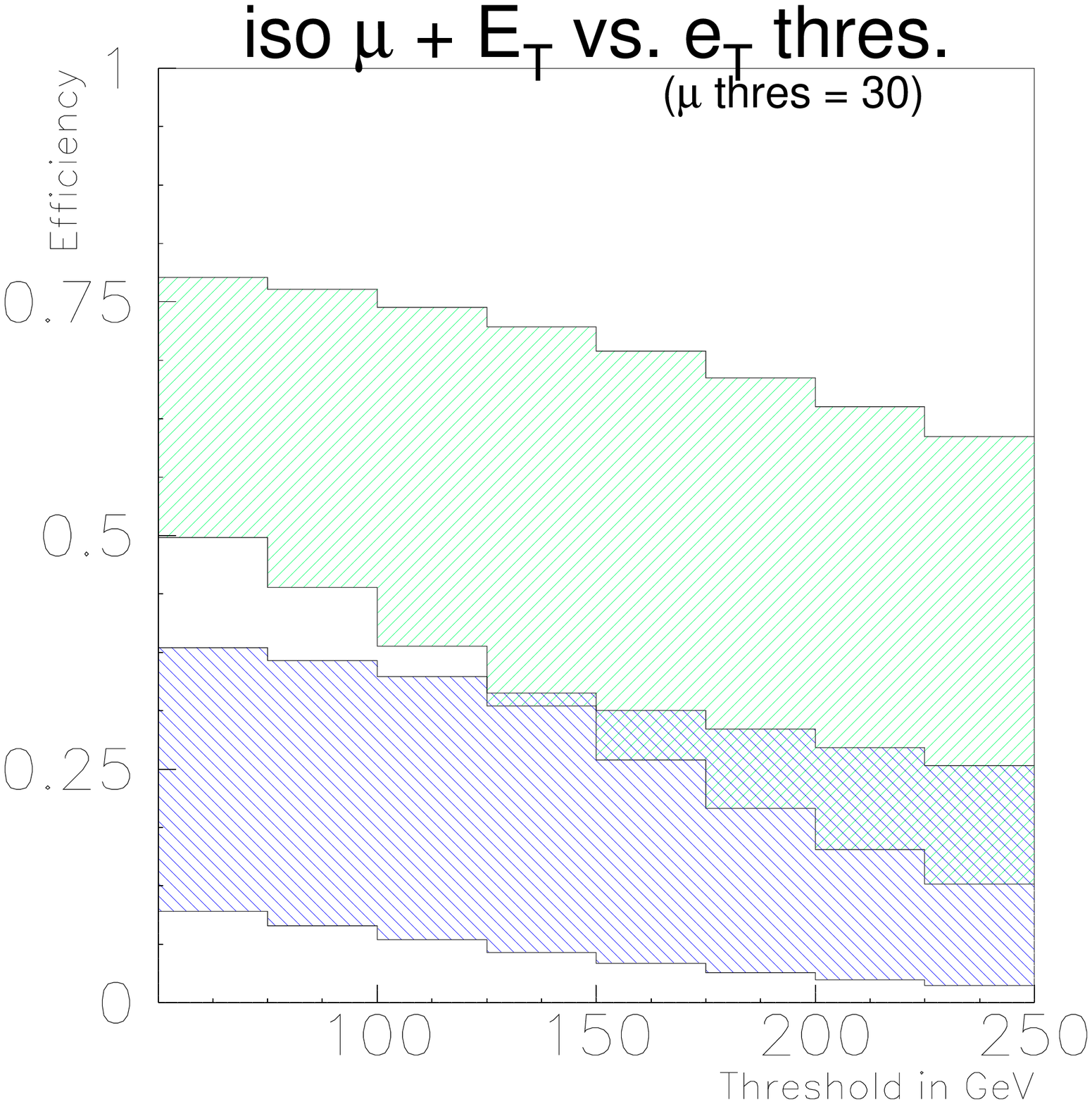} \hspace*{-3mm} 
\includegraphics*[scale=0.25]{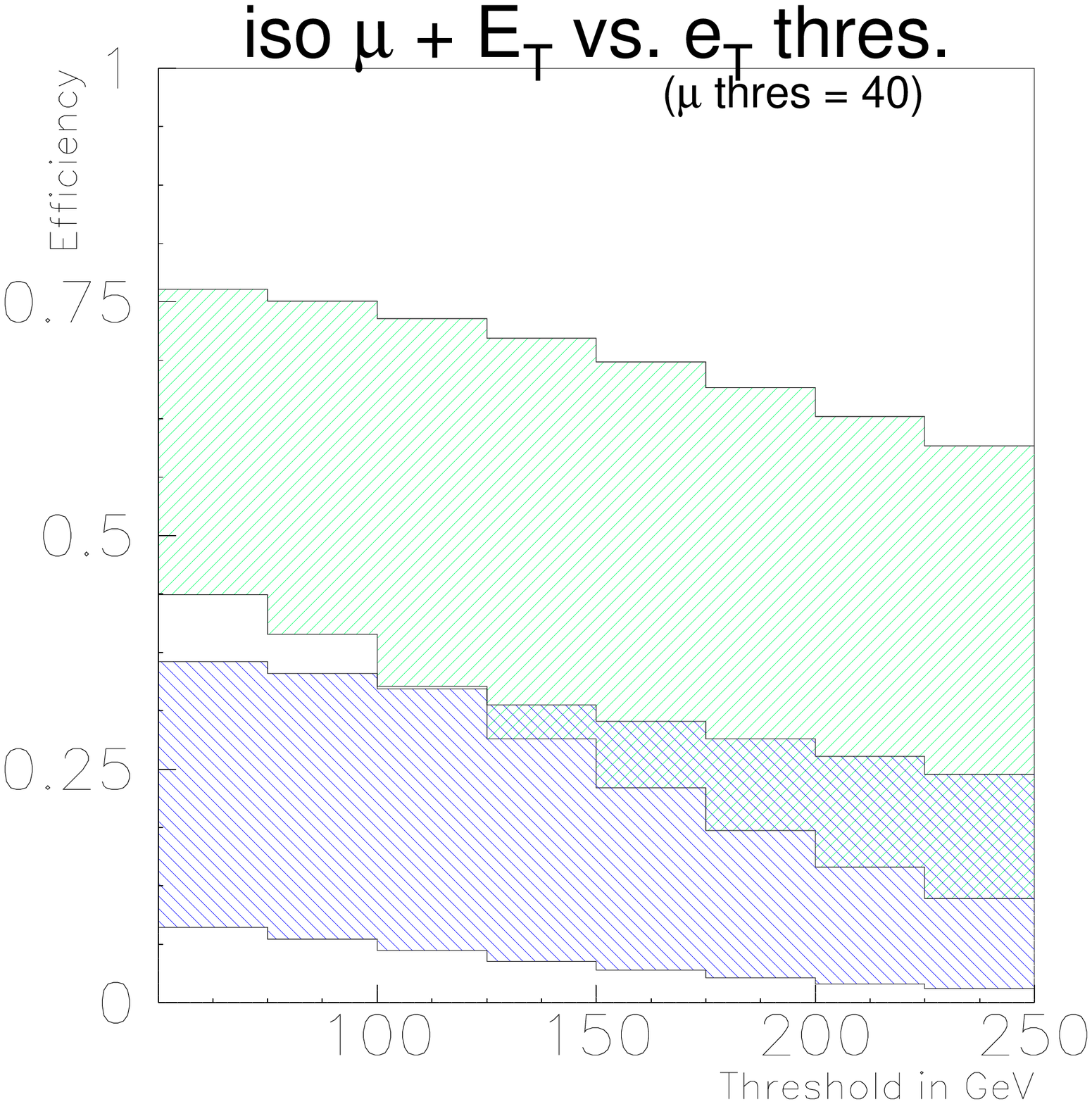} \hspace*{-3mm} 
\includegraphics*[scale=0.25]{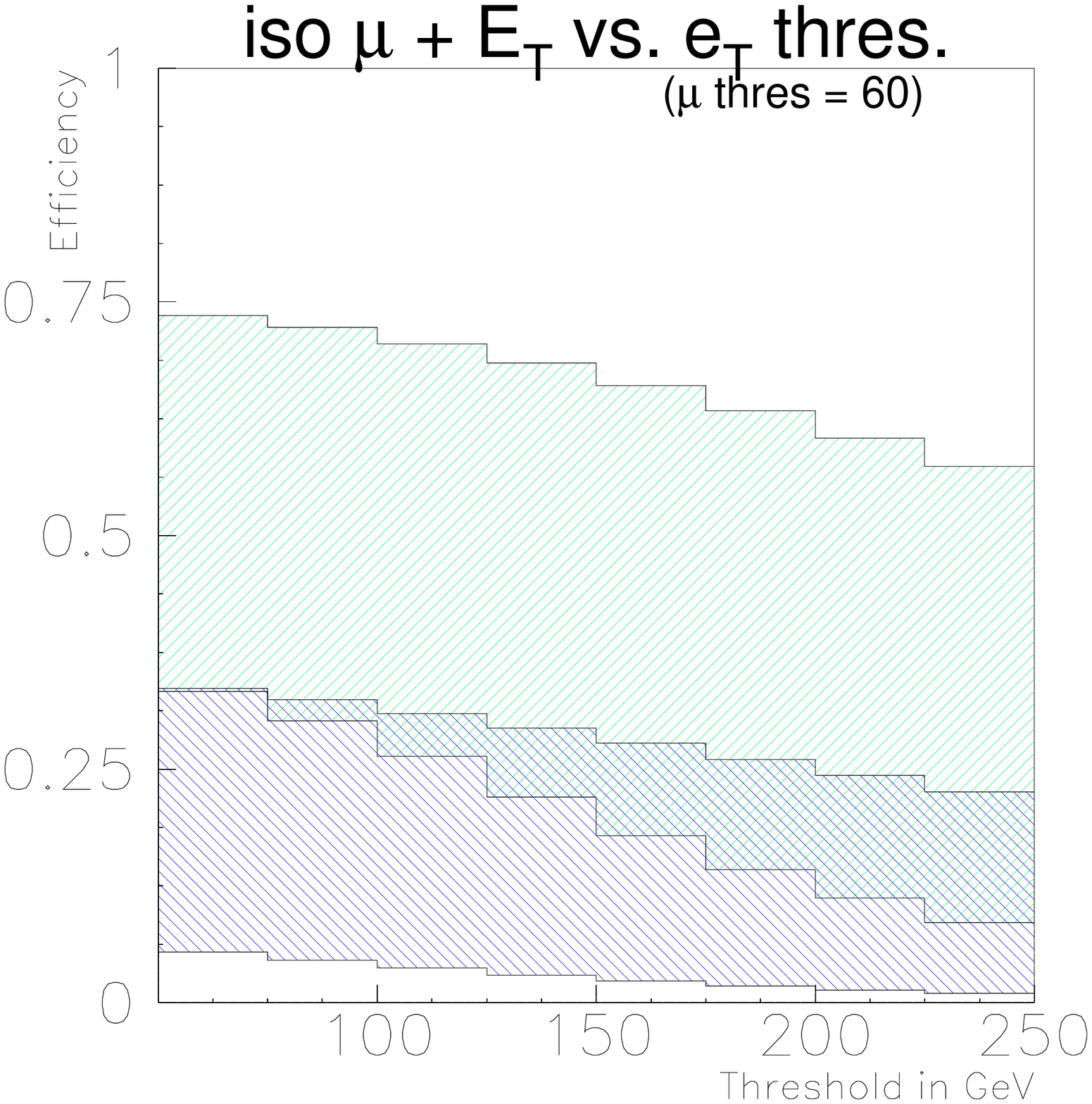} \vspace*{-1mm} \\
\includegraphics*[scale=0.25]{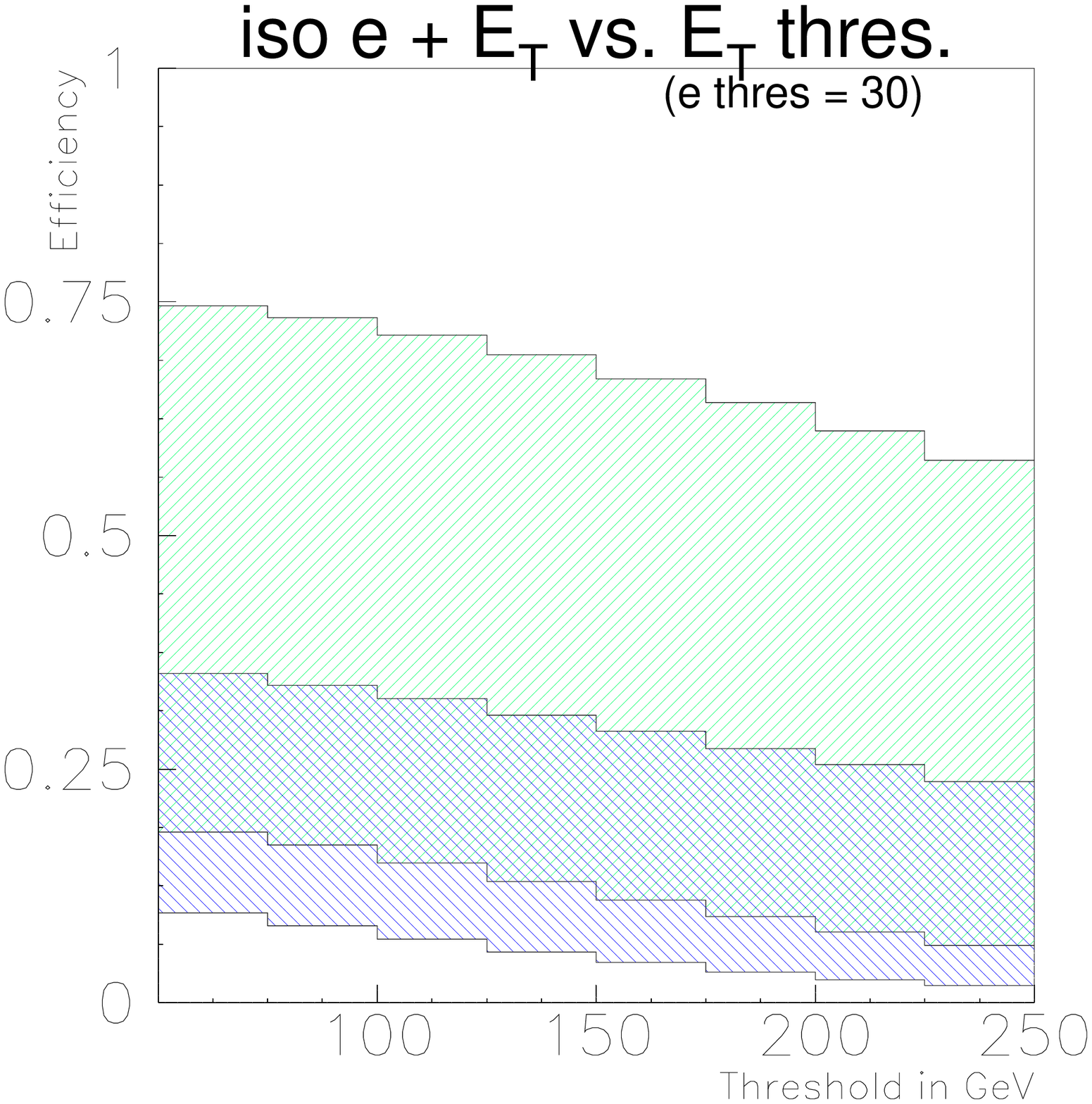} \hspace*{-3mm} 
\includegraphics*[scale=0.25]{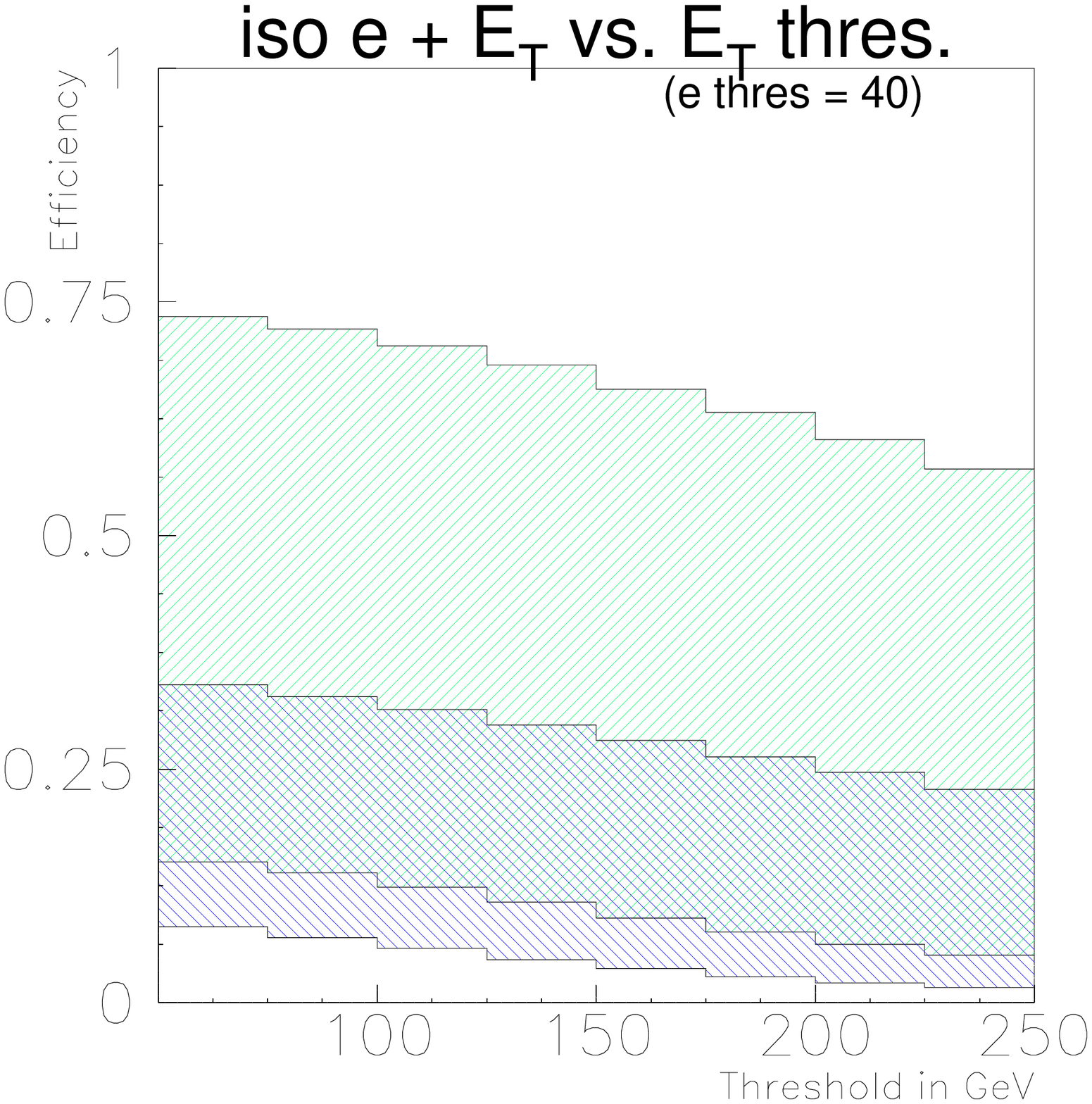} \hspace*{-3mm} 
\includegraphics*[scale=0.25]{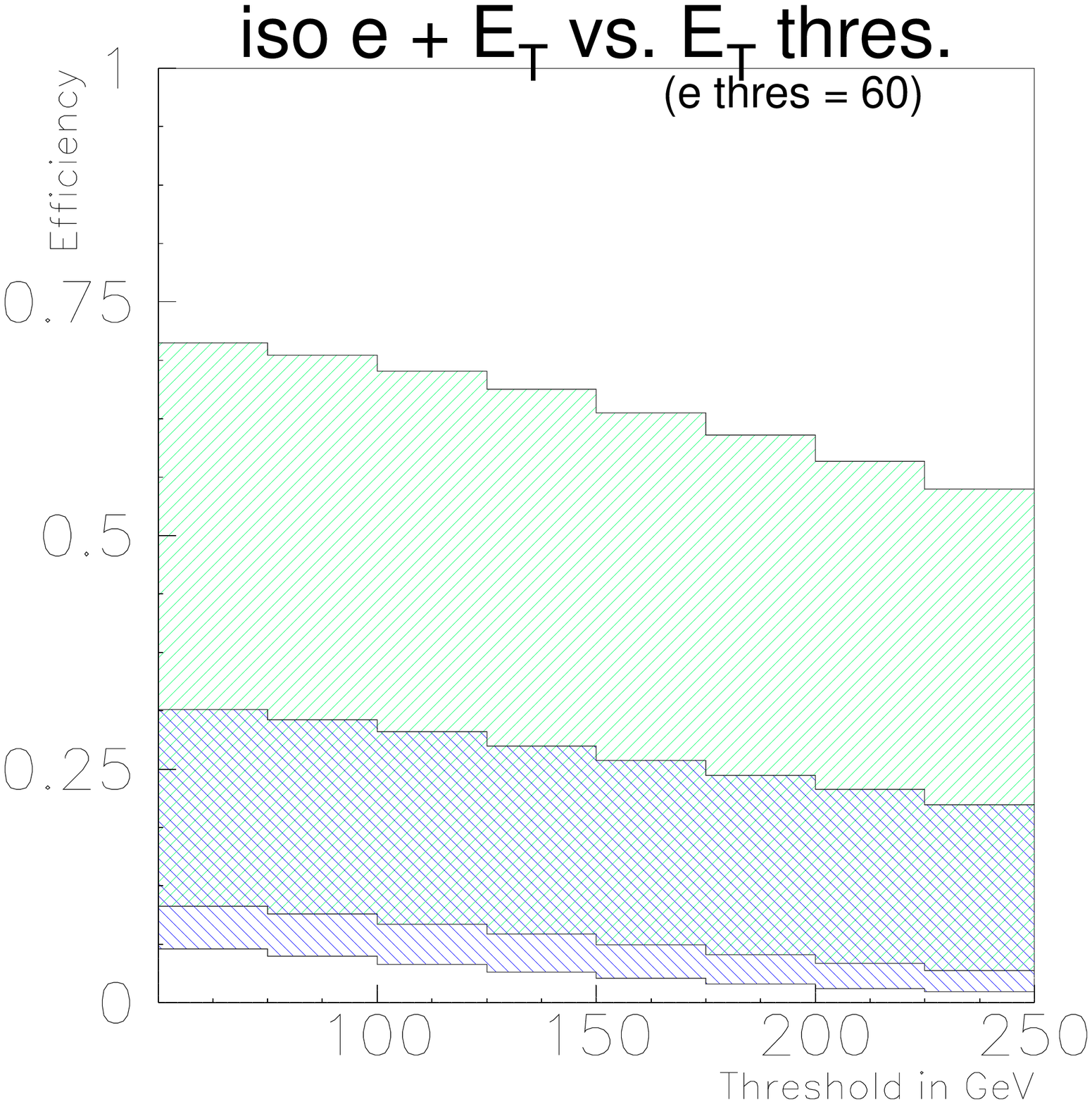} \vspace*{-1mm} \\
\includegraphics*[scale=0.25]{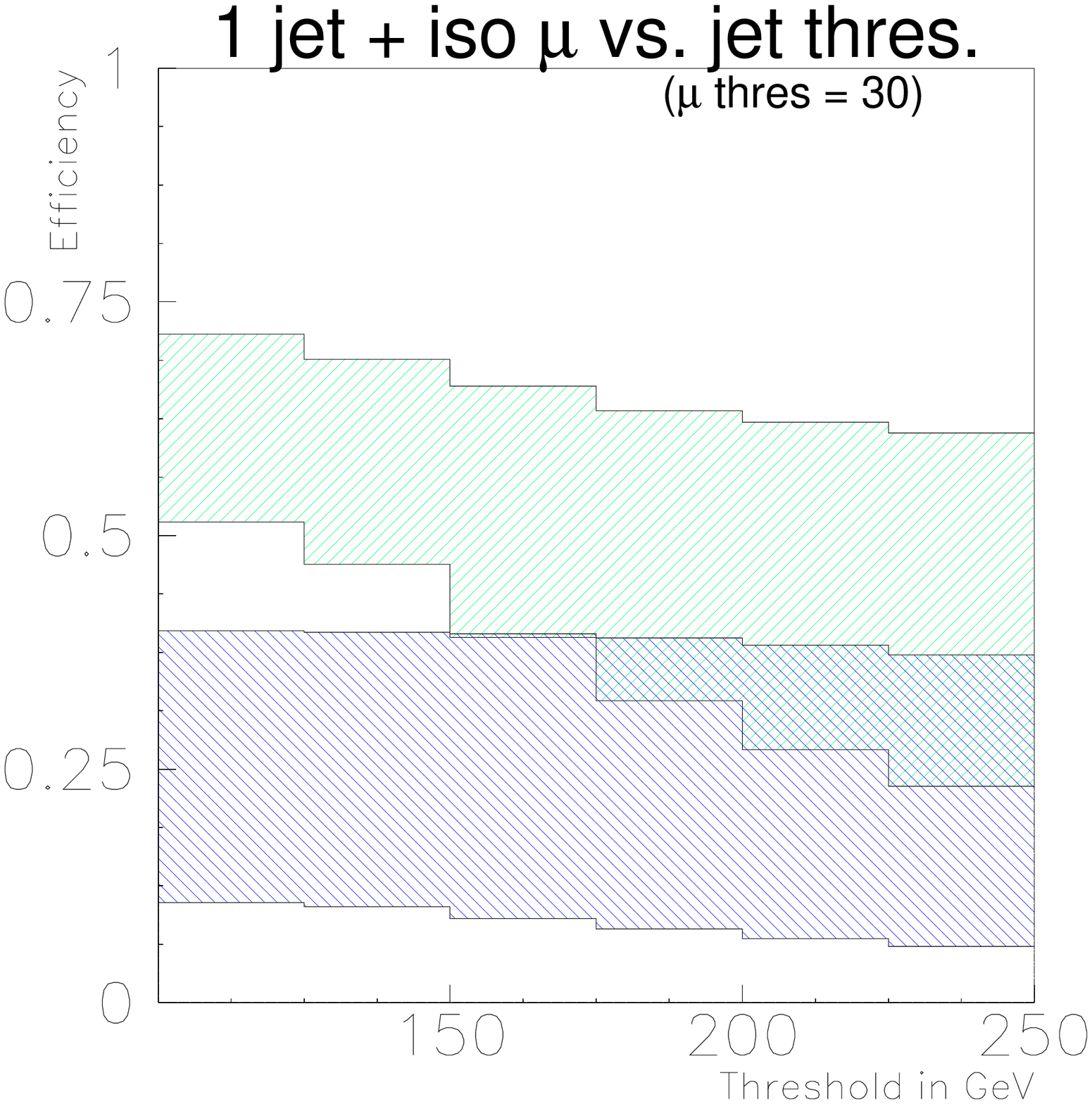} \hspace*{-3mm} 
\includegraphics*[scale=0.25]{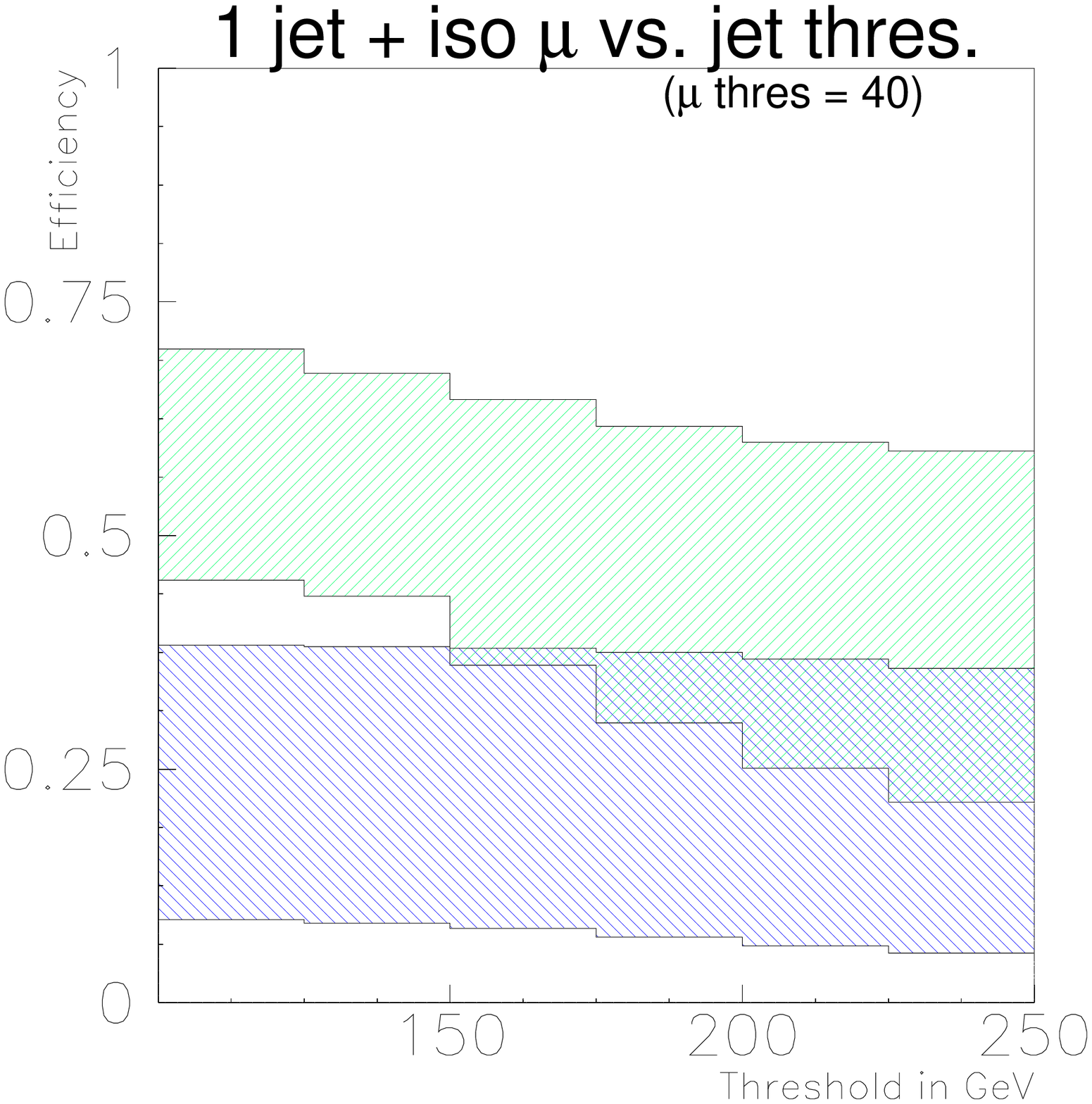} \hspace*{-3mm} 
\includegraphics*[scale=0.25]{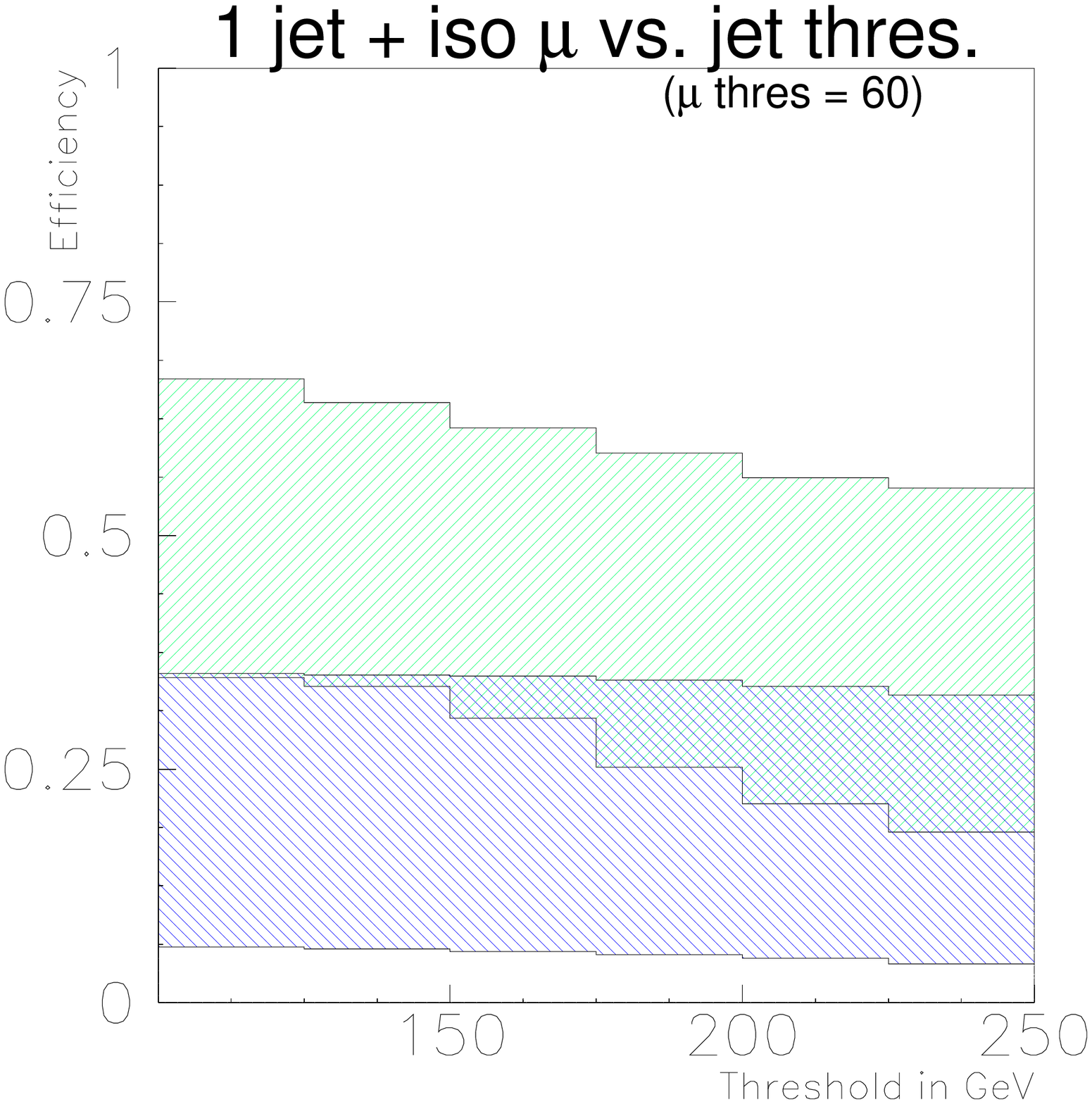} \vspace*{-1mm} \\
\includegraphics*[scale=0.25]{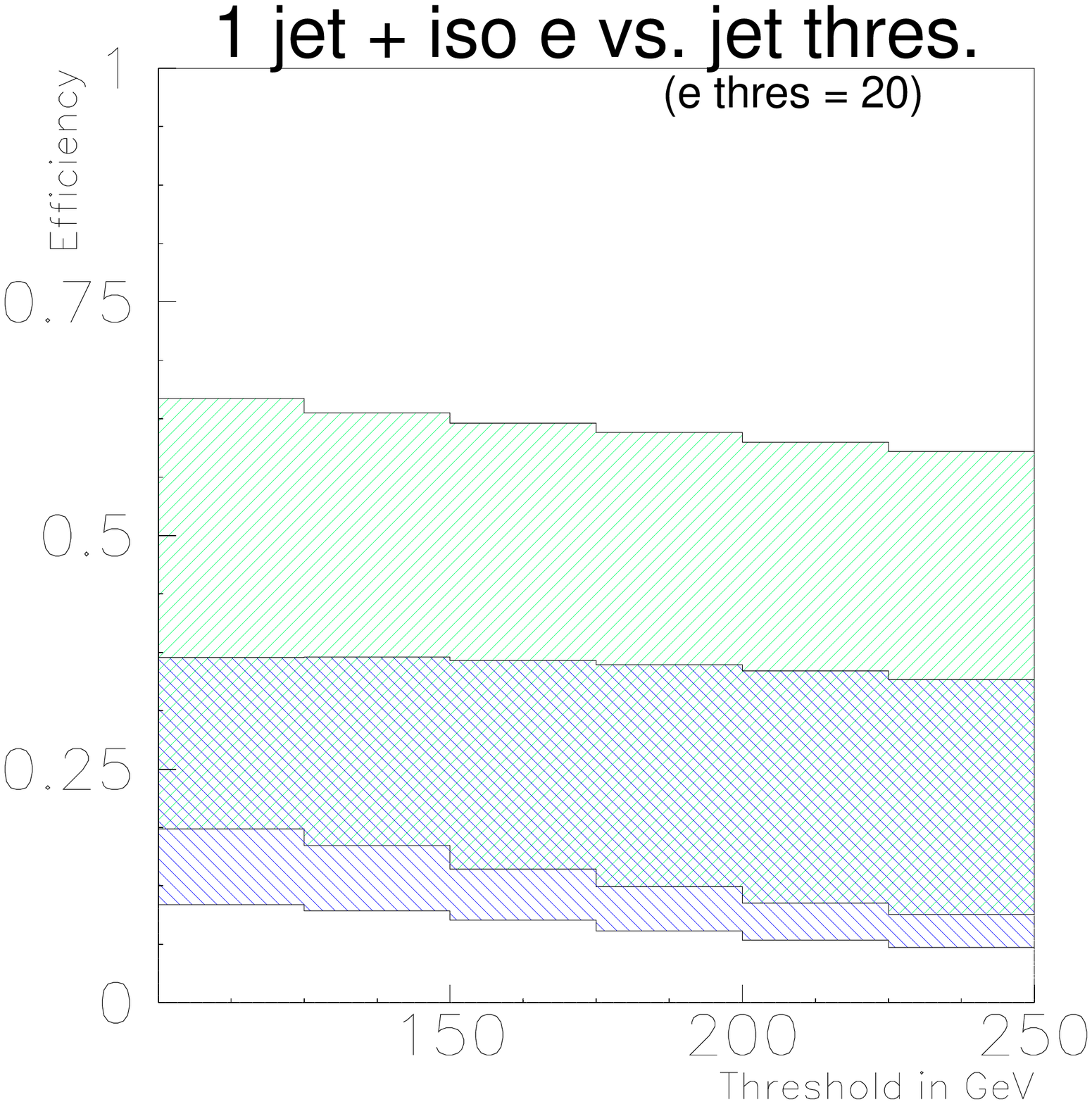} \hspace*{-3mm} 
\includegraphics*[scale=0.25]{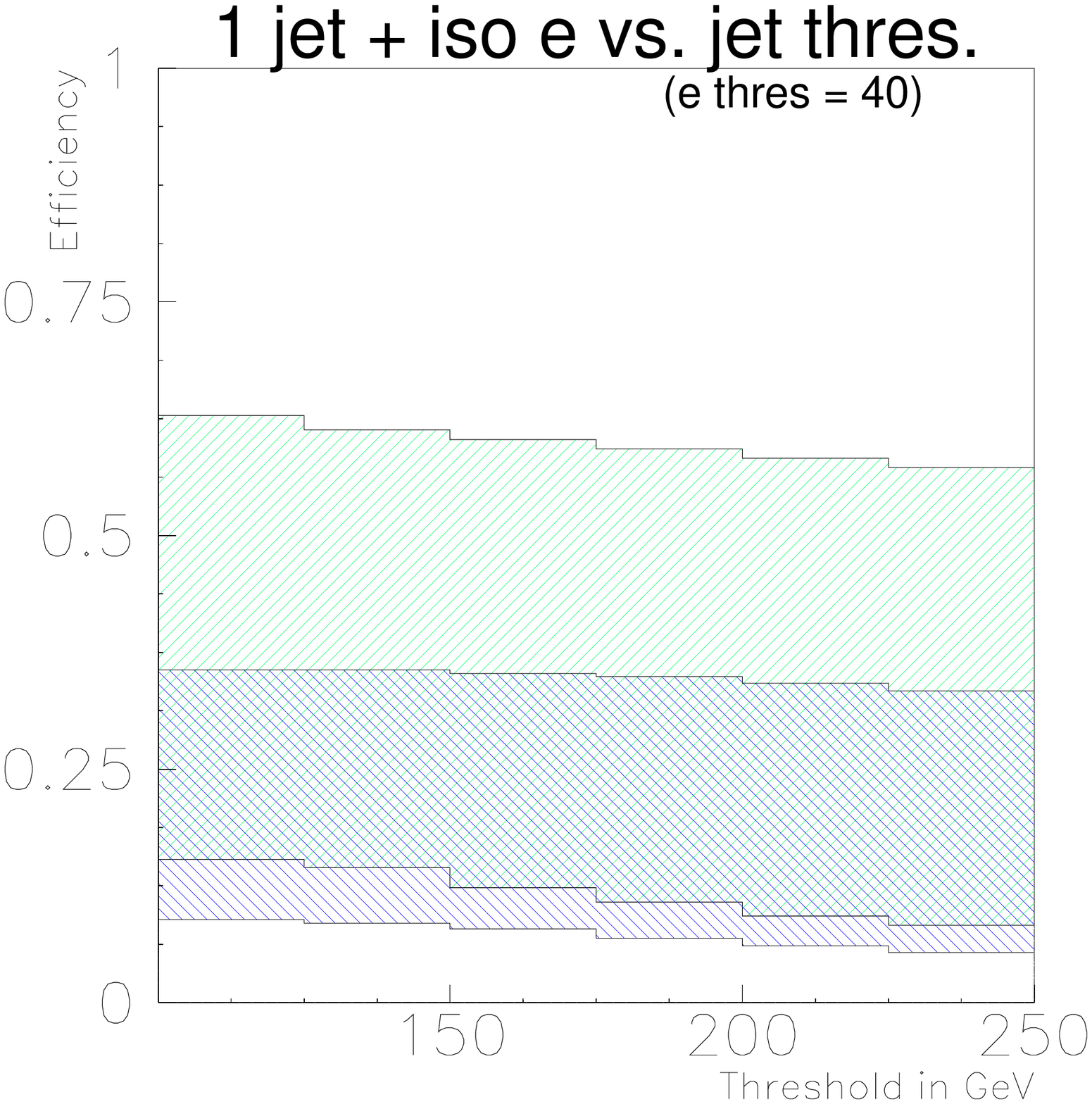} \hspace*{-3mm} 
\includegraphics*[scale=0.25]{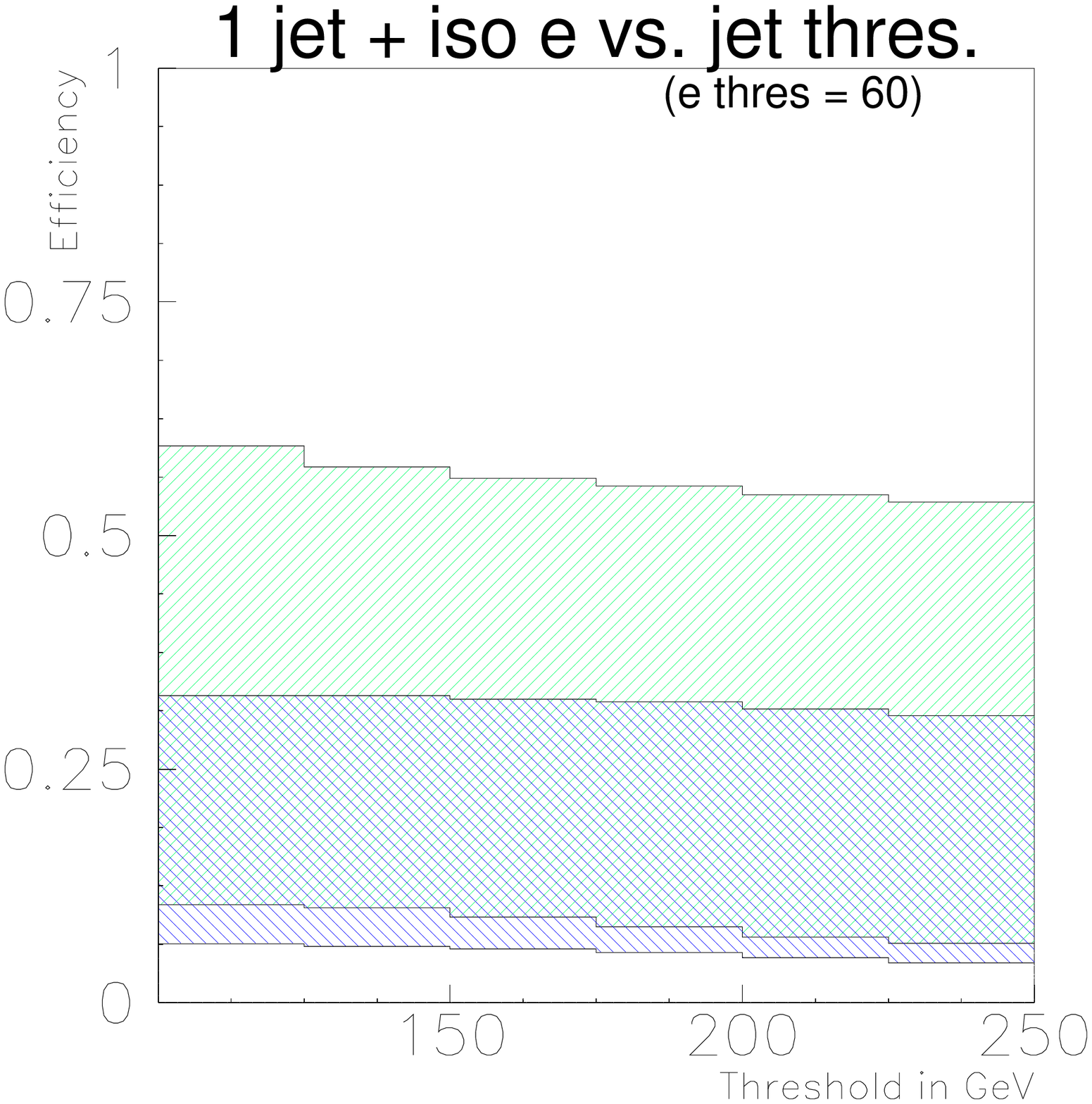} \vspace*{-1mm} \clearpage
\includegraphics*[scale=0.25]{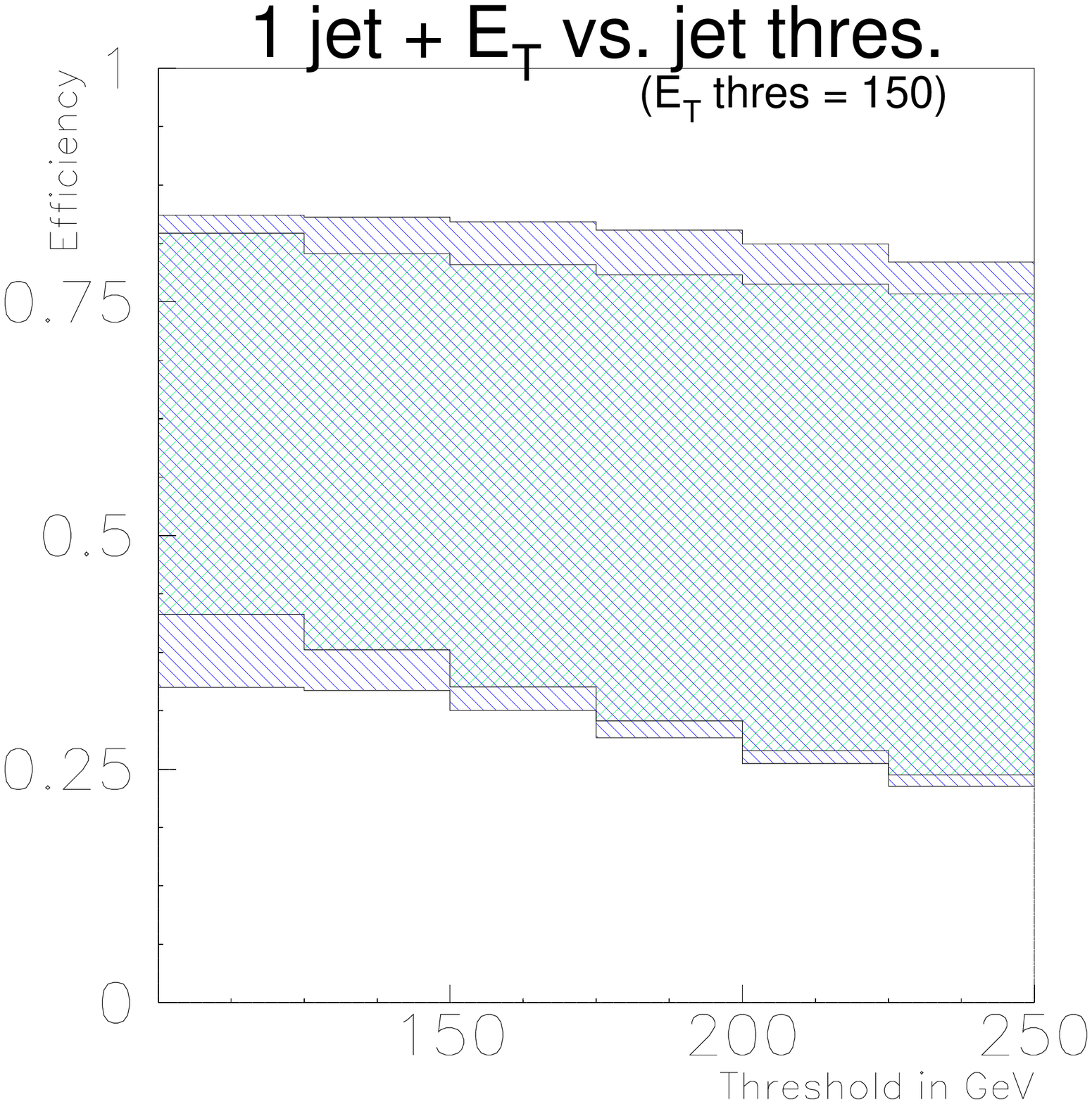} \hspace*{-3mm} 
\includegraphics*[scale=0.25]{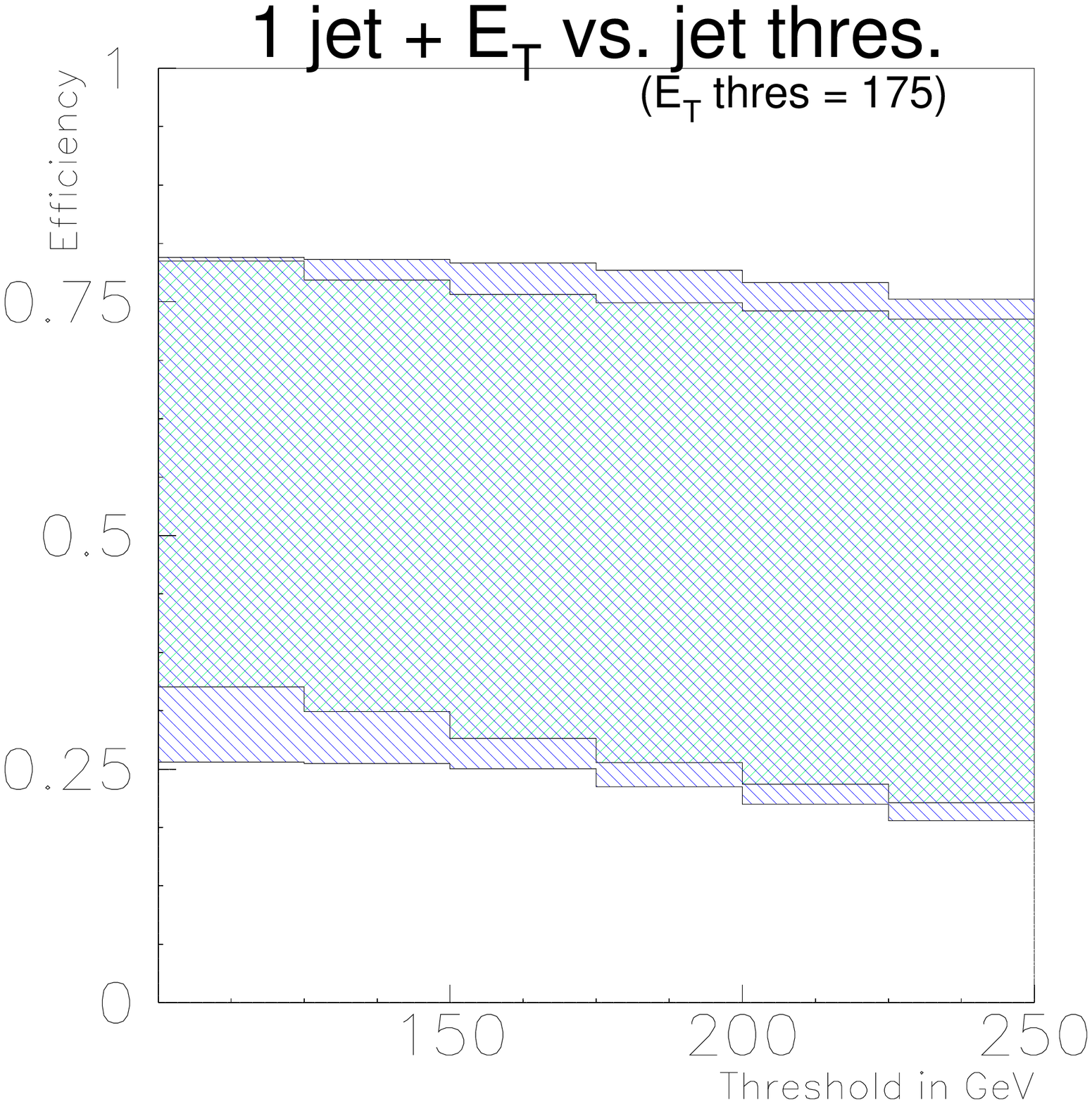} \hspace*{-3mm} 
\includegraphics*[scale=0.25]{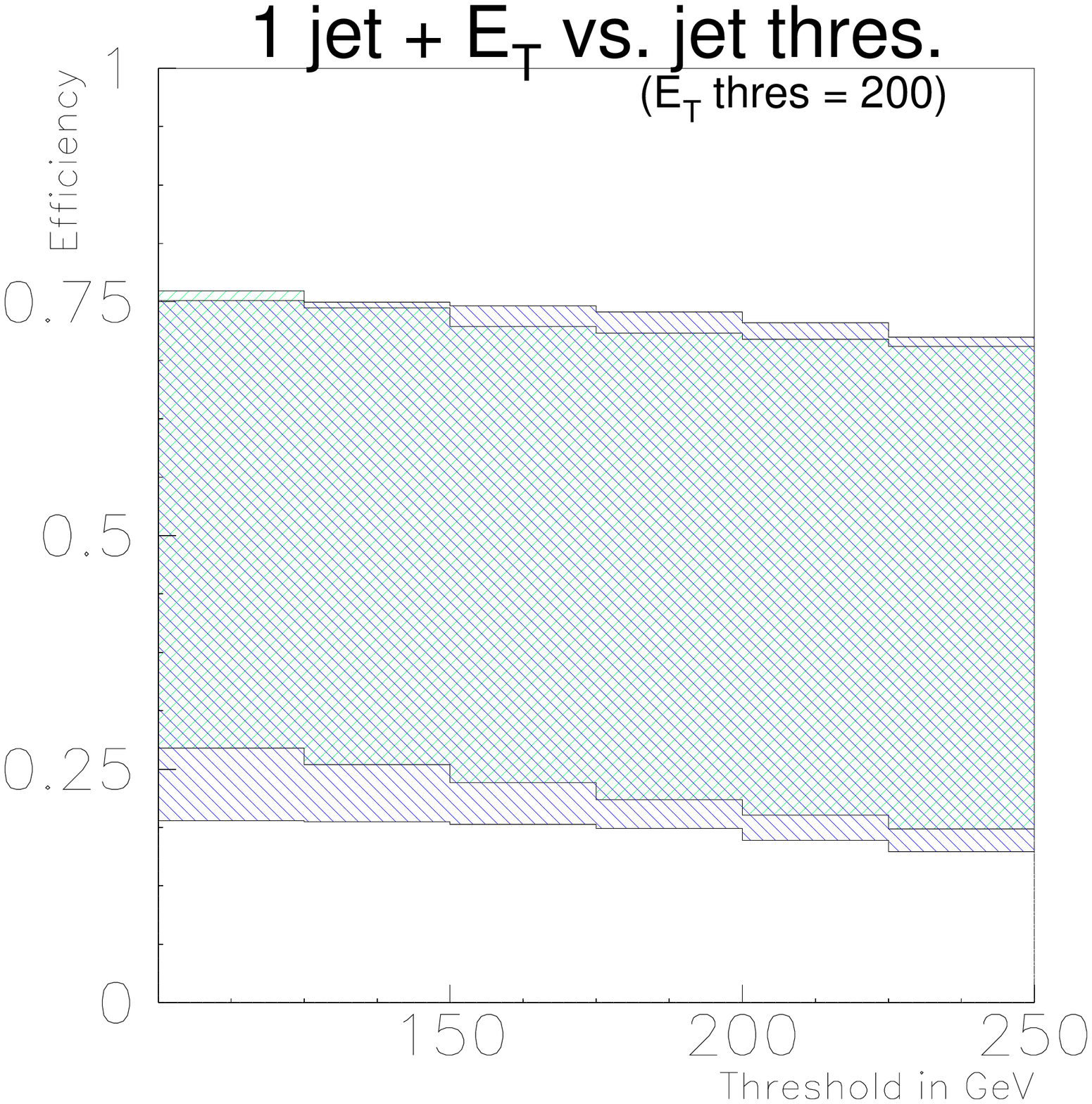} \vspace*{-1mm} \\
\includegraphics*[scale=0.25]{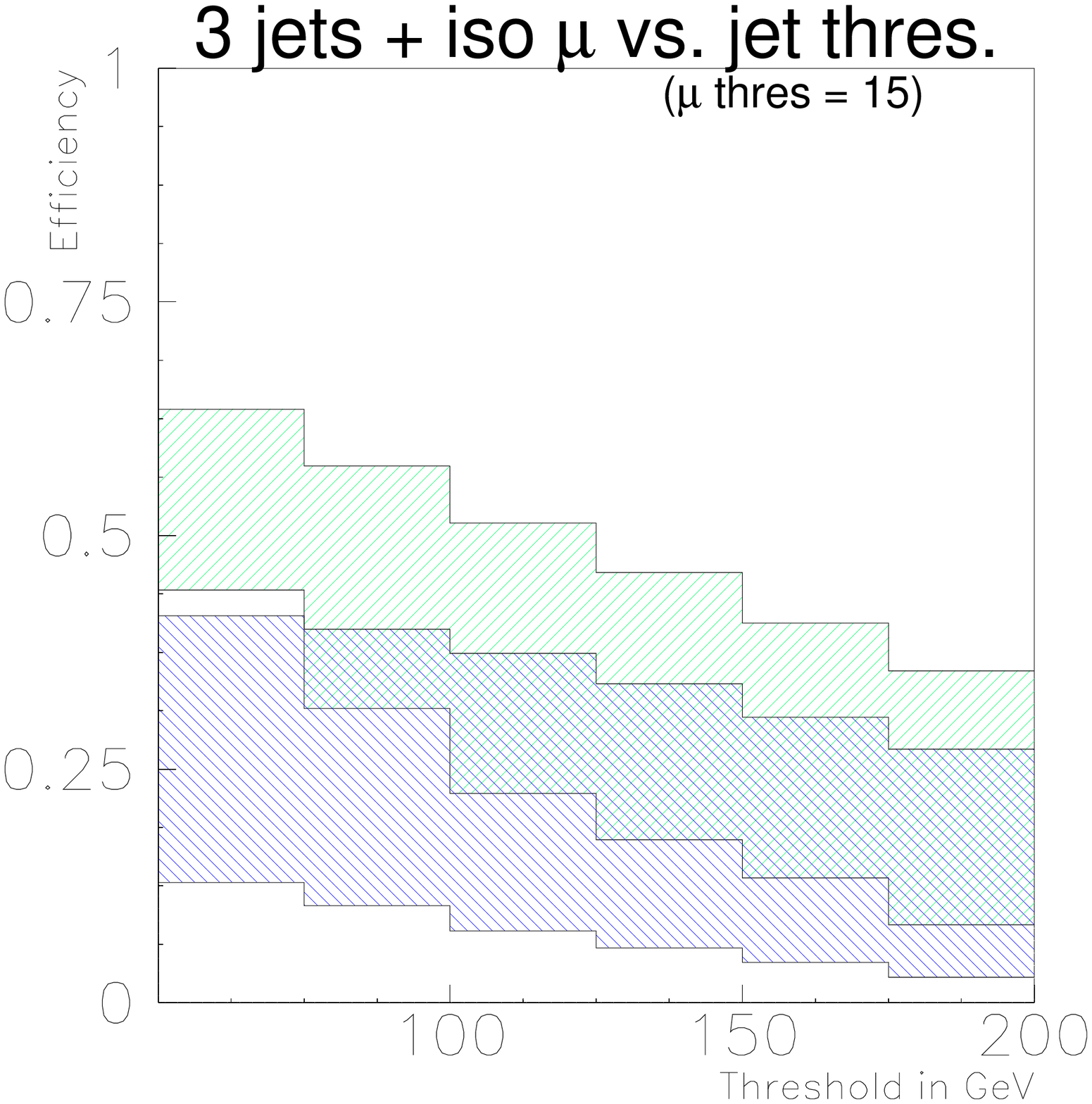} \hspace*{-3mm} 
\includegraphics*[scale=0.25]{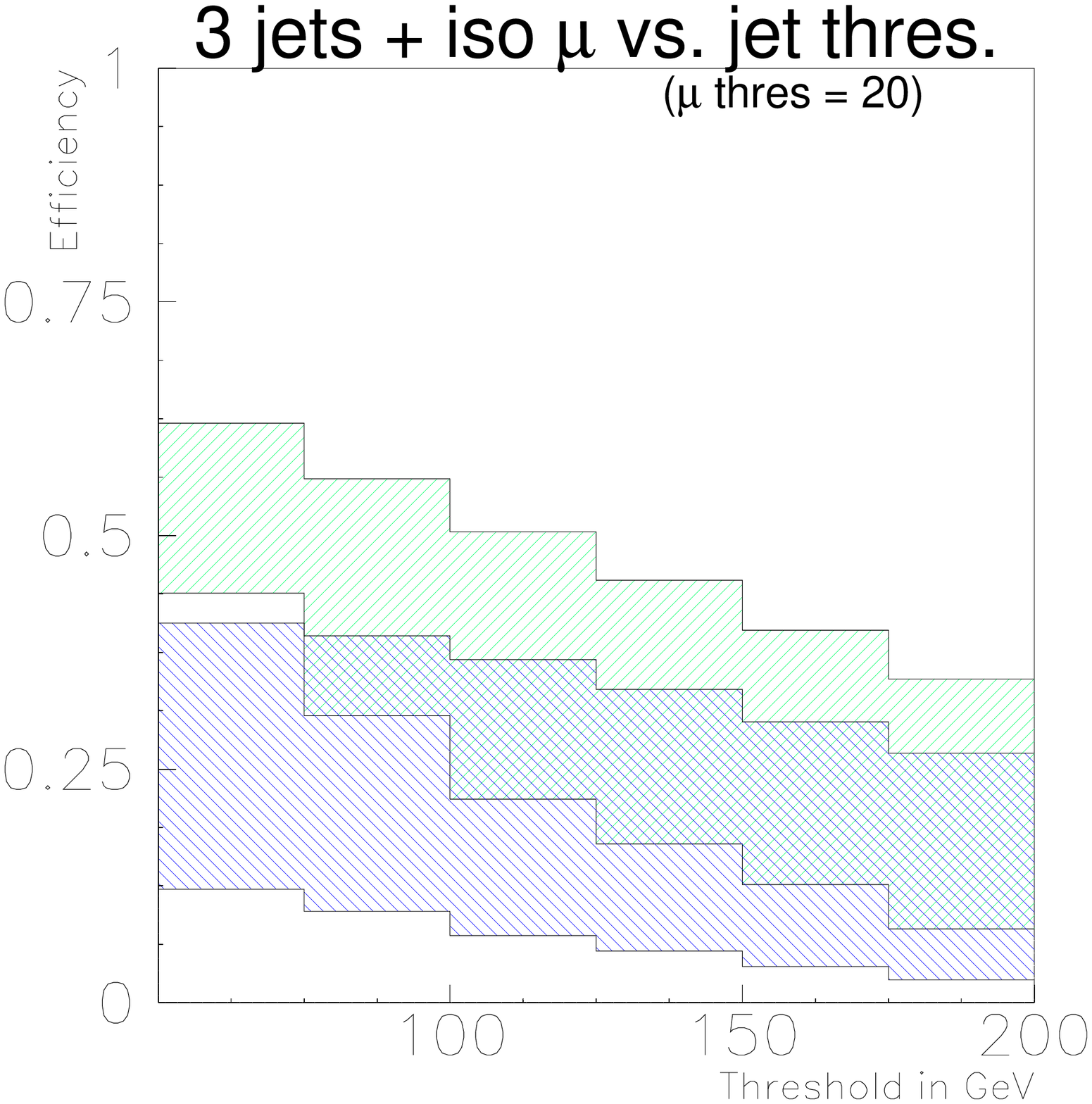} \hspace*{-3mm}  
\includegraphics*[scale=0.25]{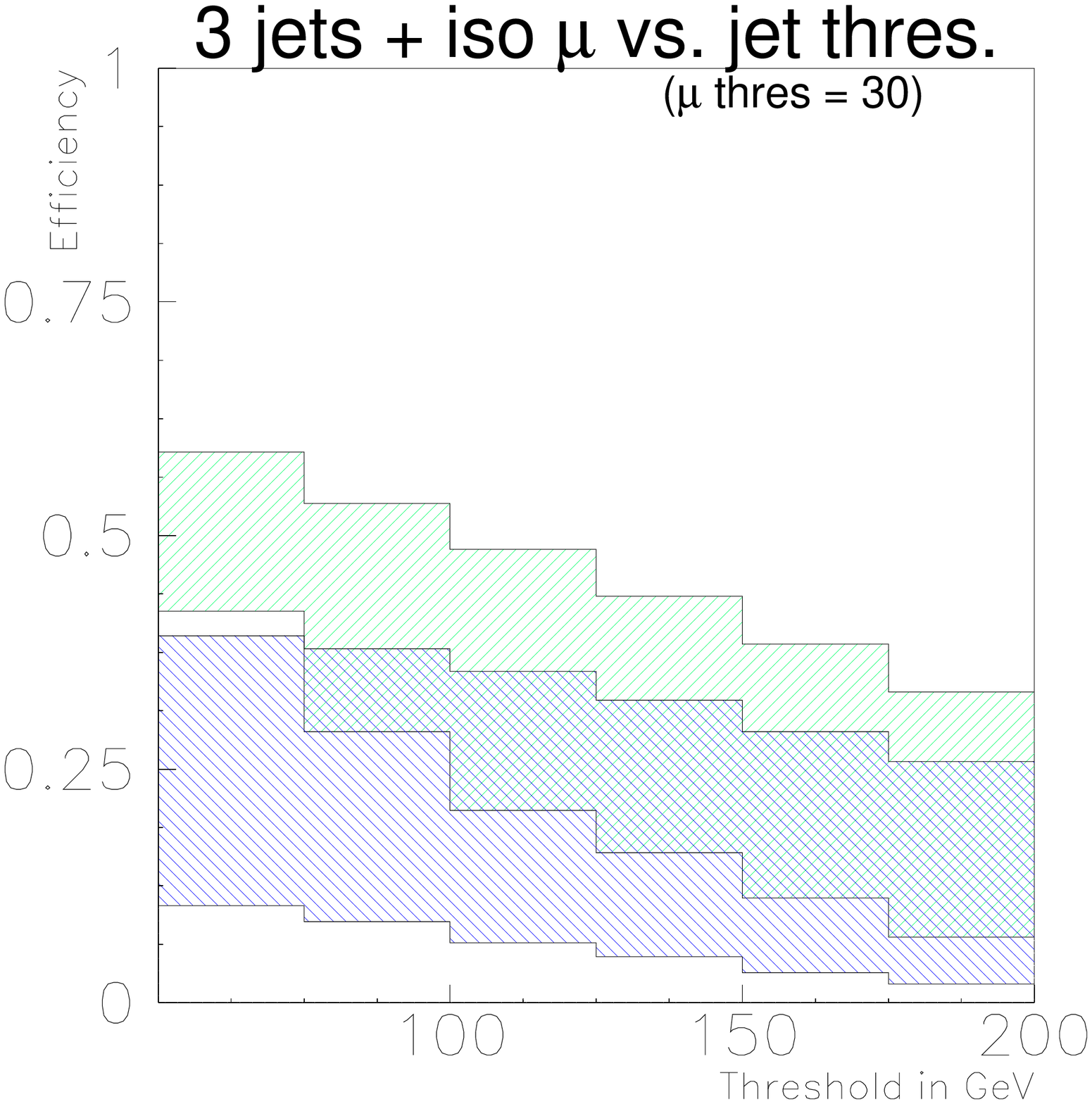} \vspace*{-1mm} \\
\includegraphics*[scale=0.25]{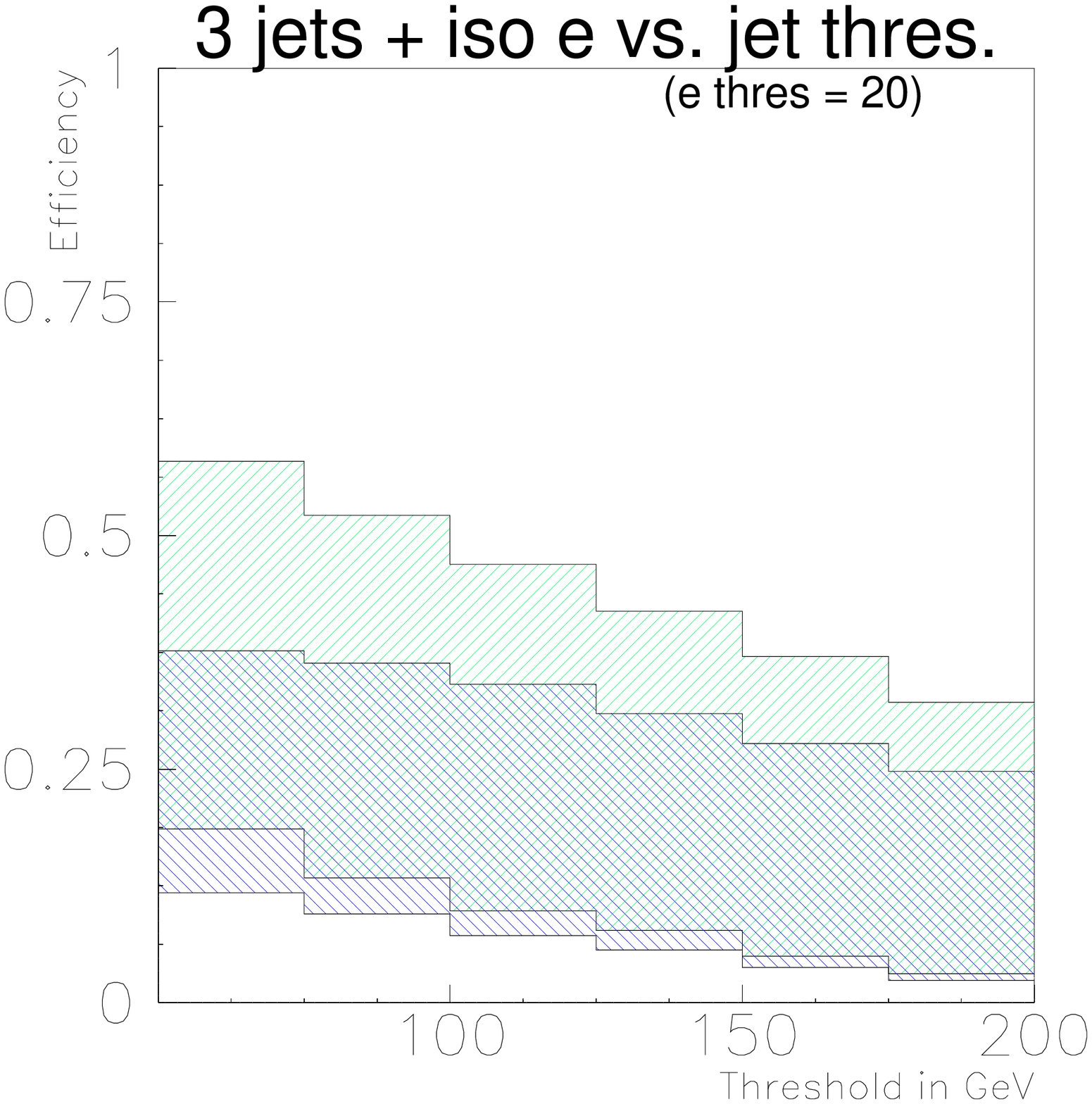} \hspace*{-3mm} 
\includegraphics*[scale=0.25]{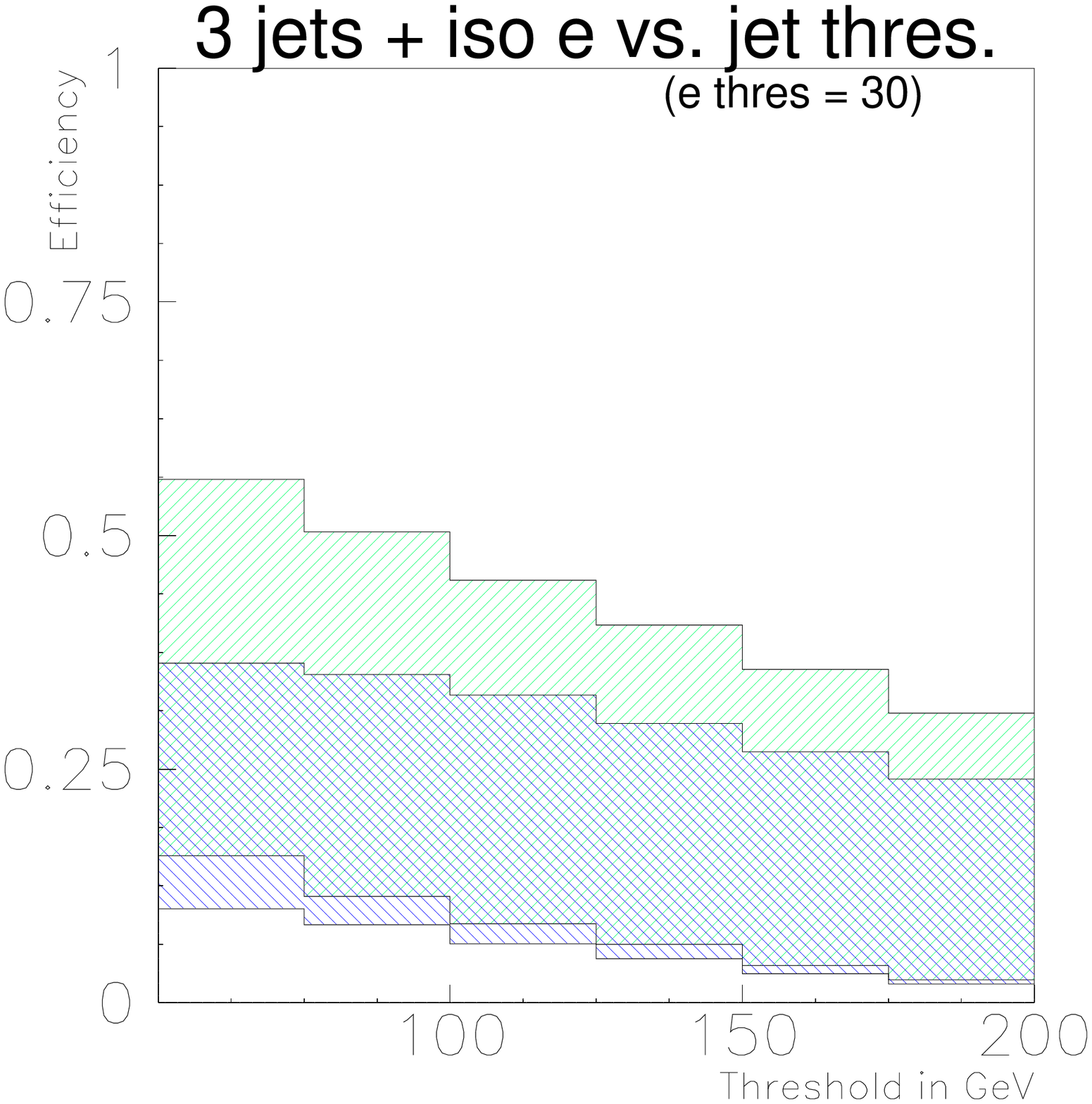} \hspace*{-3mm} 
\includegraphics*[scale=0.25]{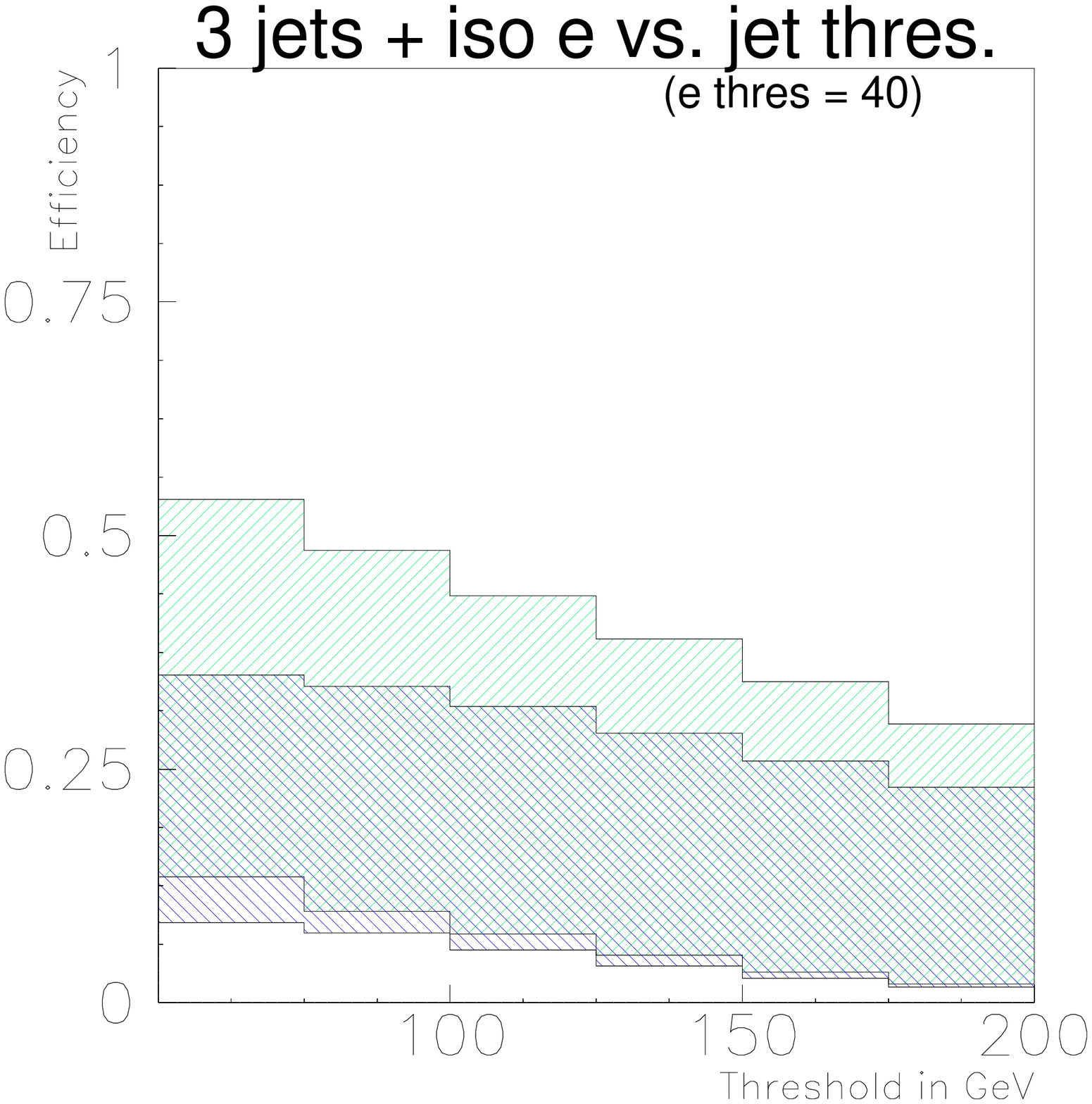} \vspace*{-1mm} \\
\includegraphics*[scale=0.25]{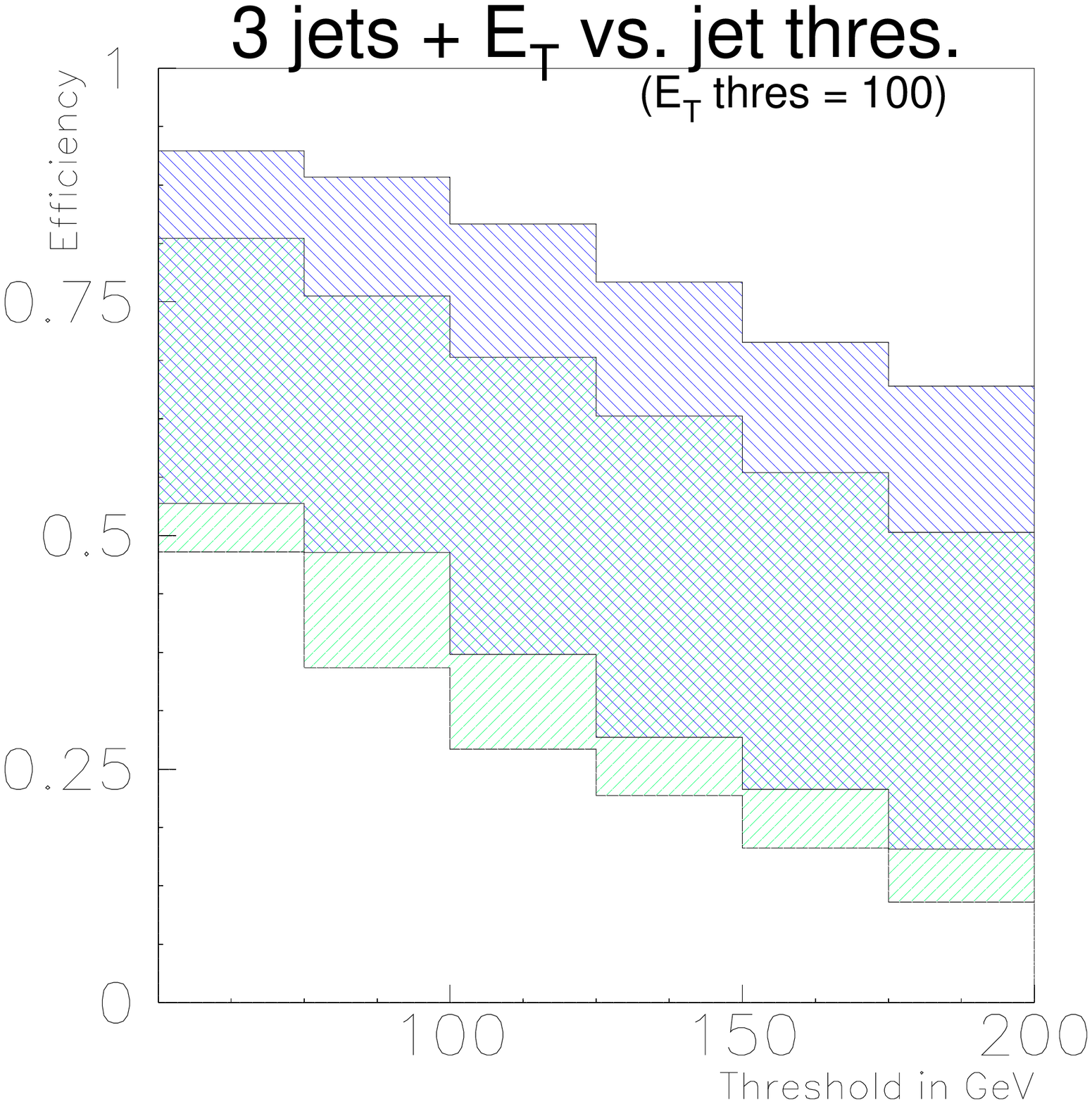} \hspace*{-3mm} 
\includegraphics*[scale=0.25]{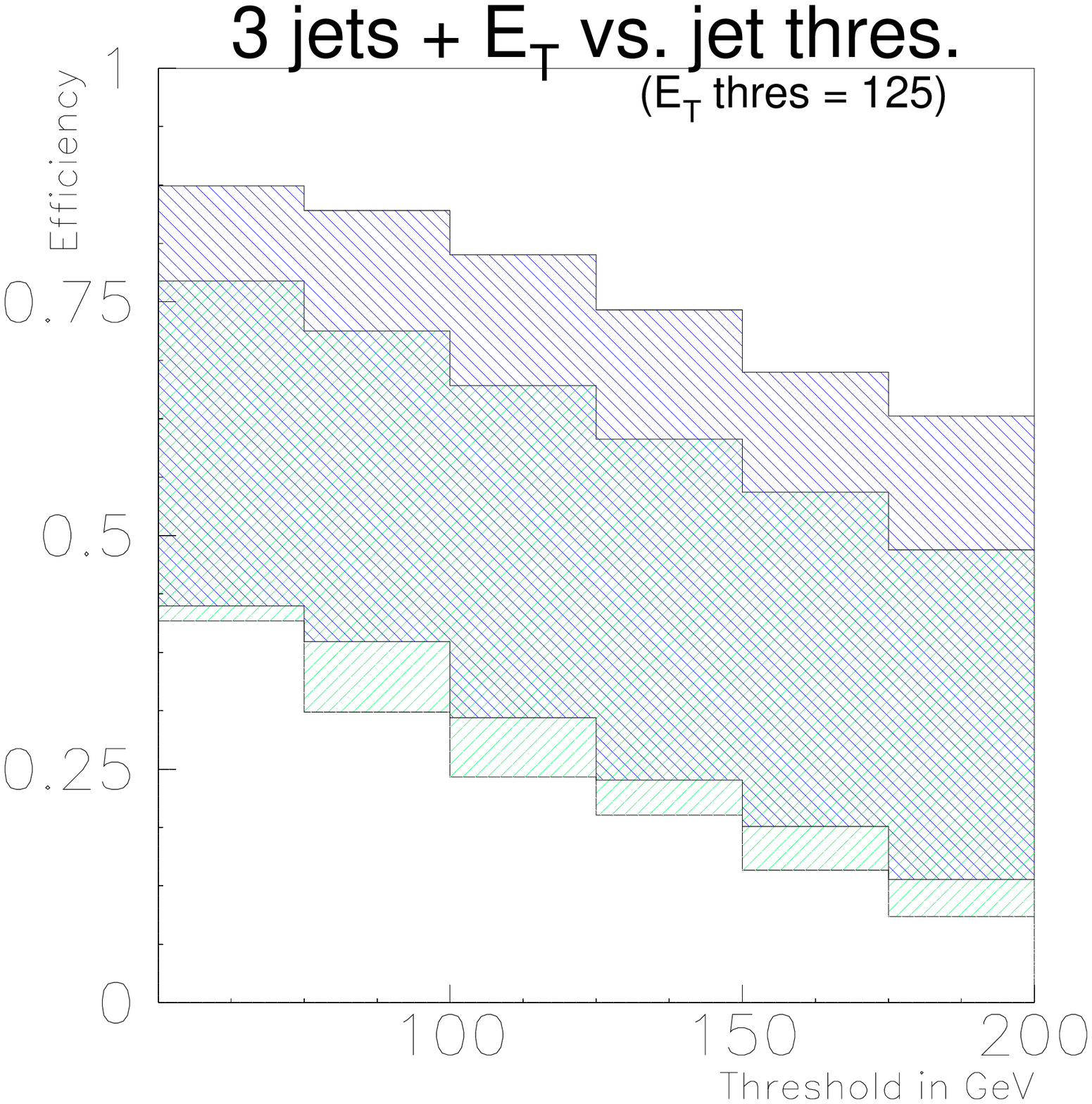} \hspace*{-3mm} 
\includegraphics*[scale=0.25]{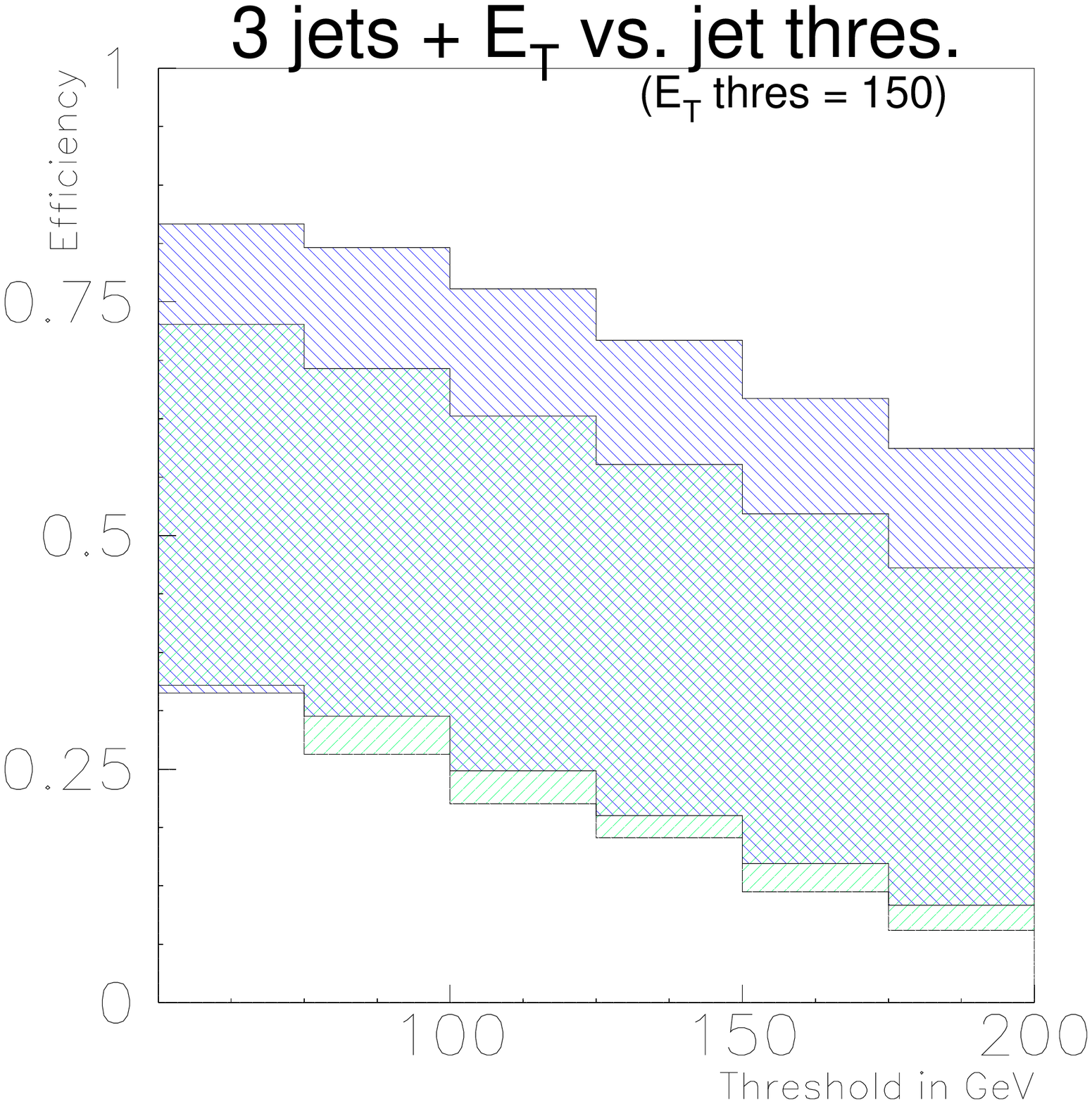} \vspace*{-1mm}
\end{center}

\clearpage
\addcontentsline{toc}{section}{References}
\bibliography{main}

\begin{thebibliography}{10}

\bibitem{atlastdr}
ATLAS Collaboration.
\newblock {ATLAS:} {Detector} and {Physics Performance} {Technical Design
  Report}.
\newblock Technical report, CERN, 1999.
\newblock CERN/LHCC 99-14.

\bibitem{coleman67}
S.~Coleman and J.~Mandula.
\newblock All possible symmetries of the {S} matrix.
\newblock {\em Phys. Rev.}, 159:1251, 1967.

\bibitem{haag75}
R.~Haag, J.~T. Lopuszanski, and M.~Sohnius.
\newblock All possible generators of supersymmetries of the {S} matrix.
\newblock {\em Nucl. Phys.}, B(88):257, 1975.

\bibitem{kane98}
G.~L. Kane, editor.
\newblock {\em Perspectives on Supersymmetry}.
\newblock World Scientific, 1998.

\bibitem{lykken96}
J.~D. Lykken.
\newblock Introduction to supersymmetry.
\newblock In {\em Talk given at TASI 96: Fields, Strings, and Duality}, June
  1996.
\newblock hep-th/9612114.

\bibitem{cahill99}
K.~Cahill.
\newblock Elements of supersymmetry, {Chp.~14}.
\newblock hep-ph/9907295, 1999.

\bibitem{wu57}
C.~S. Wu et~al.
\newblock Experimental test of parity conservation in beta decay.
\newblock {\em Phys. Rev.}, 105:1413, 1957.

\bibitem{martin98}
S.~P. Martin.
\newblock A supersymmetry primer.
\newblock In G.~Kane, editor, {\em Perspectives on supersymmetry}, chapter~1.
  World Scientific, 1998.

\bibitem{weinberg82}
S.~Weinberg.
\newblock Supersymmetry at ordinary energies. {M}asses and conservation laws.
\newblock {\em Phys. Rev.}, D(26):287, 1982.

\bibitem{dimopoulos81}
S.~Dimopoulos and H.~Georgi.
\newblock Softly broken supersymmetry and {SU}(5).
\newblock {\em Nucl. Phys.}, B(193):150, 1981.

\bibitem{buras96}
A.~J. Buras and R.~Fleischer.
\newblock Quark mixing, {CP} violation and rare decays after the top quark
  discovery.
\newblock In A.~J. Buras and M.~Linder, editors, {\em Heavy Flavours II}. World
  Scientific, 1996.
\newblock hep-ph/9704376.

\bibitem{baer94}
H.~Baer, J.~Sender, and X.~Tata.
\newblock The search for top squarks at the fermilab tevatron collider.
\newblock {\em Phys. Rev.}, D(50):4517, 1994.
\newblock hep-ph/9404342.

\bibitem{dreiner00}
H.~Dreiner and M.~H. Seymour.
\newblock Parton shower simulations of {R}-parity violating supersymmetric
  models.
\newblock {\em JHEP}, April 2000.
\newblock hep-ph/9912407.

\bibitem{diracmajorana}
E.~K. Akhmedov.
\newblock Neutrino physics.
\newblock In {\em Lectures given at ICTP Summer School in Particle Physics
  1999}, 1999.
\newblock hep-ph/0001264.

\bibitem{baltz98}
E.~A. Baltz and P.~Gondolo.
\newblock Neutralino decay rates with explicit {$R$} parity violation.
\newblock {\em Phys. Rev.}, D(57):31, 1998.
\newblock hep-ph/9709445.

\bibitem{haber85}
H.~E. Haber and G.~L. Kane.
\newblock The search for supersymmetry: Probing physics beyond the {S}tandard
  {M}odel.
\newblock {\em Phys. Rep.}, 117:76--263, 1985.
\newblock Mixing conventions can be found in App. C.

\bibitem{gunion86}
J.~F. Gunion and H.~E. Haber.
\newblock Higgs bosons in {S}upersymmetric models - {I}.
\newblock {\em Nucl. Phys.}, B(272):1, 1986.
\newblock SLAC-PUB-3404.

\bibitem{okun99}
L.~B. Okun.
\newblock {\em Leptons and Quarks}, chapter~24.
\newblock Elsevier Sciene, 1st edition, 1984.
\newblock Third impression (1999).

\bibitem{ellis00}
J.~Ellis.
\newblock The 115 {GeV} {Higgs Odyssey}.
\newblock hep-ex/0011086, 2000.

\bibitem{witten81_2}
E.~Witten.
\newblock Dynamical breaking of supersymmetry.
\newblock {\em Nucl. Phys.}, B(188):513, 1981.

\bibitem{ibanez82}
L.~E. Iba{\~{n}}ez and G.~G. Ross.
\newblock {SU(2)$_L \times$ U(1)} symmetry breaking as a radiative effect of
  supersymmetry breaking in {GUT}s.
\newblock {\em Phys. Lett.}, B(110):215, 1982.

\bibitem{drees95}
M.~Drees and S.~P. Martin.
\newblock Implications of {SUSY} model building.
\newblock In T.~L. Barklow et~al., editors, {\em Electroweak symmetry breaking
  and new physics at the {TeV} scale}, pages 146--215, 1995.
\newblock hep-ph/9504324.

\bibitem{arkani-hamed98}
N.~Arkani-Hamed, S.~Dimopoulos, and G.~Dvali.
\newblock The hierarchy problem and new dimensions at a millimeter.
\newblock {\em Phys. Lett.}, B(429):263, 1998.
\newblock hep-ph/9803315.

\bibitem{randall99}
L.~Randall and R.~Sundrum.
\newblock A large mass hierarchy from a small extra dimension.
\newblock {\em Phys. Rev. Lett.}, 83:3370--3373, 1999.
\newblock hep-ph/9905221.

\bibitem{kanti00}
P.~Kanti, K.~A. Olive, and M.~Pospelov.
\newblock Solving the hierarchy problem in two-brane cosmological models.
\newblock hep-ph/0005146, 2000.

\bibitem{randall99_2}
L.~Randall and R.~Sundrum.
\newblock Out of this world supersymmetry breaking.
\newblock {\em Nucl. Phys.}, B(557):79, 1999.
\newblock hep-th/9810155.

\bibitem{antoniadis98}
I.~Antoniadis, N.~Arkani-Hamed, S.~Dimopoulos, and G.~Dvali.
\newblock New dimensions at a millimeter to a fermi and superstrings at a
  {TeV}.
\newblock {\em Phys. Lett.}, B(436):257, 1998.
\newblock hep-ph/9804398.

\bibitem{gunion90}
J.~F. Gunion, H.~E. Haber, G.~Kane, and S.~Dawson.
\newblock {\em The Higgs hunter's guide}, chapter~7.
\newblock Addison-Wesley, 1990.

\bibitem{europhys}
{Particle Data Group,}~D. Groom et~al.
\newblock Review of {P}article {P}hysics.
\newblock {\em Eur. Phys. J.}, C(15), 2000.

\bibitem{llewellyn81}
C.~H.~Llewellyn Smith, G.~G. Ross, and J.~F. Wheather.
\newblock Low-energy predictions from grand unified theories.
\newblock {\em Nucl. Phys.}, B(177):263, 1981.

\bibitem{ibanez81}
L.~E. Ib{\'{a}\~{n}}ez and G.~G. Ross.
\newblock Low energy predictions in supersymmetric grand unified theories.
\newblock {\em Phys. Lett.}, 105B(6):439, 1981.

\bibitem{ellis00_2}
J.~Ellis and D.~Ross.
\newblock A light higgs boson would invite supersymmetry.
\newblock hep-ph/0012067, 2000.

\bibitem{dreiner98}
H.~Dreiner.
\newblock An introduction to explicit {$R$}-parity violation.
\newblock In G.~Kane, editor, {\em Perspectives on supersymmetry}, chapter~20.
  World Scientific, 1998.

\bibitem{ibanez92}
L.~E. Ib{\'{a}\~{n}}ez and G.~G. Ross.
\newblock Discrete gauge symmetries and the origin of baryon and lepton number
  conservation in supersymmetric versions of the standard model.
\newblock {\em Nucl. Phys.}, B368:3--37, 1992.

\bibitem{hinchliffe93}
I.~Hinchliffe and T.~Kaeding.
\newblock {$B$ and $L$-violating} couplings in the minimal supersymmetric
  standard model.
\newblock {\em Phys. Rev.}, D(47):279, 1993.

\bibitem{farrar78}
G.~R. Farrar and P.~Fayet.
\newblock Phenomenology of the production, decay, and detection of new hadronic
  states associated with supersymmetry.
\newblock {\em Phys. Lett.}, B76:575--579, 1978.

\bibitem{takayama00}
F.~Takayama and M.~Yamaguchi.
\newblock Gravitino dark matter without {$R$}-parity.
\newblock {\em Phys. Lett.}, B(485):388, 2000.
\newblock hep-ph/0005214.

\bibitem{superK99}
SuperKamiokande Collaboration (Y.~Fukuda et~al.).
\newblock Measurement of the flux and zenith angle distribution of upward
  through going muons by superkamiokande.
\newblock {\em Phys. Rev. Lett.}, 82:2644, 1999.
\newblock hep-ex/9812014.

\bibitem{sno01}
SNO Collaboration (Q. R.~Ahmad et~al.).
\newblock Measurement of the charged current interactions produced by b-8 solar
  neutrinos at the sudbury neutrino observatory.
\newblock {\em Submitted to Phys.Rev.Lett.}, June 2001.
\newblock nucl-ex/0106015.

\bibitem{drees98}
M.~Drees, S.~Pakvasa, X.~Tata, and T.~ter Veldhuis.
\newblock A supersymmetric resolution of solar and atmosperic neutrino puzzles.
\newblock {\em Phys. Rev.}, D(57):5335, 1998.
\newblock hep-ph/9712392.

\bibitem{abada00}
A.~Abada and G.~Bhattacharyya.
\newblock Can {$R$}-parity violation explain the {LSND} data as well?
\newblock hep-ph/0007016, 2000.

\bibitem{adloff97}
H1~Collaboration{,}~C. Adloff et~al.
\newblock Observation of events at very high {$Q^2$} in $ep$ collisions at
  {HERA}.
\newblock {\em Z. Phys.}, C(74):191, 1997.
\newblock hep-ex/9702012.

\bibitem{breitweg97}
ZEUS Collaboration{,}~J. Breitweg et~al.
\newblock Comparison of {ZEUS} data with standard model predictions for
  {$e^+p\to e^+X$} scattering at high {$x$ and $Q^2$}.
\newblock {\em Z. Phys.}, C(74):207, 1997.
\newblock hep-ex/9702015.

\bibitem{altarelli97_1}
G.~Altarelli, J.~Ellis, G.~F. Giudice, et~al.
\newblock Pursuing interpretations of the {HERA} large-{$Q^2$} data.
\newblock {\em Nucl. Phys.}, B(506):3, 1997.
\newblock hep-ph/9703276.

\bibitem{perez00}
E.~Perez (on behalf of~the H1 and ZEUS collaborations).
\newblock Searches for exotica at {HERA}.
\newblock In K.~Huitu et~al., editors, {\em International Europhysics
  Conference on High Energy Physics 99}, page 818, 2000.
\newblock hep-ex/9911037.

\bibitem{oehler00}
C.~Oehler.
\newblock Analysis of the {KARMEN} time-anomaly.
\newblock {\em Nucl. Phys. B, Proc. Suppl.}, 85:101, 2000.

\bibitem{zimmerman00}
E.~Zimmerman.
\newblock Recent results adressing the {KARMEN} timing anomaly.
\newblock In {\em CIPANP2000}, 2000.
\newblock hep-ex/0009008.

\bibitem{choudhury00}
D.~Choudhury, H.~Dreiner, P.~Richardson, and S.~Sarkar.
\newblock A supersymmetric solution to the {KARMEN} time anomaly.
\newblock {\em Phys. Rev.}, D(61):95, 2000.
\newblock hep-ph/9911365.

\bibitem{choudhury96}
D.~Choudhury and S.~Sarkar.
\newblock A supersymmetric resolution of the {KARMEN} anomaly.
\newblock {\em Phys. Lett. B.}, 374:87, 1996.
\newblock hep-ph/9511357.

\bibitem{richardson00}
P.~Richardson.
\newblock {\em Simulations of {R}-parity violating {SUSY} models}.
\newblock PhD thesis, University of Oxford, 2000.
\newblock hep-ph/0101105.

\bibitem{ibanez91}
L.~E. Ib{\'{a}\~{n}}ez and G.~G. Ross.
\newblock Discrete gauge symmetry anomalies.
\newblock {\em Phys. Lett.}, B(260):291--295, 1991.

\bibitem{gilbert88}
G.~Gilbert.
\newblock Wormhole-induced proton decay.
\newblock {\em Nuc. Phys.}, B(328):159--170, 1989.

\bibitem{skands01}
P.~Z. Skands.
\newblock Did {$R$}-conservation kill the proton?. {T}alk given at 3{${^{rd}}$}
  {N}ordic {LHC} {W}orkshop.
\newblock Slides at http://www.fys.uio.no/epf/nordic-network/programme.htm.

\bibitem{wilczek89}
L.~M. Krauss and F.~Wilczek.
\newblock {Discrete Gauge Symmetry in Continuum Theories}.
\newblock {\em Phys. Rev. Lett.}, 62(11):1221--1223, 1989.

\bibitem{weinberg79}
S.~Weinberg.
\newblock Baryon- and lepton-nonconserving processes.
\newblock {\em Phys. Rev. Lett.}, 43(21):1566, 1979.

\bibitem{murayama98}
H.~Murayama.
\newblock Probing physics at short distances with supersymmetry.
\newblock In G.~Kane, editor, {\em Perspectives on supersymmetry}, chapter~13.
  World Scientific, 1998.

\bibitem{carlos96}
B.~de~Carlos and P.~L. White.
\newblock R-parity violation effects through soft supersymmetry breaking terms
  and the renormalisation group.
\newblock {\em Phys. Rev. D}, 54(5):3427, 1996.
\newblock hep-ph/9602381.

\bibitem{peskin95}
M.~E. Peskin and D.~V. Schroeder.
\newblock {\em An Introduction to Quantum Field Theory}.
\newblock Perseus Books, 1995.

\bibitem{pythia5.7}
T.~Sj{\"{o}}strand.
\newblock {\sc Pythia 5.7} and {\sc jetset 7.4} {Physics and Manual}.
\newblock hep-ph/9508391, 1993.

\bibitem{james80}
F.~James.
\newblock Monte {C}arlo theory and practice.
\newblock {\em Rep. Prog. Phys.}, 43:73, 1980.

\bibitem{atlas_inner}
ATLAS Inner~Detector Community.
\newblock {Inner Detector Technical Design Report}.
\newblock Technical report, CERN, 1997.
\newblock CERN/LHCC 97-16.

\bibitem{atlas_calo}
ATLAS Collaboration.
\newblock {Calorimeter Performance Technical Design Report}.
\newblock Technical report, CERN, 1997.
\newblock CERN/LHCC 96-40.

\bibitem{atlas_tile}
ATLAS Tile~Calorimeter Collaboration.
\newblock {Tile Calorimeter Technical Design Report}.
\newblock Technical report, CERN, 1996.
\newblock CERN/LHCC 96-42.

\bibitem{atlas_lar}
ATLAS Collaboration.
\newblock {Liquid Argon Calorimeter Technical Design Report}.
\newblock Technical report, CERN, 1996.

\bibitem{atlas_tp}
ATLAS Collaboration.
\newblock {ATLAS Technical Proposal}.
\newblock Technical report, CERN, 1994.
\newblock CERN/LHCC 94-43.

\bibitem{atlfast2.0}
E.~Richter-Was et~al.
\newblock {\texttt{ATLFAST 2.0}} a fast simulation package for {ATLAS}.
\newblock ATLAS Internal Note ATL-PHYS-98-131, 1998.

\bibitem{stirling96}
R.~K. Ellis, W.~J. Stirling, and B.~R. Webber.
\newblock {\em {QCD} and Collider Physics}.
\newblock Cambridge University Press, 1996.

\bibitem{abott01}
{D\O} Collaboration (B.~Abbot et~al.).
\newblock Determination of the absolute jet energy scale in the {D\O}
  calorimeters.
\newblock {\em Nucl. Instr. Meth.}, A424:352, 1999.
\newblock hep-ex/9805009.

\bibitem{atlas_muon}
ATLAS~Muon Collaboration.
\newblock {ATLAS Muon Spectrometer Technical Design Report}.
\newblock Technical report, CERN, 1997.
\newblock CERN/LHCC 97-22.

\bibitem{feldman98}
G.~Feldman and R.~Cousins.
\newblock Unified approach to the classical statistical analysis of small
  signals.
\newblock {\em Phys.Rev.D.}, 57:3873, 1998.

\bibitem{rvbounds}
B.~C. Allanach, A.~Dedes, and H.~K. Dreiner.
\newblock Bounds on {R}-parity violating couplings at the weak scale and at the
  {GUT} scale.
\newblock {\em Phys. Rev.}, D60:075014, 1999.
\newblock hep-ph/9906209.

\bibitem{hinchliffe_private}
I.~Hinchliffe.
\newblock Private communication.
\newblock 2001.

\bibitem{webber00}
B.~Webber.
\newblock Improving {QCD} event generators.
\newblock Talk Given at the Durham Workshop, 2000. Slides can be found at
  http://www.hep.phy.cam.ac.uk/theory/webber.

\bibitem{forshaw99}
J.~Forshaw and M.~Seymour.
\newblock Subjet rates in hadron collider jets.
\newblock {\em JHEP}, 09:009, 1999.
\newblock hep-ph/9908307.

\bibitem{seymour00}
M.~Seymour.
\newblock Jets in hadron collisions.
\newblock In {\em QCD and High Energy Hadronic Interactions}, Les Arcs, Savoie,
  France, March 2000.
\newblock hep-ph/0007051.

\bibitem{bystricky96}
J.~Bystricky et~al.
\newblock {ATLAS} trigger menus at luminosity
  {$10^{33}\mathrm{cm^{-2}s^{-1}}$}.
\newblock {\em ATLAS Internal Note}, 1996.
\newblock DAQ-NO-054.

\bibitem{mcclelland89}
J.~McClelland and D.~Rumelhart.
\newblock {\em Explorations in Parallel Distributed Processing}.
\newblock MIT Press, 1989.
\newblock Fourth print.

\bibitem{lecun90}
Y.~LeCun et~al.
\newblock Optimal brain damage.
\newblock In D.~S. Touretsky et~al., editors, {\em Advances in Neural
  Information Processing Systems 2}, pages 396--404. MIT Press, 1990.

\bibitem{baer97}
H.~Baer, C.~Chen, and X.~Tata.
\newblock Impact of hadronic decays of the lightest neutralino on the reach of
  the  {CERN LHC}.
\newblock {\em Phys. Rev.}, D55:1466--1470, 1997.

\bibitem{mrenna97}
S.~Mrenna.
\newblock {\texttt{SPYTHIA}}, a supersymmetric extension {of \texttt{PYTHIA
  5.7}}.
\newblock {\em Comp. Phys. Comm.}, 101:232, 1997.

\bibitem{herwig6}
G.~Corcella et~al.
\newblock Herwig 6: An event generator for hadron emission reactions with
  interfering gluons (including supersymmetric processes).
\newblock {\em JHEP}, 01:010, 2001.
\newblock hep-ph/0011363.

\bibitem{baer87}
H.~Baer et~al.
\newblock Higgs boson signals in superstring inspired models at hadron
  supercolliders.
\newblock {\em Phys. Rev.}, D(36):96, 1987.

\end{thebibliography}
\end{document}